\documentclass{article}

\usepackage{arxiv}
\usepackage{array}
\usepackage[inkscapelatex=false]{svg}
\usepackage[utf8]{inputenc} % allow utf-8 input
\usepackage[T1]{fontenc}    % use 8-bit T1 fonts
\usepackage{hyperref}       % hyperlinks
\usepackage{url}            % simple URL typesetting
\usepackage{booktabs}       % professional-quality tables
\usepackage{amsfonts}       % blackboard math symbols
\usepackage{amsmath}
\usepackage{nicefrac}       % compact symbols for 1/2, etc.
\usepackage{microtype}      % microtypography
\usepackage{cleveref}       % smart cross-referencing
\usepackage{lipsum}         % Can be removed after putting your text content
\usepackage{graphicx}
\usepackage[export]{adjustbox}
\usepackage{subcaption}
\usepackage[permil]{overpic}
\usepackage{caption}
\usepackage{tikz}
\usetikzlibrary{arrows.meta}

\newcommand\shshimage[1]{\adjustbox{valign=m, vspace=1pt}{\includegraphics[width=.09\linewidth]{#1}}}

\usepackage[framemethod=TikZ]{mdframed}

\newenvironment{labeledbox}[1][]{%
    \mdfsetup{
        innertopmargin=10pt,
        linecolor=blue!20,
        linewidth=2pt,
        topline=true,
        frametitleaboveskip=\dimexpr-\ht\strutbox\relax,
    }
    \ifstrempty{#1}%
    {\mdfsetup{frametitle={}}}
    {\mdfsetup{%
        frametitle={%
            \tikz[baseline=(current bounding box.east),outer sep=0pt]
            \node[anchor=east,rectangle,fill=blue!20]
            {\strut #1};}}
    }%
    \begin{mdframed}[]\relax%
}{\end{mdframed}}

\usepackage{tabularx}
\usepackage{xcolor}
\usepackage[numbers,sort&compress]{natbib}
\usepackage{doi}

\usepackage{algorithm}
\usepackage{algpseudocode}

\usepackage{authblk}

\title{Generative reconstruction of 2D and 3D polycrystalline microstructures using symmetrized hyperspherical harmonics}

\author[1]{Ali R. Safi}
\author[2]{Paul Seibert}
\author[3]{Santiago Benito}
\author[2]{Alexander Raßloff}
\author[2]{Markus Kästner}
\author[1,4]{Benjamin Klusemann}

\affil[1]{Helmholtz-Zentrum Hereon, Institute of Material and Process Design, Solid State Materials Processing\\ Max–Planck-Straße 1, 21502 Geesthacht, Germany}
\affil[2]{TU Dresden, Institute of Solid Mechanics, Chair of Computational and Experimental Solid Mechanics\\ George-Bähr-Straße 3c, 01069 Dresden, Germany}
\affil[3]{Ruhr University Bochum, Institute for Materials, Chair of Materials Technology \\ Universitätsstraße 150, 44801 Bochum, Germany}
\affil[4]{Leuphana University Lüneburg, Institute for Production Technology and Systems\\ Universitätsallee 1, 21335 Lüneburg, Germany}

\date{}

\renewcommand{\headeright}{}
\renewcommand{\undertitle}{}
\renewcommand{\shorttitle}{}
\begin{document}
\maketitle

\begin{abstract}
Establishing structure-property linkages in polycrystalline materials requires representative two- (2D) and three- (3D) dimensional microstructural inputs for full-field simulations. A core objective of microstructure characterization and reconstruction is the generative synthesis of 2D and 3D microstructures that reflect a target statistical ensemble using limited 2D data as a reference. This work introduces an orientation-based differentiable microstructure characterization and reconstruction framework, implemented in MCRpy, to perform reconstructions of voxelized images. Unit quaternions in combination with symmetrized hyperspherical harmonics are utilized to derive a continuous, symmetry-invariant representation of crystallographic orientations to overcome the numerical singularities and discontinuities associated with traditional Euler-based methods.

The descriptor-based reconstructions are driven by a set combining two-point spatial correlations, a novel hybrid three-point variogram, and a mean variation regularizer to capture both global texture and local interfacial topology. The framework's efficiency is demonstrated by reconstructing 3D realizations from 2D orientation data of an aluminum alloy after thermo-mechanical processing, successfully recovering both morphological features and crystallographic distribution. Systematic benchmarking indicates that second-order gradient-based optimization, utilizing the L-BFGS-B algorithm, effectively navigates the complex loss landscape to generate high-fidelity realizations with minimal residuals. This methodology provides a versatile, open-source framework for the digital synthesis of polycrystalline representative volume elements to facilitate the rapid development of microstructure-informed materials design workflows.
\end{abstract}

% keywords can be removed

\section{Introduction}
The homogenized material properties of polycrystalline materials emerge from complex interactions between microstructural constituents across multiple length scales. Manufacturing and processing routes inherently produce heterogeneous microstructures whose characteristics must be rigorously quantified to establish structure-property relationships and enable microstructure-informed materials design. Computational modeling frameworks have become indispensable tools for efficiently exploring these structure-property relationships and can be broadly classified into full-field and mean-field approaches. Mean-field methods provide valuable insights through the prediction of statistical distribution functions \cite{lebensohn_self-consistent_1993} that characterize microstructural features and properties as well as their evolution. In contrast, full-field methods, such as phase-field \cite{steinbachGeneralizedFieldMethod1999, chenPhaseFieldModelsMicrostructure2002} or crystal plasticity \cite{roters_damask_2019, yaghoobi_prisms-plasticity_2019}, resolve the local material response at the voxel or element level, explicitly accounting for spatial heterogeneity and thereby requiring detailed three-dimensional (3D) microstructural realizations as computational input. This fundamental requirement for representative microstructural geometries motivates the development of robust characterization and reconstruction techniques that can capture the essential features governing material behavior.

Explicit geometric microstructure definitions for full-field simulation frameworks typically rely on either two-dimensional (2D) micrographs extended via columnar subsurface assumptions \cite{hestroffer_subsurface_2025} or the generative reconstruction of synthetic 3D volume elements that reproduce target statistical descriptors. While experimental 3D characterization has become increasingly feasible through combined serial sectioning and imaging \cite{groeber_3d_2006, wirth_focused_2009, calcagnotto_orientation_2010} or the use of high-energy X-ray diffraction microscopy \cite{li_three-dimensional_2012, gustafson_quantifying_2020, pagan_graph_2022}, the cost and complexity of these techniques make them largely inaccessible for high-throughput data generation required by the Integrated Computational Materials Engineering \cite{yi_wang_integrated_2019} paradigm. This bottleneck has strongly motivated the domain of Microstructure Characterization and Reconstruction (MCR) \cite{yeong_reconstructing_1998, torquato_optimal_2010, bostanabad_computational_2018, bargmannGeneration3DRepresentative2018}, which seeks to generate statistically equivalent synthetic microstructures to serve as representative inputs for computational analyses.

The fundamental premise of MCR is based on the definition of microstructural descriptors that are capable of quantifying features and statistics of interest in random heterogeneous media. Consequently, generating a representative volume element (RVE) from these descriptors constitutes an ill-posed inverse problem, where the objective is to identify a geometric realization that belongs to the target statistical ensemble. Polycrystalline microstructures have been an extensive subject of research in the context of MCR, where the reconstruction methodologies are predominantly categorized into geometric tessellation techniques of grains and discrete voxel-based stochastic optimization of microstructural indicator functions such as crystallographic orientations. Tessellation-based approaches, such as Voronoi tessellations and their generalized weighted counterparts known as Laguerre tessellations, partition space based on low-parametric statistical models of seed point distributions. In contrast, voxel-based techniques employ iterative algorithms that update local microstructural states pointwise to minimize an objective function to converge to a configuration that is statistically equivalent \cite{ghosh_statistically_2023} to the target experimental data.

Several widely adopted open-source software frameworks have been developed in the recent years, including DREAM.3D \cite{groeber_framework_2008, groeber_dream3d_2014}, Neper \cite{quey_large-scale_2011}, Kanapy \cite{prasad_kanapy_2019}, and DRAGen \cite{henrich_novel_2020, henrich_dragen_2023} as well as approaches coupling tessellation with spatial stochastic modeling frameworks \cite{furat_artificial_2021,fuchs_generating_2025}. While the computational efficiency and low-parametric representation of Voronoi-based tessellations remain distinct advantages, these methods are geometrically constrained by the inherent convexity of standard Voronoi cells. Although advanced variants such as multi-level Voronoi tessellations \cite{yadegari_analysis_2014} can introduce non-convexity to a limited extent, these geometric abstractions often fail to capture the intragranular substructures characteristic that arise during dynamic recrystallization of metallic alloys \cite{humphreys_recrystallization_1995}. Consequently, they are frequently insufficient for representing the hierarchical heterogeneity observed in materials subjected to severe thermomechanical loading \cite{suhuddin_microstructure_2023} or the rapid solidification conditions of additive manufacturing \cite{motaman_anisotropic_2020}.

In contrast, voxel-based techniques offer extended topological flexibility, as the state of each discretization point is independently modifiable, which effectively maximizes the degrees of freedom within the design space for a given resolution. Within this domain, data-driven machine learning methodologies have gained significant attention, ranging from classification-based approaches like support vector machines \cite{sundararaghavan_classification_2005} to deep generative frameworks such as generative adversarial networks \cite{kench_generating_2021, zhang_da-vegan_2024, murgas_generative_2024} and denoising diffusion probabilistic models \cite{dureth_conditional_2023, lee_denoising_2024, lee_multi-plane_2024, buzzy_statistically_2024}. Given the inherently spatial nature of microstructural data, Convolutional Neural Networks (CNNs) represent a natural choice as architecture for reconstruction tasks trough utilizing translation-invariant filters to automatically extract hierarchical feature maps instead of relying on manually designed descriptors. The utility of CNNs can be extended significantly beyond the application of individually trained architectures on specific microstructure classes. Deep architectures that are pre-trained on large-scale generic image datasets, such as VGG19 \cite{simonyan_very_2015} or ResNet \cite{he_deep_2015}, can be effectively applied to materials characterization \cite{li_transfer_2018, lubbers_inferring_2017, bostanabad_reconstruction_2020}. This is attributed to lower-level convolutional filters that learn edge and texture detection and are, thus, applicable across different domains. By employing Gram matrices to correlate the outputs of selected convolutional layers, it becomes possible to quantify and reconstruct complex microstructural features \cite{seibert_descriptor-based_2022, blumer_generative_2025}, even when such features were absent from the original training dataset. However, while these models have demonstrated the capacity to generate plausible realizations, their non-transparent nature makes the physical interpretability difficult.

Parallel to geometric and deep learning strategies, stochastic reconstruction algorithms, such as texture synthesis \cite{kopf_solid_2007,liu_random_2015}, have emerged as a powerful alternative class of methods. These techniques typically model the microstructure as a realization of a stationary random field. A prominent methodology within this domain utilizes gaussian random fields \cite{robertson_efficient_2021}, which synthesize microstructures by thresholding continuous fields constructed from randomized Fourier modes. To recover higher-order topological features, other texture synthesis algorithms have been adapted, employing patch-based sampling \cite{kopf_solid_2007} or Markov Random Fields \cite{senthilnathan_markov_2021, javaheri_polycrystalline_2020} to sequentially grow 3D volumes that locally preserve morphologies of 2D exemplars. 

Descriptor-based voxel techniques have recently garnered significant traction due to their versatility, particularly in generating statistically equivalent 3D realizations from limited 2D input data. While the seminal stochastic reconstruction algorithm of Yeong and Torquato \cite{yeong_reconstructing_1998} established the benchmark for this inverse problem, its reliance on discrete pixel-swapping and simulated annealing inherently suffers from slow convergence, therefore, significantly limiting scalability for large reconstructions \cite{seibert_reconstructing_2021}. To overcome these computational bottlenecks, the Differentiable Microstructure Characterization and Reconstruction (DMCR) framework \cite{seibert_descriptor-based_2022} has emerged as a robust alternative, substituting stochastic sampling with gradient-based optimization to guide the microstructure generation efficiently towards statistical minima. MCRpy \cite{seibert_microstructure_2022} is an open-source framework that allows for a modular and extendable implementation of various DMCR strategies. It enables the flexible integration of custom loss functions and optimizers. However, despite its proven ability in reconstructing multiphase systems \cite{seibert_two-stage_2023} and complex morphologies like Ti-6Al-4V basketweave structures \cite{blumer_generative_2025}, current implementations remain largely restricted to scalar phase fractions. Consequently, extending the MCRpy framework to directly accommodate the crystallographic lattice orientations remains a significant challenge.

In this work, a gradient-based optimization framework, implemented in MCRpy, for the reconstruction of polycrystalline microstructures is employed \cite{seibert_microstructure_2025}, utilizing direct crystallographic orientation data, as acquired via Electron Backscatter Diffraction (EBSD). Central to this approach is the derivation of a symmetry-invariant formulation that strictly accounts for the crystal symmetry group of the lattice. Instead of binning the high-dimensional orientation space, a continuous representation of local states based on Symmetrized Hyperspherical Harmonics (SHSH) is employed. The statistical fidelity of the reconstruction is governed by a descriptor set combining two-point spatial correlations with a hybrid three-point variogram, which have been demonstrated to be suitable classes for capturing crystallographic textures \cite{paulson_reduced-order_2017} and morphological features \cite{benito_statistical_2023} of microstructures. Furthermore, regularization through variation is incorporated as an auxiliary descriptor to enforce smoothness and mitigate noise. Collectively, these developments constitute a significant extension of MCRpy \cite{seibert_microstructure_2022}, establishing a methodology for the direct, orientation-based reconstruction of polycrystalline materials.

\section{Methodology}\label{2}

This section establishes the theoretical and methodological foundations for the proposed gradient-based microstructure reconstruction framework using lattice orientation fields. First MCR is formalized as an inverse optimization problem, where the objective is to determine a spatial and crystallographic configuration that minimizes the statistical divergence from a target descriptor. Central to this formulation is the definition of the local microstructural state. Therefore, standard representations of crystallographic orientation are critically reviewed to address the inherent topological limitations and associated singularities that pose significant challenges to numerical optimizers. To resolve these issues, unit quaternions are utilized as a mathematically singularity-free alternative for orientation space parameterization. Building upon this, a continuous description that is invariant to symmetrically equivalent orientations is employed by utilizing SHSH. These harmonic basis functions subsequently serve as the fundamental components for constructing the differentiable descriptors that are required to perform the reconstruction process.

\subsection{Microstructure characterization and reconstruction of random heterogeneous media}

Mathematically, a microstructure $\mathbf{M}$ can be represented as a collection of field variables defined over an $N_d$-dimensional spatial domain $\Omega \in \mathbb{R}^{N_d}$, where the field values at each material point may encode phase composition and crystallographic orientation in a unified description. A general representation of a polycrystalline microstructure with $N_p$ phases is given by

\begin{equation}
\label{eq:micr_def}
\begin{aligned}
&\mathbf{M}(\boldsymbol{x}) = \{ I_1(\boldsymbol{x}),\dots,I_{N_p}(\boldsymbol{x}),\, \boldsymbol{g}(\boldsymbol{x}) \}, \quad \boldsymbol{x}\in\Omega,
\end{aligned}
\end{equation}

where $I_q(\boldsymbol{x})$ denotes the phase indicator function for phase $q$, subject to the constraint $\sum^{N_p}_{q=1} I_q(\boldsymbol{x})=1$ to ensure a complete domain coverage. For the simplest case of a heterogeneous two-phase system, the indicator function $I_q(\boldsymbol{x})$ assumes a value of one at every material point occupied by phase $q$ and vanishes elsewhere. In polycrystalline materials, the microstructure description is further augmented by an orientation field $\boldsymbol{g}(\boldsymbol{x})$ that characterizes the local crystallographic lattice orientation as a continuous spatial distribution. This orientation field captures the morphological grain structure and crystallographic texture, and is experimentally commonly determined in 2D via EBSD.

In practical applications, the continuous field representation defined in Eq.~\eqref{eq:micr_def} is discretized into a voxel-based grid. This discretization maps the microstructure onto a regular array where each voxel represents the local phase and orientation state. To enable quantitative analysis and subsequent reconstruction, the high-dimensional information contained within this discrete domain is characterized by a set of statistical descriptors, $\mathbf{D}$. These descriptors are designed to be statistically stationary and translation-invariant. Consequently, the task of reconstructing a synthetic microstructure $\mathbf{M}^\text{recon}$ that reproduces the desired statistics becomes an inverse problem and is formulated as the following optimization problem

\begin{equation}
\label{eq:micr_opt}
\begin{aligned}
&\mathbf{M}^\text{recon}=\underset{\mathbf{M}}{\arg \min} \,\mathcal{L}(\mathbf{M)} ,
\end{aligned}
\end{equation}

where $\mathcal{L}(\mathbf{M})$ denotes a scalar microstructure loss function that quantifies the statistical divergence between the candidate microstructure and a prescribed reference state. This loss is formally expressed in terms of a distance metric between the microstructural descriptors of the candidate and the reference, given by

\begin{equation}
\label{eq:micr_loss}
\begin{aligned}
&\mathcal{L}(\mathbf{M)} = f(\mathbf{D(\mathbf{M}), \mathbf{D}^{\text{ref}}}),
\end{aligned}
\end{equation}

where $\mathbf{D}^{\text{ref}}$ denotes the target descriptor set derived from experimental observations or synthetic reference data, and $f$ represents a suitable distance metric, such as the Mean-Squared Error (MSE), which penalizes statistical deviations from the target. Historically, stochastic optimization algorithms such as simulated annealing \cite{rintoul_reconstruction_1997} have been employed to solve Eq.~\eqref{eq:micr_loss}. However, the convergence rate of these methods scales poorly with the dimensionality of the voxel grid making them computationally challenging for high-resolution 2D images or performing 3D reconstructions \cite{seibert_descriptor-based_2022}. These scalability limitations motivate the adoption of DMCR. By formulating the entire characterization scheme as a differentiable computation, DMCR utilizes gradient information to efficiently navigate through the high-dimensional microstructural design space. Consequently, the reconstruction problem is recast as a differentiable optimization, where the iterative evolution of the microstructure is driven directly by the sensitivity of the loss function with respect to the local state variables

\begin{equation}
\label{eq:micr_update}
\begin{aligned}
&\mathbf{M}^{i+1} = \mathbf{M}^{i} + \Delta \mathbf{M}\left(\frac{d\mathcal{L}}{d\mathbf{M}^{i}}\right).
\end{aligned}
\end{equation}

Here, $\mathbf{M}^{i}$ denotes the microstructural state at iteration $i$, and $\Delta \mathbf{M}$ represents the update increment computed by the optimizer, which is derived from the gradient of the loss function, $\frac{d\mathcal{L}}{d\mathbf{M}^{i}}$. In practice, optimizers may also utilize higher-order information, such as approximations of the Hessian, to compute the update $\Delta \mathbf{M}$. While the formulation detailed in Eqs.~\eqref{eq:micr_opt}-\eqref{eq:micr_update} is sufficient for direct 2D-to-2D reconstruction tasks, the extension to 2D-to-3D reconstruction requires further stereological assumptions. Since the direct acquisition of full 3D volumetric descriptors is associated with significant experimental difficulties, the loss function can be generalized to enforce statistical consistency across the volume by evaluating orthogonal planar 2D sections. Specifically, during optimization the loss function aggregates the discrepancies computed across all of the 2D cross-sections oriented along orthogonal directions to ensure that the reconstructed volume reproduces the target statistics along all principal axes

\begin{equation}
\label{eq:slice_loss}
\begin{aligned}
&\mathcal{L}(\mathbf{M)} = \sum_{i=1}^{N_d}\sum_{s=1}^{N^i_s} f(\mathbf{D}^s_i(\mathbf{M}), \mathbf{D}_i^{\text{ref}}),
\end{aligned}
\end{equation}

where $N^i_s$ denotes the number of parallel slices sampled along the $i$-th direction. The term $\mathbf{D}^s_i(\mathbf{M})$ refers to the descriptors computed on the $s$-th discrete 2D section extracted from $\mathbf{M}$ normal to direction $i$, while $\mathbf{D}_i^{\text{ref}}$ represents the corresponding target descriptors. It is important to note that while the reference descriptors $\mathbf{D}_i^{\text{ref}}$ may vary between directions $i$ to account for global microstructural anisotropy, they are typically assumed to be constant across all slices $s$ within a given direction in assumption that they are statistically wide-sense stationary throughout the volume.

\subsection{Representation of crystallographic orientation}

The quantitative description of a material's crystallographic microstructure relies on the mathematical representation of lattice orientation $\boldsymbol{g}$. Conventionally, the transformation from the specimen reference frame to the local crystal reference frame is parameterized using a set of three Bunge-Euler \cite{bunge_texture_2013} angles $(\phi_1, \Phi, \phi_2)$. Despite their widespread use in texture analysis, Euler angles present significant theoretical and computational challenges in the context of numerical optimization. The parameterization is inherently non-unique and suffers from singularities, a phenomenon referred to as gimbal lock \cite{hemingway_perspectives_2018}. Near these singular configurations, infinitesimal physical rotations inducing large and discontinuous jumps in the angular coordinates and would destabilize gradient-based reconstruction schemes. While the full rotation matrix, $\mathbf{R}$, offers a singularity-free alternative, it is rarely employed as a primary variable in optimization frameworks due to its excessive dimensionality as it requires nine components to represent a rotation. Moreover, strictly enforcing the orthogonality constraint ($\mathbf{R}^T\mathbf{R} = \mathbf{I}$) and unit determinant during iterative updates adds significant computational complexity and make it an inefficient choice for microstructure reconstruction. This motivates the axis-angle representation, following from Euler's rotation theorem, wherein any rotation is fully parameterized by a single rotation axis $\mathbf{v} \in \mathbb{R}^3$ and a corresponding rotation angle $\omega$ about that axis as an active rotation formulation that fall under the broader class of neo-Eulerian representations. While the axis-angle representation introduces a fourth parameter, various neo-Eulerian subclasses have emerged that restore the dimensionality to three by transforming the axis and angle into a single vector through the product $f(\omega)\cdot\mathbf{v}$, where $f(\omega)$ is an elementary function. One prominent neo-Eulerian subclass is the Rodrigues vector $\boldsymbol{\rho} \in \mathbb{R}^3$, which is defined by the scaling function $f(\omega) = \tan(\omega/2)$ \cite{frank_orientation_1988, neumann_representation_1991}. The Rodrigues vector, $\boldsymbol{\rho}$, can thus be written as

\begin{equation}
\label{eq:rodrigues}
\begin{aligned}
&\boldsymbol{\rho}=\mathbf{v}\tan(\omega/2),
\end{aligned}
\end{equation}

which provides an intuitive axis-direction representation where the vector magnitude increases monotonically with $\omega$. While the Rodrigues parameterization is associated with favorable properties, such as the unique definition of orientations near the center of the orientation space and its direct relation to the geometry of misorientations, it is also limited by singularities at $\omega = -\pi$ and $\omega = \pi$, where the tangent function diverges. Furthermore, the Rodrigues space does not naturally support harmonic analysis or series expansions for the formulation of symmetry-invariant descriptors. Numerous further valid neo-Eulerian forms for the scaling function $f(\omega)$ exist, out of which the specific choice of $f(\omega)= \sin(\omega/2)$ leads to the derivation of quaternions \cite{altmann_rotations_2005}. By embedding the rotation in a four-dimensional (4D) hypersphere, quaternions provide a globally singularity-free description that uniquely characterizes any 3D rotation

\begin{equation}
\label{eq:quats}
\begin{aligned}
&\mathbf{q} = \left(\cos(\omega/2), \mathbf{v} \sin(\omega/2)\right) = \left(q_0, \mathbf{q}_v\right).
\end{aligned}
\end{equation}

The four-component quaternion is formally defined as a hypercomplex number consisting of a real part $q_0$ and an imaginary vector part $\mathbf{q}_{v} = (q_1, q_2, q_3)$ for which the Hamilton rules apply \cite{hamilton_new_1840}. In the context of spatial rotations, pure rotational transformations are represented exclusively by unit quaternions and lie on the surface of the 4D unit hypersphere $\mathbf{S}^3$. Thus, it is crucial to enforce the normalization constraint, $\|\mathbf{q}\| = 1$, as non-normalized quaternions also infer scaling. Consequently, all subsequent references to \textit{quaternions} in this work will implicitly denote unit quaternions. Due to their shared geometric origin, neo-Eulerian representations can be easily converted to each other. For example, $\boldsymbol{\rho}$ can be recovered from $\mathbf{q}$ via the projection

\begin{equation}
\label{eq:quat2rodrigues}
\begin{aligned}
&\boldsymbol{\rho}=\frac{\mathbf{q}_v}{q_0}.
\end{aligned}
\end{equation}

Conversely, $\mathbf{q}$ is related to $\boldsymbol{\rho}$ as follows

\begin{equation}
\label{eq:rodrigues2quat}
\begin{aligned}
&q_0=\frac{1}{\sqrt{1+\|\boldsymbol{\rho}\|^2}},\qquad \mathbf{q}_v=\frac{\boldsymbol{\rho}}{\sqrt{1+\|\boldsymbol{\rho}\|^2}}.
\end{aligned}
\end{equation}

Beyond these numerical advantages, quaternions are particularly attractive for DMCR because they naturally allow for a spectral representation via the Hyperspherical Harmonics (HSH) \cite{mason_hyperspherical_2008}. This property is analogous to the established use of Generalized Spherical Harmonics (GSH) for texture analysis in Euler space \cite{bunge_texture_2013}, but it avoids the associated singularities. To exploit this, the unit quaternion is mapped onto the surface of $\mathbf{S}^3$ and can be parameterized by a set of three hyperspherical angles $(\omega, \theta, \phi)$. Defined within the domains $0 \le \omega \le 2\pi$, $0 \le \theta \le \pi$, and $0 \le \phi < 2\pi$, these angular coordinates allow the transformation from the Cartesian quaternion components to the hyperspherical domain via the following relations

\begin{equation}
\label{eq:hyper2quats}
\begin{aligned}
q_0 &= \cos(\omega/2), \\
q_1 &= \sin(\omega/2)\sin(\theta)\cos(\phi), \\
q_2 &= \sin(\omega/2)\sin(\theta)\sin(\phi), \\
q_3 &= \sin(\omega/2)\cos(\theta).
\end{aligned}
\end{equation}

\begin{figure}
\begin{labeledbox}[Eulerian representation]
\begin{subfigure}[c]{.2\textwidth}
  \centering
  \includegraphics[width=.8\linewidth]{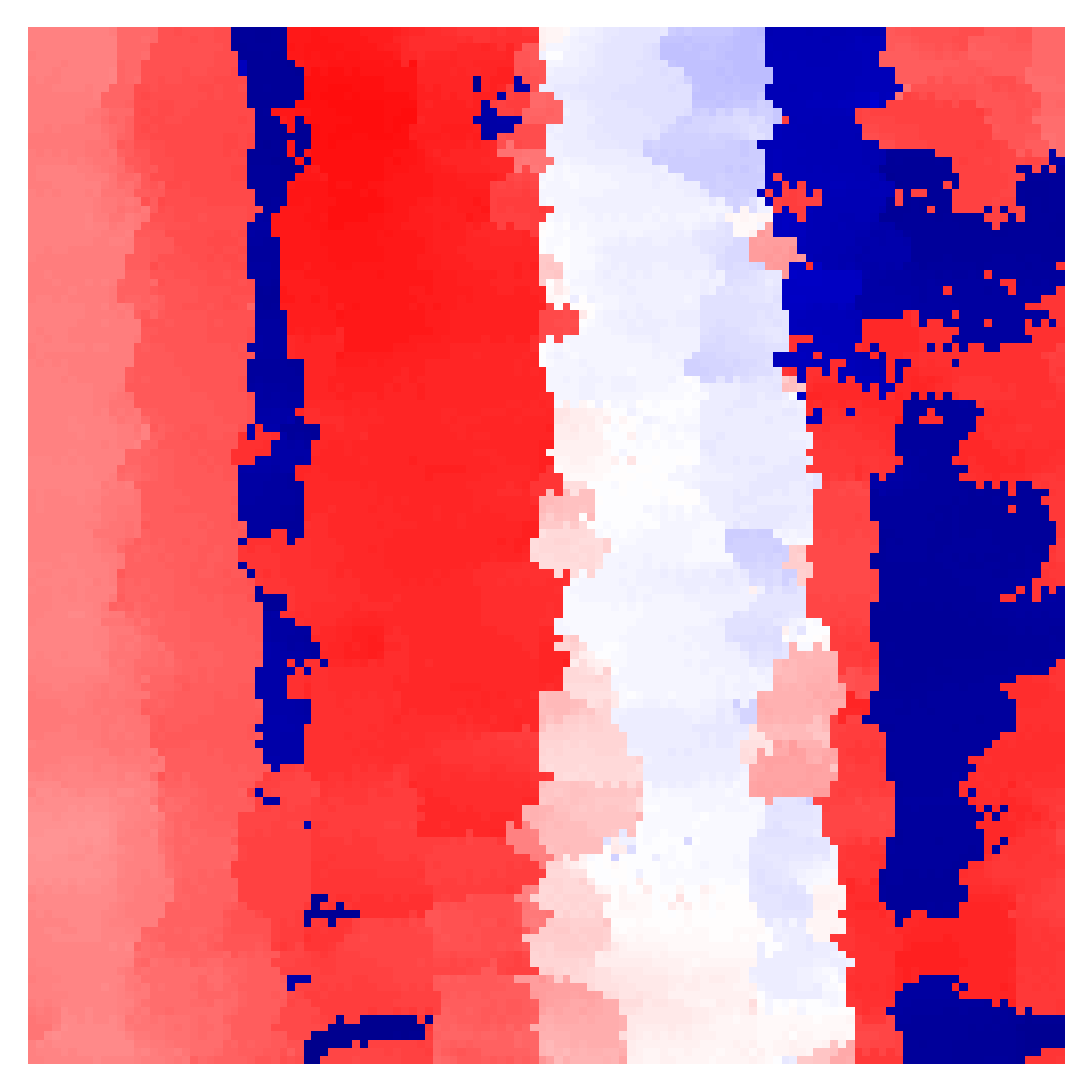}
  \caption*{$\varphi_1$}
  \label{fig:phi1}
\end{subfigure}%
\begin{subfigure}[c]{.2\textwidth}
  \centering
  \includegraphics[width=.8\linewidth]{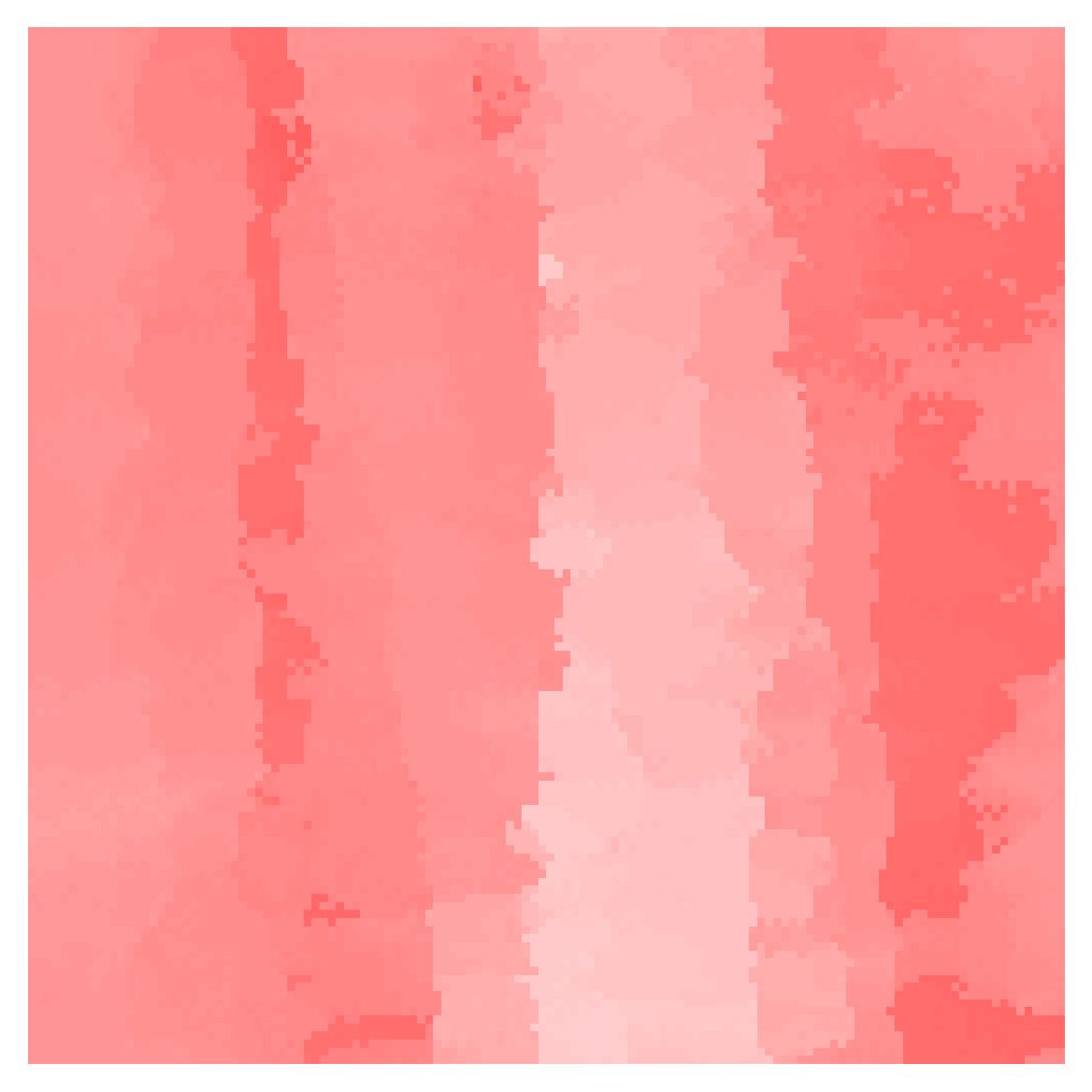}
  \caption*{$\Phi$}
  \label{fig:Phi}
\end{subfigure}
\begin{subfigure}[c]{.2\textwidth}
  \centering
  \includegraphics[width=.8\linewidth]{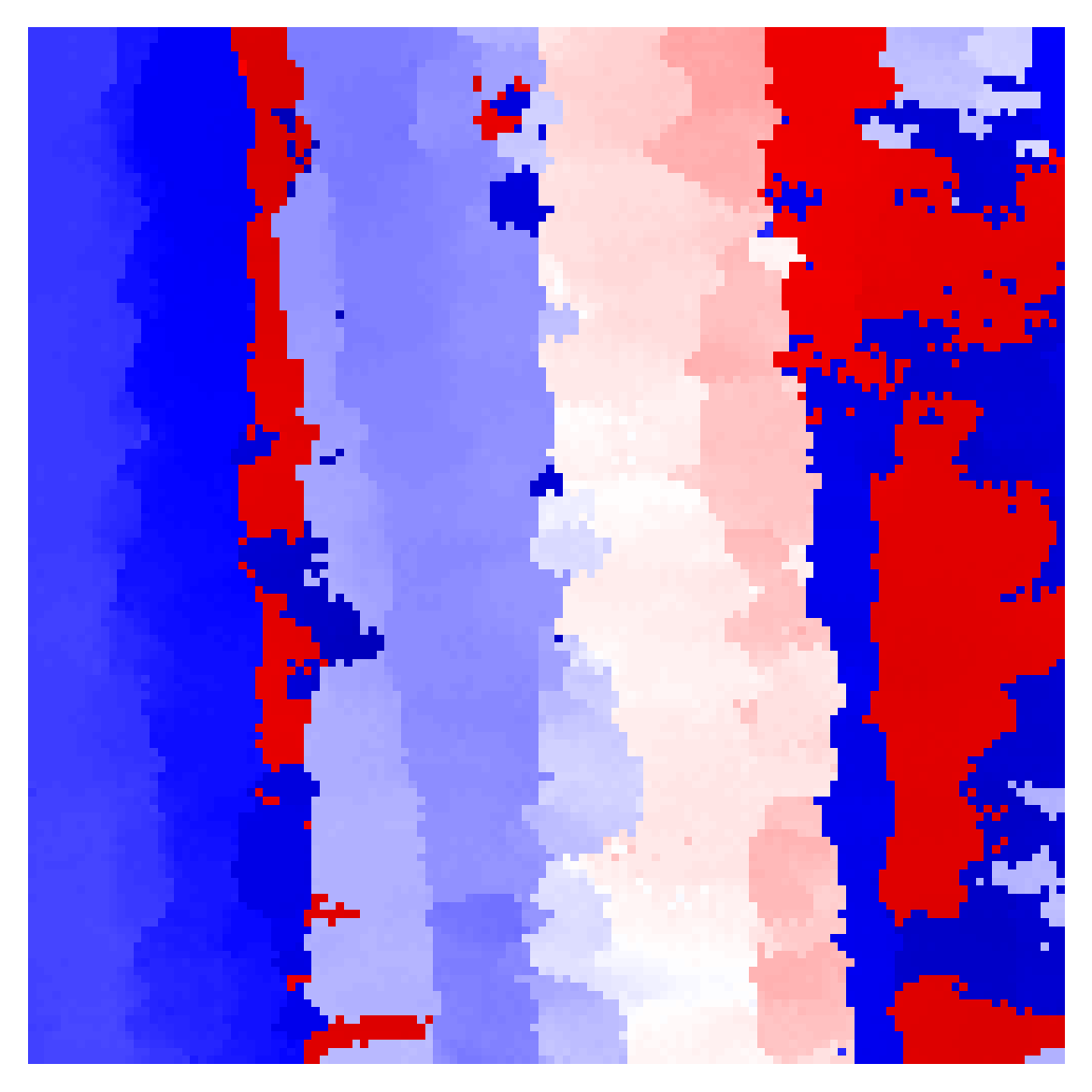}
  \caption*{$\varphi_2$}
  \label{fig:phi2}
\end{subfigure}%
\begin{subfigure}[c]{.08\textwidth}
  \centering
  \includegraphics[height=2cm]{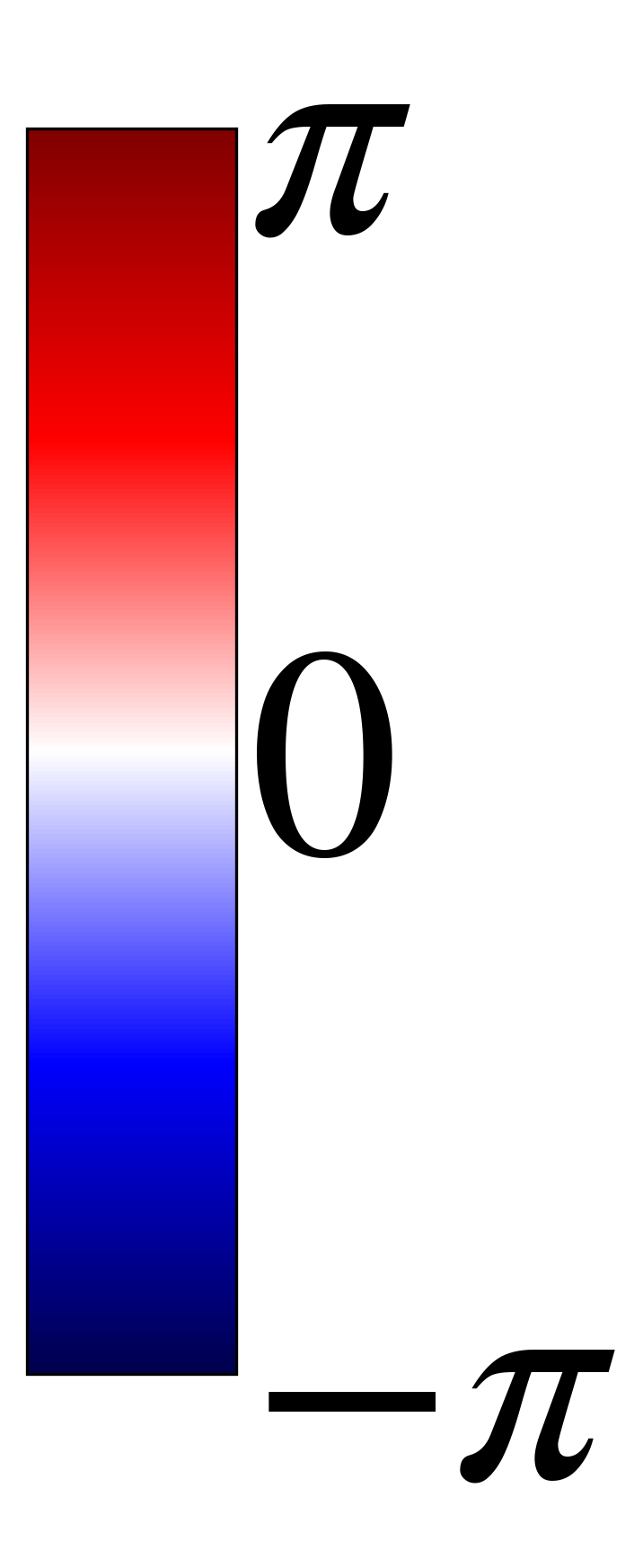}
  \caption*{}
\end{subfigure}%
\end{labeledbox}
\begin{labeledbox}[Neo-Eulerian representation]
 \begin{subfigure}[c]{.2\textwidth}
  \centering
  \includegraphics[width=.8\linewidth]{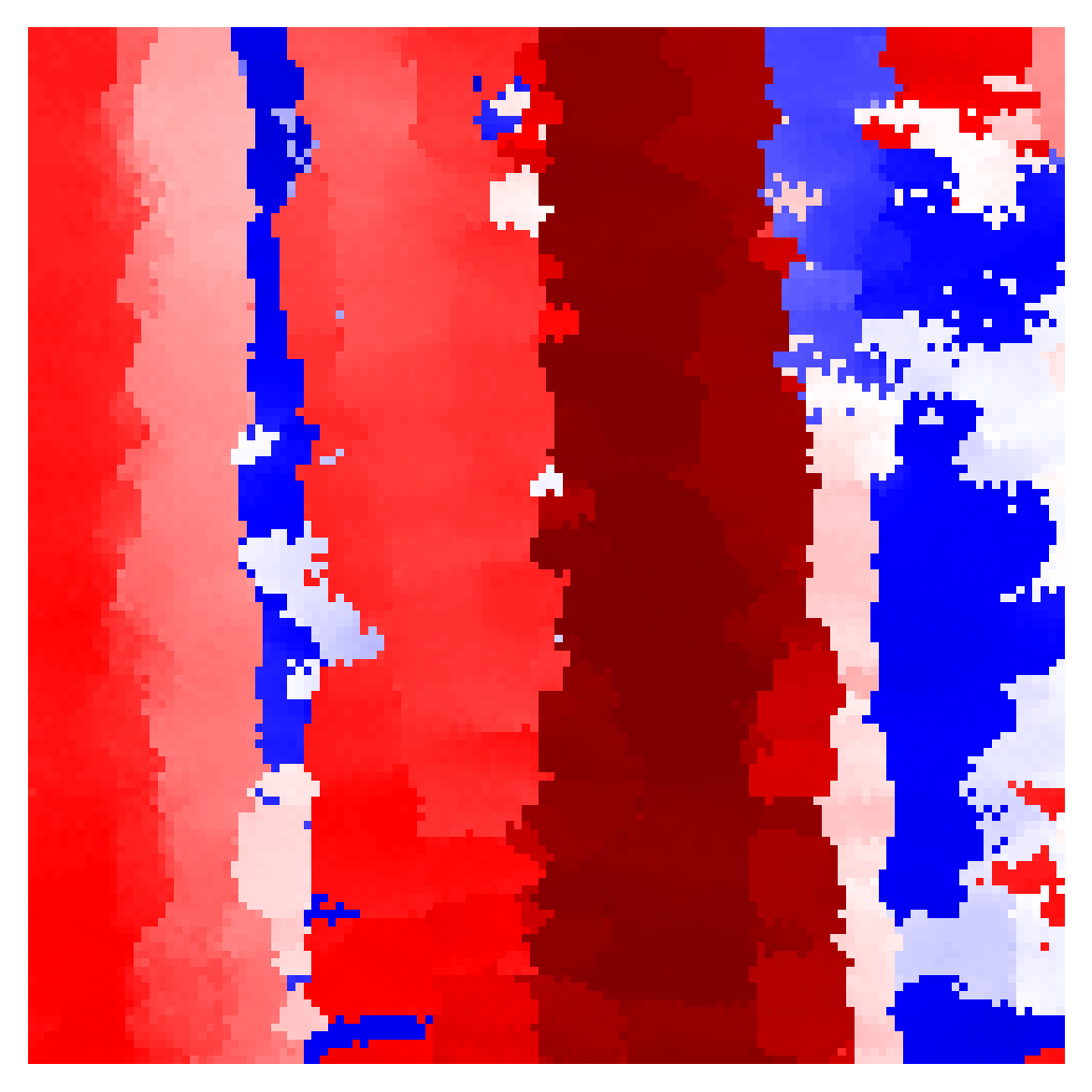}
  \caption*{$v_1$}
  \label{fig:v1}
\end{subfigure}%
\begin{subfigure}[c]{.2\textwidth}
  \centering
  \includegraphics[width=.8\linewidth]{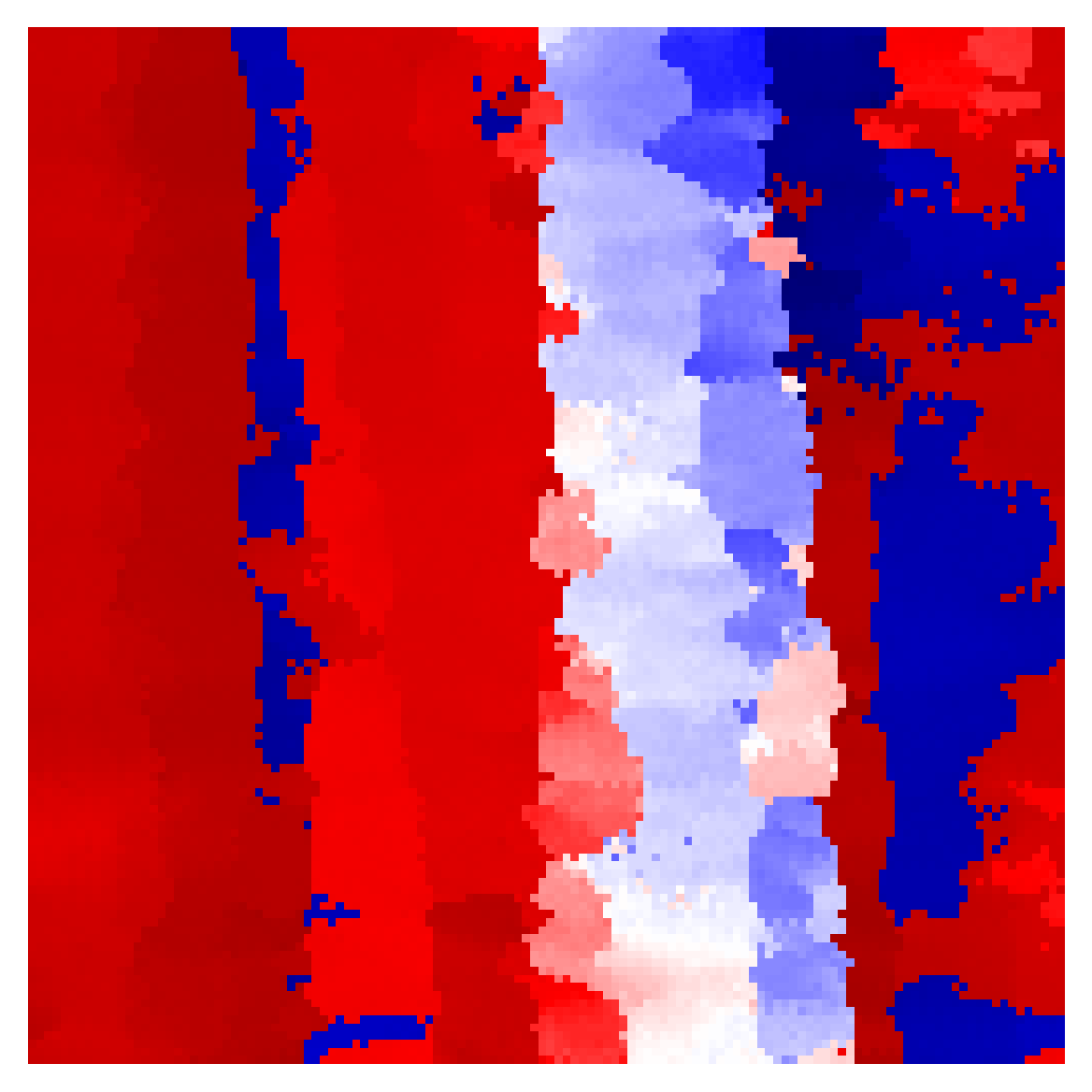}
  \caption*{$v_2$}
  \label{fig:v2}
\end{subfigure}
\begin{subfigure}[c]{.2\textwidth}
  \centering
  \includegraphics[width=.8\linewidth]{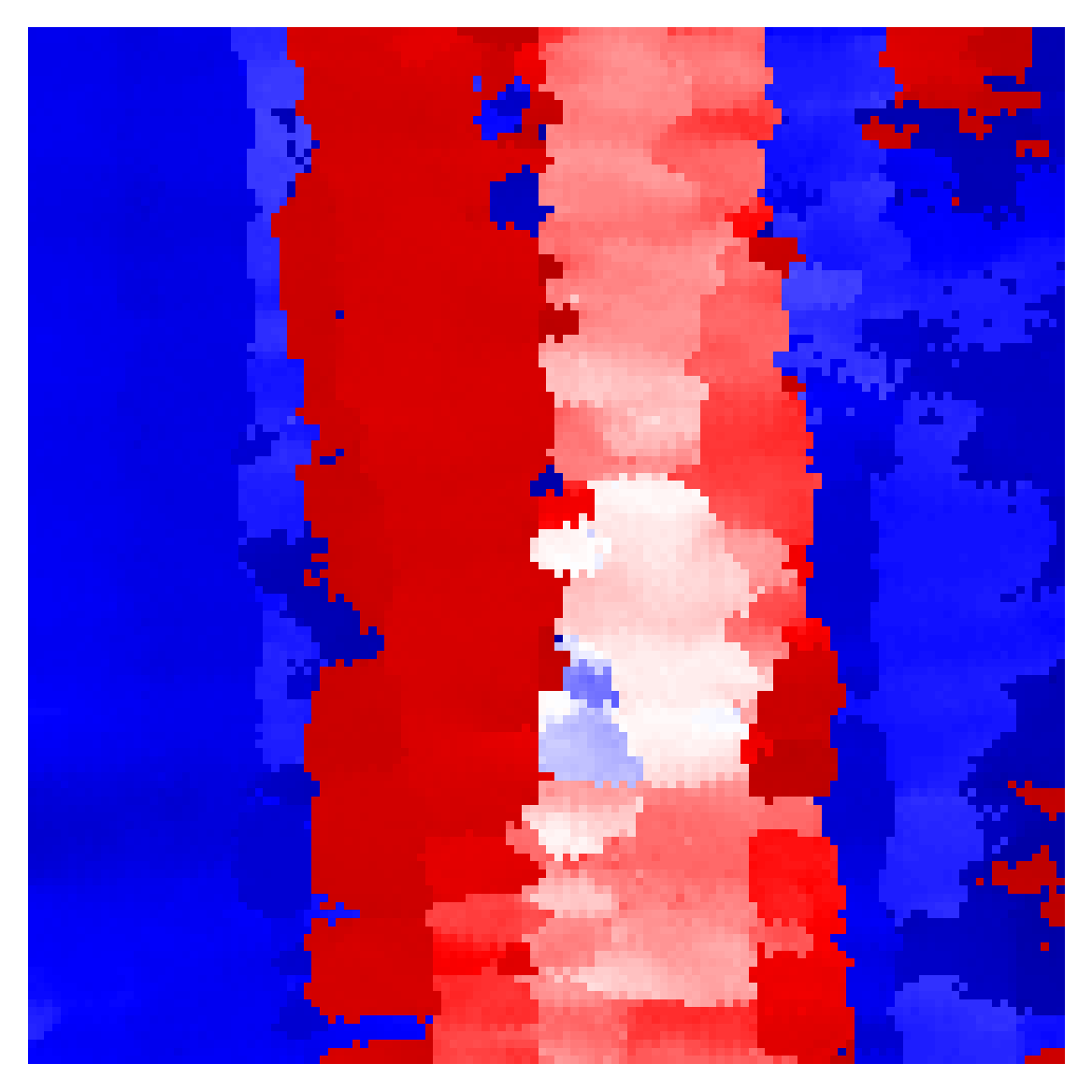}
  \caption*{$v_3$}
  \label{fig:v3}
\end{subfigure}%
\begin{subfigure}[c]{.2\textwidth}
  \centering
  \includegraphics[width=.8\linewidth]{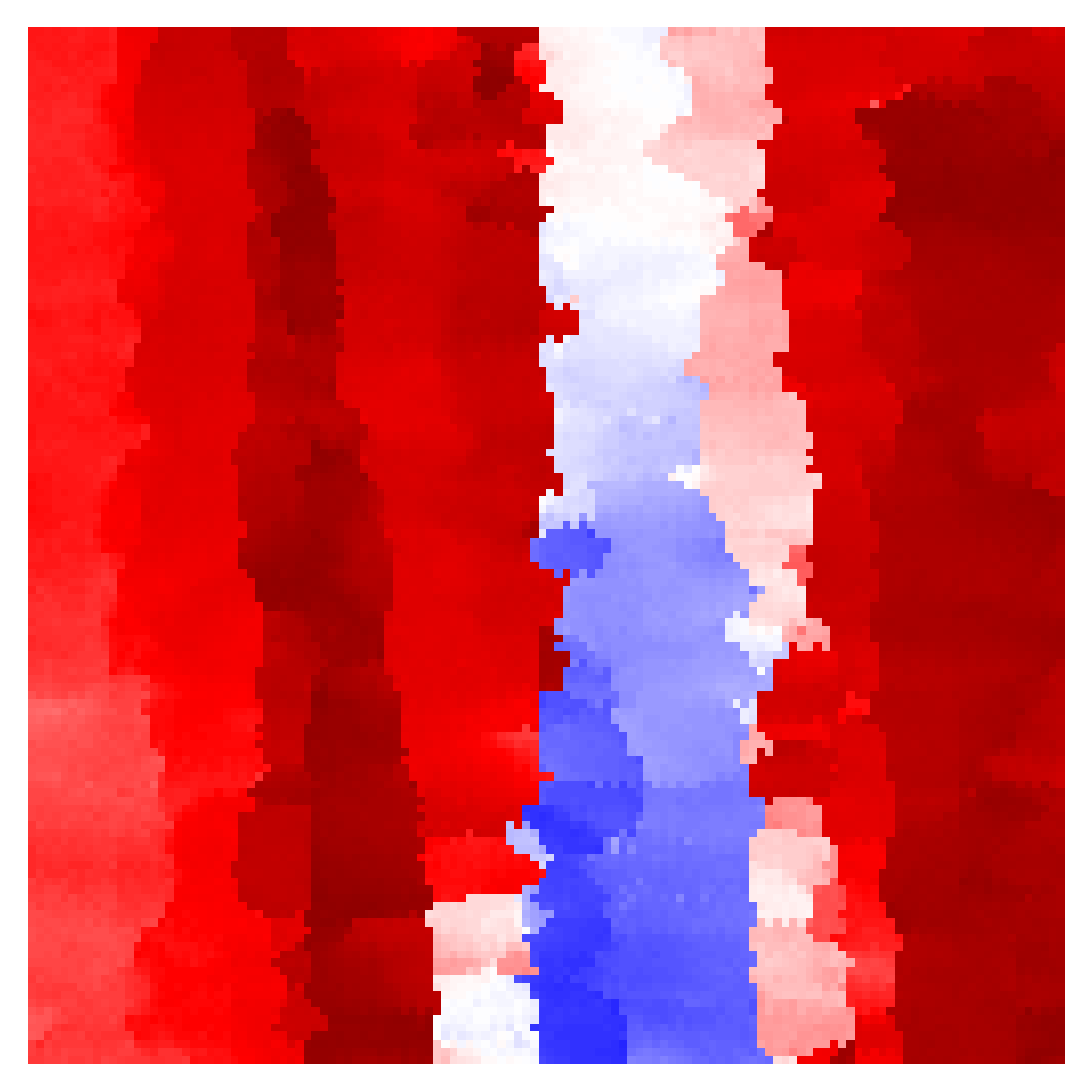}
  \caption*{$\omega$}
  \label{fig:omega}
\end{subfigure}%
\begin{subfigure}[c]{.2\textwidth}
  \centering
  \begin{tabular}{m{0.2cm} m{0.5cm} m{0.7cm} m{0.3cm}}
    \includegraphics[height=1.8cm]{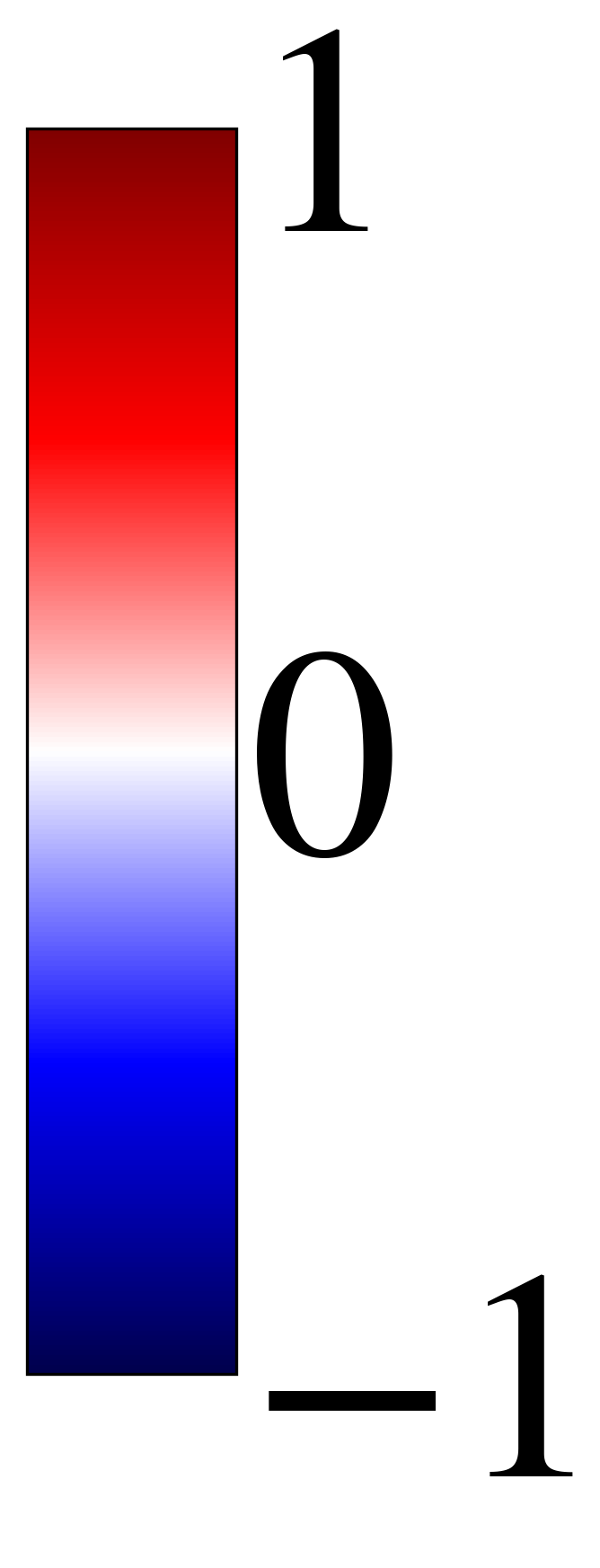} &
    \rotatebox[origin=c]{90}{\footnotesize $v_i$} & 
    \includegraphics[height=1.8cm]{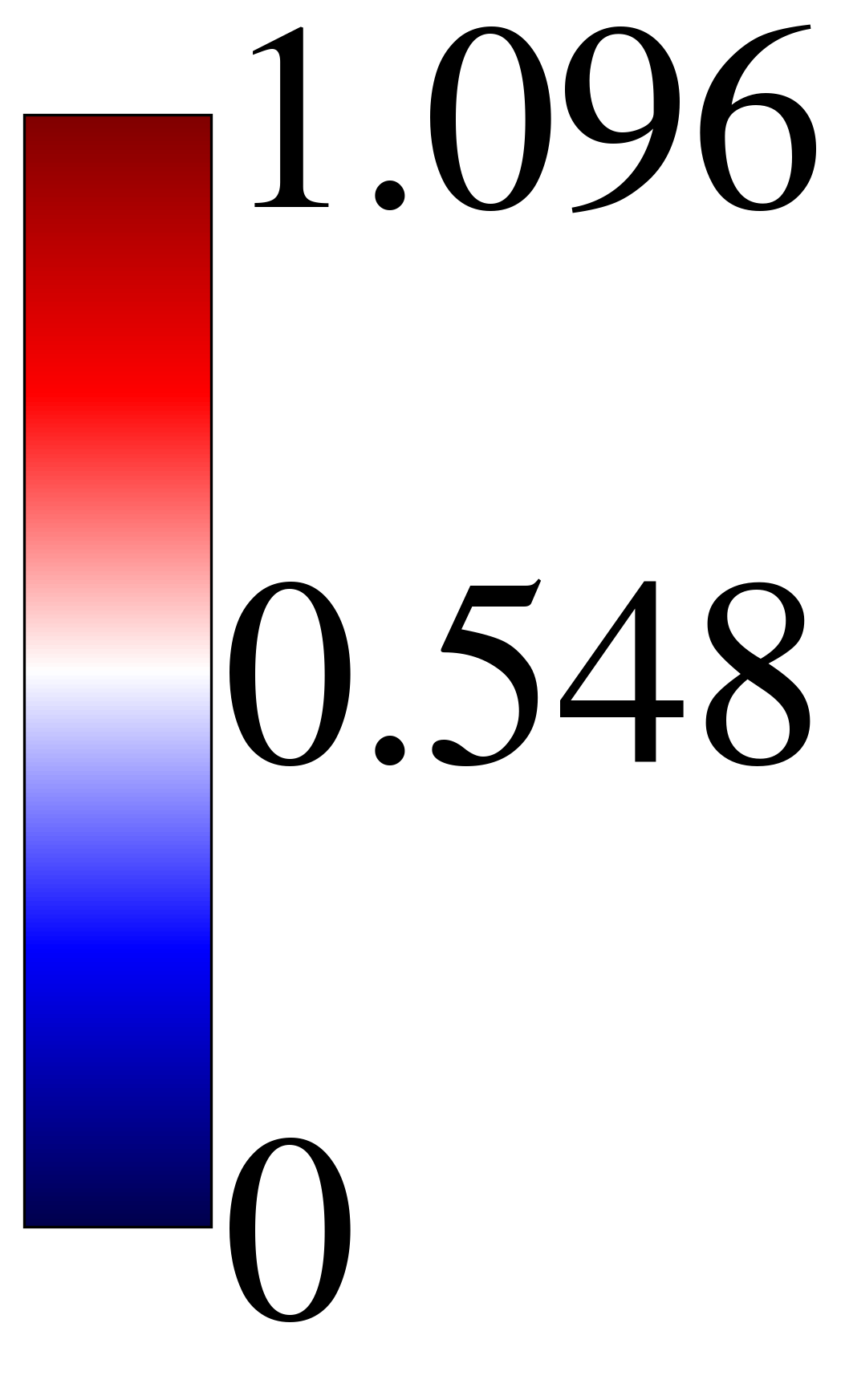}& 
    \rotatebox[origin=c]{90}{\footnotesize $\omega$} \\
  \end{tabular}
  \caption*{}
\end{subfigure}
\\
\begin{subfigure}[c]{.2\textwidth}
  \centering
  \includegraphics[width=.8\linewidth]{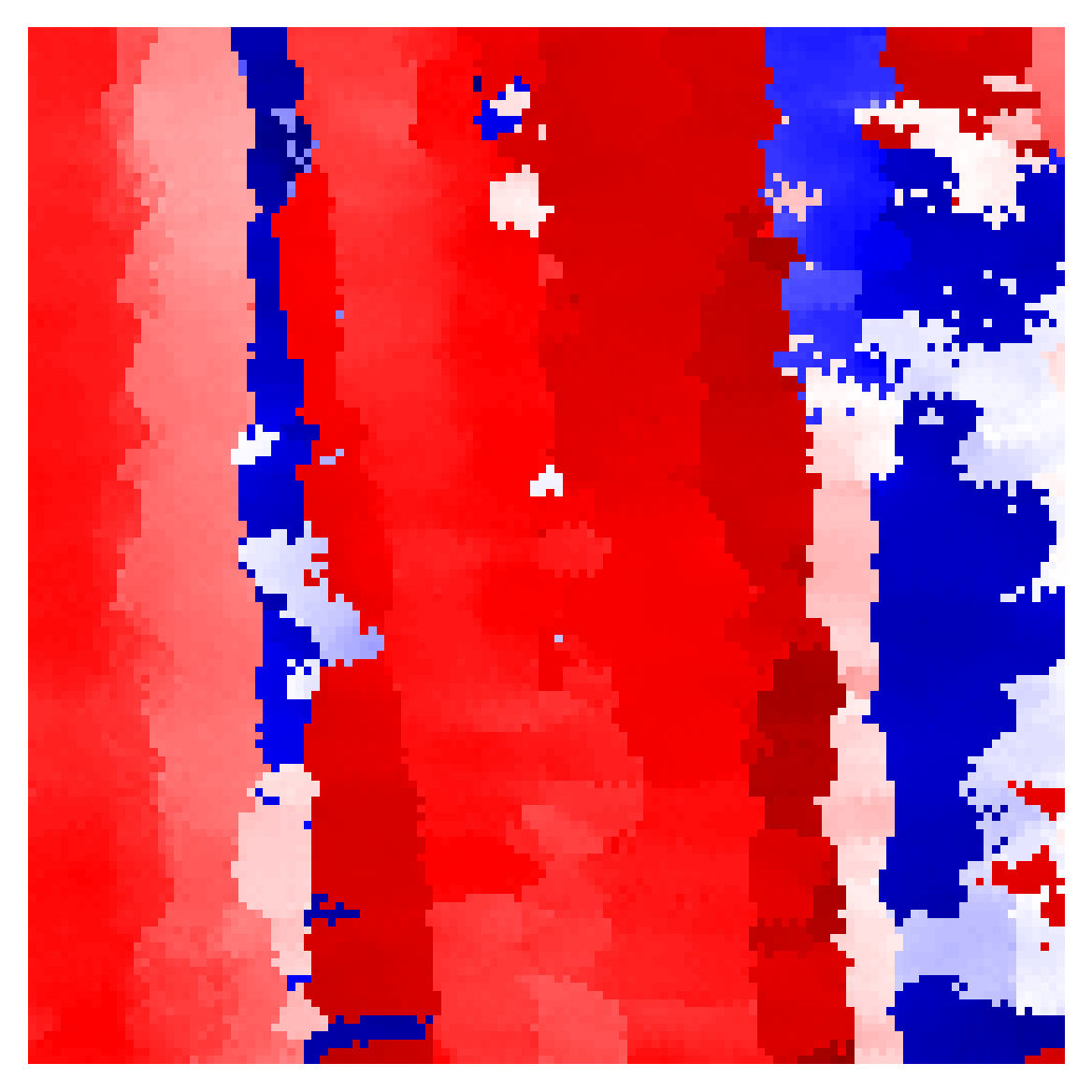}
  \caption*{$\rho_1$}
  \label{fig:rho1}
\end{subfigure}%
\begin{subfigure}[c]{.2\textwidth}
  \centering
  \includegraphics[width=.8\linewidth]{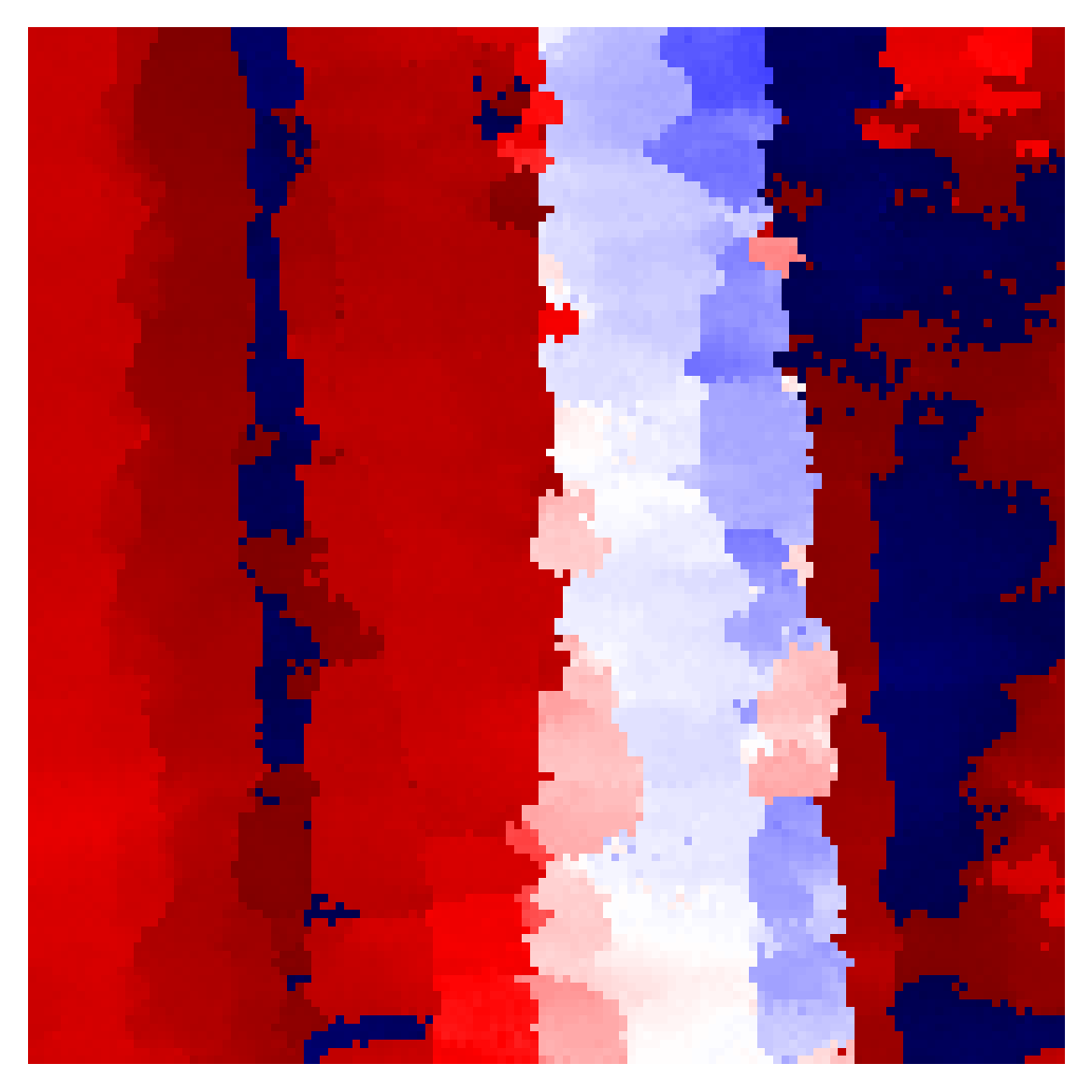}
  \caption*{$\rho_2$}
  \label{fig:rho2}
\end{subfigure}
\begin{subfigure}[c]{.2\textwidth}
  \centering
  \includegraphics[width=.8\linewidth]{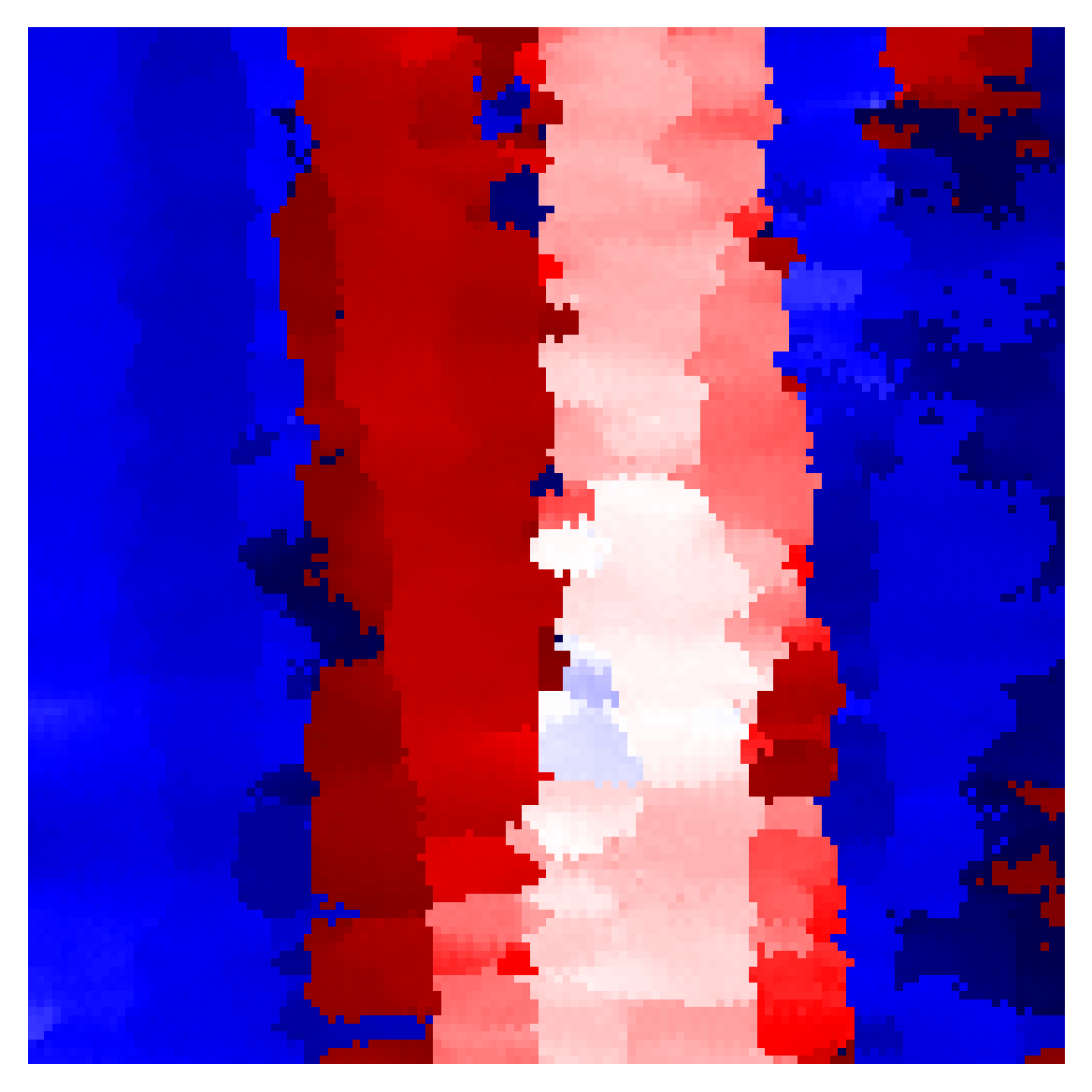}
  \caption*{$\rho_3$}
  \label{fig:rho3}
\end{subfigure}%
\begin{subfigure}[c]{.1\textwidth}
  \centering
  \includegraphics[height=2cm]{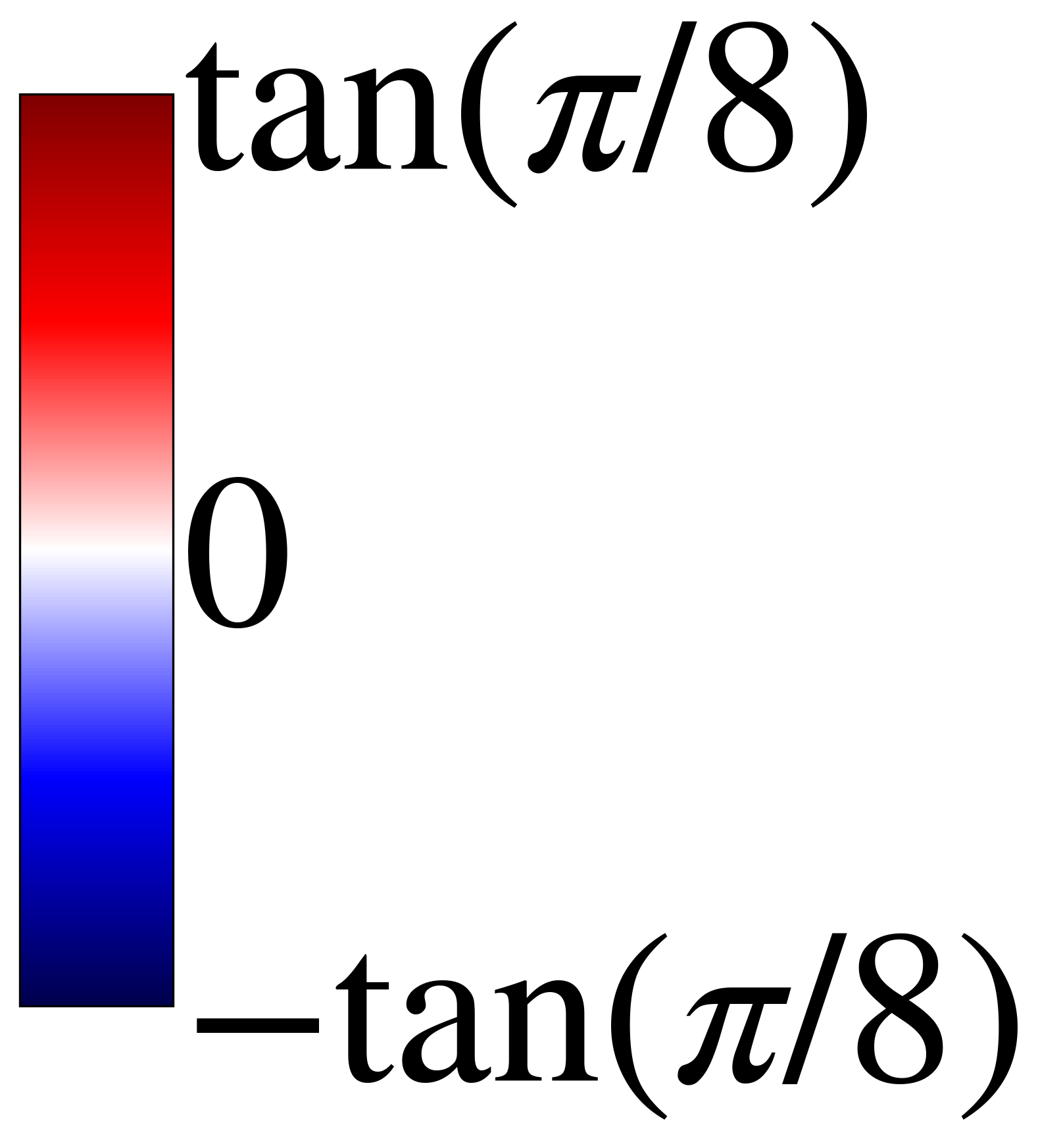}
  \caption*{}
\end{subfigure}%
\\
 \begin{subfigure}[c]{.2\textwidth}
  \centering
  \includegraphics[width=.8\linewidth]{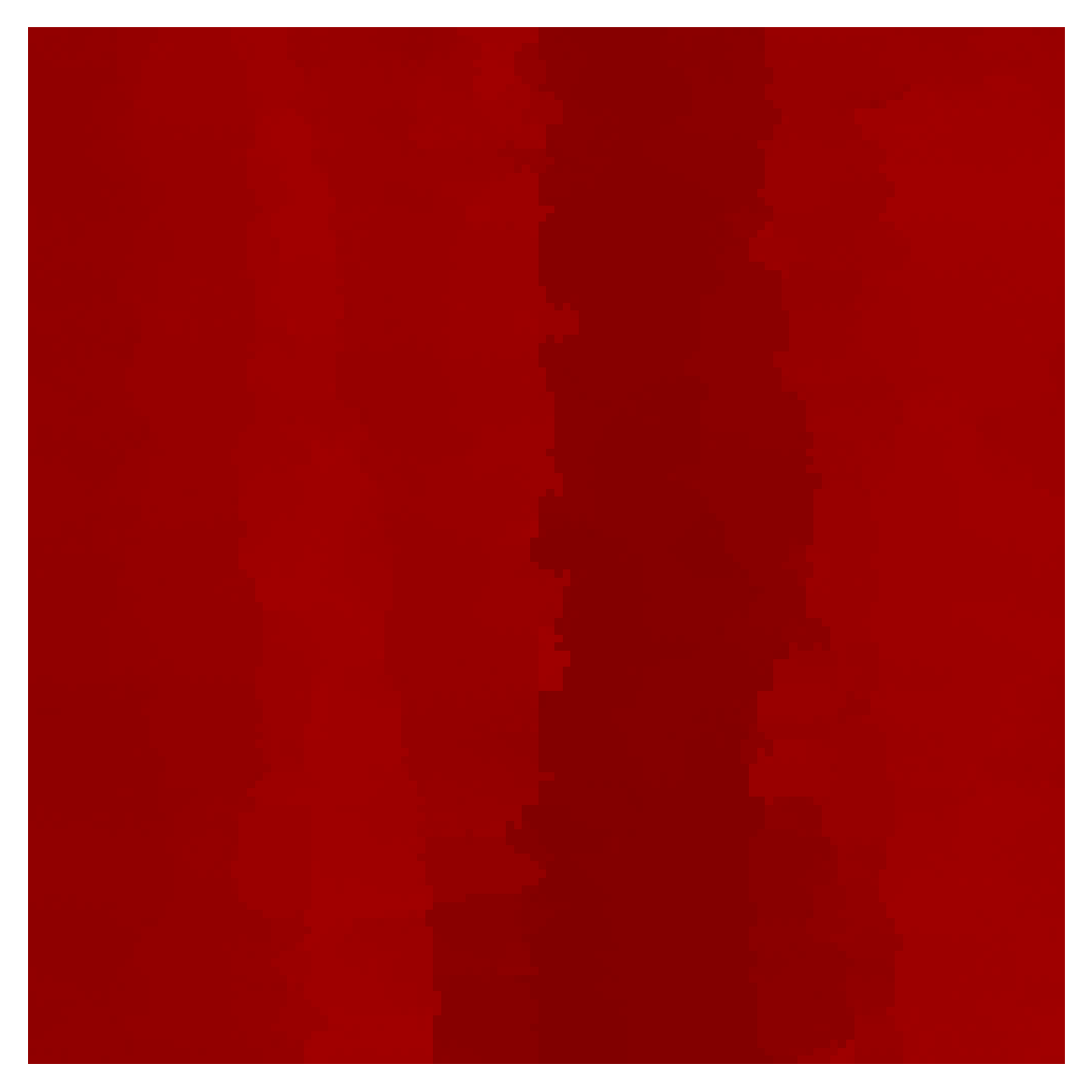}
  \caption*{$q_0$}
  \label{fig:q0}
\end{subfigure}%
\begin{subfigure}[c]{.2\textwidth}
  \centering
  \includegraphics[width=.8\linewidth]{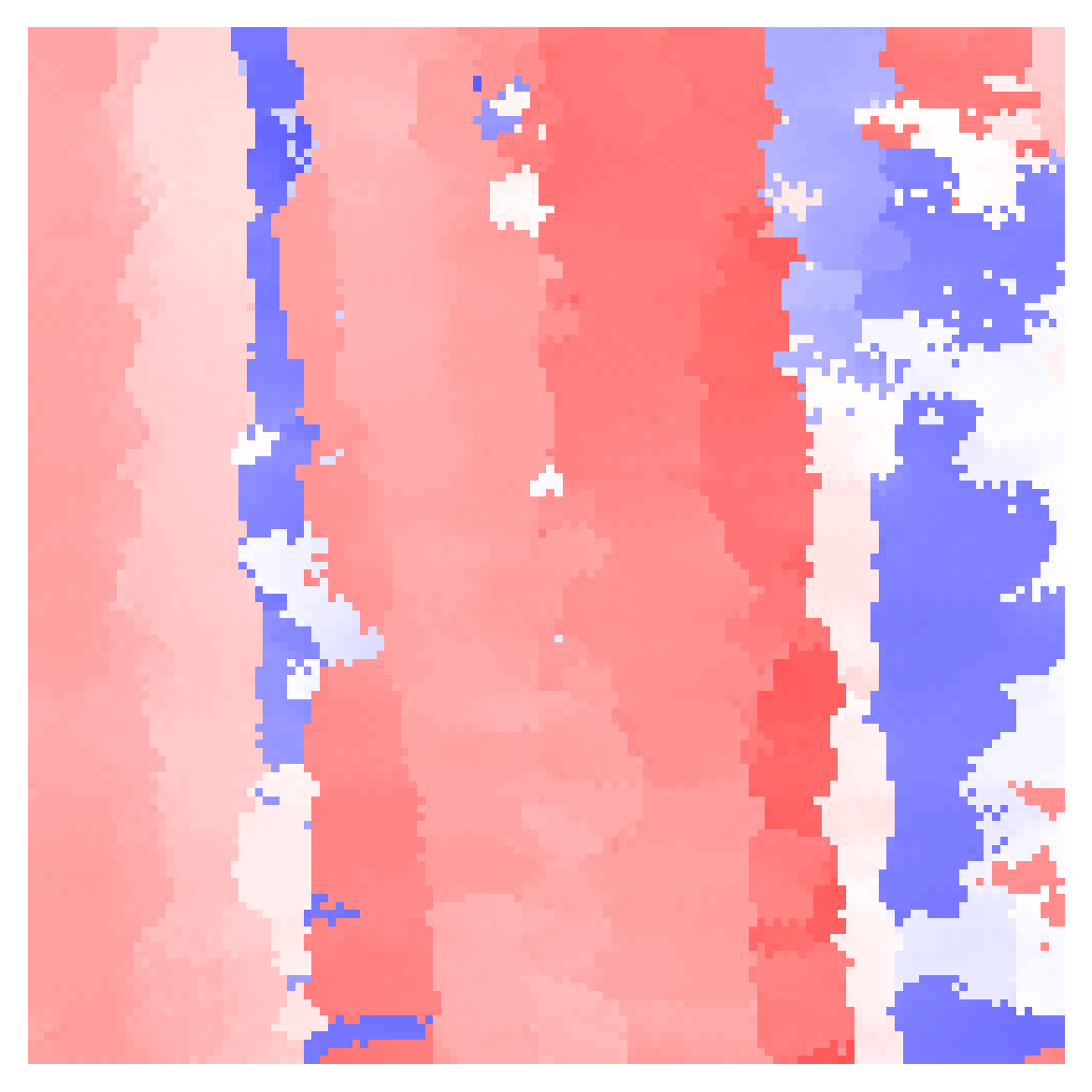}
  \caption*{$q_1$}
  \label{fig:q1}
\end{subfigure}
\begin{subfigure}[c]{.2\textwidth}
  \centering
  \includegraphics[width=.8\linewidth]{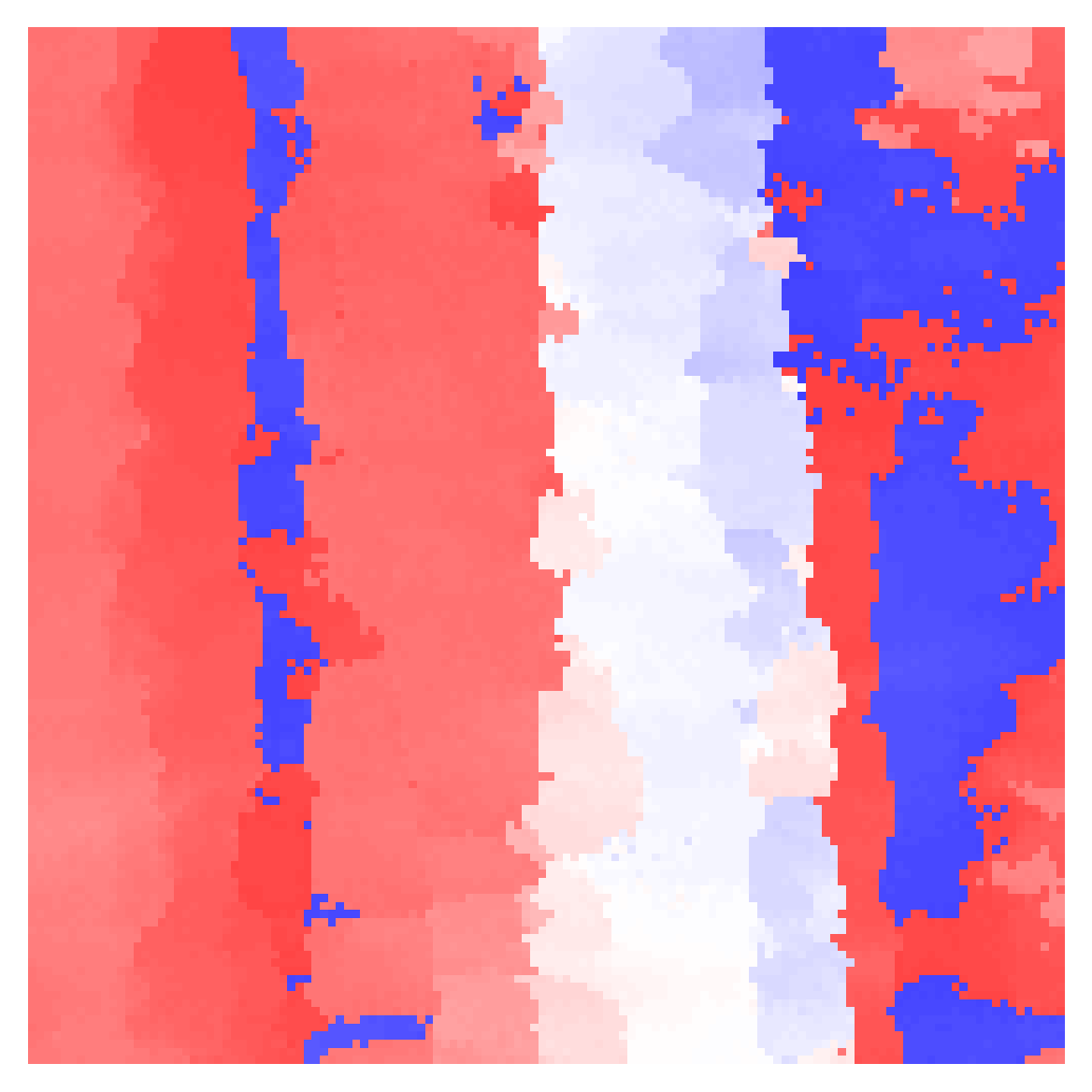}
  \caption*{$q_2$}
  \label{fig:q2}
\end{subfigure}%
\begin{subfigure}[c]{.2\textwidth}
  \centering
  \includegraphics[width=.8\linewidth]{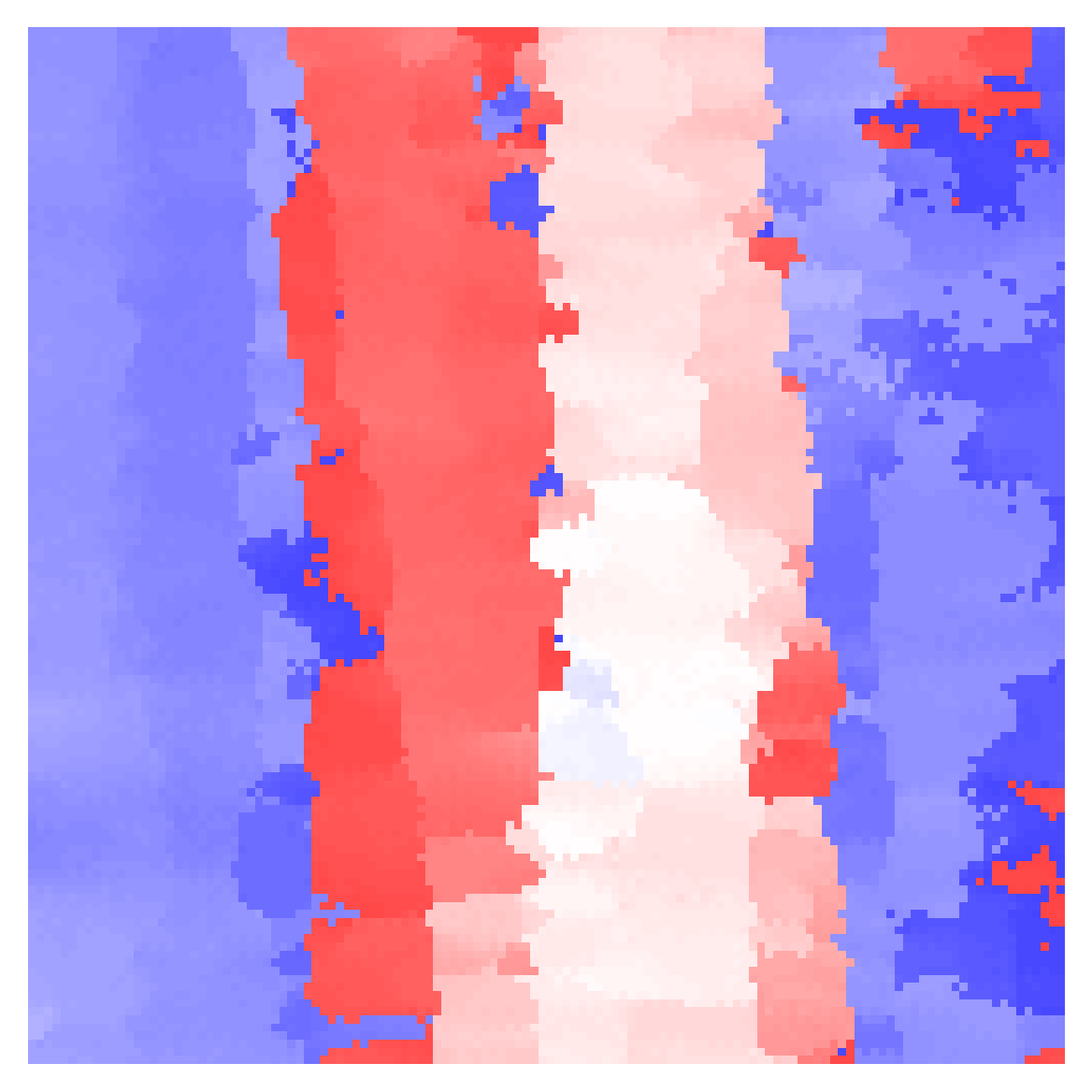}
  \caption*{$q_3$}
  \label{fig:q3}
\end{subfigure}%
\begin{subfigure}[c]{.1\textwidth}
  \centering
  \includegraphics[height=2cm]{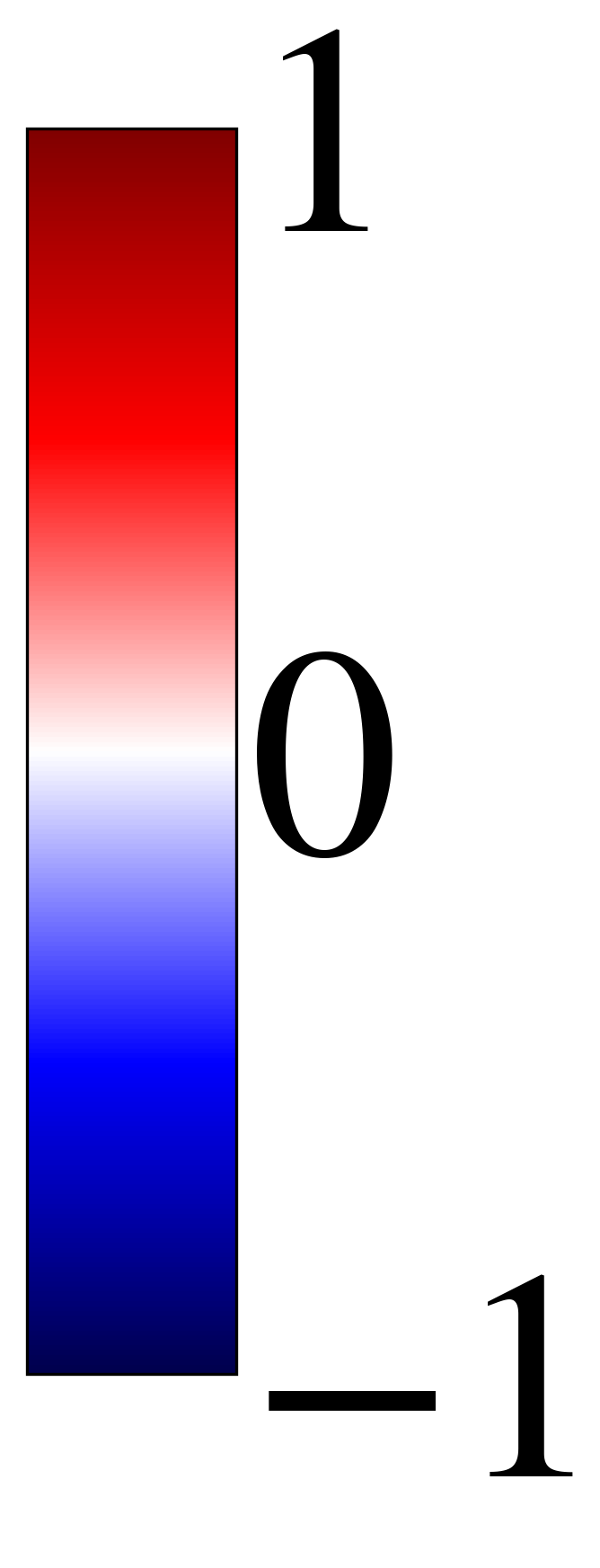}
  \caption*{}
\end{subfigure}%
\end{labeledbox}
\caption{Comparative visualization of the local crystallographic orientation state across a polycrystalline microstructure using Eulerian and neo-Eulerian mappings. The top panel displays the Eulerian representation using the three Bunge-Euler angles $(\phi_1, \Phi, \phi_2)$. The bottom panel illustrates three neo-Eulerian parametrizations: the axis-angle formulation, defined by the rotation-axis components $v_i$ and the rotation angle $\omega$, which reaches a maximum value of approximately 1.096 for cubic symmetry; the Rodrigues vector components $\rho_i$, which scale the axis by $\tan(\omega/2)$; and the unit quaternion components $q_i$, which map the rotation onto the hypersphere.}
\label{fig:comp_ori_repr}
\end{figure}

To illustrate how various orientation parameterizations transform a physical microstructural state into distinct numerical signals, Fig.~\ref{fig:comp_ori_repr} maps a polycrystalline microstructure into each respective coordinate space. The representations show distinct variation in contrast across the domain. While certain orientation formulations clearly differentiate adjacent grains, others appear nearly uniform, which offers no discriminatory signal for those regions. This visualization demonstrates that adjacent regions displayed with differing numerical values may still be physically indistinguishable when crystal lattice symmetry is considered. None of these standard representations alone accounts for the non-uniqueness of orientations, meaning that numerical discontinuities or transitions shown in the maps do not necessarily correlate with true physical misorientations, which highlights the inherent ambiguity in using raw orientation parameters for reconstruction and emphasizes the necessity of a symmetry-invariant framework to accurately characterize the microstructural state.

A fundamental topological property of the quaternion group is its two-to-one homomorphism with the rotation group $SO(3)$, meaning that the antipodal quaternions $\mathbf{q}$ and $-\mathbf{q}$ map to the identical physical orientation. Consequently, any statistical description must be additionally invariant under this sign inversion. HSH provide the natural framework for this, serving as an orthonormal basis for functions defined on $\mathbf{S}^3$ and are analogous to the role of GSH for rotations on $\mathbf{S}^2$, while being convertible to each other \cite{mason_expressing_2009}. While general definitions of HSH are often complex-valued, Mason et al. \cite{mason_hyperspherical_2008} established a real-valued convention utilizing a linear combination of cosine-type $Z^{nc}_{l,m}(\mathbf{q})$ and sine-type $Z^{ns}_{l,m}(\mathbf{q})$ basis functions. Here, the integer indices $n, l, m$ determine the harmonic order and frequency, while the superscripts $c$ and $s$ denote the specific trigonometric contribution. This spectral decomposition enables the Orientation Distribution Function (ODF), $f_\text{ODF}$, to be expressed as a series expansion, where the corresponding coefficients encode the crystallographic texture. The generalized form of this expansion is given by

\begin{equation}
\label{eq:odf}
\begin{aligned}
f_\text{ODF}(\omega, \theta, \phi) = \sum_{n=0}^{\infty} \sum_{l=0}^{n} \left[ f^{nc}_{l0} Z^{nc}_{l0} + \sum_{m=1}^{l} \left( f^{nc}_{lm} Z^{nc}_{lm}+ f^{ns}_{lm} Z^{ns}_{lm} \right) \right].
\end{aligned}
\end{equation}

In this expansion, the scalar weights $f^{nc}_{lm}$ and $f^{ns}_{lm}$ constitute the spectral coefficients of the ODF and can be determined from experimental datasets. In practice, the infinite series is truncated at a finite maximum harmonic degree, $n_{\max}$, that depends on the feature complexity of the ODF and is usually low in strongly textured representations and high when describing weak textures. The explicit forms of the basis functions are defined as follows

\begin{equation}
\label{eq:cosine_HSH}
\begin{aligned}
Z^{nc}_{lm} =
\begin{cases}
Y^{n}_{lm} \cos(m\phi) & \text{if } m > 0, \\
Y^{n}_{lm} & \text{if } m = 0,
\end{cases}
\end{aligned}
\end{equation}
and
\begin{equation}
\label{eq:sine_HSH}
\begin{aligned}
Z^{ns}_{lm} &=
\begin{cases}
Y^{n}_{lm} \sin(m\phi) & \text{if } m > 0, \\
0 & \text{if } m = 0,
\end{cases}
\end{aligned}
\end{equation}

where the function $Y^{n}_{lm}(\omega,\theta)$ takes the form

\begin{equation}
\label{eq:HSH_prefactor}
\begin{aligned}
Y^{n}_{lm}(\omega,\theta) &= (-1)^m \frac{2^l l!}{\pi}
\sqrt{(2l + 1)\frac{(l - m)!}{(l + m)!}
\frac{(n + 1)!(n - l)!}{(n + l + 1)!}
}
\sin\left(\frac{\omega}{2}\right)
C_{n - l}^{l + 1} \left( \cos\left( \frac{\omega}{2} \right) \right)
P_l^m(\cos\theta).
\end{aligned}
\end{equation}

The ultraspherical Gegenbauer polynomials, $C_n^{\nu}(x)$, are orthogonal functions that are defined as

\begin{equation}
\label{eq:Gegenbauer}
\begin{aligned}
C_\kappa^{\nu}(x) &=
\frac{(-2)^n (\nu + \kappa - 1)! (2\nu + \kappa - 1)!}{\kappa!(\nu - 1)! (2\nu + 2\kappa - 1)!}
\left(1 - x^2\right)^{\frac{1}{2} - \nu}
\frac{d^\kappa}{dx^\kappa} \left[ \left(1 - x^2\right)^{\nu + \kappa - \frac{1}{2}} \right],
\end{aligned}
\end{equation}

whereas the associated Legendre polynomials, $ P_\eta^\mu(x) $, are given by the Rodrigues' formula

\begin{equation}
\label{eq:Legendre}
\begin{aligned}
P_\eta^\mu(x) &= \frac{(-1)^\mu}{2^\eta \eta!} (1 - x^2)^{\frac{\mu}{2}}
\frac{\operatorname{d}^{\eta+\mu}}{\operatorname{d}x^{\eta+\mu}} \left[ (x^2 - 1)^\eta \right].
\end{aligned}
\end{equation}

To effectively deploy this description in DMCR, invariance to symmetrically equivalent orientations must be ensured. An advantage of the HSH is its ability to handle these constraints through the construction of SHSH \cite{mason_hyperspherical_2008}. Mathematically, the imposition of symmetry conditions establishes a system of linear dependencies among the primitive coefficients. This allows to project the expansion onto a reduced subspace spanned by a basis set. The general expansion in Eq.~\eqref{eq:odf} is transformed into a compact symmetrized form, where the summation runs only over even harmonic orders $n$ and a reduced index $\lambda$

\begin{equation}
\label{eq:odf_symm}
\begin{aligned}
f_\text{ODF}(\omega, \theta, \phi) &= \sum_{n=0,2\dots}^{\infty} \sum_{\lambda=1}^{\Lambda(n)} f^n_{\lambda} \dot{\ddot{Z}}^n_{\lambda},
\end{aligned}
\end{equation}

where $\Lambda(n)$ denotes the number of linearly independent SHSH. $\dot{\ddot{Z}}^n_{\lambda}$ refers to the SHSH basis functions that results from a linear combination of the previously defined HSH in the following manner

\begin{equation}
\label{eq:SHSH}
\begin{aligned}
\dot{\ddot{Z}}_\lambda^n &= \sum_{l=0}^{n} \left[\dot{\ddot{a}}_{l0\lambda}^n Z_{l0}^{nc} + \sum_{m=1}^{l} \left(\dot{\ddot{a}}_{lm\lambda}^n Z_{lm}^{nc} + \dot{\ddot{b}}_{lm\lambda}^n Z_{lm}^{ns} \right) \right].
\end{aligned}
\end{equation}

\begin{figure}
\centering
    \begin{subfigure}[c]{.2\textwidth}
      \centering
      \includegraphics[width=.8\linewidth]{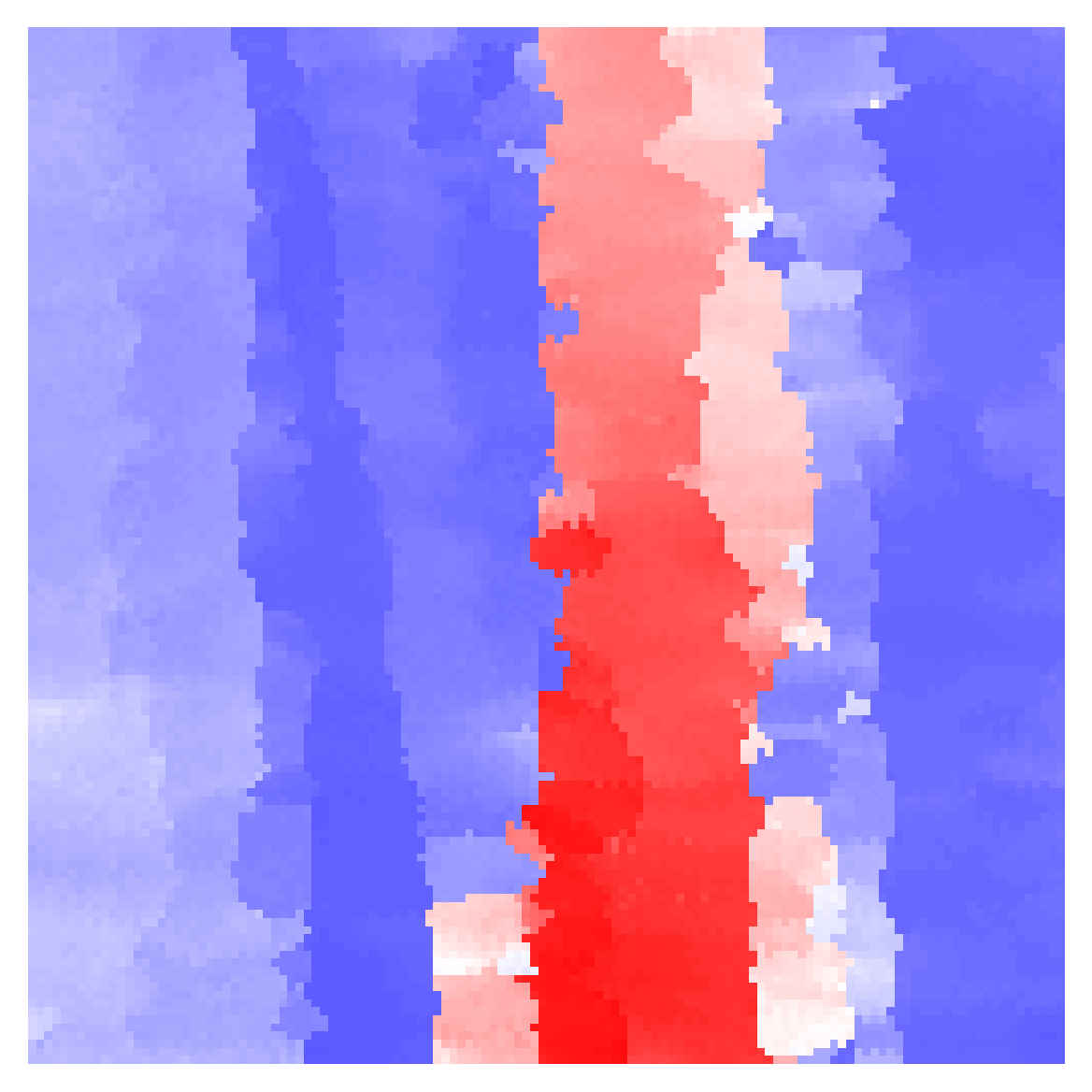}
      \caption*{$\dot{\ddot{Z}}_1^8$}
      \label{fig:Z81}
    \end{subfigure}%
    \begin{subfigure}[c]{.2\textwidth}
      \centering
      \includegraphics[width=.8\linewidth]{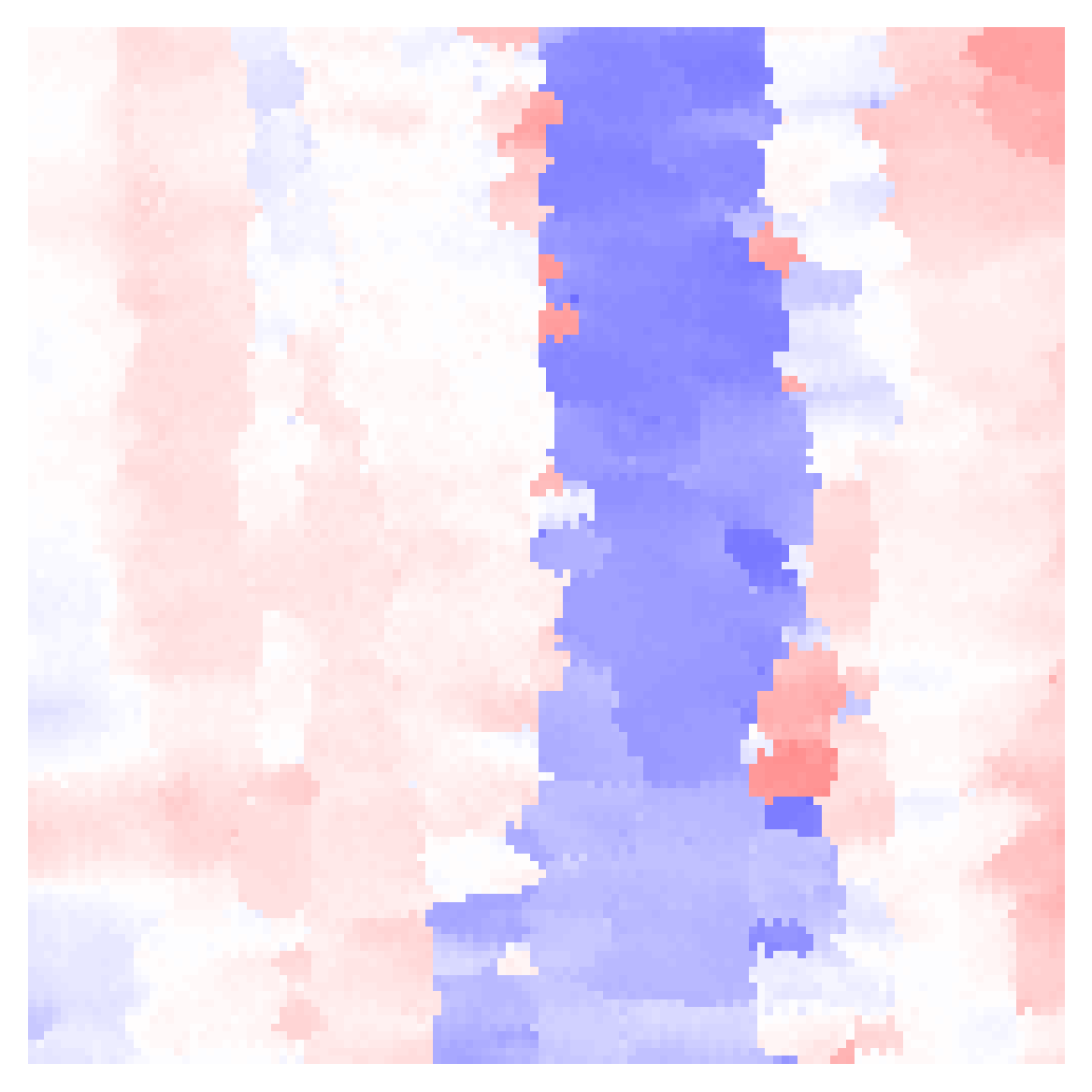}
      \caption*{$\dot{\ddot{Z}}_5^8$}
      \label{fig:Z85}
    \end{subfigure}%
    \begin{subfigure}[c]{.2\textwidth}
      \centering
      \includegraphics[width=.8\linewidth]{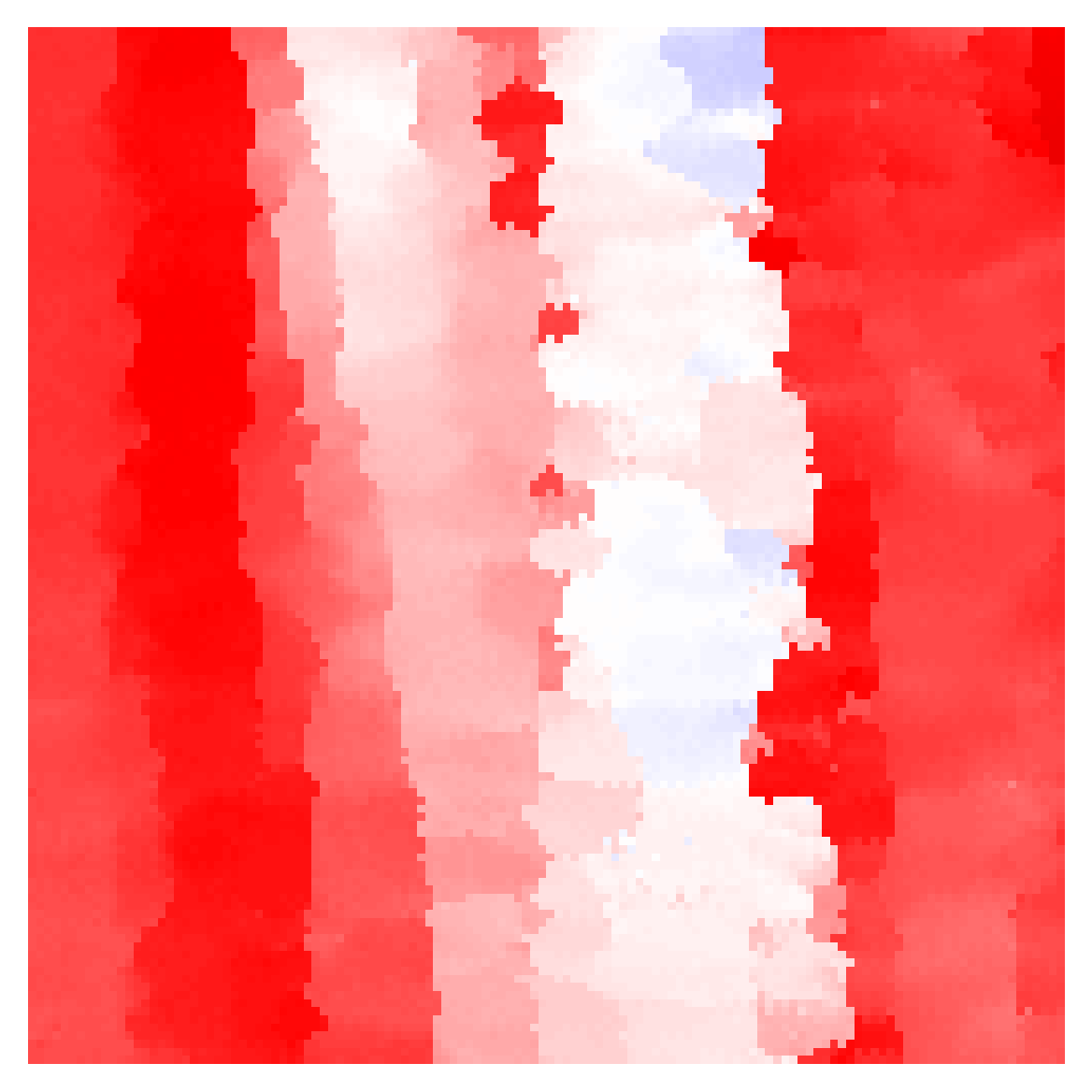}
      \caption*{$\dot{\ddot{Z}}_9^8$}
      \label{fig:Z89}
    \end{subfigure}%
    \begin{subfigure}[c]{.1\textwidth}
      \centering
      \includegraphics[height=2cm]{QuaternionCbar.png}
      \caption*{}
    \end{subfigure}%
    \caption{Spatial distribution of selected SHSH basis function evaluations mapped onto the polycrystalline microstructure. The basis functions $\dot{\ddot{Z}}^n_{\lambda}$ for harmonic degree $n=8$ and distinct symmetry indices $\lambda = \{1, 5, 9\}$ are displayed in a range of $[-1, 1]$.}
    \label{fig:micro_SHSH}
\end{figure}

Here, $\dot{\ddot{a}}_{lm\lambda}^n$ and $\dot{\ddot{b}}_{lm\lambda}^n$ are the coefficients that are fully determined by the symmetry group, known a priori, that define the reduced set of basis functions. The coefficients are computable via the implementation provided by Mason \cite{mason_quaternionic_2012}. Visualized stereographic projections of the first set $\dot{\ddot{Z}}_\lambda^n$ for cubic symmetry is shown in Fig.~\ref{fig:SHSH_basefunction} of Appendix~\ref{app1}. The efficacy of the reduced spectral basis is visually demonstrated in Fig.~\ref{fig:micro_SHSH}, which maps the microstructure presented in Fig.~\ref{fig:comp_ori_repr} to a selection of SHSH basis functions $\dot{\ddot{Z}}^n_{\lambda}$ at harmonic degree $n=8$ for $\lambda=\{1,5,9\}$. A striking benefit of this representation is the elimination of the discontinuities observed in the raw Eulerian and neo-Eulerian spaces shown in Fig.~\ref{fig:comp_ori_repr}. The resulting fields show distinct orientation regions with physically consistent transitions and clearly indicate regions with similar orientations. However, the transition to the SHSH space introduces specific requirements for unique orientation recovery. While algebraic relations suggest that as few as three SHSH components are sufficient to reconstruct a set of hyperspherical angles, trigonometric ambiguities can yield unintended solutions that are not covered by the symmetry considerations. In practice, the SHSH signals for active and passive rotations or orientations belonging to closely related symmetry groups, can be part of solutions if the SHSH subset is too small. Consequently, a minimum set of basis functions is required to ensure that distinct orientations are uniquely distinguishable \cite{seibert_microstructure_2025} and to prevent the optimizer from converging to valid but unintended states. By mapping the large redundant orientation space onto a compact symmetrized subset, the constructed SHSH enable an efficient parameterization of the local microstructural state, which is essential for subsequent optimization tasks.

\subsection{Orientation-based differentiable spatial microstructure descriptors}

\begin{figure}[ht]
\centering
\renewcommand{\arraystretch}{1.5} % Adjust row height
\begin{tabular}{m{0.5cm} m{4cm} m{4cm} m{4cm}}
    % --- Header Row ---
    & \centering $\dot{\ddot{Z}}_1^8$ & \centering $\dot{\ddot{Z}}_5^8$ & \centering $\dot{\ddot{Z}}_9^8$ \tabularnewline
    
    % --- Row 1 ---
    \rotatebox[origin=c]{90}{$\dot{\ddot{Z}}_1^8$} & 
    \begin{subfigure}{4cm} \includegraphics[width=\linewidth]{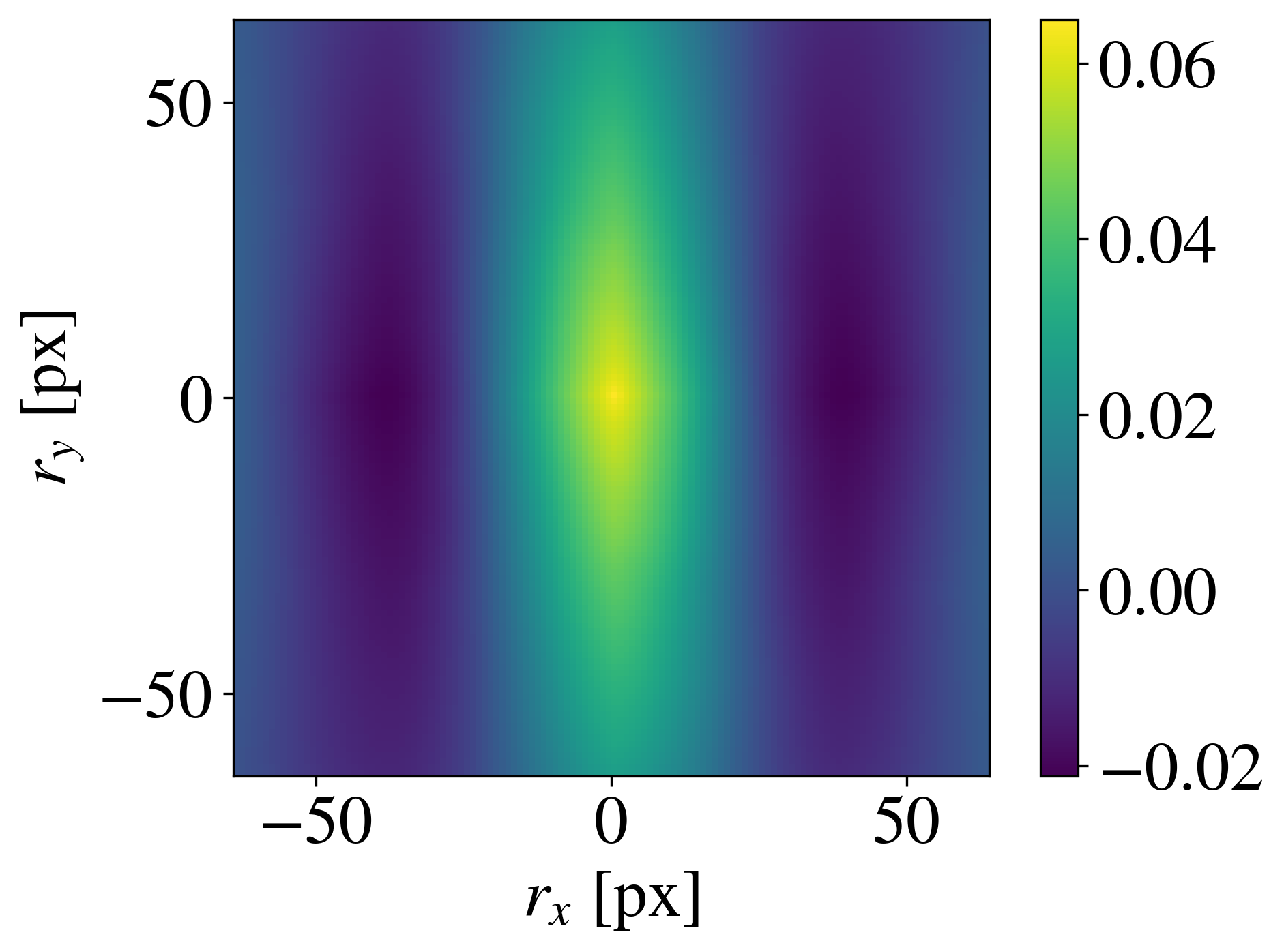} \caption*{$\mathbf{S}\left(\dot{\ddot{Z}}_1^8, \dot{\ddot{Z}}_1^8|\mathbf{r}\right)$} \end{subfigure} & 
    &
    \rotatebox[origin=c]{-45}{\textbf{Symmetric}} \tabularnewline
    
    % --- Row 2 ---
    \rotatebox[origin=c]{90}{$\dot{\ddot{Z}}_5^8$} & 
    \begin{subfigure}{4cm} \includegraphics[width=\linewidth]{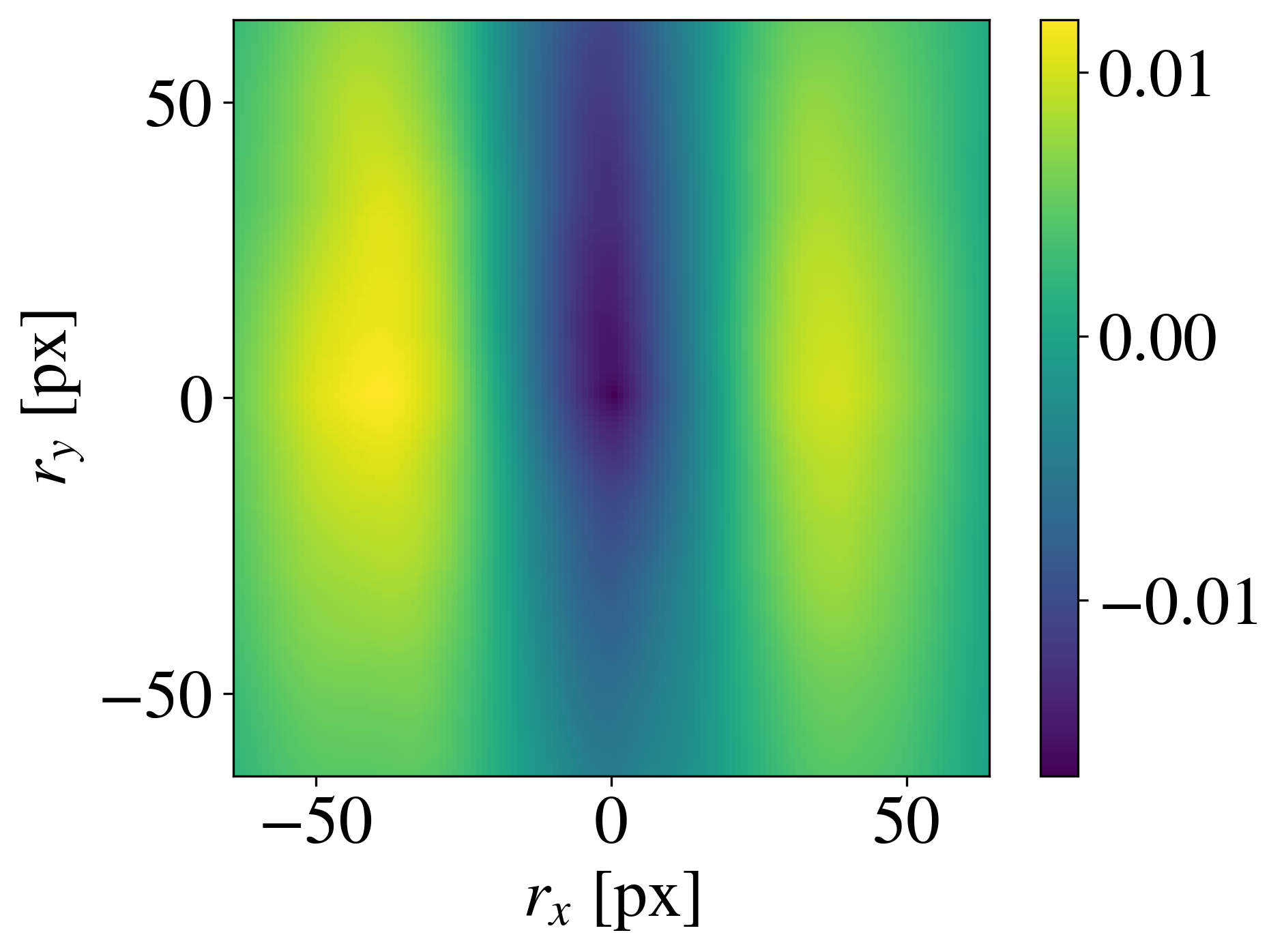} \caption*{$\mathbf{S}\left(\dot{\ddot{Z}}_5^8, \dot{\ddot{Z}}_1^8|\mathbf{r}\right)$} \end{subfigure} & 
    \begin{subfigure}{4cm} \includegraphics[width=\linewidth]{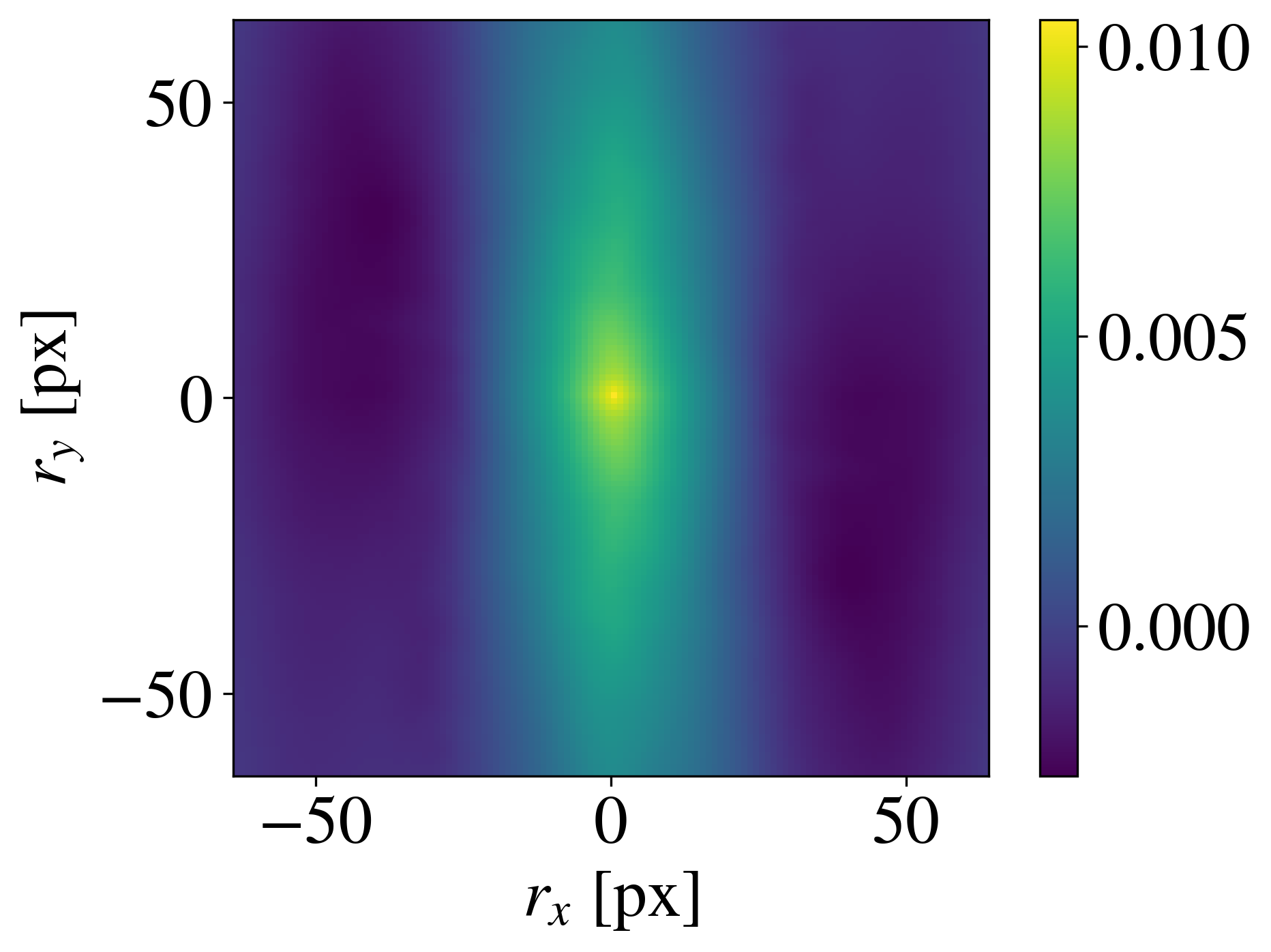} \caption*{$\mathbf{S}\left(\dot{\ddot{Z}}_5^8, \dot{\ddot{Z}}_5^8|\mathbf{r}\right)$} \end{subfigure} & 
    \tabularnewline
    
    % --- Row 3 ---
    \rotatebox[origin=c]{90}{$\dot{\ddot{Z}}_9^8$} & 
    \begin{subfigure}{4cm} \includegraphics[width=\linewidth]{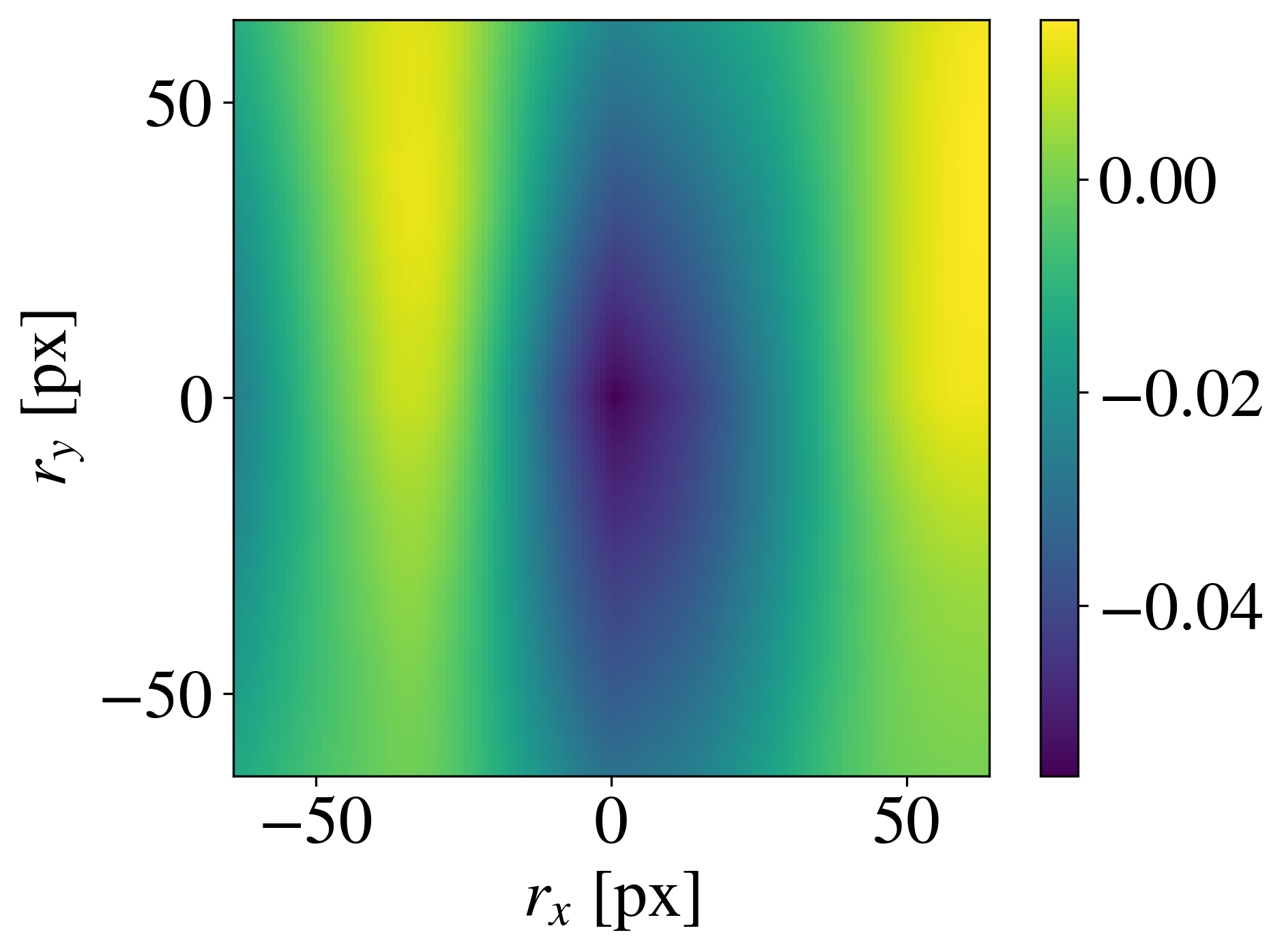} \caption*{$\mathbf{S}\left(\dot{\ddot{Z}}_9^8, \dot{\ddot{Z}}_1^8|\mathbf{r}\right)$} \end{subfigure} & 
    \begin{subfigure}{4cm} \includegraphics[width=\linewidth]{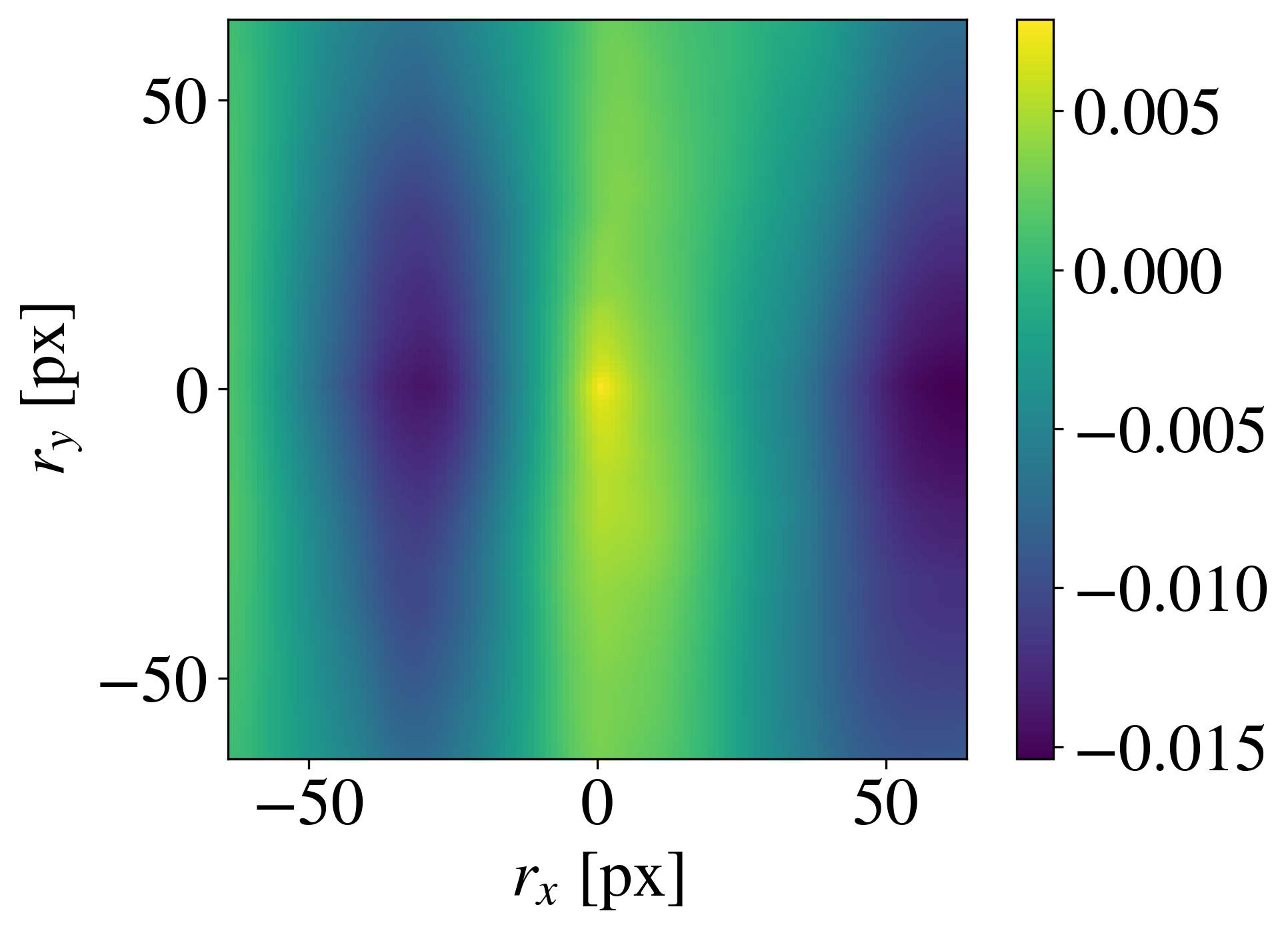} \caption*{$\mathbf{S}\left(\dot{\ddot{Z}}_9^8, \dot{\ddot{Z}}_5^8|\mathbf{r}\right)$} \end{subfigure} & 
    \begin{subfigure}{4cm} \includegraphics[width=\linewidth]{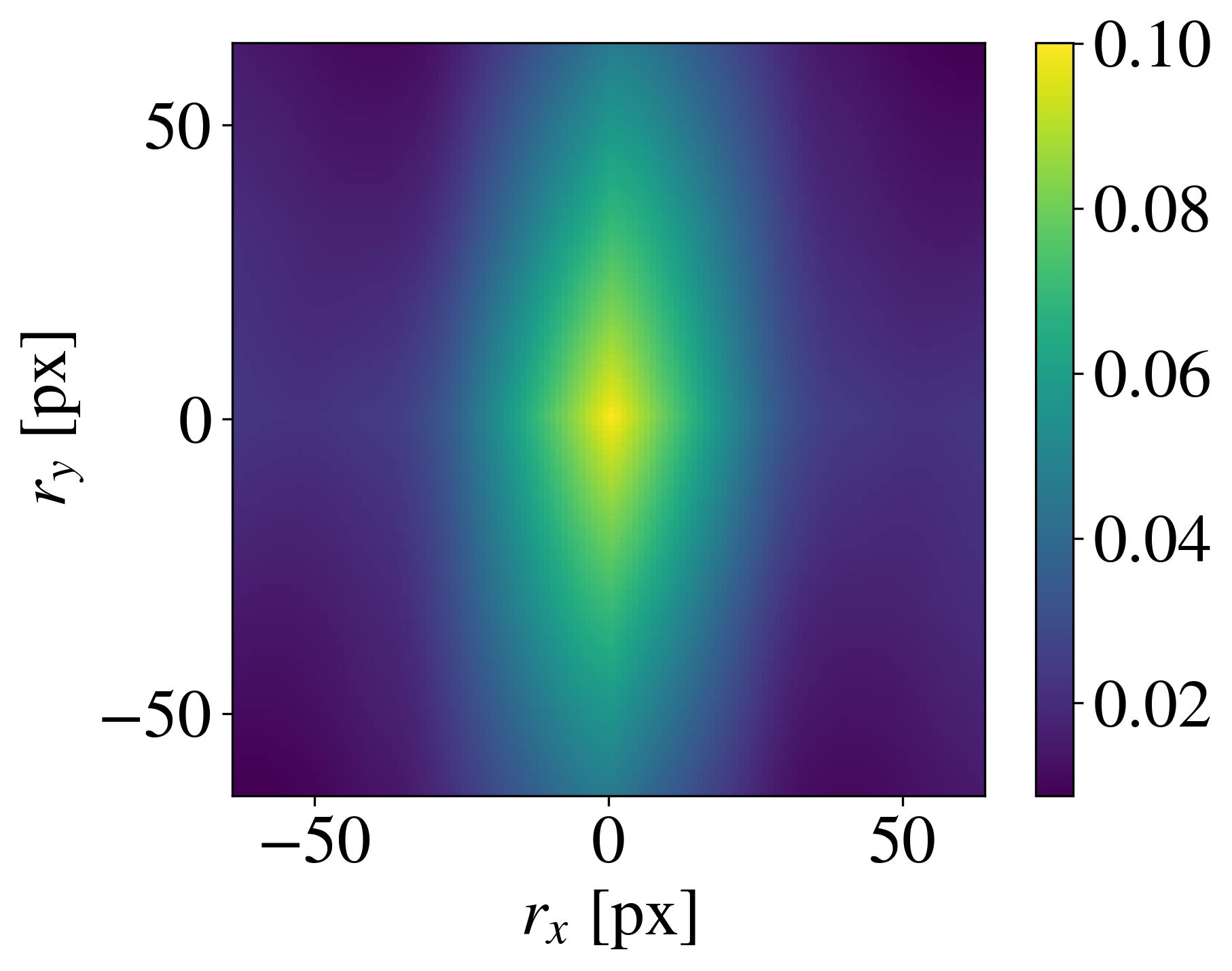} \caption*{$\mathbf{S}\left(\dot{\ddot{Z}}_9^8, \dot{\ddot{Z}}_9^8|\mathbf{r}\right)$} \end{subfigure} \tabularnewline

\end{tabular}
\caption{Grid of auto- and cross-correlation maps for selected SHSH basis functions. The plots visualize the two-point spatial correlation function $\mathbf{S}(\dot{\ddot{Z}}^n_{\lambda}, \dot{\ddot{Z}}^n_{\lambda'} | \mathbf{r})$ for harmonic degree $n=8$ and symmetry indices $\lambda, \lambda' \in \{1, 5, 9\}$. Diagonal elements display the auto-correlations $\mathbf{S}(\dot{\ddot{Z}}^n_{\lambda}, \dot{\ddot{Z}}^n_{\lambda} | \mathbf{r})$, revealing the characteristic morphological anisotropy and correlation length of each individual mode. Off-diagonal elements display the cross-correlations $\mathbf{S}(\dot{\ddot{Z}}^n_{\lambda}, \dot{\ddot{Z}}^n_{\lambda'} | \mathbf{r})$ that quantify the spatial coupling between SHSH components.}
\label{fig:shsh_corr_grid}
\end{figure}

While the ODF, as introduced in Eq.~\eqref{eq:odf_symm}, serves as a reliable statistical descriptor for the global crystallographic texture, it does not account for spatially heterogeneity and fails to capture the morphological grain properties and neighborhood relationships of the polycrystals. To resolve this limitation and quantify the spatial heterogeneity essential for structure-property modeling, higher-order statistical descriptors are employed that are capable of encoding spatial correlations. Specifically, the two-point spatial correlation function \cite{jiao_modeling_2007,jiao_modeling_2008}, $\mathbf{S}(\mathbf{r})$, is used which acts as a metric for morphological characterization. In the context of the derived SHSH framework, this function is defined to correlate a specific pair of SHSH modes $(n, \lambda)$ and $(n', \lambda')$ as the ensemble average of the product of their respective spatial coefficients separated by a lag vector $\mathbf{r}$

\begin{equation}
\label{eq:two-point_SHSH}
\begin{aligned}
\mathbf{S}(\dot{\ddot{Z}}^n_{\lambda},\dot{\ddot{Z}}^{n'}_{\lambda'}|\mathbf{r}) = \frac{1}{\Omega} \int_\Omega \dot{\ddot{Z}}^n_{\lambda}(\mathbf{x}) \dot{\ddot{Z}}^{n'}_{\lambda'}(\mathbf{x} + \mathbf{r}) \, d\Omega,
\end{aligned}
\end{equation}

where $\mathbf{x}$ is the position vector. It is pointed out that, for a statistically homogeneous material, the $\mathbf{S}(\mathbf{r})$ (and any of the higher-order n-point correlation functions) is independent of the absolute position and characterizes only relationships relative to $\mathbf{r}$. Eq.~\eqref{eq:two-point_SHSH} takes into account the auto-correlation function, for $n = n'$ and $\lambda=\lambda'$, and assembles the cross-correlations for every other case. The spatial correlation grid presented in Fig.~\ref{fig:shsh_corr_grid} illustrates how these descriptors decompose the microstructural topology. The auto-correlations shown along the diagonal capture the characteristic morphology associated with each SHSH mode. The elongated central features in these maps directly reflect the directional anisotropy and mean grain dimensions prevalent in the original microstructure shown in Figs.~\ref{fig:comp_ori_repr}~and~\ref{fig:micro_SHSH}. Meanwhile, the cross-correlations, shown in the off-diagonal entries, quantify the statistical coupling between different SHSH basis functions across the spatial domain $\mathbf{r}$.

Alternatively to $\mathbf{S}(\mathbf{r})$, the variogram $\boldsymbol{\gamma}(\mathbf{r})$ of order $\alpha$ is a complementary metric of spatial correlations. It measures the absolute $\alpha$-powered difference of the increments between coefficient values at two locations separated by $\mathbf{r}$ and is formally defined as

\begin{equation}
\label{eq:variogram_standard}
\begin{aligned}
\boldsymbol{\gamma}(\dot{\ddot{Z}}^n_{\lambda}|\mathbf{r}) = \frac{1}{2\Omega} \int_\Omega \bigg| \dot{\ddot{Z}}^n_{\lambda}(\mathbf{x})) - \dot{\ddot{Z}}^{n}_{\lambda}(\mathbf{x} + \mathbf{r}) \bigg|^\alpha \operatorname{d}\Omega.
\end{aligned}
\end{equation}

\begin{figure}[ht]
\centering
\renewcommand{\arraystretch}{1.5} % Adjust row height
\begin{tabular}{m{4cm} m{4cm} m{4cm}}
    % --- Header Row ---
    \begin{subfigure}{4cm} \caption*{$\boldsymbol{\gamma}\left(\dot{\ddot{Z}}_1^8|\mathbf{r}\right)$} \includegraphics[width=\linewidth]{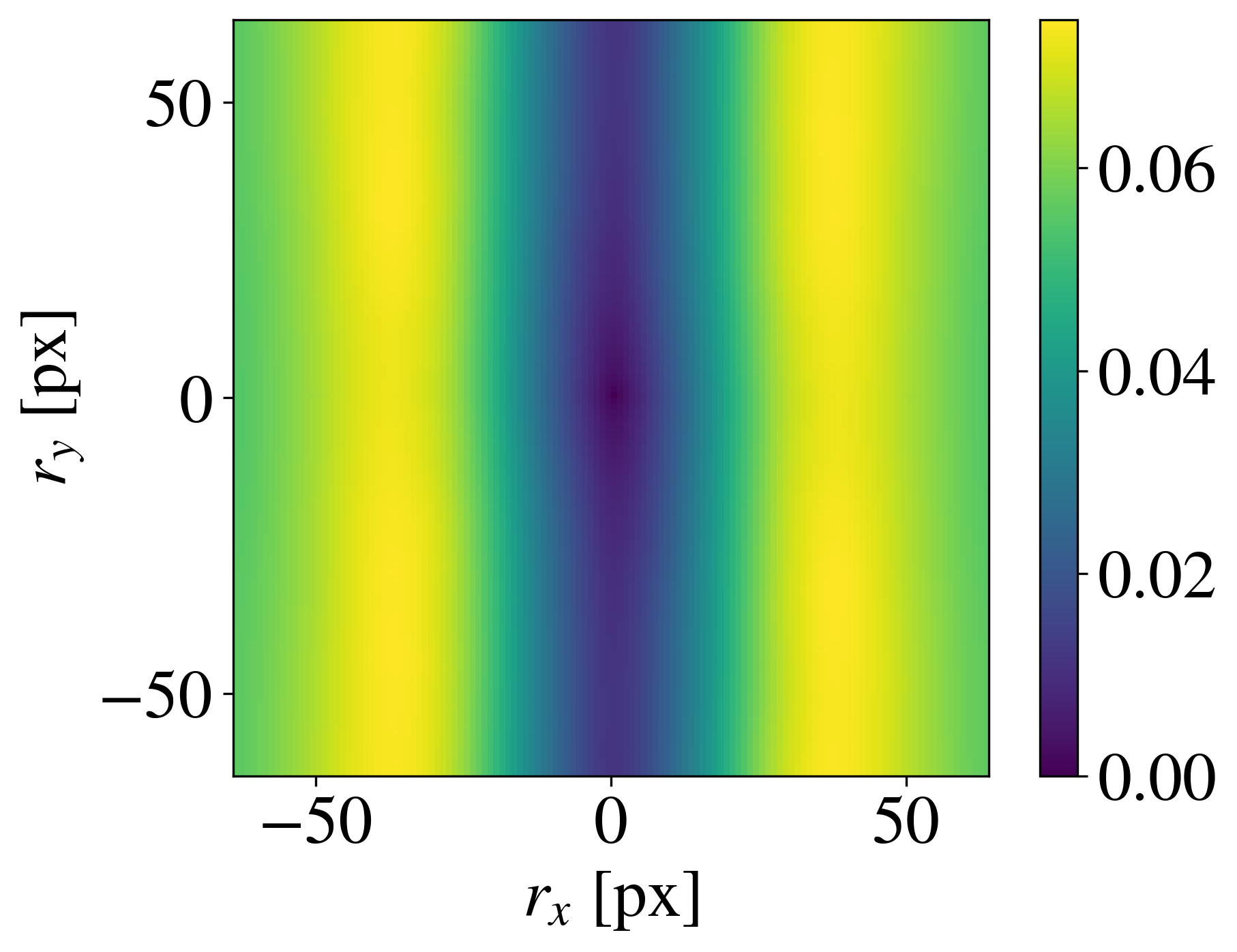} \end{subfigure} & 
    \begin{subfigure}{4cm} \caption*{$\boldsymbol{\gamma}\left(\dot{\ddot{Z}}_5^8|\mathbf{r}\right)$} \includegraphics[width=\linewidth]{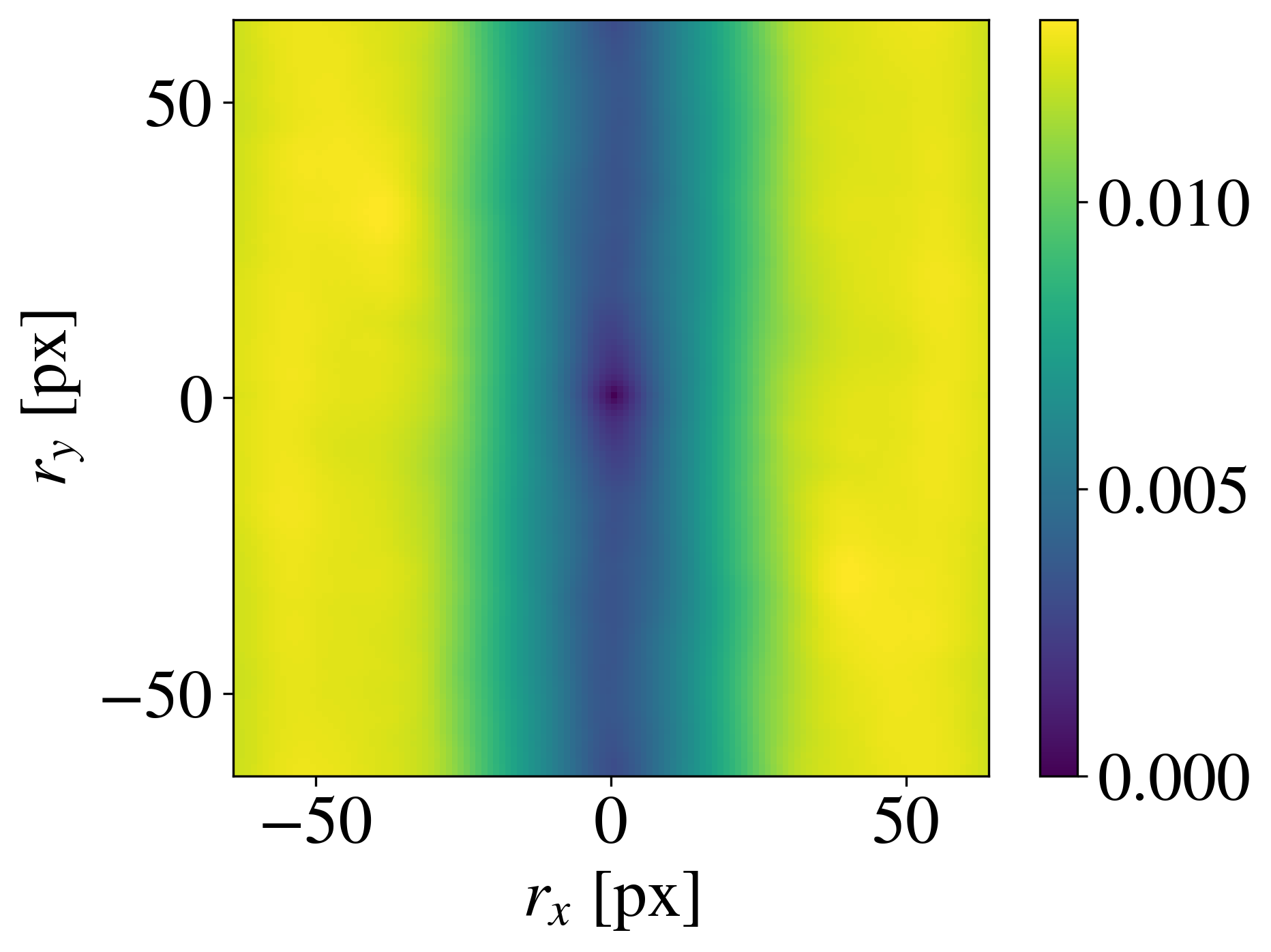} \end{subfigure} & 
    \begin{subfigure}{4cm} \caption*{$\boldsymbol{\gamma}\left(\dot{\ddot{Z}}_9^8|\mathbf{r}\right)$} \includegraphics[width=\linewidth]{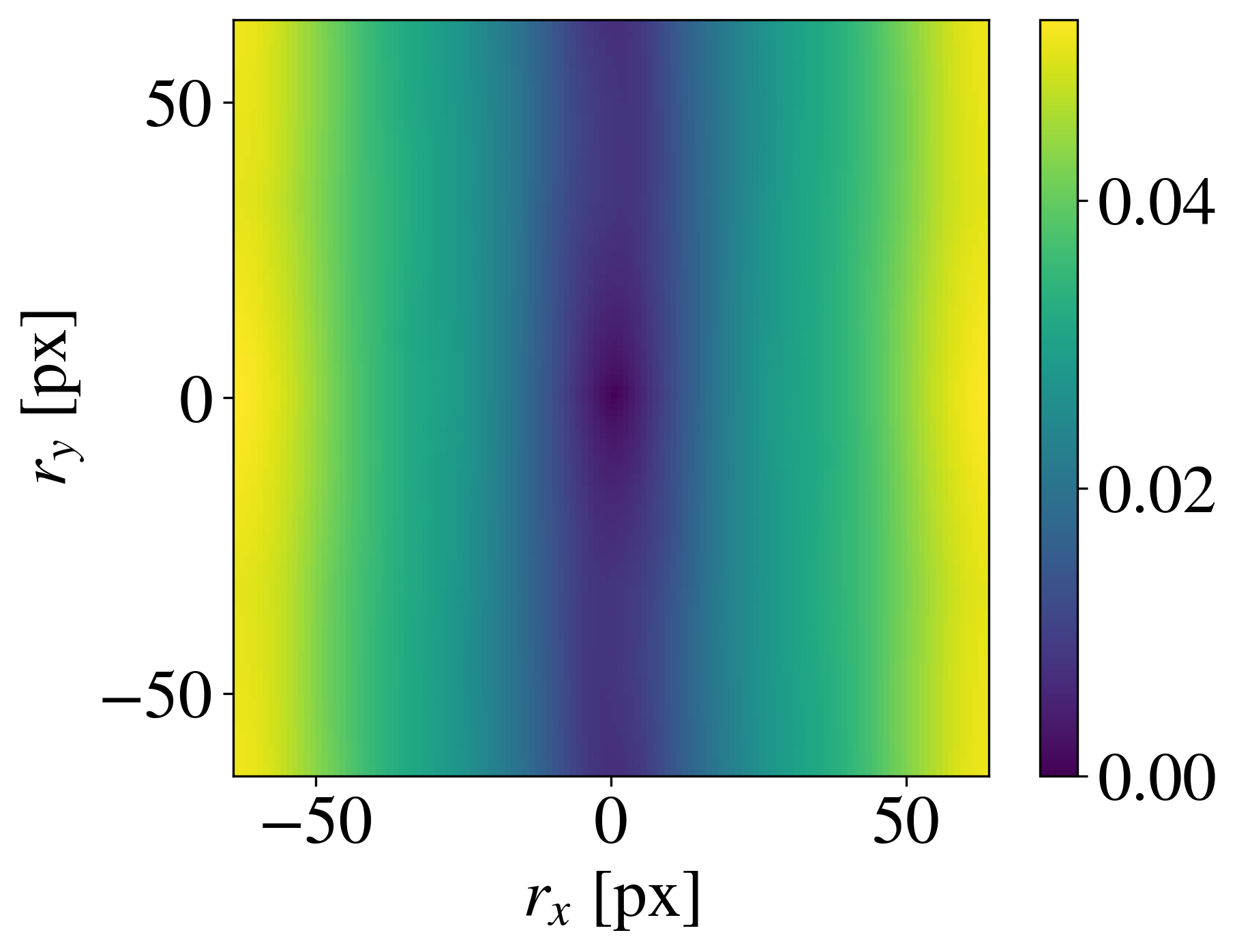} \end{subfigure}  \tabularnewline
    \end{tabular}
    \caption{Variogram maps for selected SHSH basis modes. The plots display the first-order variogram $\boldsymbol{\gamma}(\dot{\ddot{Z}}^n_{\lambda} | \mathbf{r})$ for harmonic degree $n=8$ and symmetry indices $\lambda \in \{1, 5, 9\}$. Each map quantifies the expected absolute difference in the respective SHSH coefficient field for a given lag vector $\mathbf{r}$. The pronounced anisotropy, characterized by lower values (darker regions) extending vertically along the $r_y$ axis, signifies that the microstructural features are more continuous and exhibit longer correlation lengths in the vertical direction compared to the horizontal direction.}
\label{fig:shsh_var}
\end{figure}

In the context of this work, a first-order variogram is assumed by setting $\alpha=1$, a form widely referred to in geostatistics as the madogram \cite{chiles_geostatistics_1999}. The resulting variogram maps for the SHSH components, presented in Fig.~\ref{fig:shsh_var}, provide a direct measure of spatial dissimilarity across $\mathbf{r}$. While these maps share morphological ties with $\mathbf{S}(\mathbf{r})$ functions, they diverge fundamentally in their statistical sensitivity.By relying on first-order variograms related to the absolute differences rather than (co)-variance, this approach eliminates the strong weighing of large differences characteristic of the squared terms in $\mathbf{S}(\mathbf{r})$. This, combined with a mean-free formulation, allows the new metric to robustly describe orientation fields even in the presence of non-stationarity.

To capture higher-order spatial dependencies, a hybrid three-point variogram is introduced, denoted as $\boldsymbol{\gamma}_3$. Unlike the standard variogram, which evaluates the magnitude of increments for a single lag vector, this generalized descriptor quantifies the correlation between two distinct spatial increments separated by different lag vectors $\mathbf{r}_1$ and $\mathbf{r}_2$. The formulation is given by

\begin{equation}
\label{eq:variogram_hybrid}
\begin{aligned}
\boldsymbol{\gamma}_3(\dot{\ddot{Z}}^n_{\lambda}|\mathbf{r}_1,\mathbf{r}_2) = \frac{1}{2\Omega} \int_\Omega \left|\dot{\ddot{Z}}^n_{\lambda}(\mathbf{x}) - \dot{\ddot{Z}}^{n}_{\lambda}(\mathbf{x} + \mathbf{r}_1) \right| \cdot \left|\dot{\ddot{Z}}^n_{\lambda}(\mathbf{x}) - \dot{\ddot{Z}}^{n}_{\lambda}(\mathbf{x} + \mathbf{r}_2) \right| \operatorname{d}\Omega.
\end{aligned}
\end{equation}

When $\mathbf{r}_1 = \mathbf{r}_2$, this definition recovers the semi-variogram. For differing lag vectors it generalizes to a three-point measure that probes how orientation increments co-vary along different directions. In this sense, $\boldsymbol{\gamma}_3$ complements $\mathbf{S}(\mathbf{r})$ by characterizing curvature information and directional coupling in the orientation gradient fields.

To govern the local smoothness and characteristic interface density of the reconstructed fields, the mean variation $\mathcal{V}$ is incorporated as an explicit statistical descriptor. While total variation $\mathcal{V^\text{tot}}$ is conventionally employed in image processing \cite{chambolle_image_1997} as a regularization term to suppress noise, it may also eliminate local features that are physically relevant. Alternatively, this work utilizes it as a target statistic to be reproduced as detailed by Seibert et al. \cite{seibert_descriptor-based_2022}. The mean variation for the microstructure field is defined as the total variation normalized by the domain volume ensuring resolution invariance

\begin{equation}
\label{eq:mean_variation}
\begin{aligned}
\mathcal{V}(\dot{\ddot{Z}}^n_{\lambda}) = \frac{1}{\Omega} \int_\Omega  \left| \nabla \dot{\ddot{Z}}^n_{\lambda}(\mathbf{x}) \right| \operatorname{d}\Omega.
\end{aligned}
\end{equation}

By treating $\mathcal{V}(\dot{\ddot{Z}}^n_{\lambda})$ as a descriptor, the framework allows to preserve essential features, such as intragranular substructures, which would otherwise be smoothed out by standard total variation denoising schemes. The term $\nabla \dot{\ddot{Z}}^n_{\lambda}(\mathbf{x})$ in Eq.~\eqref{eq:mean_variation} is evaluated via first-order numerical differentiation. To account for the presented descriptors in a unified framework, the individual contributions are weighted with prefactors, $w_\mathbf{S}, w_{\boldsymbol{\gamma}_3}$ and $w_\mathcal{V}$. The total descriptor set can now be assembled as
 
\begin{equation}
\label{eq:final_loss}
\begin{aligned}
\mathbf{D} = \{w_\mathbf{S}\mathbf{S}, w_{\boldsymbol{\gamma}_3} \boldsymbol{\gamma}_3, w_\mathcal{V}\mathcal{V} \}.
\end{aligned}
\end{equation}

\begin{figure}
  \centering
  \includegraphics[width=1\textwidth]{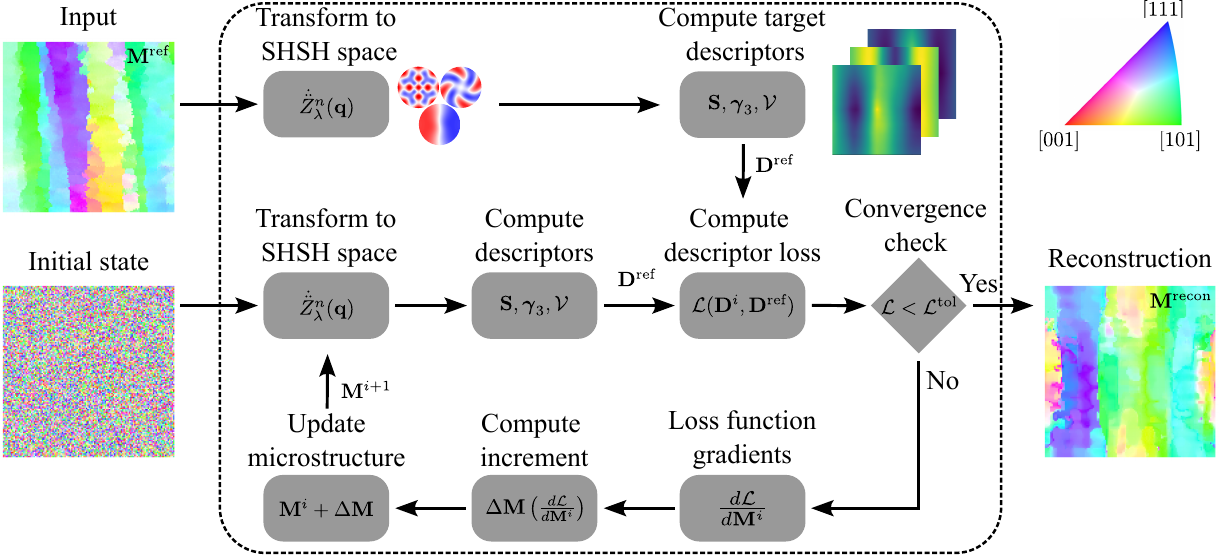}
  \caption{Schematic workflow of the DMCR framework expanded to orientation information. The pipeline begins with the extraction of a target descriptor set $\mathbf{D}^{\text{ref}}$ from a reference EBSD map $\mathbf{M}^{\text{ref}}$ by utilizing the SHSH transform as microstructural local state. The reconstruction is initialized from a random noise field $\mathbf{M}^{0}$ and proceeds through an iterative optimization loop. In each iteration $i$, the current microstructure $\mathbf{M}^i$ is transformed into SHSH space to compute its descriptors. The discrepancy $\mathcal{L}$ between the current and target descriptors is evaluated, and its gradients $\nabla_{\mathbf{M}} \mathcal{L}$ are backpropagated to compute the update increment $\Delta \mathbf{M}$. The process repeats until the loss falls below a specified tolerance $\mathcal{L}^{\text{tol}}$ and resulting in the final reconstructed microstructure $\mathbf{M}^{\text{recon}}$.}
  \label{fig:DMCR_algo}
\end{figure}

The complete reconstruction algorithm is visually summarized in Fig.~\ref{fig:DMCR_algo}. The process commences with the preprocessing of the experimental input, where the reference EBSD map $\mathbf{M}^{\text{ref}}$ is transformed into the continuous SHSH spectral space $\dot{\ddot{Z}}^n_{\lambda}(\mathbf{q})$. From this spectral representation, the target statistical signature $\mathbf{D}^{\text{ref}}$ is compiled, comprising the two-point spatial correlations $\mathbf{S}$, the hybrid three-point variogram $\boldsymbol{\gamma}_3$, and the mean variation $\mathcal{V}$.

The reconstruction is initialized as a spatially uncorrelated randomized field $\mathbf{M}^0$. During each iteration, the current microstructural state $\mathbf{M}^i$ is first mapped into the SHSH domain to compute its descriptor set $\mathbf{D}^i$ via the differentiable operations defined in Eqs.~(\ref{eq:two-point_SHSH}) - (\ref{eq:mean_variation}). A scalar objective $\mathcal{L}$ is then calculated as the weighted sum of squared residuals between $\mathbf{D}^i$ and $\mathbf{D}^{\text{ref}}$ to aggregate the statistical discrepancies across all selected SHSH functions and spatial lags. Benefiting from the underlying computational graph of DMCR, the gradient of this loss with respect to each voxel-wise orientation component, $\nabla_{\mathbf{M}} \mathcal{L}$, is computed via automatic differentiation in TensorFlow. Finally, a gradient-based optimizer utilizes these values to determine the update step $\Delta \mathbf{M}$ to iteratively adjust the noise field into a reconstructed microstructure that minimizes the statistical error. This cycle repeats until the loss converges below a predefined tolerance $\mathcal{L}^{\text{tol}}$ and yielding a final synthesized microstructure $\mathbf{M}^{\text{recon}}$ that is statistically similar to the prescribed descriptors obtained from the experimental reference.

\section{Results}\label{3}
This section presents the numerical results starting from controlled descriptor studies to practical reconstructions of experimentally acquired microstructures. First, the effect of the individual loss terms are isolated to identify the information captured by each descriptor, then examine their combined performance under varying weight ratios and regularization strength. Building on these benchmark studies, statistically homogeneous and inhomogeneous experimental orientation maps are reconstructed in 2D and 3D. The section concludes by discussing the current limitations of the framework, including descriptor realizability, optimization cost, and sensitivity to initialization.

\subsection{Numerical studies}\label{3.1}
\begin{figure*}
  \centering
  \begin{subfigure}{0.25\textwidth}
    \centering
    \includegraphics[width=0.76\textwidth]{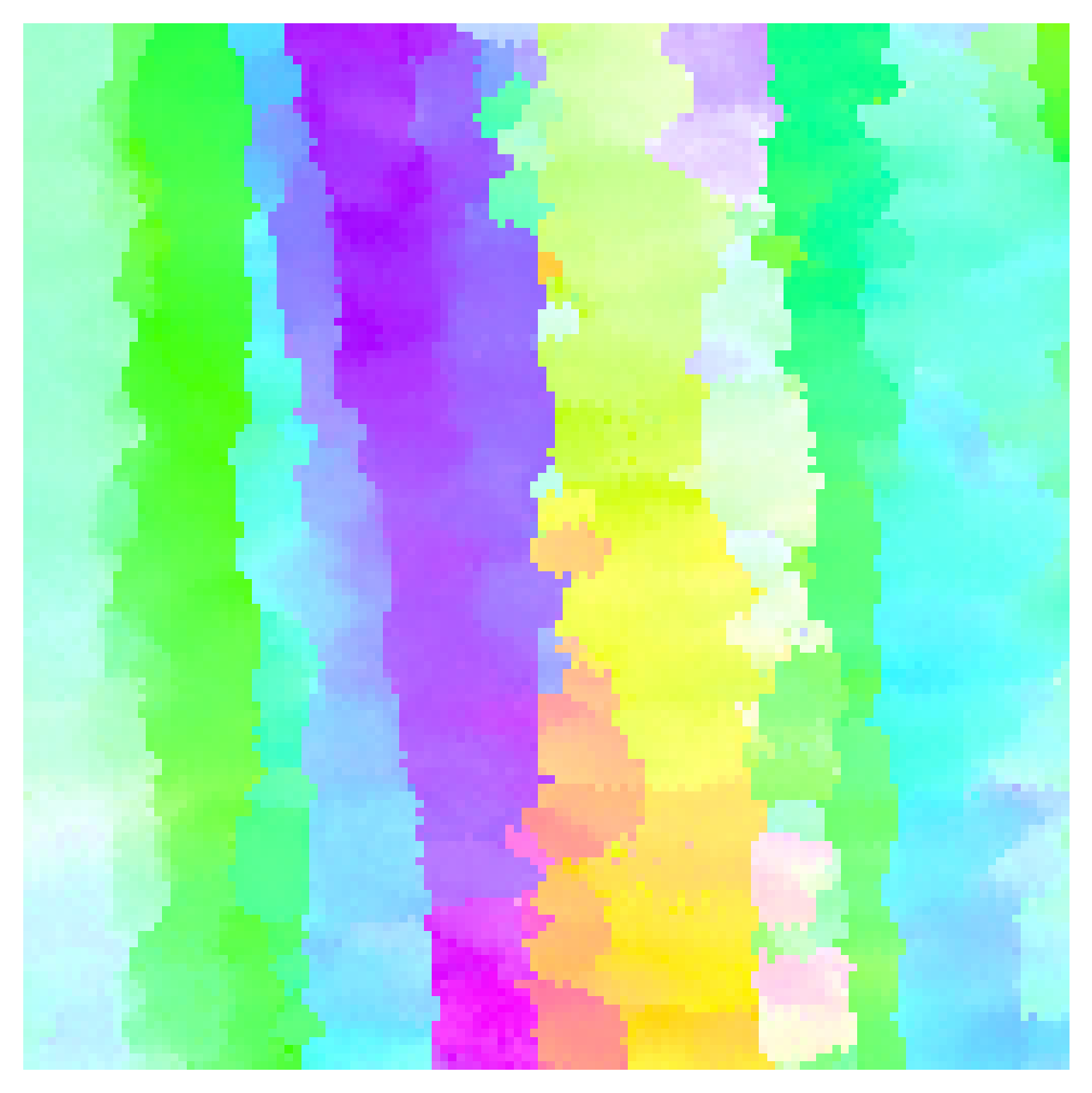}
    \caption*{$\mathbf{M}^{\text{ref}}$}
  \end{subfigure}%
  \begin{subfigure}{0.25\textwidth}
    \centering
    \includegraphics[width=0.76\textwidth]{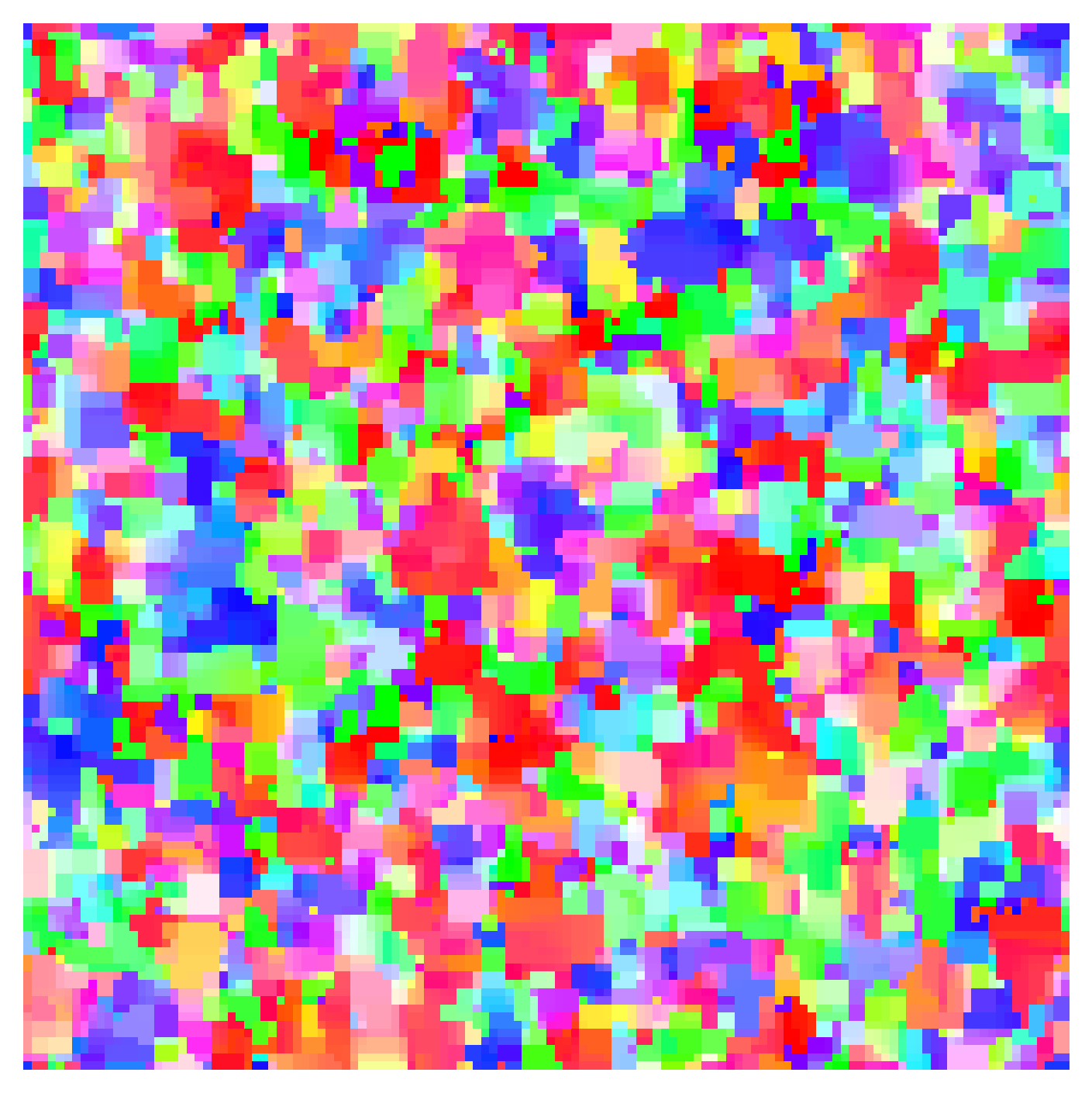}
    \caption*{$\mathbf{M}^{\text{recon}} = \arg \min \mathcal{L}_\mathcal{V}$}
  \end{subfigure}%
  \begin{subfigure}{0.25\textwidth}
    \centering
    \includegraphics[width=0.76\textwidth]{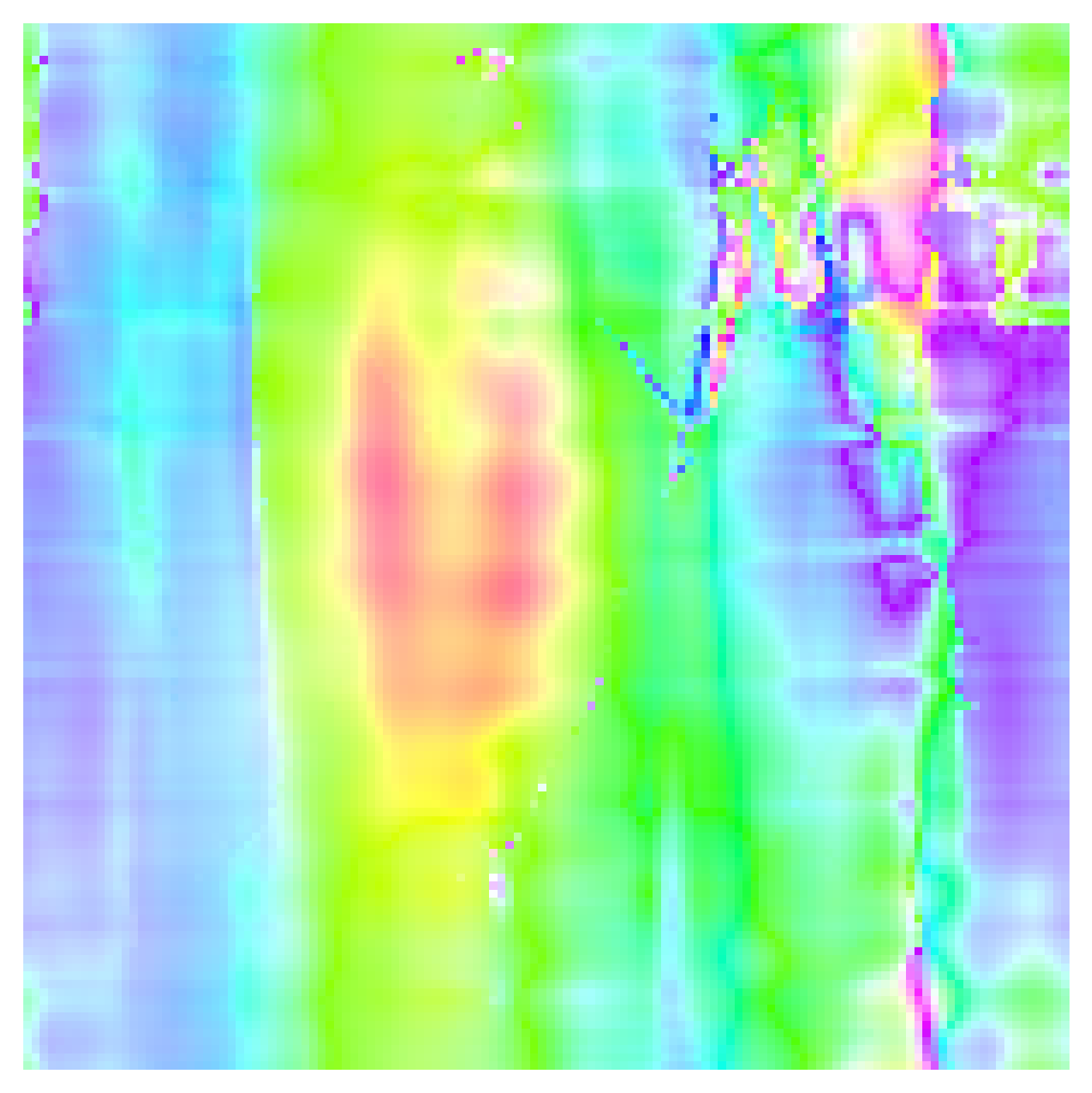}
    \caption*{$\mathbf{M}^{\text{recon}} = \arg \min \mathcal{L}_\mathbf{S}$}
  \end{subfigure}%
  \begin{subfigure}{0.25\textwidth}
    \centering
    \includegraphics[width=0.76\textwidth]{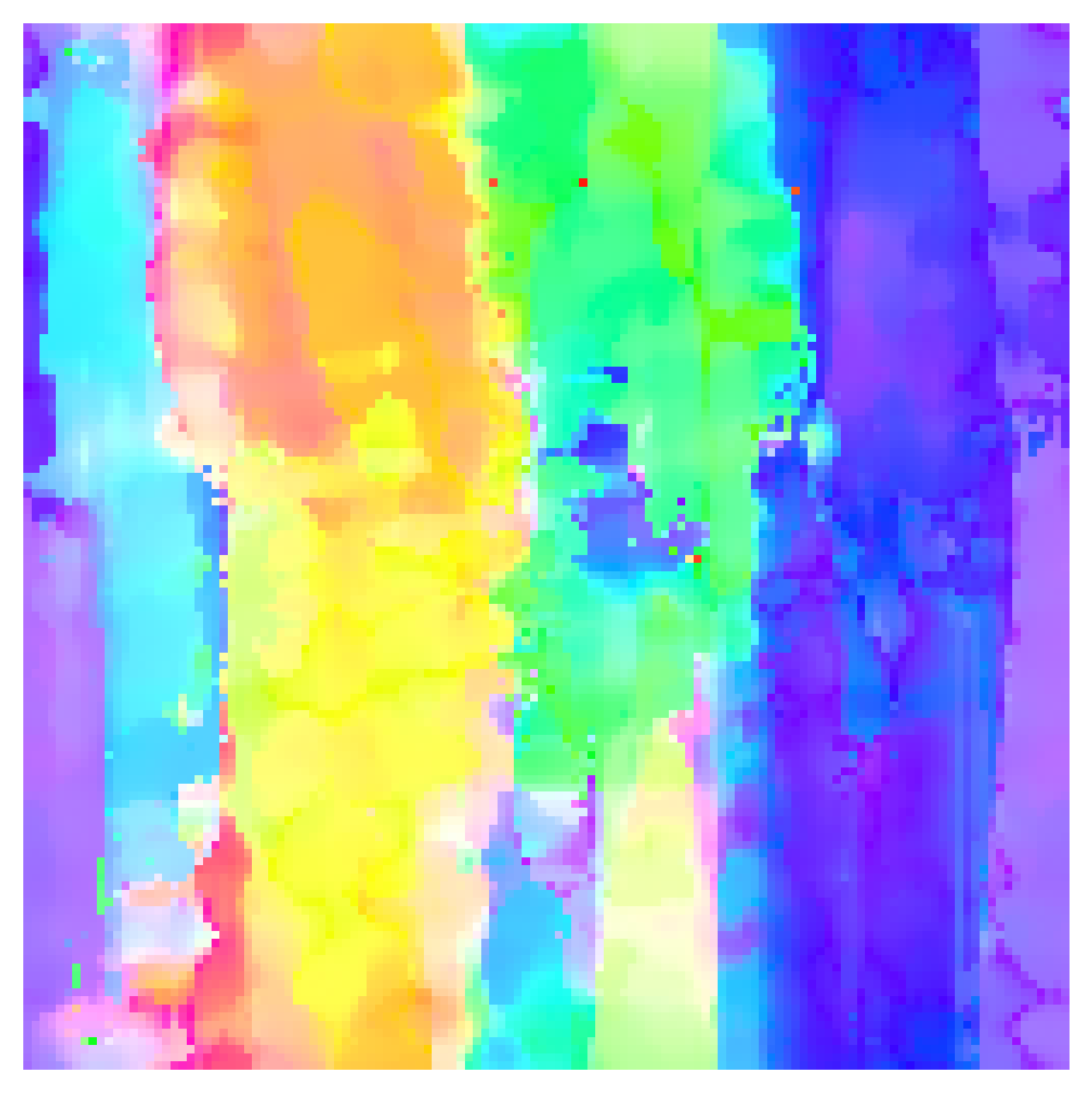}
    \caption*{$\mathbf{M}^{\text{recon}} = \arg \min \mathcal{L}_{\boldsymbol{\gamma}_3}$}
  \end{subfigure}%
  \caption{Comparison of reconstructed microstructure $\mathbf{M}^{\text{recon}}$ that optimize with respect to a single descriptor by minimizing $\mathcal{L}_{\mathcal{V}}$, $\mathcal{L}_{\mathbf{S}}$ and $\mathcal{L}_{\boldsymbol{\gamma}_3}$, respectively.}
  \label{fig:Comparison_Descriptors}
\end{figure*}

A central question in the DMCR framework concerns the unique informational contribution of each descriptor and the specific microstructural features that it can, or cannot, independently recover. To isolate these biases, a study is performed where reconstructions are driven solely by minimizing the individual loss terms for mean variation ($\mathcal{L}_\mathcal{V}$), two-point correlation ($\mathcal{L}_\mathbf{S}$), and the hybrid three-point variogram ($\mathcal{L}_{\boldsymbol{\gamma}_3}$), utilizing the bounded limited-memory Broyden–Fletcher–Goldfarb–Shanno (L-BFGS-B) optimizer \cite{fletcher_practical_2000}. The first nine SHSH components for cubic crystal symmetry are selected to compute the descriptors and spatial descriptors are cropped to optimize with respect to a $17\times17\,\text{px}$ window. The resulting microstructures are compared against the reference microstructure shown in Fig.~\ref{fig:Comparison_Descriptors}.

When the optimization is governed exclusively by $\mathcal{L}_\mathcal{V}$, the algorithm effectively penalizes the average magnitude of orientation gradients and denoises $\mathbf{M}^0$ to achieve local smoothness. While this reconstruction succeeds in suppressing high-frequency noise, it fails to reproduce any meaningful grain morphology as $\mathcal{V}$ relies on first-order pixel neighborhoods and, therefore, restricts the descriptor to a purely local scope. Conversely, minimizing $\mathcal{L}_{\mathbf{S}}$ successfully captures long-range spatial dependencies and generates vertically elongated structures that recover the reference anisotropy. However, the reconstruction suffers from overly diffuse interfaces and a lack of sharp grain boundaries, which indicates that $\mathbf{S}$ prioritizes large-scale features over local interfacial details. In contrast, the reconstruction driven by $\mathcal{L}_{\boldsymbol{\gamma}_3}$ yields more distinct interfaces that aligns grain widths in a plausible manner, although artificial residual noise remains to a certain extent. Although the local information captured by $\mathcal{V}$ is implicitly contained within $\boldsymbol{\gamma}_3$, including it as a separate term allows for the explicit and controlled adjustment of interfacial and intragranular smoothness. Additionally, the choice of $\boldsymbol{\gamma}_3$ instead of a standard three-point correlation function $\mathbf{S}_3$ is motivated by its numerically advantageous behavior due to the mean-free formulation and reliance on absolute differences that relaxes the constraints when reconstructing orientations. To complement this, $\mathbf{S}$ is retained to preserve mean-sensitive information and to account for the cross-correlations between SHSH signals. The results shown in Fig.~\ref{fig:Comparison_Descriptors} demonstrate that while $\mathcal{L}_{\mathcal{V}}$ regulates gradient density, $\mathcal{L}_{\mathbf{S}}$ enforces crystallographic constraints through cross-correlation of SHSH signals, and $\mathcal{L}_{\boldsymbol{\gamma}_3}$ improves sharper local features, no single descriptor can independently recover the complexity of the polycrystalline state. This motivates the necessity of the composing a a combined set of descriptor that simultaneously drive the loss function.

\begin{figure*}
  \centering
  \begin{tikzpicture}
    % Place all subfigures in a single node
    \node (img) [inner sep=0] {
      \begin{subfigure}{0.19\textwidth}
        \includegraphics[width=\textwidth]{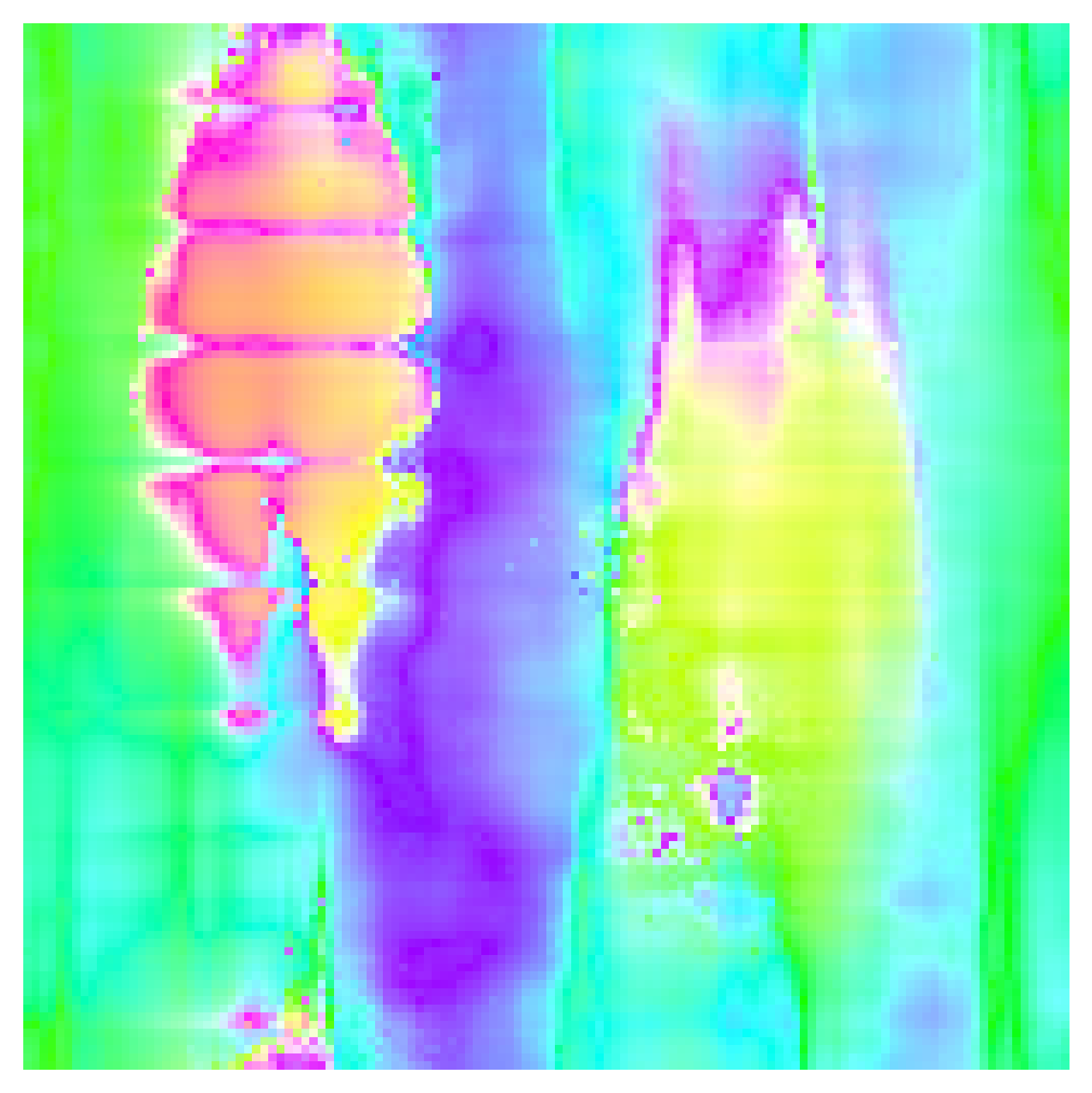}
        \caption*{$w_{\boldsymbol{\gamma}_3}/w_{\mathbf{S}} = 10^{-2}$}
      \end{subfigure}%
      \begin{subfigure}{0.19\textwidth}
        \includegraphics[width=\textwidth]{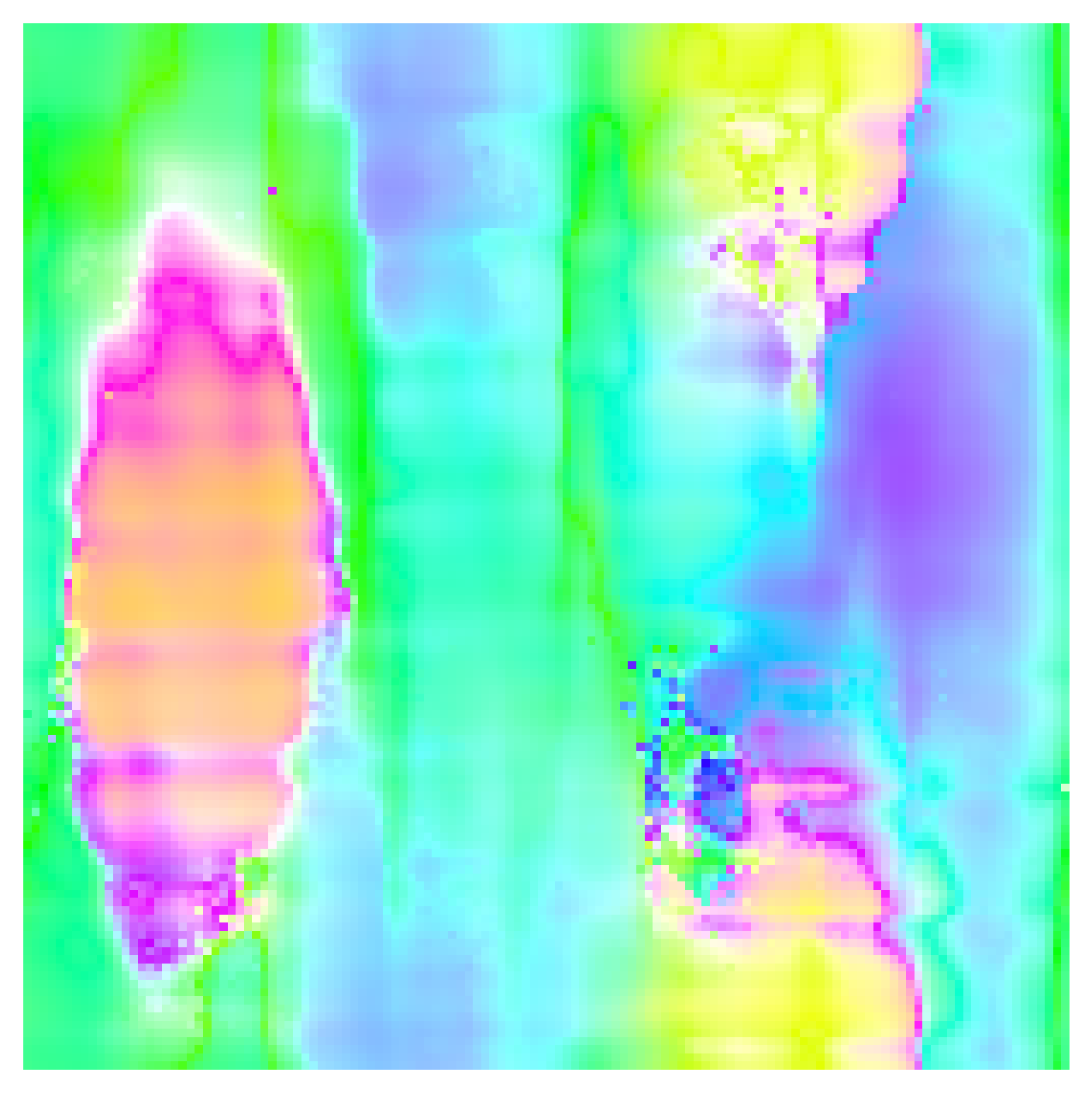}
        \caption*{$w_{\boldsymbol{\gamma}_3}/w_{\mathbf{S}} = 10^{-1}$}
      \end{subfigure}%
      \begin{subfigure}{0.19\textwidth}
        \includegraphics[width=\textwidth]{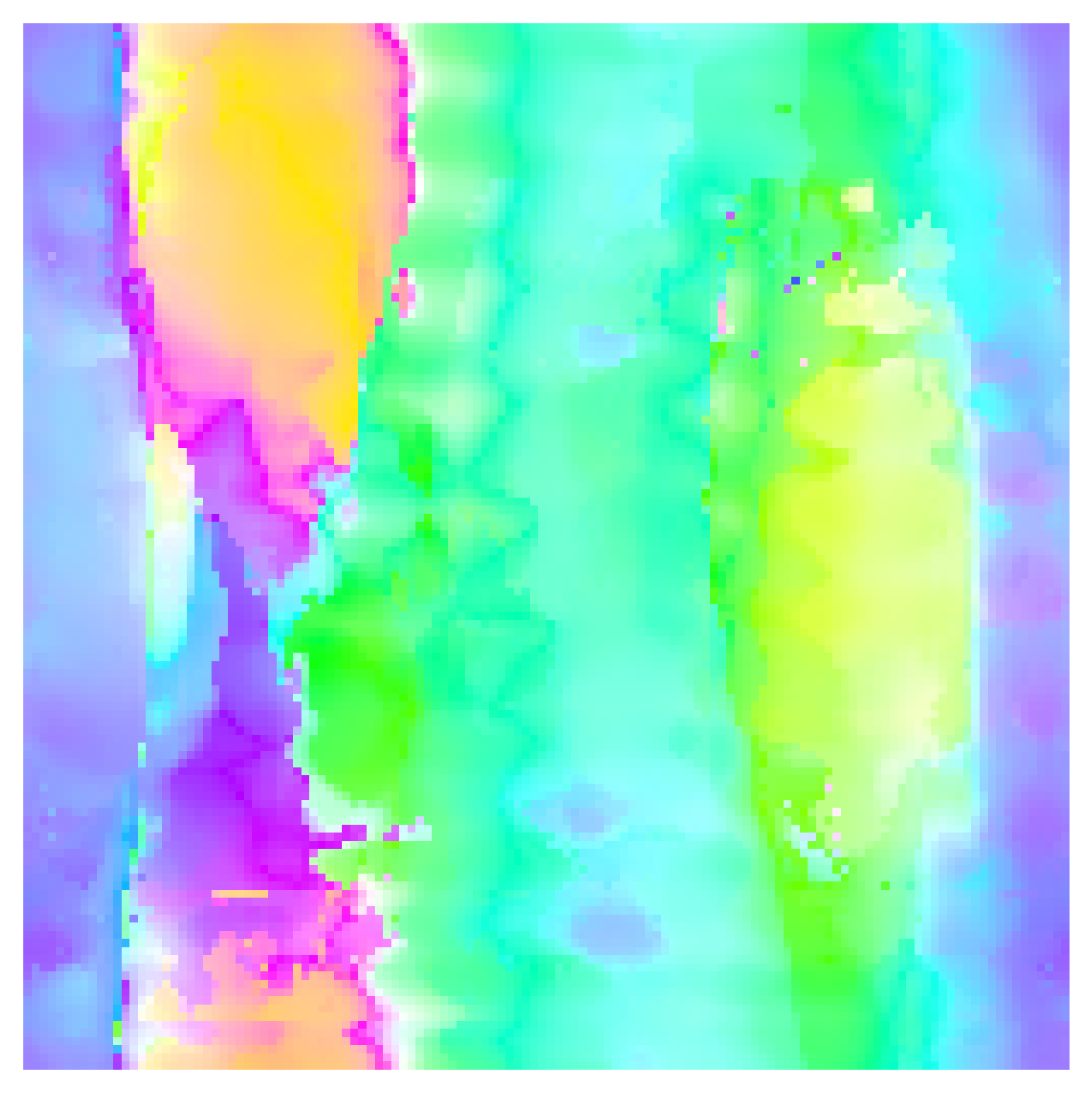}
        \caption*{$w_{\boldsymbol{\gamma}_3}/w_{\mathbf{S}} = 10^{0}$}
      \end{subfigure}%
      \begin{subfigure}{0.19\textwidth}
        \includegraphics[width=\textwidth]{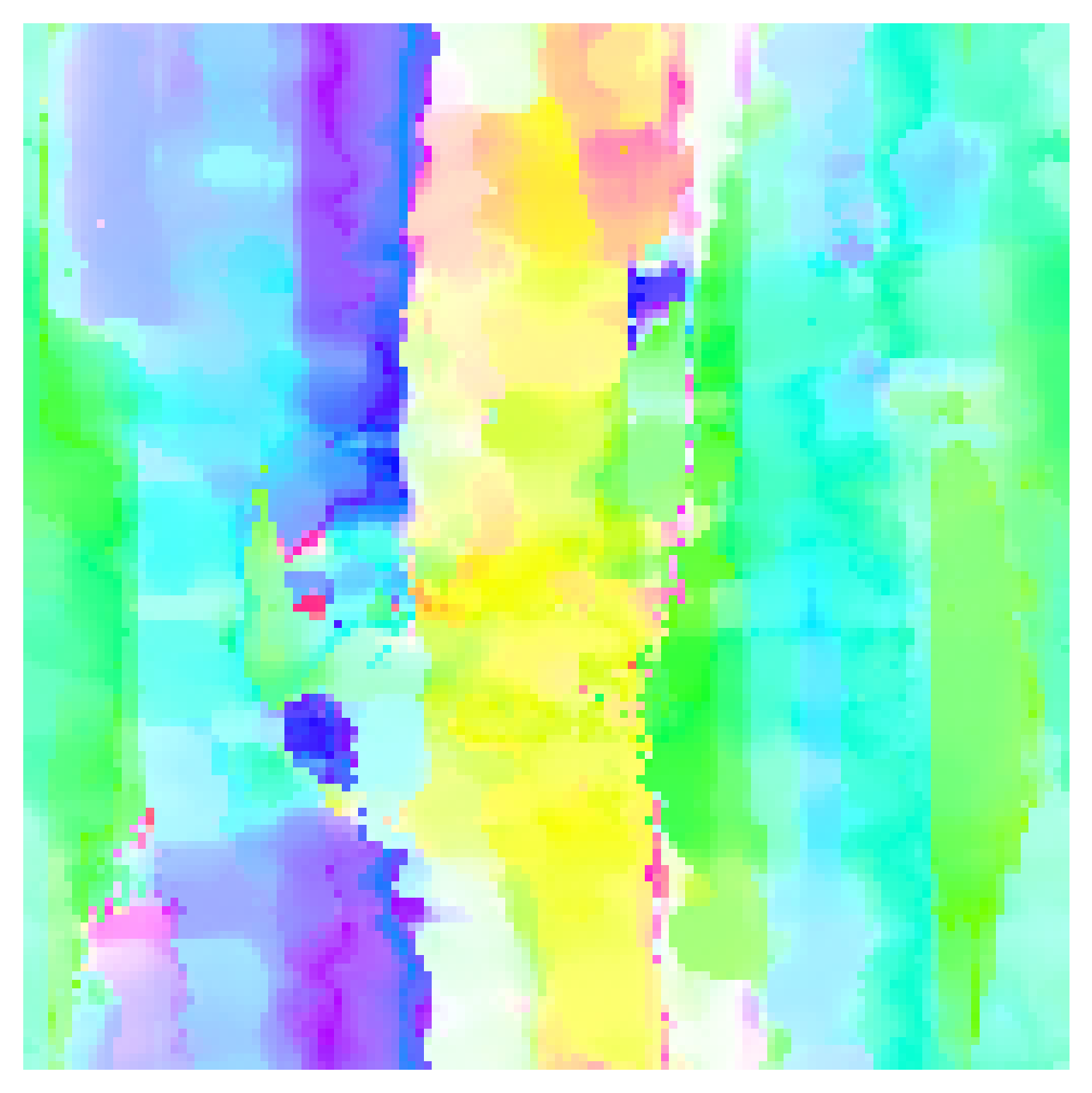}
        \caption*{$w_{\boldsymbol{\gamma}_3}/w_{\mathbf{S}} = 10^{1}$}
      \end{subfigure}%
      \begin{subfigure}{0.19\textwidth}
        \includegraphics[width=\textwidth]{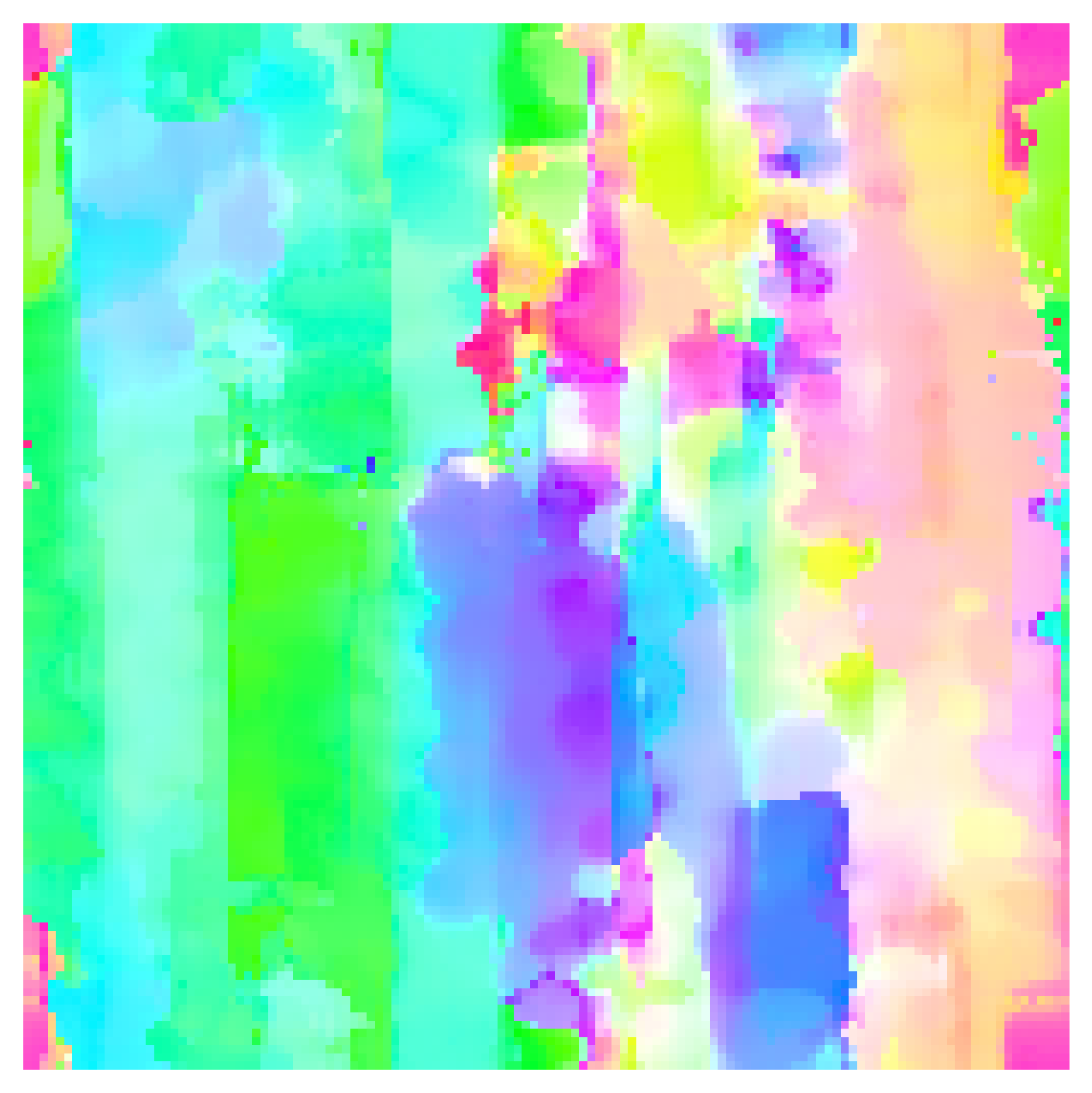}
        \caption*{$w_{\boldsymbol{\gamma}_3}/w_{\mathbf{S}} = 10^{2}$}
      \end{subfigure}
    };

    \begin{scope}[shift={(img.north)}, yshift=-0.1cm]
      % Arrow to the left
      \draw[Stealth-, thick] (-7, 0.5) -- (-1.5, 0.5) node[midway, above, font=\small] {emphasizing $\mathbf{S}$};
      % Arrow to the right
      \draw[-Stealth, thick] (1.5, 0.5) -- (7, 0.5) node[midway, above, font=\small] {emphasizing $\boldsymbol{\gamma}_3$};
    \end{scope}
  \end{tikzpicture}
  \caption{Influence of the relative weighting between $\boldsymbol{\gamma}_3$ and $\mathbf{S}$ on the reconstruction.}
  \label{fig:OC_vs_OVG}
\end{figure*}

Building upon the isolated descriptor analysis, the competitive interplay between $\mathbf{S}$ and $\boldsymbol{\gamma}_3$ is investigated by varying their relative weighting ratios, $w_{\boldsymbol{\gamma}_3}/w_{\mathbf{S}}$, as shown in Fig.~\ref{fig:OC_vs_OVG}. Increasing the influence of $\boldsymbol{\gamma}_3$ results in orientation fields that appear progressively less diffuse, with grain boundaries becoming more distinct. In regimes where the weighting of $\mathbf{S}$ is dominant, the reconstruction captures the broad columnar morphology of the reference, though the boundaries remain poorly defined with visible orientation fluctuations within the grains. Conversely, high weighting of $\boldsymbol{\gamma}_3$ leads to microstructures with high local contrast but fragmented global features, as the columnar arrangement seen in the reference data is not maintained. At intermediate ratios where both columnar traits and sharper interfaces are present, the maps continue to exhibit localized spatial noise and irregular boundary segments, which indicates that while $\mathbf{S}$ and $\boldsymbol{\gamma}_3$ constrain the global arrangement and local contrast differently, their combined application does not readily yield the noise-free structures similar to experimental orientation maps.

\begin{figure*}
  \centering
  \begin{subfigure}{0.19\textwidth}
    \includegraphics[width=1\textwidth]{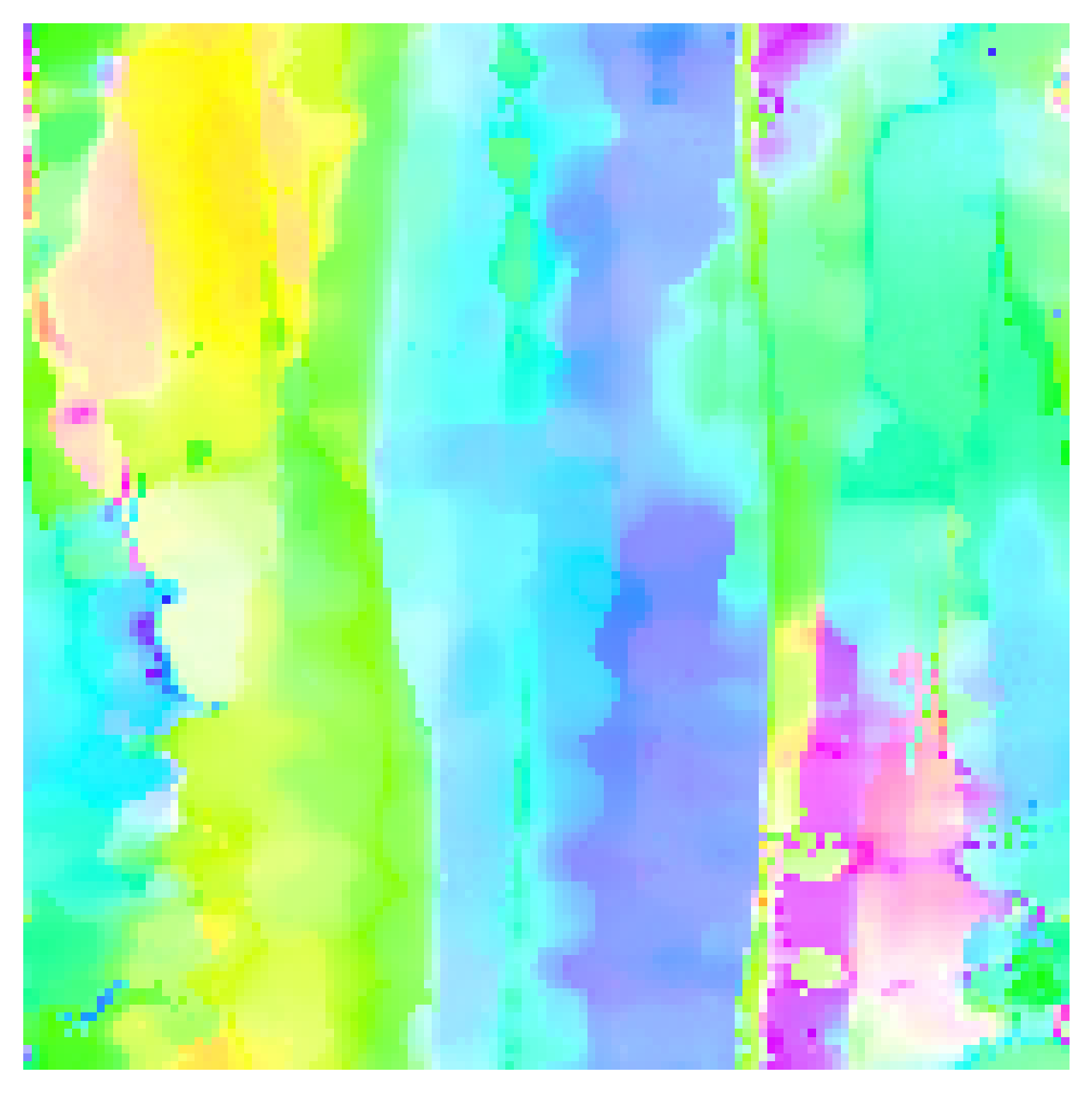}
    \caption*{$w_{\mathcal{V}} = 10^{0}$}
  \end{subfigure}%
  \begin{subfigure}{0.19\textwidth}
    \includegraphics[width=1\textwidth]{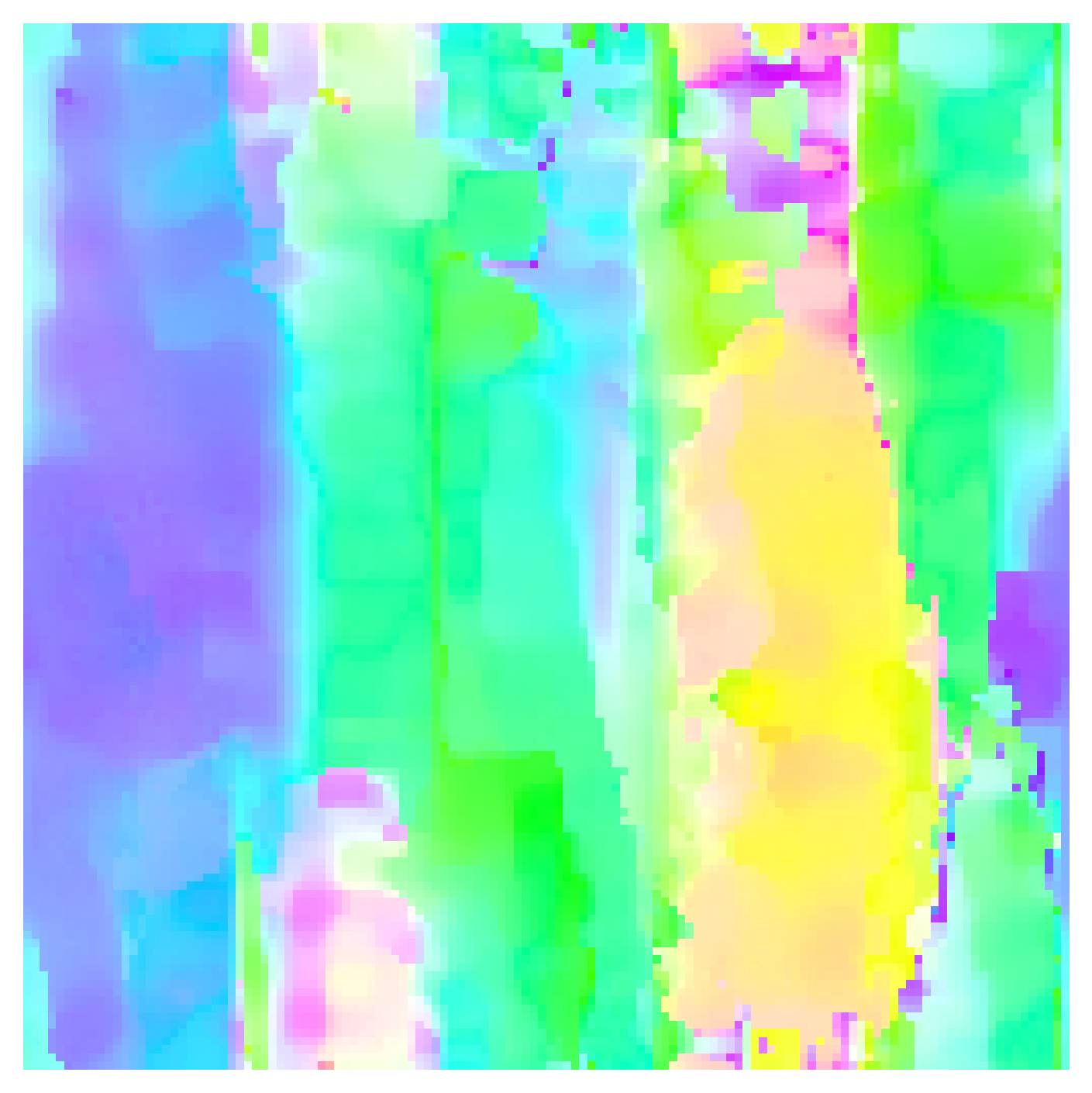}
    \caption*{$w_{\mathcal{V}} = 10^{1}$}
  \end{subfigure}%
  \begin{subfigure}{0.19\textwidth}
    \includegraphics[width=1\textwidth]{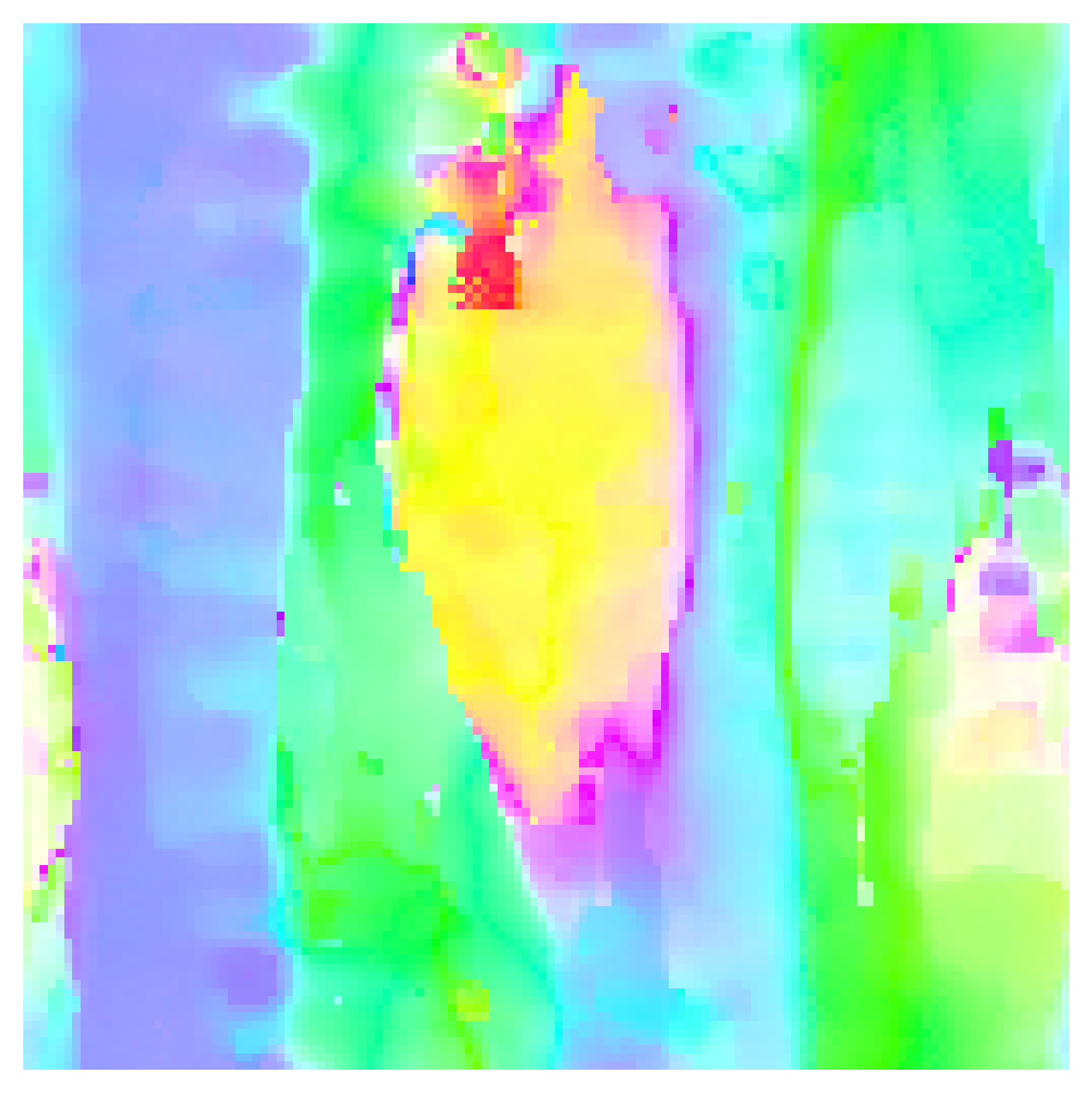}
    \caption*{$w_{\mathcal{V}} = 10^{2}$}
  \end{subfigure}%
  \begin{subfigure}{0.19\textwidth}
    \includegraphics[width=1\textwidth]{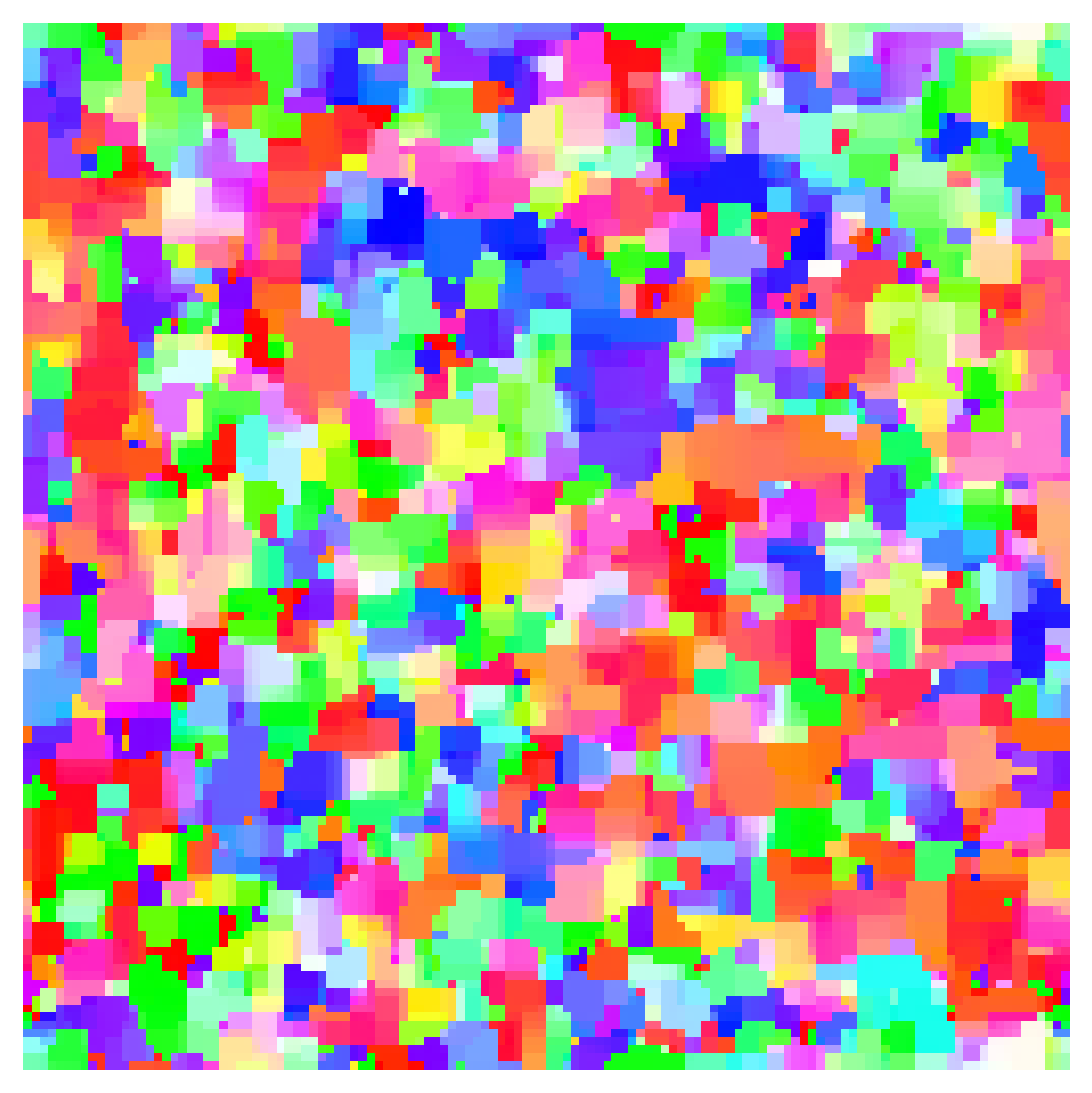}
    \caption*{$w_{\mathcal{V}} = 10^{3}$}
  \end{subfigure}%
  \caption{Influence of the weight $w_\mathcal{V}$ on a reconstruction with fixed spatial descriptor weights ($w_{\mathbf{S}}=1$, $w_{\boldsymbol{\gamma}_3}=10$). Increasing regularization progresses from noise suppression (left) to morphological oversmoothing and eventual structural collapse (right).}
  \label{fig:OV_study}
\end{figure*}

To refine the microstructures to physically more realistic grains, the regularizing influence of $\mathcal{V}$ is investigated by applying it alongside fixed baselines of $w_{\mathbf{S}}=1$ and $w_{\boldsymbol{\gamma}_3}=10$, with results shown across four orders of magnitude in Fig.~\ref{fig:OV_study}. The baseline was held constant to isolate the effect of the regularizer. While other weight combinations may yield similar visual trends, this choice provides a consistent reference for comparison. At lower weights ($w_{\mathcal{V}} = 10^0$ to $10^1$), the $\mathcal{V}$ acts as a stabilizing constraint that reduces nonphysical intragranular fluctuations and helps to distinctly define grain regions while preserving the general columnar arrangement. However, as the weight increases to $w_{\mathcal{V}} = 10^3$, the regularization competes with the spatial descriptors and the optimization appears to be dominated by the gradient penalty. As a result, the columnar structure is largely lost and replaced by the fragmented orientation field that was visible in Fig.~\ref{fig:Comparison_Descriptors}. These observations suggest that while $\mathcal{V}$ is useful for suppressing local noise, its influence must be carefully balanced against the $\mathbf{S}$ and $\boldsymbol{\gamma}_3$ descriptors to maintain the underlying morphological features of the target microstructure.

\begin{figure*}
  \centering
    \includegraphics[width=0.8\textwidth]{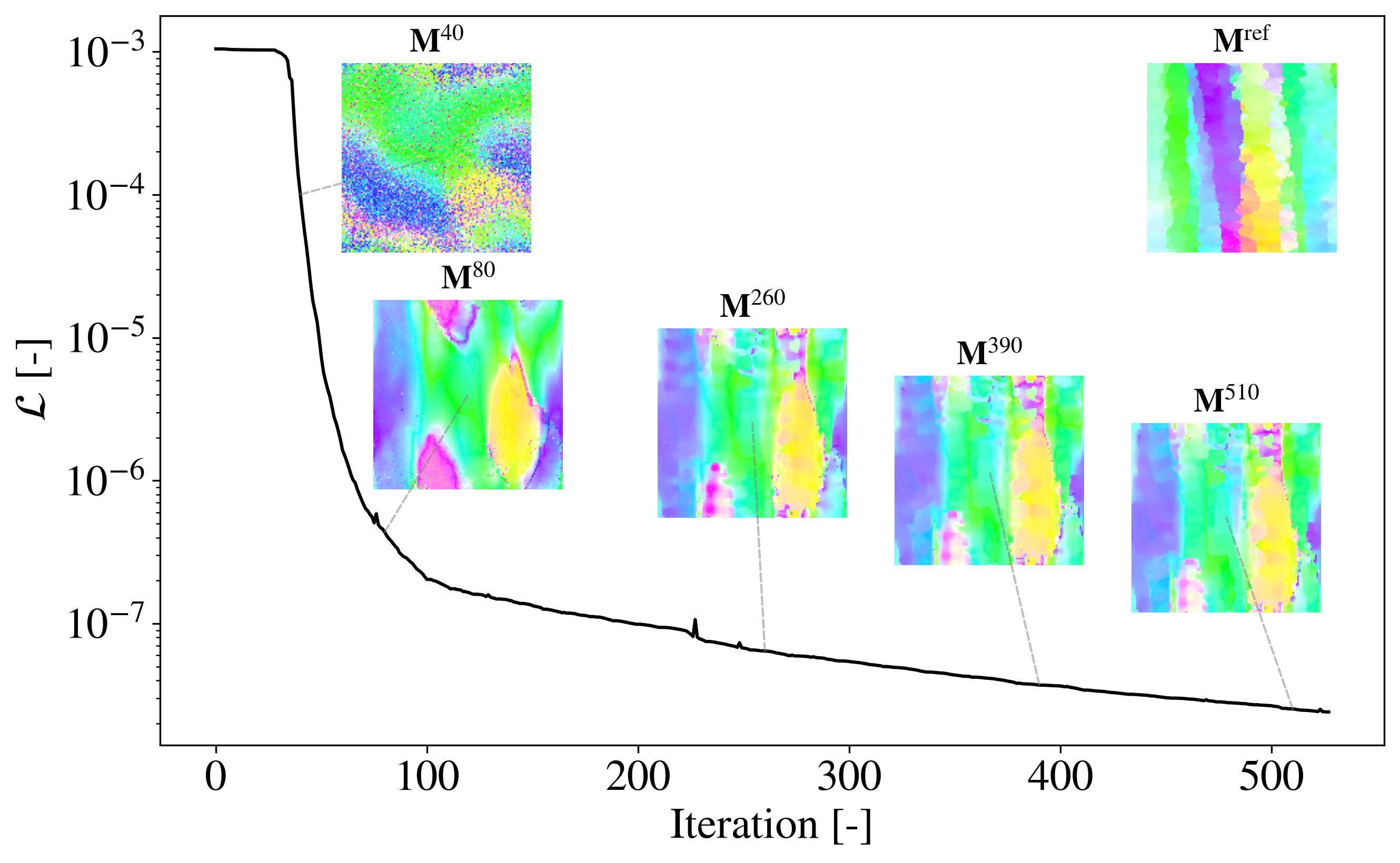}
    \caption{Convergence trajectory of the multi-objective loss function $\mathcal{L}$ over 530 iterations. Inset snapshots ($\mathbf{M}^{40}$ to $\mathbf{M}^{510}$) illustrate the evolution from initial noise-dominated states to a refined reconstructed morphology that statistically matches the reference ($\mathbf{M}^{\mathrm{ref}}$).}
  \label{fig:micro_vs_loss}
\end{figure*}

The stability and efficiency of the orientation-based DMCR framework are quantified by monitoring the evolution of the composite ($w_{\mathbf{S}}=1$, $w_{\boldsymbol{\gamma}_3}=10$, $w_{\mathcal{V}}=100$) loss function $\mathcal{L}$ throughout the optimization process. Fig.~\ref{fig:micro_vs_loss} depicts the convergence trajectory for the reconstruction of orientation field of the microstructure introduced in Fig.~\ref{fig:comp_ori_repr}, showing the progressive emerging of microstructural features and directly correlates it to the logarithmic reduction of the loss function. The optimization trajectory reveals three distinct phases of structural development: (i) During the initial descent, the loss drops steeply to $10^{-4}$, yielding a reconstruction ($\mathbf{M}^{40}$) where first diffuse clusters of orientations appear while significant high-frequency noise persists. (ii) As the loss reaches the levels below the $10^{-6}$ regime, the microstructure shows a morphological feature transition where the columnar grain architecture emerges. (iii) The final stage consist of continuous migration of the interfaces that have emerged in (ii), showing little visual difference between $\mathbf{M}^{260}$ and $\mathbf{M}^{510}$. In general, the smooth logarithmic descent across four orders of magnitude confirms that the SHSH-based descriptor set yields a continuous loss landscape in which the DMCR framework enables efficient gradient-based updates.

\subsection{Reconstruction of experimental microstructures}\label{3.2}

\begin{figure*}
  \centering
  \begin{subfigure}{0.4\textwidth}
    \includegraphics[width=1\textwidth]{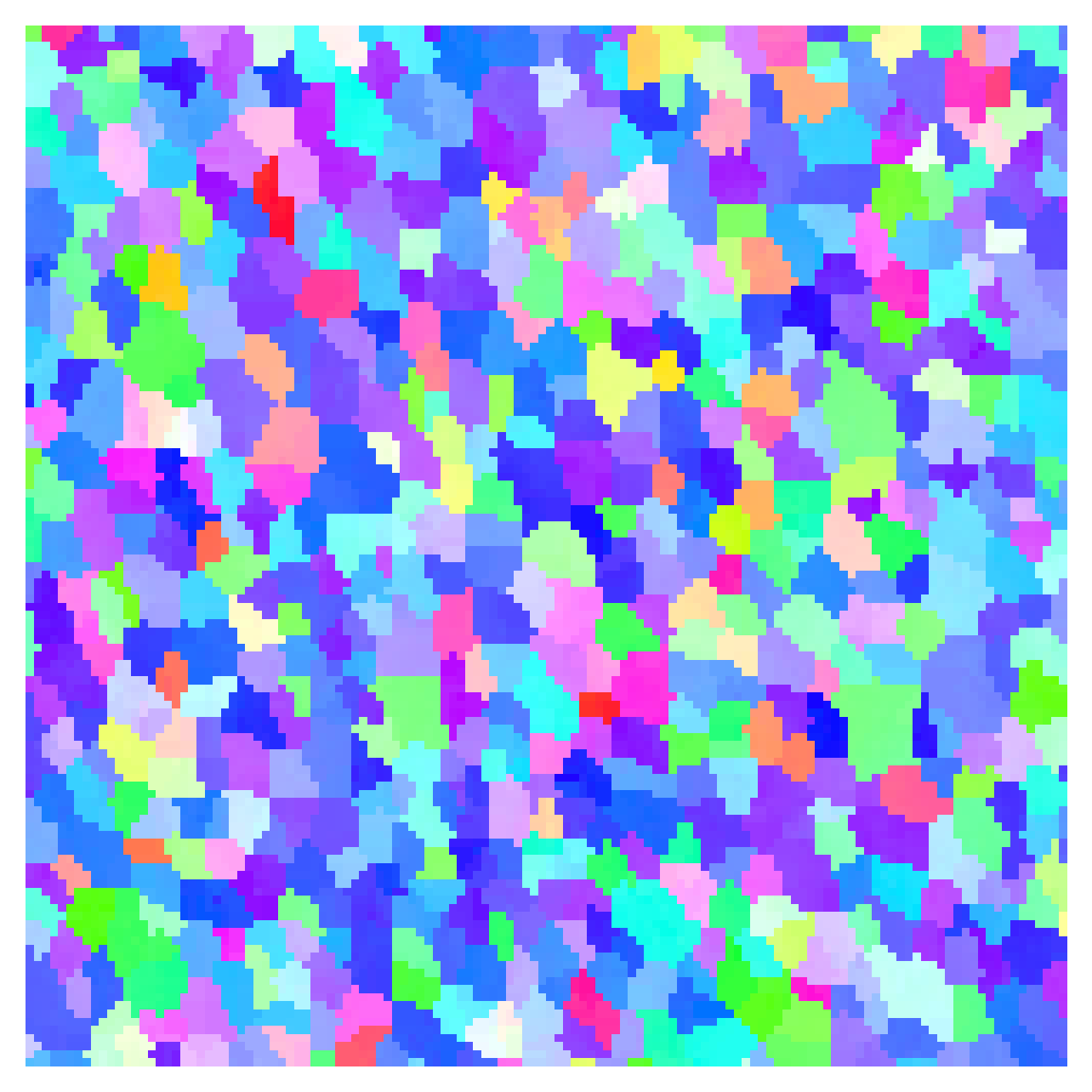}
    \caption{}
    \label{fig:figure1}
  \end{subfigure}
  \begin{subfigure}{0.4\textwidth}
    \begin{overpic}[width=1\textwidth]{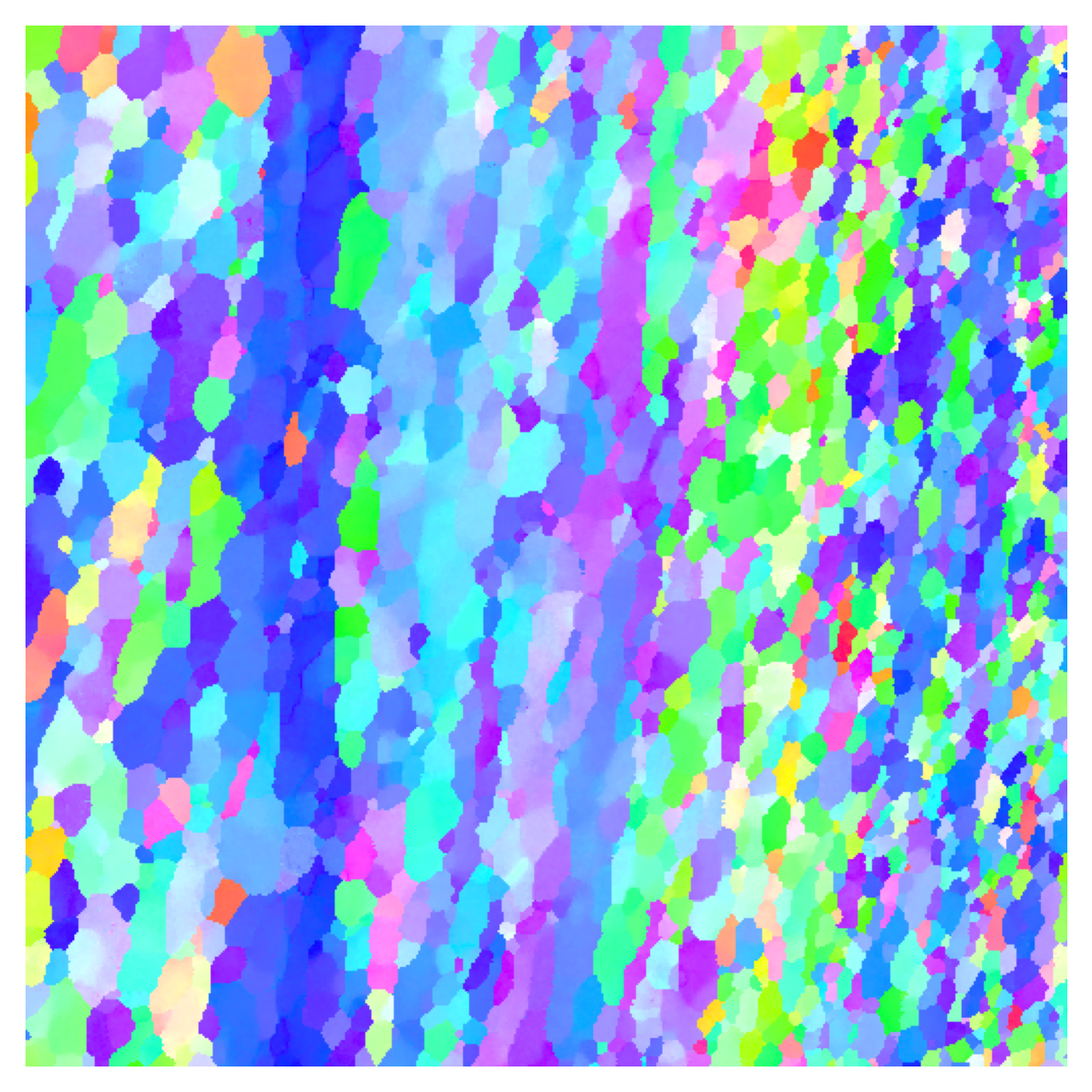}%
    \put(750,780){\includegraphics[width=0.2\textwidth]{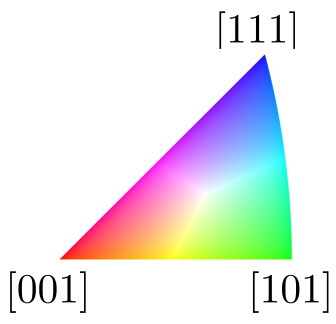}}%
    \end{overpic}
    \caption{}
  \end{subfigure}
  \caption{Inverse pole figure (IPF) maps, exemplarily for aluminum microstructures produced via thermo-mechanical processing. (a) Statistically homogeneous microstructure showing uniform grain distribution. (b) Statistically inhomogeneous region characterized by significant spatial gradients in grain size and orientation.}
  \label{fig:FE_micros}
\end{figure*}

To evaluate the versatility and robustness of the DMCR framework, the analysis is extended to orientation maps derived after thermo-mechanical processing of metallic alloys, such as Friction Extrusion (FE) of an aluminum alloy \cite{suhuddin_microstructure_2023}. Two distinct regions of interest are selected, as illustrated in Fig.~\ref{fig:FE_micros}, which pose different challenges for statistical reconstruction. The first case study, Fig.~\ref{fig:FE_micros}(a), shows a $128 \times 128\,  \text{px}$ region that demonstrated statistical homogeneity, where the distributions of grain morphology and crystallographic orientation remain spatially invariant across the observation window. In contrast, the second case study, Fig.~\ref{fig:FE_micros}(b), presents a $512 \times 512\, \text{px}$ statistically inhomogeneous field exhibiting a pronounced macro-scale texture and grain size gradients, resulting from the localized thermomechanical history. The reconstructions are performed with $w_{\mathbf{S}}=1$, $w_{\boldsymbol{\gamma}_3}=10$ and $w_{\mathcal{V}}=100$.

\subsubsection{Reconstruction of statistically homogeneous microstructures}\label{3.2.1}

\begin{figure*}
\centering
  \begin{subfigure}{0.32\textwidth}
    \includegraphics[width=\linewidth]{Case2_128_orig.png}
    \caption{}
  \end{subfigure}
  \hfill
   \begin{subfigure}{0.32\textwidth}
    \includegraphics[width=\linewidth]{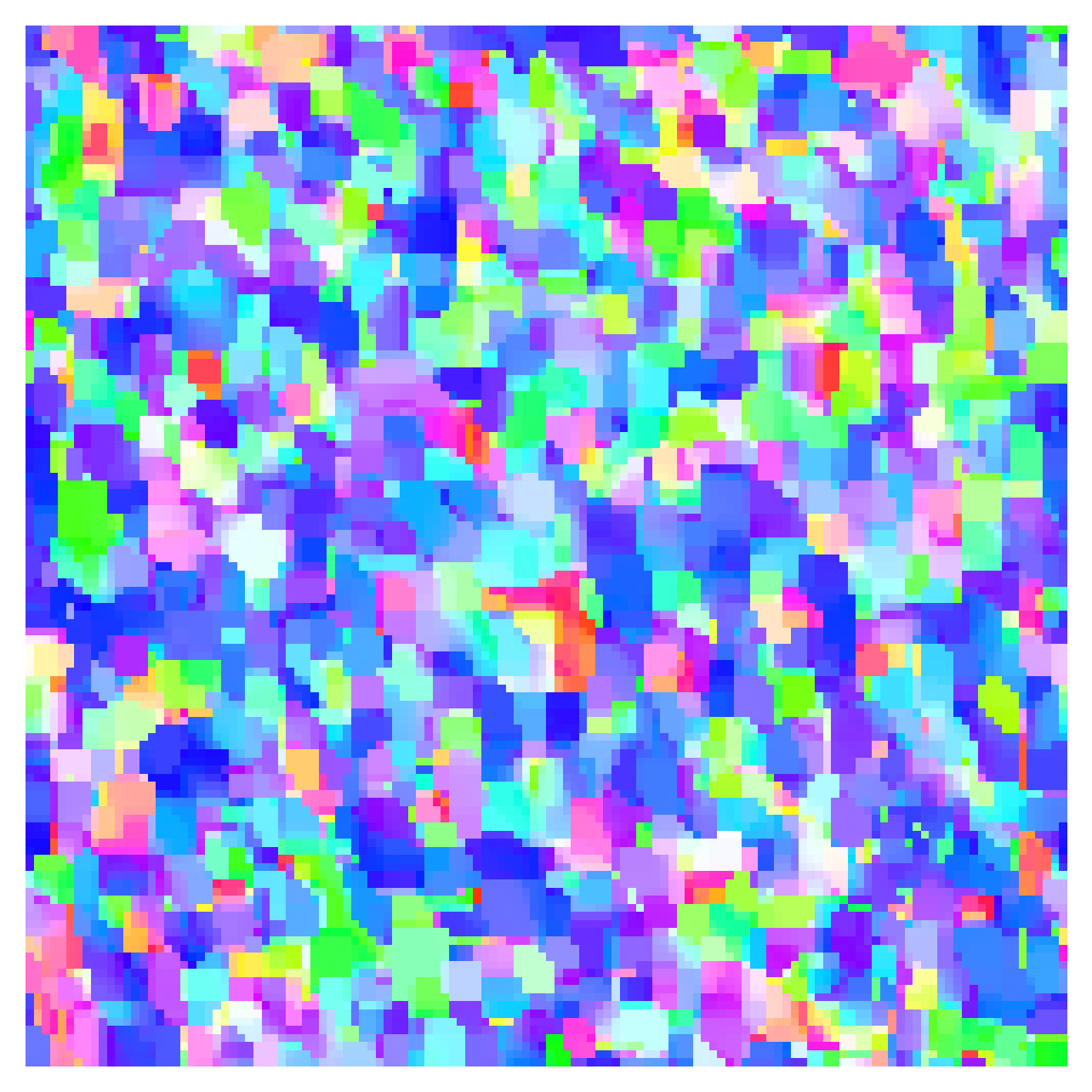}
    \caption{}
  \end{subfigure}
  \hfill
\begin{subfigure}{0.32\textwidth}
    \includegraphics[width=\linewidth]{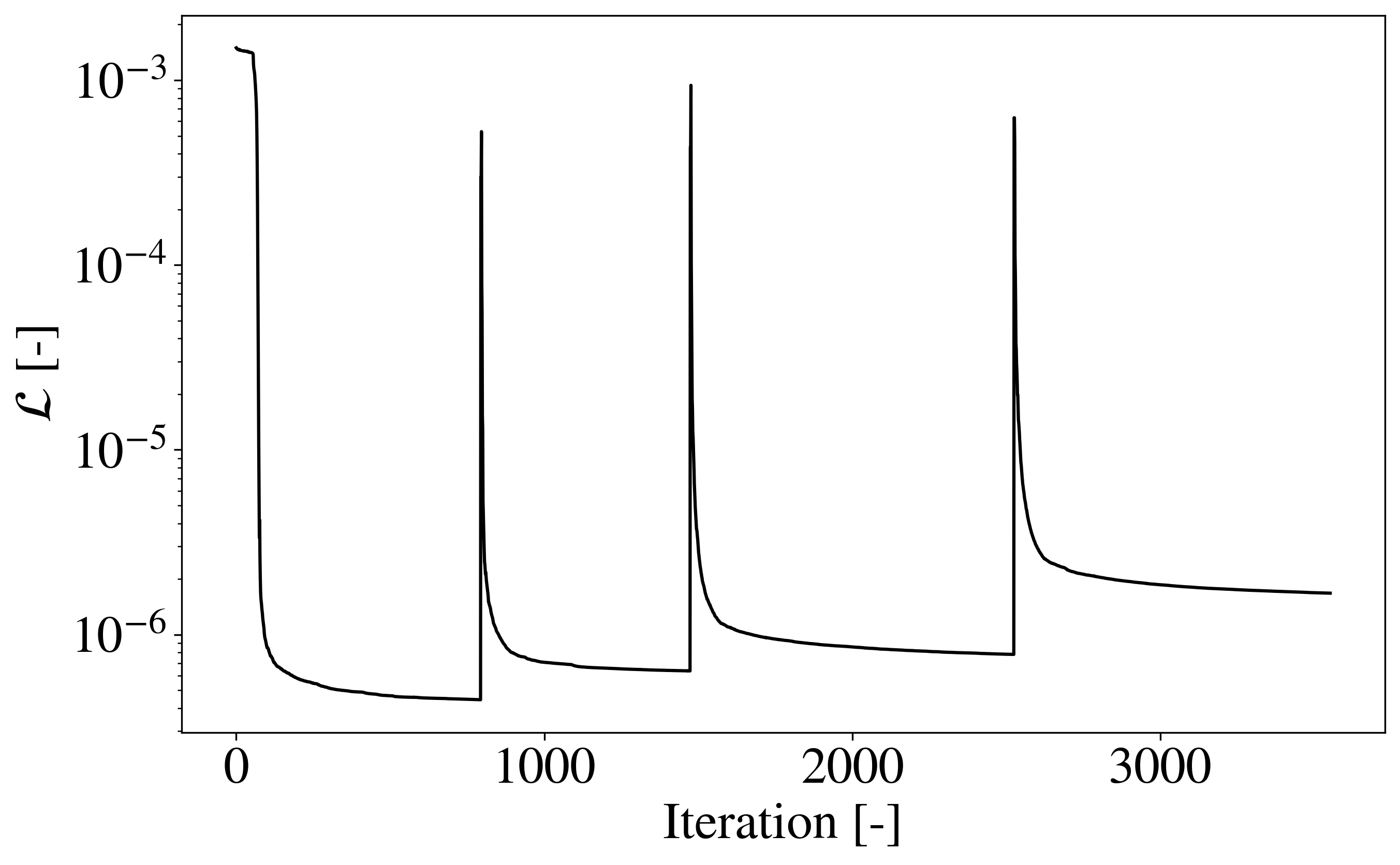}
    \caption{}
  \end{subfigure}
  \hfill
\begin{subfigure}{0.8\textwidth}
    \begin{overpic}[width=\linewidth]{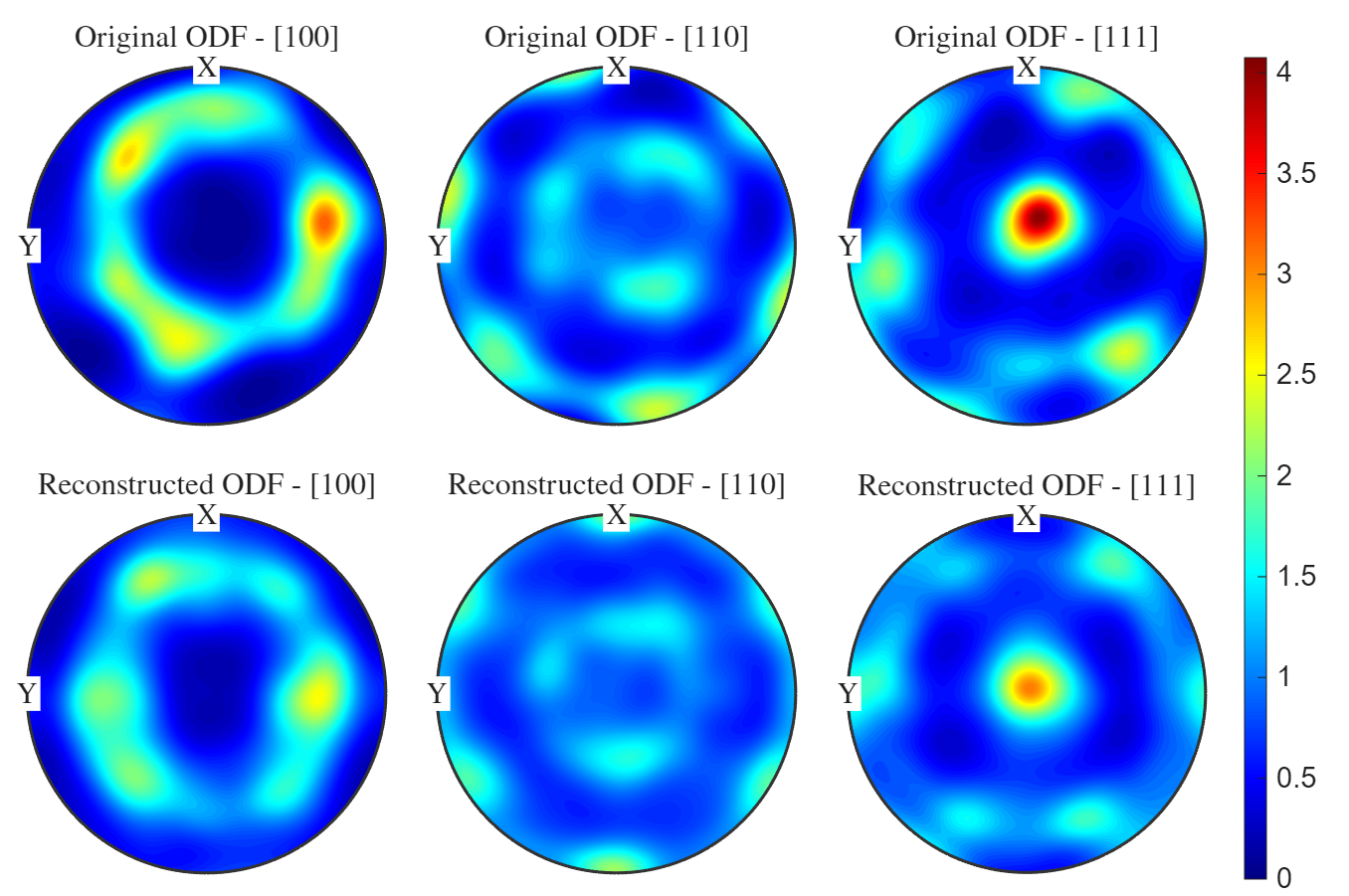}
      \put(1000,180){\rlap{\ \rotatebox{90}{Texture intensity (m.r.d.)}}}
    \end{overpic}
    \caption{}
  \end{subfigure}
  \hfill
  \caption{Reconstruction results for the statistically homogeneous microstructure, highlighting (a) the original EBSD map, (b) the synthetic reconstruction, (c) the convergence of the loss function during multigrid reconstruction, and (d) the comparison of original and reconstructed $\{001\}$, $\{011\}$, and $\{111\}$ pole figures. The loss jumps in (c) arise from the refinement steps of the multigrid algorithm where additional, and so far unseen, terms are introduced to the loss function as detailed in \cite{seibert_descriptor-based_2022}.}
  \label{fig:Case2_recon}
\end{figure*}

In the following, reconstructions of the cubic microstructure presented in Fig.~\ref{fig:FE_micros} are performed using the L-BFGS-B algorithm together with a multigrid strategy, as described by Seibert et al. \cite{seibert_reconstructing_2021}, to improve the convergence behavior. The statistical target is defined by the first nine SHSH modes, with spatial correlations captured over a $17 \times 17\,  \text{px}$ cropped window to ensure sufficient range for $\mathbf{S}$ and $\boldsymbol{\gamma}_3$. The reconstruction of the statistically homogeneous microstructure demonstrates the framework's ability in resolving both topological and crystallographic features. Fig.~\ref{fig:Case2_recon} presents a comprehensive analysis comparing the target experimental EBSD map (Fig.~\ref{fig:Case2_recon}(a)) against the synthetic realization (Fig.~\ref{fig:Case2_recon}(b)). Visually, the reconstructed orientation field exhibits excellent agreement with the original EBSD map as the algorithm successfully reproduces the characteristic equiaxed grain morphology, with the combined descriptor constraints yielding distinct boundaries and low-noise. The evaluation of the crystallographic features is shown in Fig.~\ref{fig:Case2_recon}(d), where the reconstructed pole figures, computed using MTEX \cite{bachmann_texture_2010} for the $\{001\}$, $\{011\}$, and $\{111\}$ planes show reasonable alignment with the reference data. While many peaks are covered within the synthetic texture, some peaks coalesce due to remaining diffuse boundaries. In general, the DMCR framework can successfully characterize and reconstruct microstructure using the SHSH representation, while recovering lower-order descriptors like the ODF without explicit prescription. Furthermore, the loss trajectory Fig.~\ref{fig:Case2_recon}(c) confirms the efficiency of the multigrid L-BFGS-B scheme, showing a stable, monotonic convergence within each multigrid step to the statistical target.

\begin{figure*}
\centering
    \begin{subfigure}{0.3\textwidth}
        \caption*{$\boldsymbol{\gamma}^{\text{ref}}\left(\dot{\ddot{Z}}^8_{1}|\mathbf{r}\right)$}
        \includegraphics[width=\textwidth]{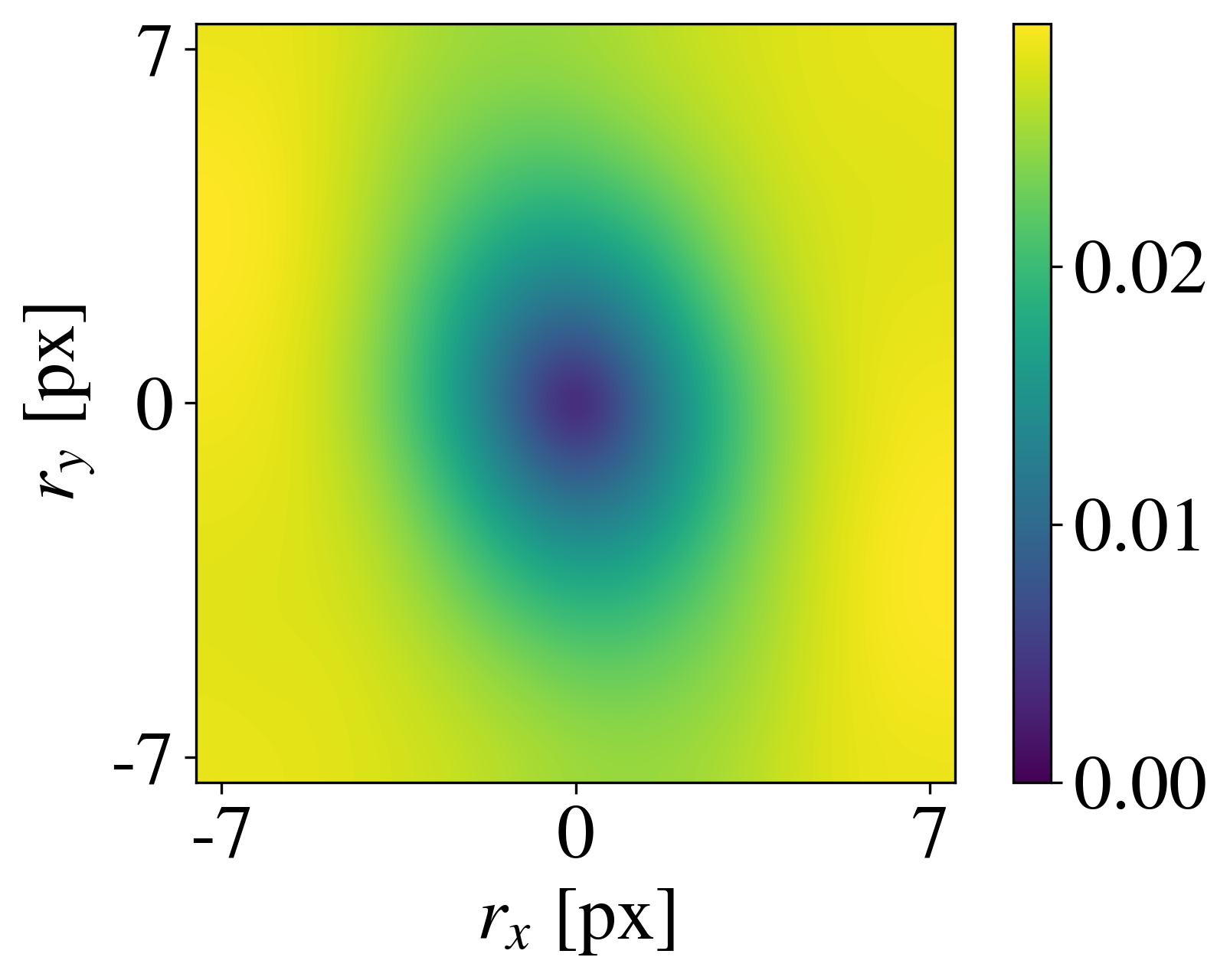}
    \end{subfigure}
    \begin{subfigure}{0.3\textwidth}
        \caption*{$\boldsymbol{\gamma}^{\text{ref}}\left(\dot{\ddot{Z}}^8_{5}|\mathbf{r}\right)$}
        \includegraphics[width=\textwidth]{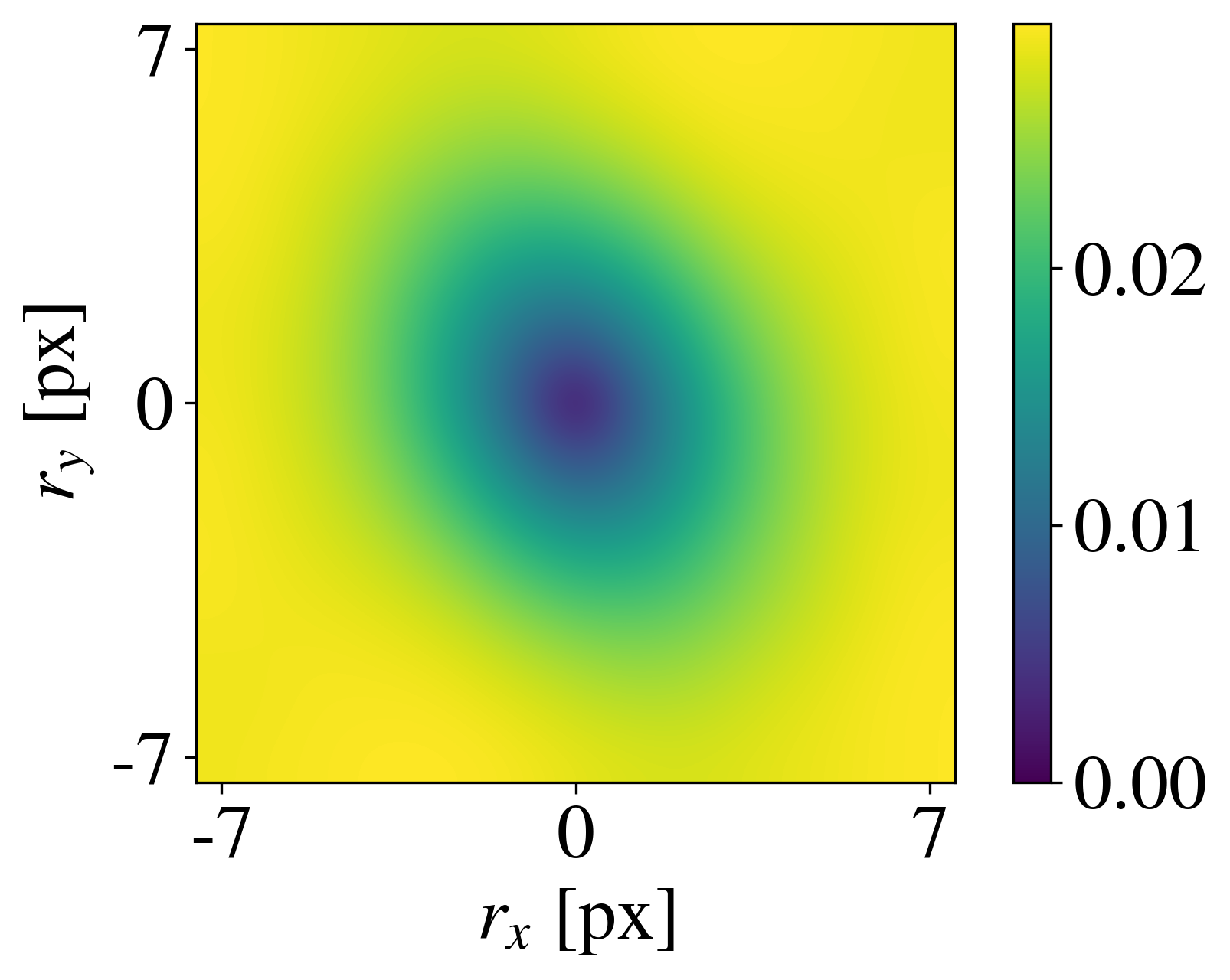}
    \end{subfigure}
    \begin{subfigure}{0.3\textwidth}
        \caption*{$\boldsymbol{\gamma}^{\text{ref}}\left(\dot{\ddot{Z}}^8_{9}|\mathbf{r}\right)$}
        \includegraphics[width=\textwidth]{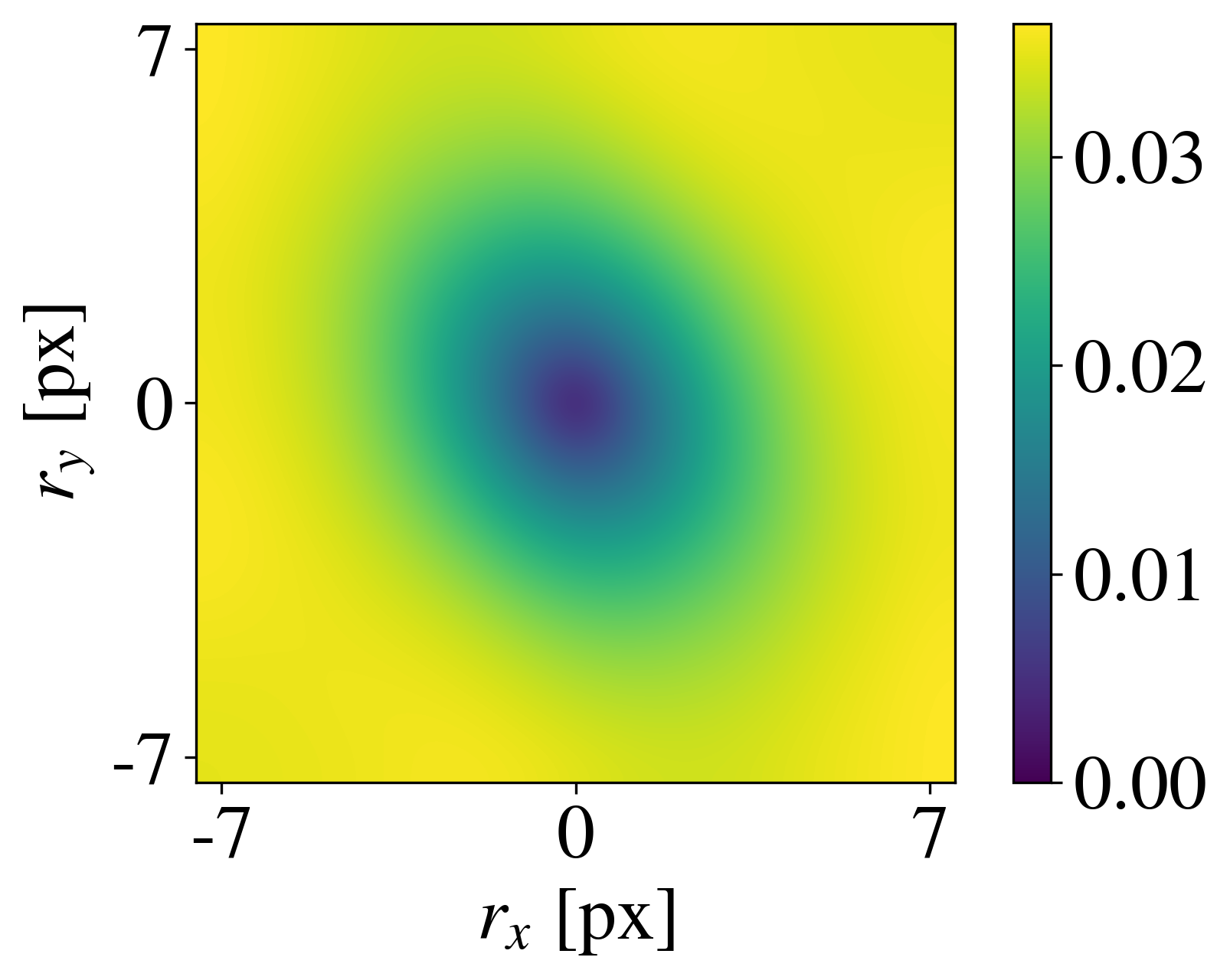}
    \end{subfigure} \\[1ex]
    \begin{subfigure}{0.3\textwidth}
        \caption*{$\boldsymbol{\gamma}^{\text{recon}}\left(\dot{\ddot{Z}}^8_{1}|\mathbf{r}\right)$}
        \includegraphics[width=\textwidth]{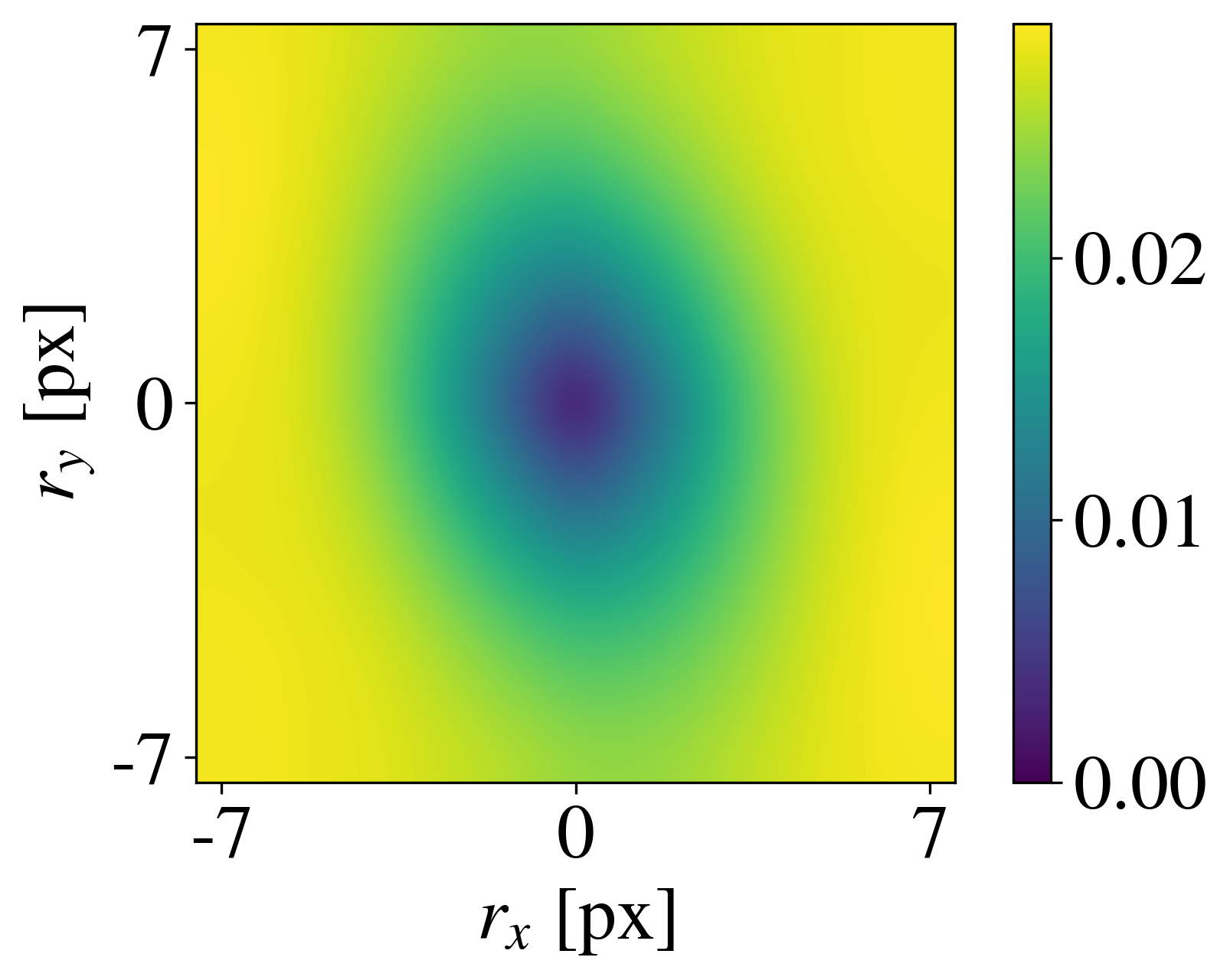}
    \end{subfigure}
    \begin{subfigure}{0.3\textwidth}
        \caption*{$\boldsymbol{\gamma}^{\text{recon}}\left(\dot{\ddot{Z}}^8_{5}|\mathbf{r}\right)$}
        \includegraphics[width=\textwidth]{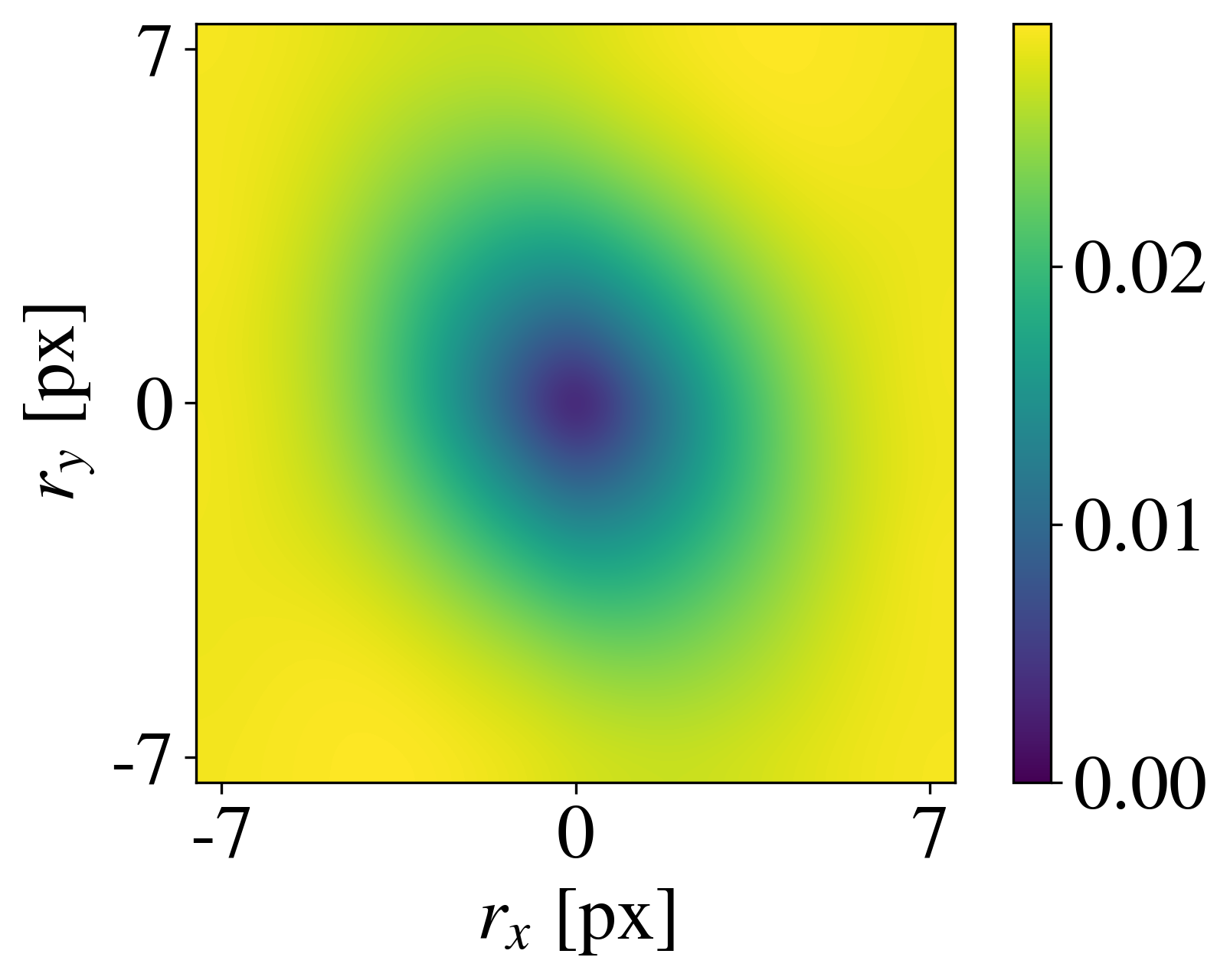}
    \end{subfigure}
    \begin{subfigure}{0.3\textwidth}
        \caption*{$\boldsymbol{\gamma}^{\text{recon}}\left(\dot{\ddot{Z}}^8_{9}|\mathbf{r}\right)$}
        \includegraphics[width=\textwidth]{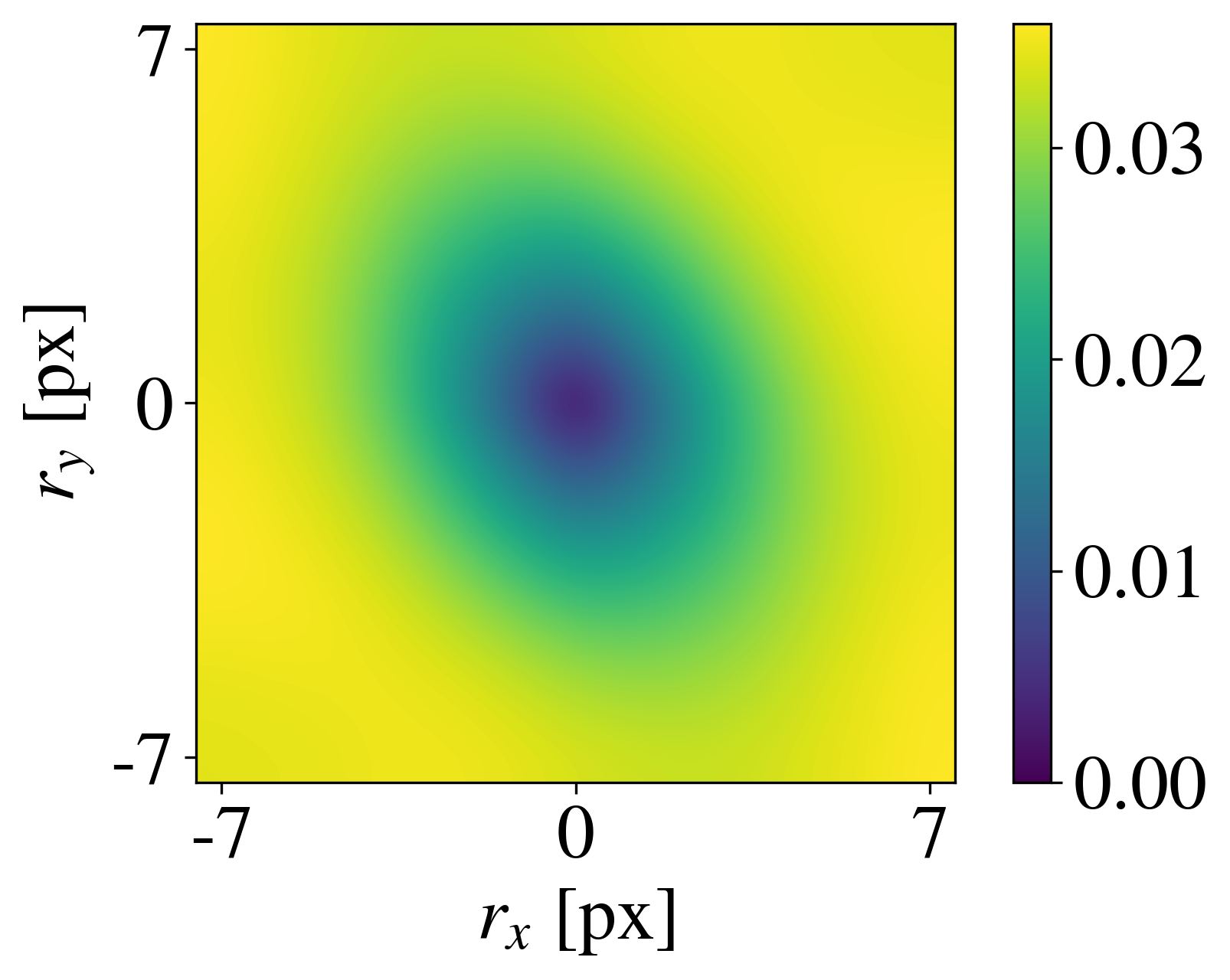}
    \end{subfigure} \\[1ex]
    \begin{subfigure}{0.3\textwidth}
        \caption*{$\frac{\boldsymbol{\gamma}^{\text{ref}}\left(\dot{\ddot{Z}}^8_{1}|\mathbf{r}\right) - \boldsymbol{\gamma}^{\text{recon}}\left(\dot{\ddot{Z}}^8_{1}|\mathbf{r}\right)}{\boldsymbol{\gamma}^{\text{ref}}\left(\dot{\ddot{Z}}^8_{1}|\mathbf{r}\right)}$}
        \includegraphics[width=\textwidth]{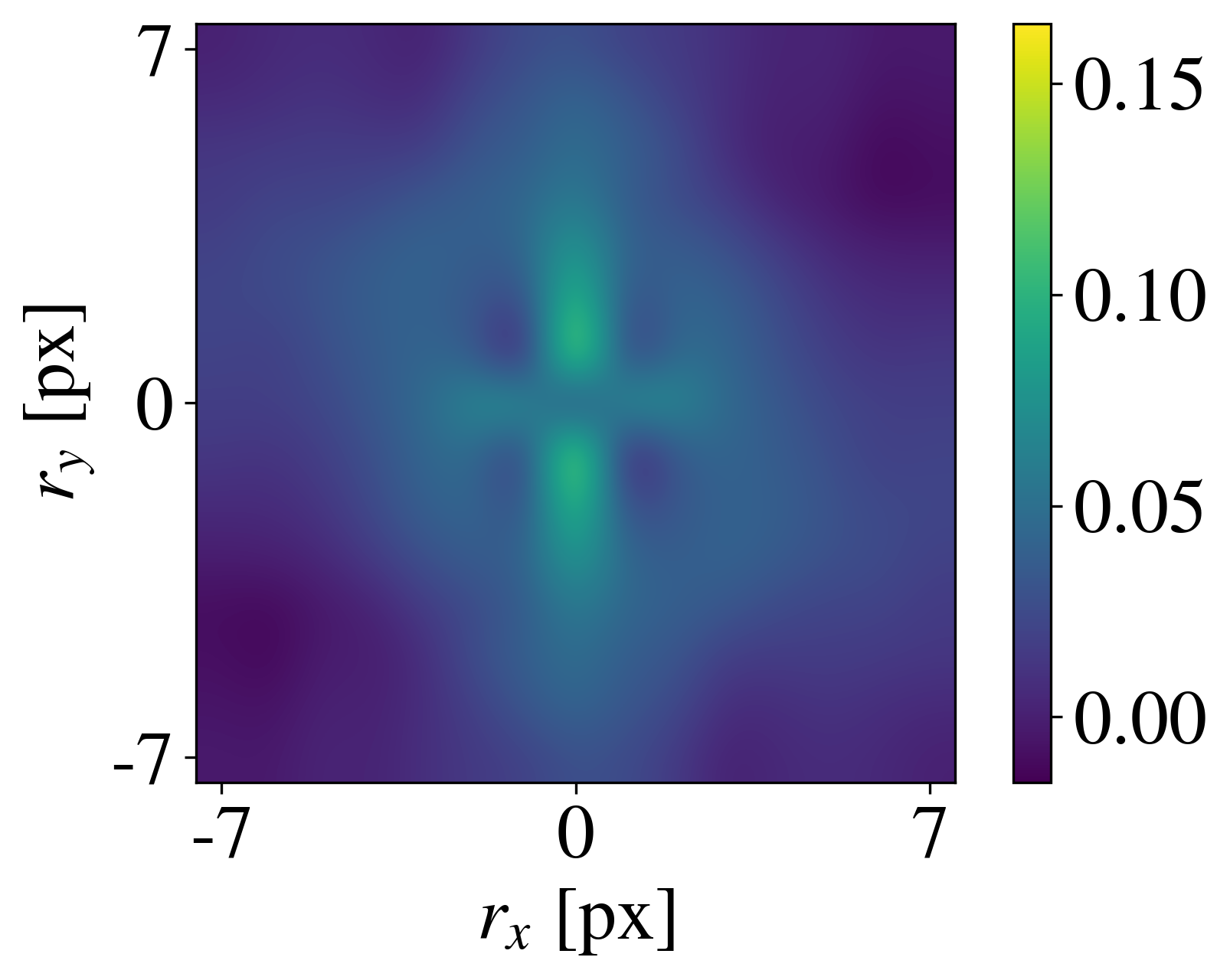}
    \end{subfigure}
    \begin{subfigure}{0.3\textwidth}
        \caption*{$\frac{\boldsymbol{\gamma}^{\text{ref}}\left(\dot{\ddot{Z}}^8_{5}|\mathbf{r}\right) - \boldsymbol{\gamma}^{\text{recon}}\left(\dot{\ddot{Z}}^8_{5}|\mathbf{r}\right)}{\boldsymbol{\gamma}^{\text{ref}}\left(\dot{\ddot{Z}}^8_{5}|\mathbf{r}\right)}$}
        \includegraphics[width=\textwidth]{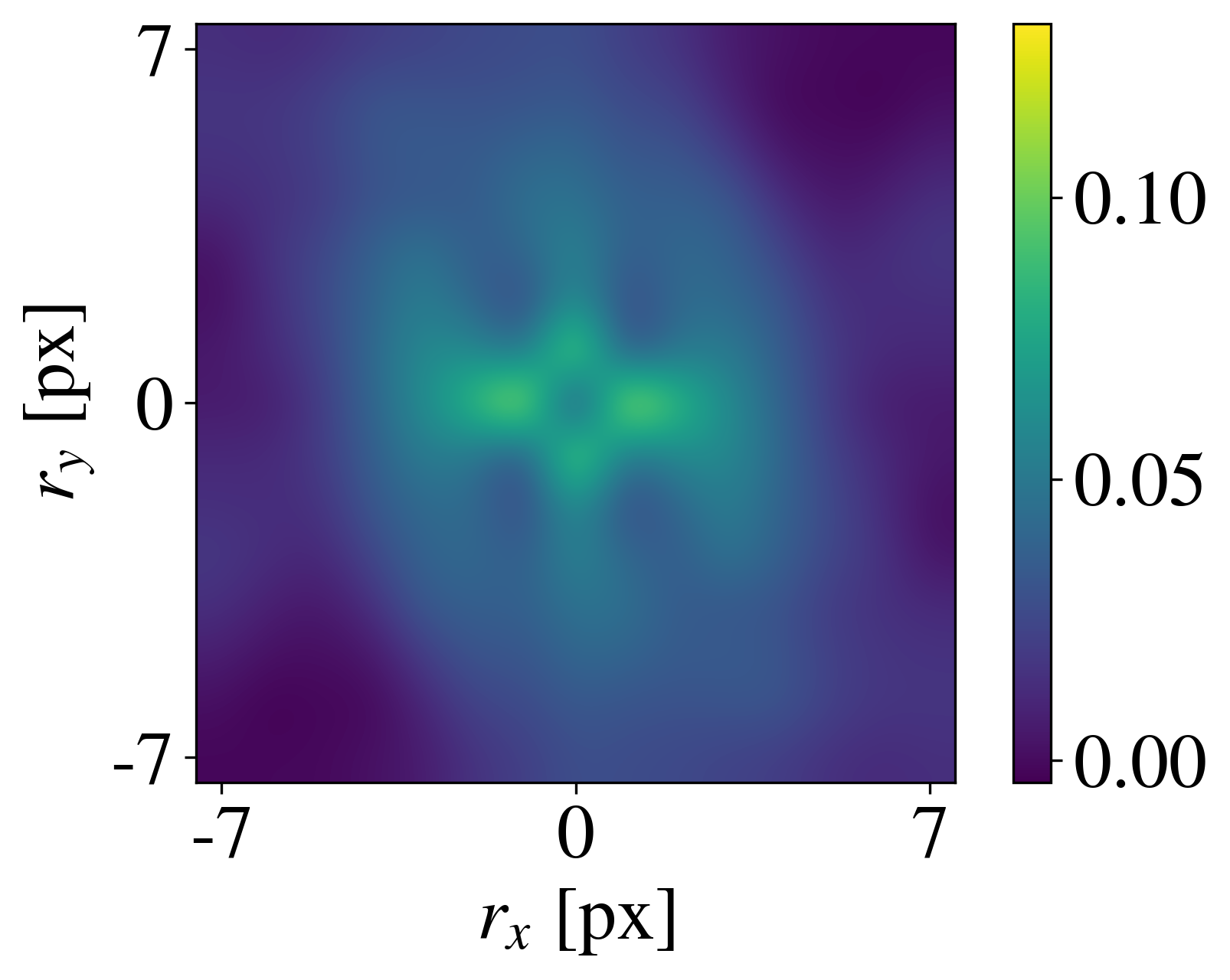}
    \end{subfigure}
    \begin{subfigure}{0.3\textwidth}
        \caption*{$\frac{\boldsymbol{\gamma}^{\text{ref}}\left(\dot{\ddot{Z}}^8_{9}|\mathbf{r}\right) - \boldsymbol{\gamma}^{\text{recon}}\left(\dot{\ddot{Z}}^8_{9}|\mathbf{r}\right)}{\boldsymbol{\gamma}^{\text{ref}}\left(\dot{\ddot{Z}}^8_{9}|\mathbf{r}\right)}$}
        \includegraphics[width=\textwidth]{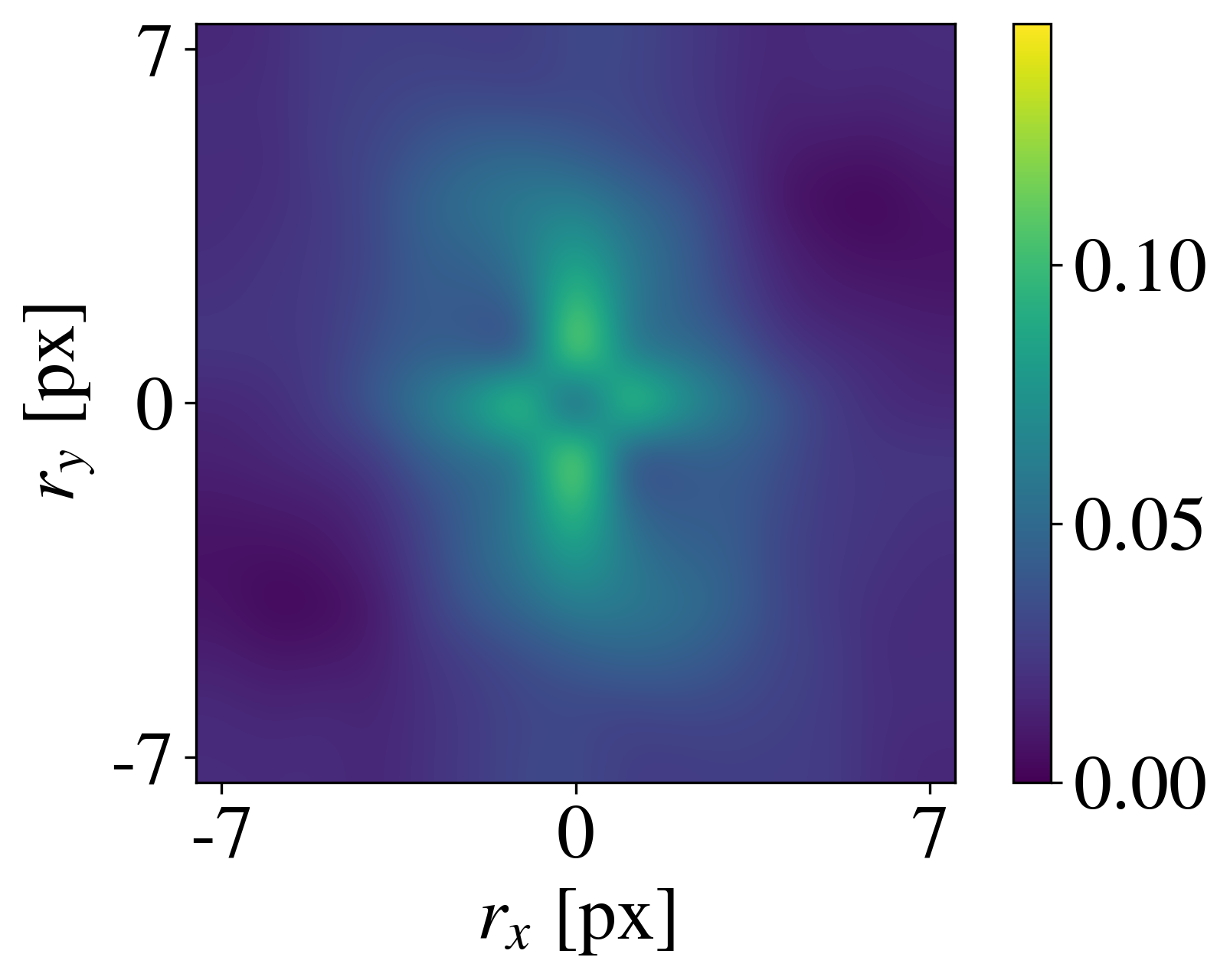}
    \end{subfigure}
    \caption{Comparative error analysis of the 2D orientation variograms for selected SHSH modes ($\dot{\ddot{Z}}_1^8, \dot{\ddot{Z}}_5^8, \dot{\ddot{Z}}_9^8$), showing the experimental reference statistics (top row), the statistics of the reconstructions (middle row), and the relative error maps (bottom row).}
    \label{fig:Case2_error}
\end{figure*}

To quantitatively assess the reconstruction fidelity for the statistically homogeneous microstructure, the variograms of the realizations are compared to the original statistics via the relative error. Fig.~\ref{fig:Case2_error} illustrates this comparison for three selected SHSH modes ($\dot{\ddot{Z}}_1^8, \dot{\ddot{Z}}_5^8, \dot{\ddot{Z}}_9^8$). The reference statistics exhibit an equiaxed pattern, characterized by a spherical minimum at the center of the variogram map, which is faithfully reproduced in the reconstructed set. No significant visual discrepancies are observed between the original and reconstructed maps for the selected modes. 

The final row of Fig.~\ref{fig:Case2_error} presents the relative error, where dominant dark regions indicate that the discrepancy remains below 5\% across the majority of the spatial domain. Minor localized deviations, approaching 15\%, are confined to the immediate vicinity of the origin and principal axes. These artifacts originate from the inherent trade-off imposed by the mean variation regularizer, which smooths the sharpest near-field gradients to ensure well defined structural geometries. Overall, the low error magnitude across diverse harmonic modes verifies that the reconstruction is not merely visually plausible but statistically consistent with the experimental reference.

\begin{figure*}
    \centering
    \begin{subfigure}{0.7\textwidth}
    \includegraphics[trim={7cm 0 14cm 2cm},clip,width=\textwidth]{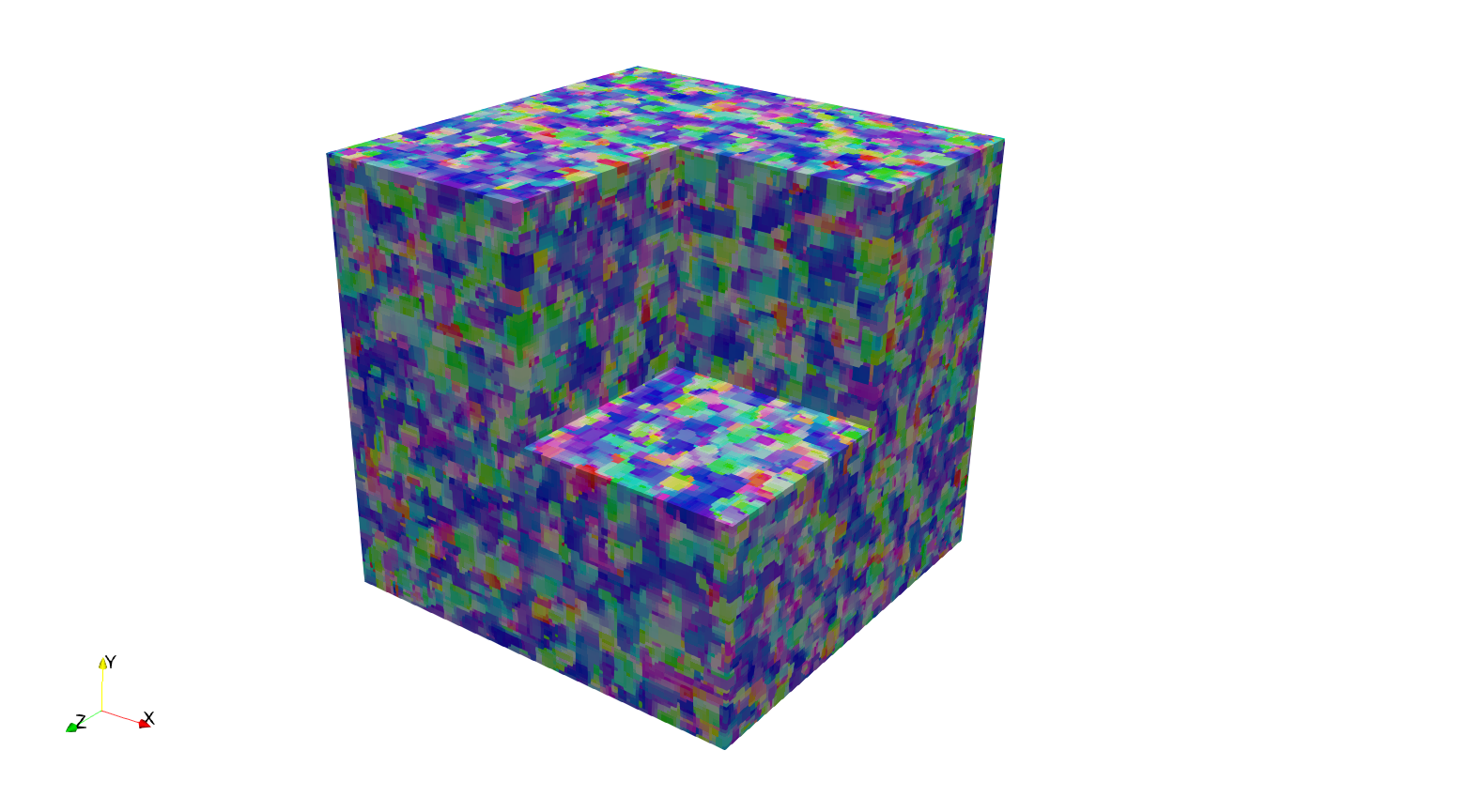}
    \caption{}
    \end{subfigure}
    \begin{subfigure}{0.7\textwidth}
    \begin{overpic}[width=\linewidth]{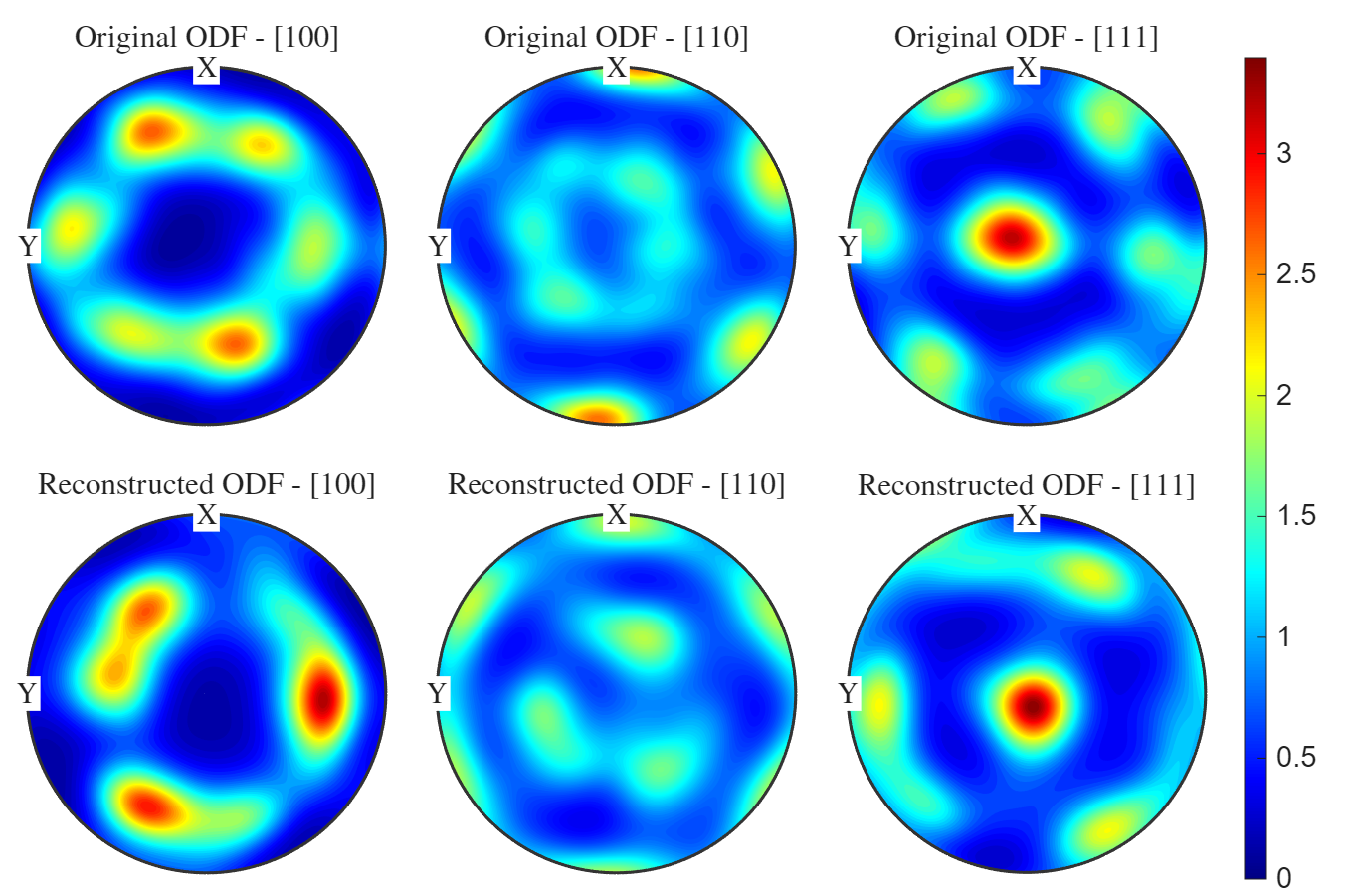}
      \put(1000,180){\rlap{\ \rotatebox{90}{Texture intensity (m.r.d.)}}}
    \end{overpic}
    \caption{}
    \end{subfigure}
    \caption{3D realization of an orientation field reconstructed from a single 2D experimental slice, showing (a) the reconstructed 3D orientation map and (b) the comparison of original and reconstructed $\{001\}$, $\{011\}$, and $\{111\}$ pole figures.}
    \label{fig:Case2_3D}
\end{figure*}

A primary advantage of the DMCR framework is its ability to generatively reconstruct 3D microstructures from a single 2D reference section. Fig.~\ref{fig:Case2_3D} illustrates this capability for the statistically homogeneous material introduced in Fig.~\ref{fig:FE_micros}(a), where the target descriptors $\mathbf{D}_i^\text{ref}$ for each orthogonal dimension are derived solely from the 2D input via the slice-loss formulation in Eq.~\eqref{eq:slice_loss}. As shown in Fig.~\ref{fig:Case2_3D}(a), the resulting 3D orientation map displays spatially coherent grain structures that successfully project the 2D morphological characteristics into the volumetric domain. The equiaxed topology is visually consistent throughout the reconstructed RVE which confirms that the SHSH-based descriptors provide sufficient constraints to infer isotropic 3D morphologies from planar data. Furthermore, despite the lack of direct volumetric measurements, the grains remain well-defined due to the mean variation regularizer’s effectiveness in suppressing numerical noise during the 3D optimization process to a certain extent.

The texture comparison in Fig.~\ref{fig:Case2_3D}(b) further validates this approach by showing good agreement between the reference and reconstructed $\{001\}$, $\{011\}$, and $\{111\}$ pole figures. Both the intensity distributions and the locations of key texture components are preserved, while the challenge of peak coalescence also persists in 3D. This correspondence confirms that the spectral coefficients extracted from a representative 2D section can accurately parameterize the ODF of the full 3D aggregate.

\subsubsection{Reconstruction of statistically inhomogeneous microstructures}\label{3.2.2}

\begin{figure*}
\centering
  \begin{subfigure}{0.32\textwidth}
    \includegraphics[width=\linewidth]{Case1-512-orig.png}
    \caption{}
  \end{subfigure}
  \hfill
   \begin{subfigure}{0.32\textwidth}
    \includegraphics[width=\linewidth]{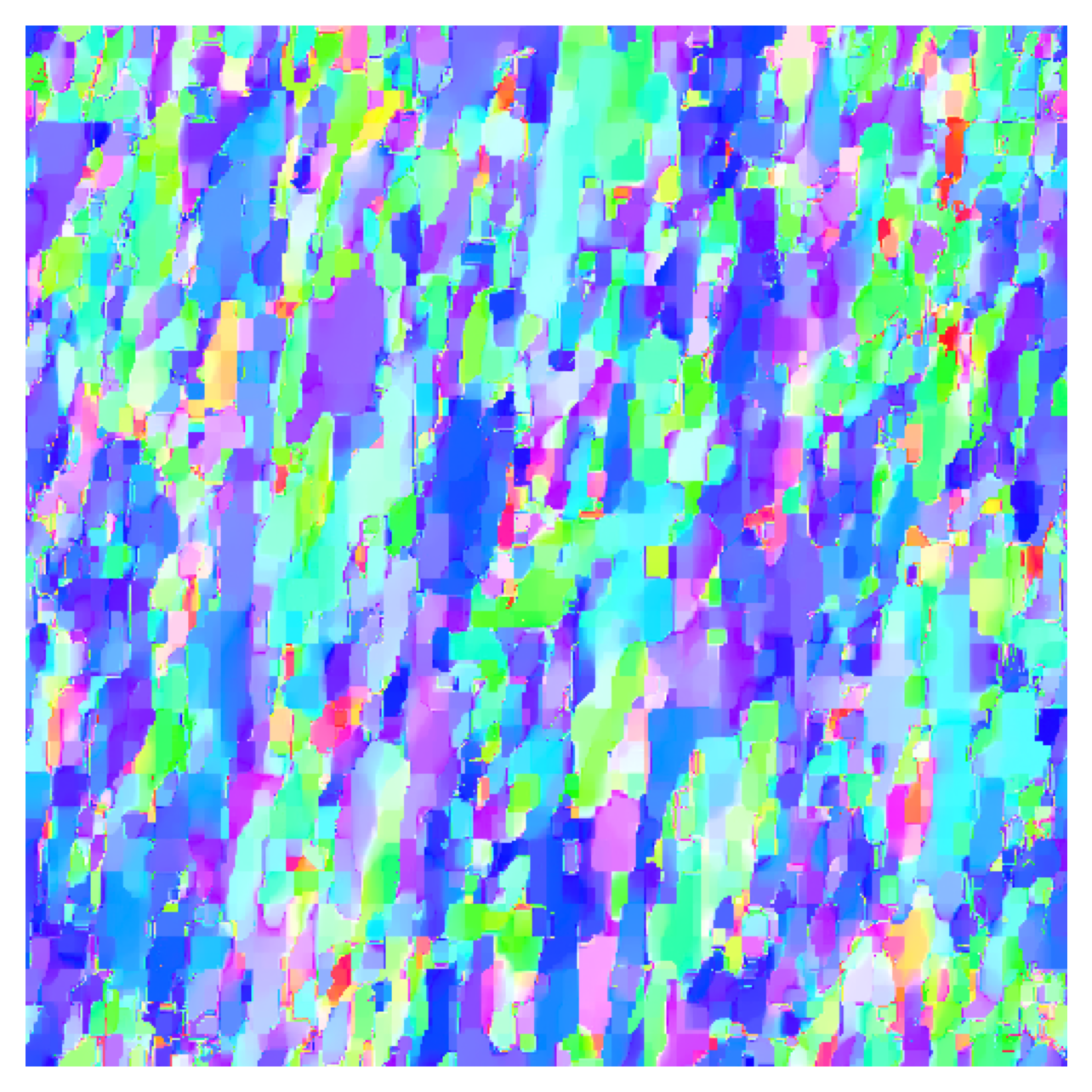}
    \caption{}
  \end{subfigure}
  \hfill
\begin{subfigure}{0.32\textwidth}
    \includegraphics[width=\linewidth]{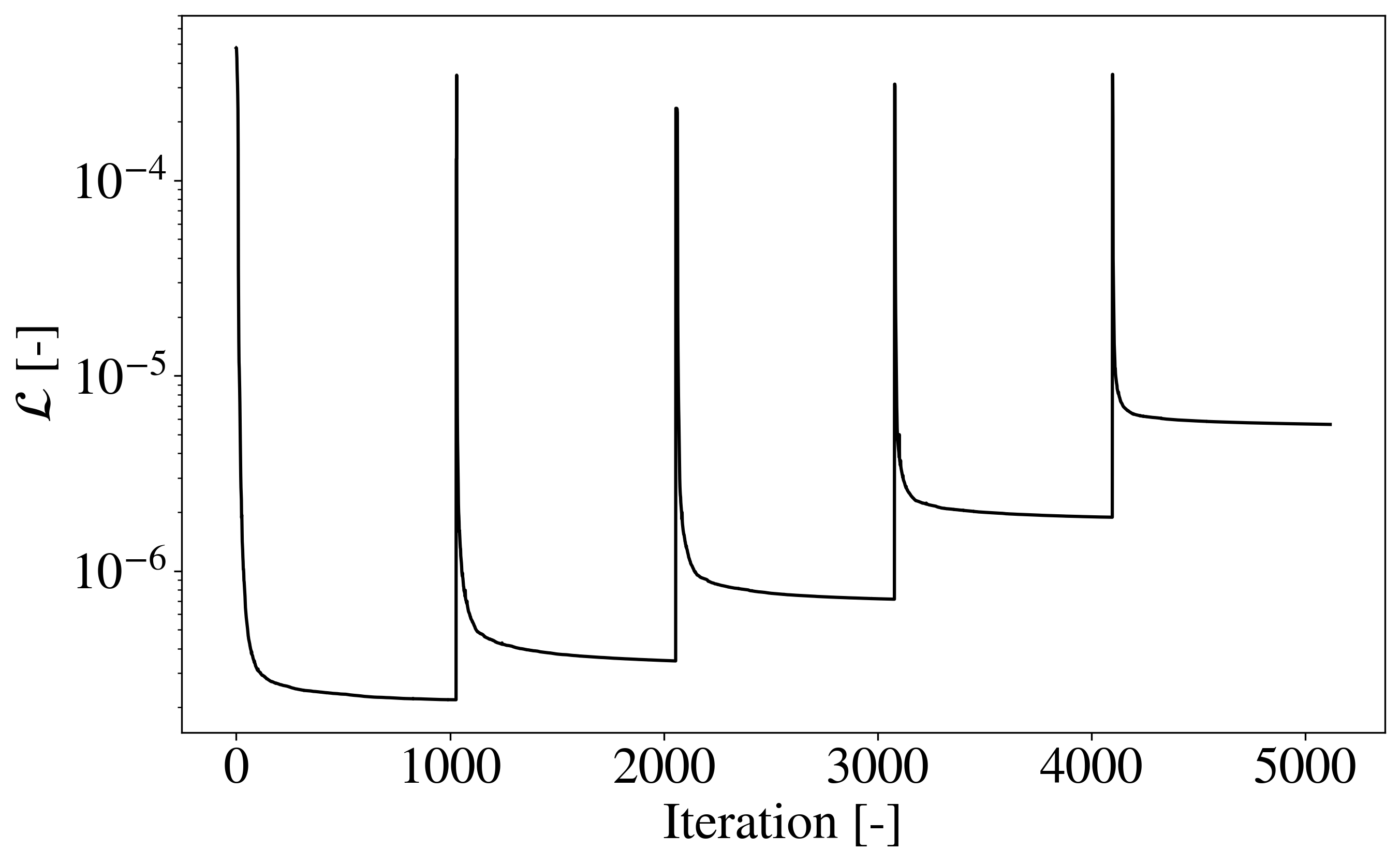}
    \caption{}
  \end{subfigure}
  \hfill
  \begin{subfigure}{0.8\textwidth}
    \begin{overpic}[width=\linewidth]{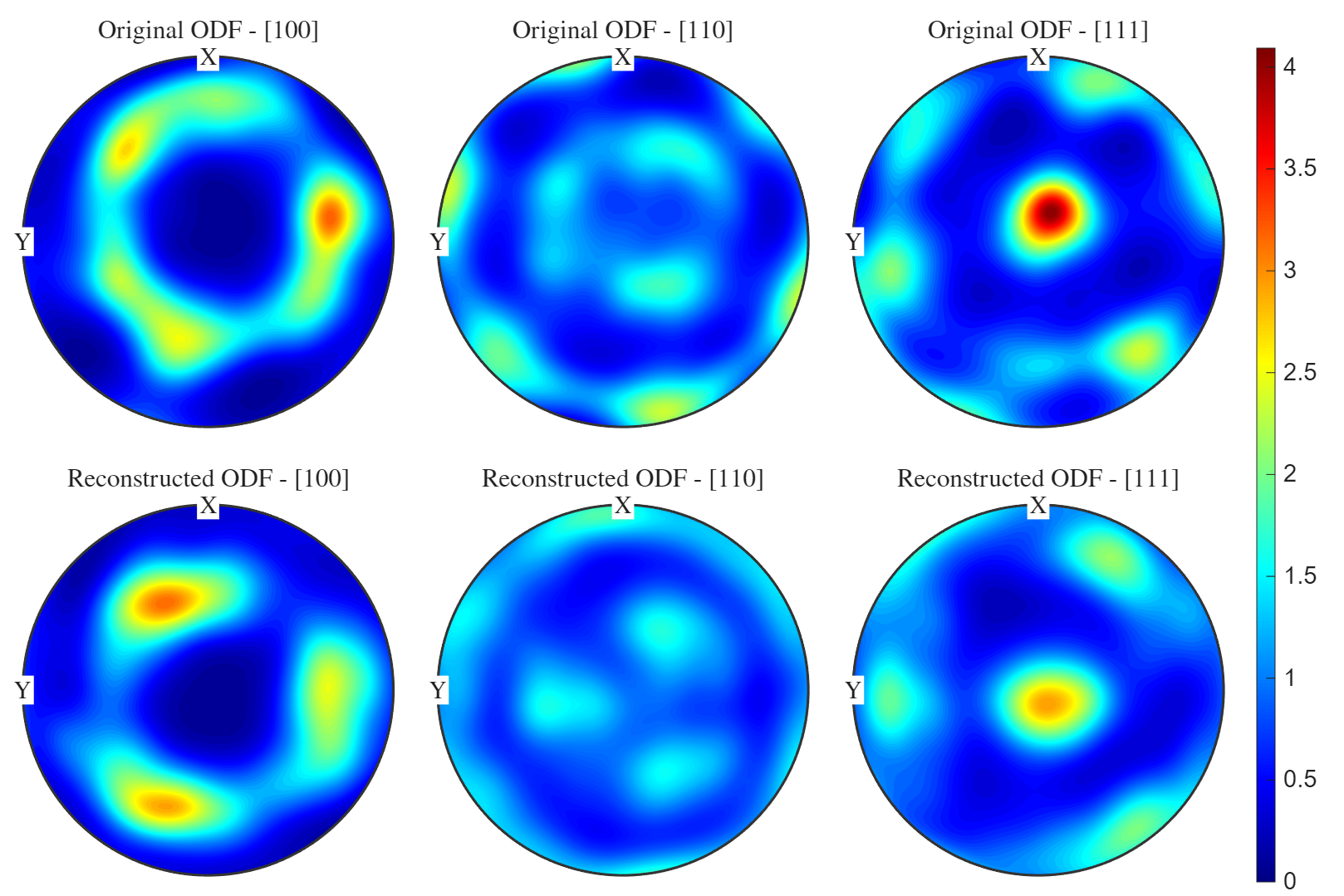}
      \put(1000,180){\rlap{\ \rotatebox{90}{Texture intensity (m.r.d.)}}}
    \end{overpic}
    \caption{}
  \end{subfigure}
  \hfill
    \caption{Reconstruction results for the statistically inhomogeneous microstructure, highlighting (a) the original EBSD map, (b) the synthetic reconstruction, (c) the convergence of the loss function during multigrid reconstruction, and (d) the comparison of original and reconstructed $\{001\}$, $\{011\}$, and $\{111\}$ pole figures.}
  \label{fig:Case1_recon}
\end{figure*}

The reconstruction of the second microstructure (Fig.~\ref{fig:FE_micros}(b)) provides a critical assessment of the framework’s performance limits when applied to statistically inhomogeneous orientation fields. Fig.~\ref{fig:Case1_recon} illustrates the results for a domain characterized by significant spatial gradients and localized grain elongation that typically arise during severe thermomechanical conditions in solid-state processing. While the DMCR framework successfully retrieves the primary global features of the target microstructure, as evidenced by the reasonable agreement of the pole figures in Fig.~\ref{fig:Case1_recon}(d), the local spatial inhomogeneities are largely homogenized. In the experimental reference, distinct vertical bands and graded transitions in grain aspect ratio are prominent. However, in the reconstruction, these localized features are replaced by a spatially stationary distribution. This occurs because the standard spatial correlation descriptors, such as the two-point correlation function and hybrid three-point variogram employed herein, are computed as ensemble averages over the entire domain. By inherently treating the microstructure as translation-invariant, these descriptors average localized gradients into a single global statistic. Consequently, the large, elongated grains are reconstructed as clusters of smaller grains that satisfy the average correlation length but fail to reproduce the specific spatial evolution of the gradient. These findings underscore a challenge of stationary statistical descriptors. While the global texture is recovered, the spatial position-dependent heterogeneity is lost.

\subsection{Current challenges}\label{3.3}

A key benefit of descriptor-based DMCR is the ability to interpolate within the descriptor space to explore novel microstructures. While this has been demonstrated for multiphase systems \cite{seibert_descriptor-based_2022}, its application to crystallographic orientation space can yield non-intuitive outcomes. Appendix~\ref{app2} illustrates this through two hypothetical, statistically homogeneous microstructures consisting of a field of circles (Fig.~\ref{fig:Interpolation_study}(a)) and a field of ellipsoids (Fig.~\ref{fig:Interpolation_study}(b)), where two distinct orientations are present, i. e. $\boldsymbol{g}_0$ for the particles and $\boldsymbol{g}_1$ for the matrix. In a purely geometric sense, an intuitive interpolation would be expected to produce intermediate morphologies, such as ellipsoids with reduced aspect ratios. However, the five equidistant interpolations shown in Fig.~\ref{fig:Interpolation_recon} reveal a more complex behavior. The realizability of hypothetical descriptors, such as those acquired from descriptor interpolation, relies on satisfying necessary conditions that ensure a valid physical configuration exists for a given set of statistical parameters. While certain necessary conditions have been established, as for instance, that microstructural states must become uncorrelated at large separation distances \cite{seibert_microstructure_2025, stillinger_pair_2004, torquato_necessary_2006}, sufficient conditions remain undefined. This challenge is amplified in the context of SHSH-based reconstruction due to the dimensionality of the spectral descriptor space. Thus, the interpolation does not move linearly between simple geometric shapes. Instead, the resulting microstructures often feature a mixture of both ellipsoidal and circular morphologies, while simultaneously interpolating through the orientation space, as highlighted by the evolving IPF coloring. These results suggest that SHSH-based interpolation traverses a high-dimensional design space where morphological and crystallographic features are fully coupled. For materials design and optimization tasks, these effects must be carefully accounted for to ensure that the generated pathways between microstructural states are physically well-behaved.

While the integration of higher-order statistical descriptors, such as the hybrid three-point variogram, enhances the recovery of morphological features, several critical challenges persist in the practical deployment of the DMCR framework. As demonstrated in the convergence analysis in Section~\ref{3.1}, the refinement of detailed grain shapes occurs predominantly in the late stages of optimization, where the global texture has already stabilized and the optimizer focuses on resolving interface curvature and sharpness. This final convergence phase is computationally demanding and requires a disproportionately large number of iterations to correct relatively small, but topologically important, features. In principle, the tensor-based implementation in TensorFlow is well suited to exploit GPU acceleration. In practice, however, the choice of optimizer becomes a major bottleneck. The SciPy implementation of the quasi-Newton method L-BFGS-B does not natively benefit from GPU resources, yet it remains the best-performing optimizer in this study, as shown in Fig.~\ref{fig:Optimizer_comp} of Appendix~\ref{app3}. First-order methods such as Adam, Adagrad, or Adamax could, in principle, be tuned to approach similar loss levels while fully utilizing GPU hardware. However, this would likely come at the cost of generality as each target microstructure defines a distinct loss landscape, so that case-specific hyperparameter tuning would be required. Other higher-order methods, such as the Barziali-Borwein algorithm \cite{barziali_two-point_1988}, show comparable performance in texture optimization as recently discussed by Krause et al. \cite{krause_generating_2026}, while maintaining higher memory efficiency compared to the L-BFGS-B algorithm, making them particularly attractive for potentially improving the efficiency of the presented DMCR of orientation maps.

Furthermore, the reconstruction problem is inherently ill-posed and non-convex, leading to significant path dependency in the optimization trajectory. Because the descriptor loss landscape has extensive local minima, the final microstructural realization is highly sensitive to the initial random seed. As detailed in Fig.~\ref{fig:Convergence_reproducibility} of Appendix~\ref{app4}, identically configured reconstructions initiated from different noise fields can converge to states that differ by an order of magnitude in terminal loss, even when utilizing the same optimizer and descriptor weights. This variance indicates that certain initialization paths may become trapped in suboptimal loss trajectories, where the solver cannot simultaneously satisfy the competing constraints within the descriptors. Consequently, achieving an optimal realizable reconstruction currently necessitates an ensemble approach of running multiple instantiations and post-selecting the realization with the minimum MSE, rather than relying on a single optimization run. This will result in a computational complexity that will be important to address in particular for high-throughput 3D reconstructions.

\section{Conclusion}

In this work, the DMCR framework in MCRpy has been extended by the gradient-based optimization approach and the generative synthesis of 2D and 3D polycrystalline microstructures from experimental EBSD data, directly using the embedded orientation information. By defining the inverse problem through the quaternion orientation representation and the descriptor assembly with respect to the SHSH, this methodology offers a versatile and modular framework for the implementation of various microstructure descriptors and optimization schemes. The key contributions and findings of this study are summarized as follows

\begin{itemize}
    \item Microstructure parametrization was performed using a continuous space of local crystallographic orientations with respect to the symmetry invariant SHSH. Unlike the Euler angle space (and its associated GSH representation) this formulation does not involve singularities, making it ideal for the use in DMCR.
    \item The integration of a novel hybrid three-point variogram alongside standard two-point correlation functions was shown to be critical for capturing complex morphological features. While the two-point correlations encode the spatial arrangement and cross-correlations between distinct SHSH modes, the hybrid three-point variogram targets higher-order interfacial configurations.
    \item Adding the mean variation as a descriptor acts as an essential regularizer to the stabilization of local features. By treating it as a target statistic rather than a minimization objective, the framework effectively suppresses noise while preserving the physically occurring intragranular features within the microstructures.
    \item Through the analysis of micrographs obtained from aluminum after thermo-mechanical processing, the framework's capability to reconstruct statistically homogeneous textures and, to a limited extent, graded microstructures was demonstrated. The results highlighted the inherent limitations of stationary descriptors in capturing local non-stationarity.
    \item Generative reconstruction of fully volumetric RVEs from a single 2D planar section was performed. By enforcing statistical consistency on cross-sectional slices, the algorithm infers isotropic 3D grain topologies that match the crystallographic texture of the reference material.
    \item Benchmarking against various optimization schemes identified the L-BFGS-B algorithm as the superior solver to converge rapidly. However, the sensitivity to initialization underscores the non-convex nature of the optimization problem and motivates ensemble-based strategies to mitigate path dependency.
\end{itemize}

Collectively, these advancements establish the orientation-based DMCR framework as a powerful and flexible extension of MCRpy for voxel-based MCR. Its modular architecture allows for the integration of novel descriptors, various optimization strategies, and the implementation of different distance metrics, offering a significant contribution to the ICME paradigm. By bridging the gap between crystallographic orientation data and gradient-based reconstruction, this work provides a foundation for the digital synthesis of polycrystalline materials.

\section*{Declaration of competing interest}
The authors declare that they have no known competing financial interests or personal relationships that could have appeared to influence the work reported in this paper.

\section*{Data availability}
The obtained data from this research will be made available on ZENODO.

\section*{Code availability}
The codes from this research will be made available on the MCRpy github page (\url{https://github.com/NEFM-TUDresden/MCRpy}).

\section*{Acknowledgements}
This project has received funding from the European Research Council (ERC) under the European Union’s Horizon 2020 research and innovation programme (grant  agreement No 101001567). This work was partially supported by German Research Foundation (DFG) within Priority Program SPP 2489 DaMic under project numbers 562138004, 562094916. The group of M. Kästner thanks the DFG which further supported this work under project number 496984632. The authors thank Jeremy Mason for providing the Matlab code to compute the symmetry coefficients to construct the SHSH.

\bibliography{MyBib}
\newpage
\appendix

\section{Stereographic projections of SHSH}\label{app1}

\begin{figure}[hb]
\centering
\begin{tabular}{ccccccc}
       & $\omega = 0 ^\circ$ & $\omega = 30^\circ$ & $\omega = 60^\circ$ & $\omega = 90^\circ$ & $\omega = 120^\circ$ & $\omega = 150^\circ$ \\
\rotatebox[origin=c]{90}{$\lambda = 1$}  & \shshimage{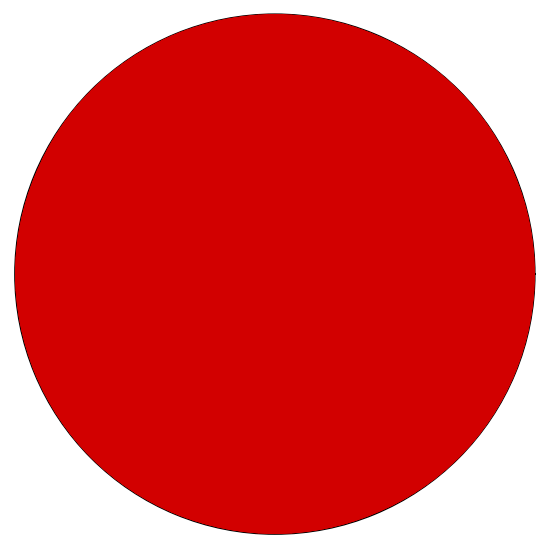} & \shshimage{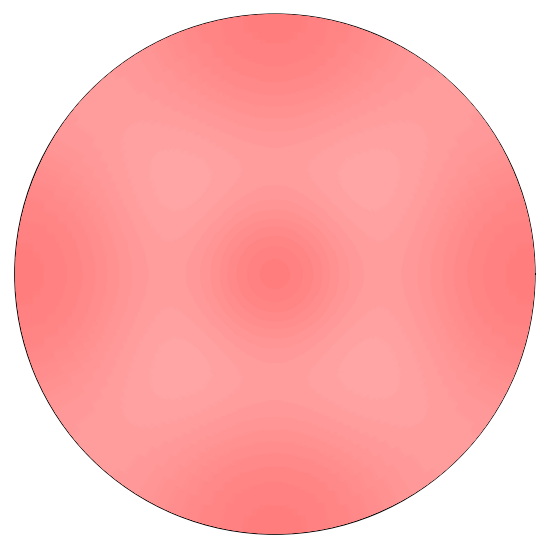} & \shshimage{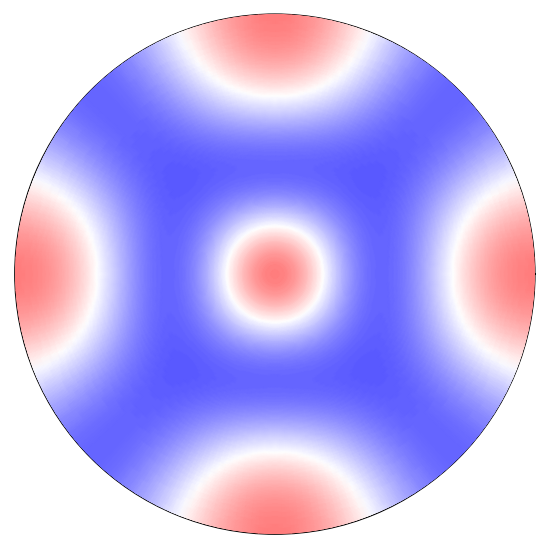} & \shshimage{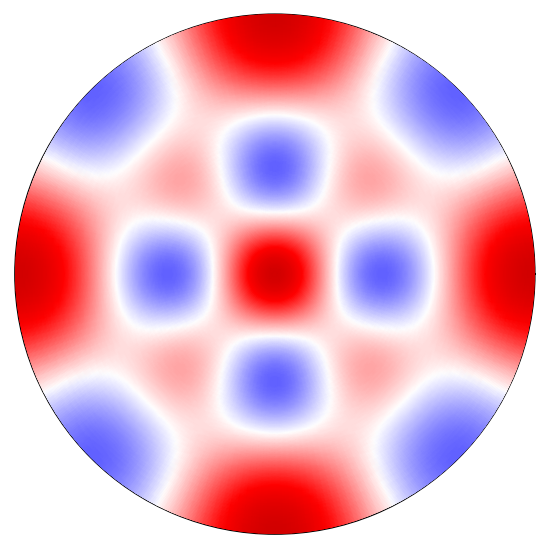} & \shshimage{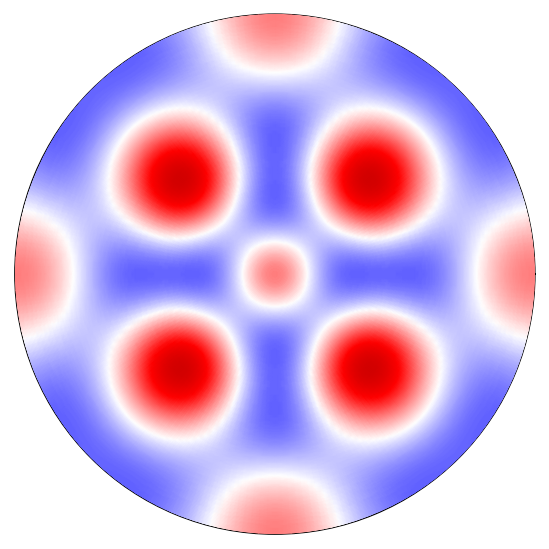} & \shshimage{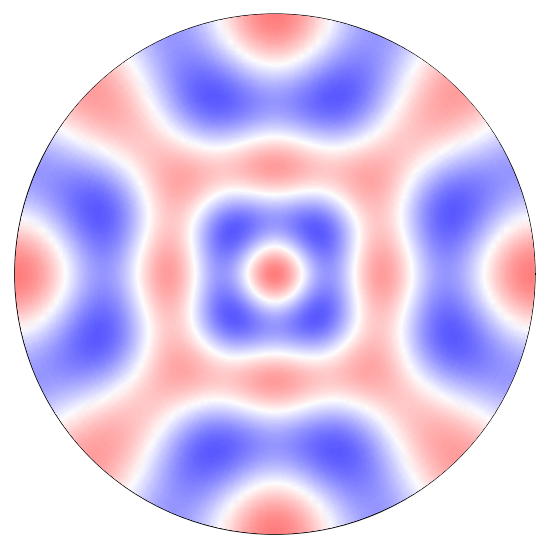} \\
\rotatebox[origin=c]{90}{$\lambda = 2$}  & \shshimage{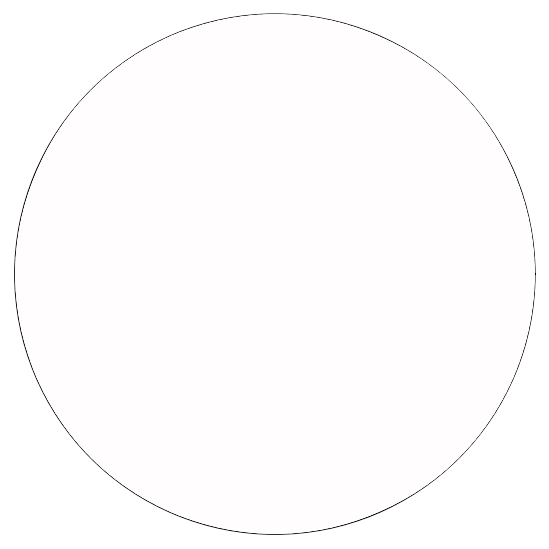} & \shshimage{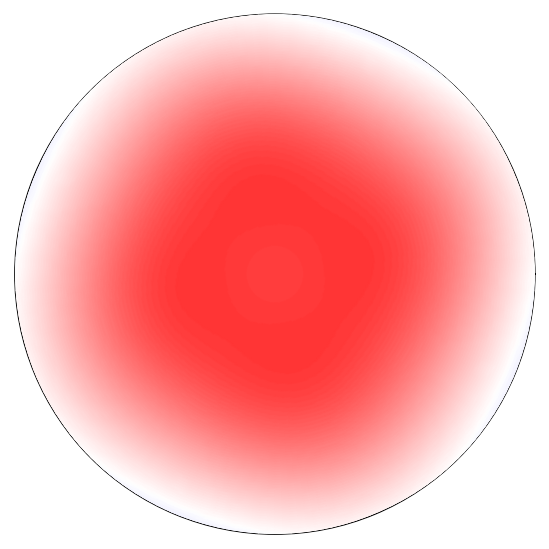} & \shshimage{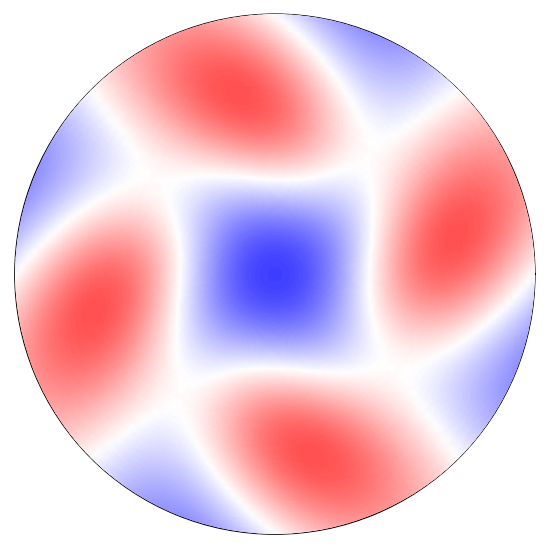} & \shshimage{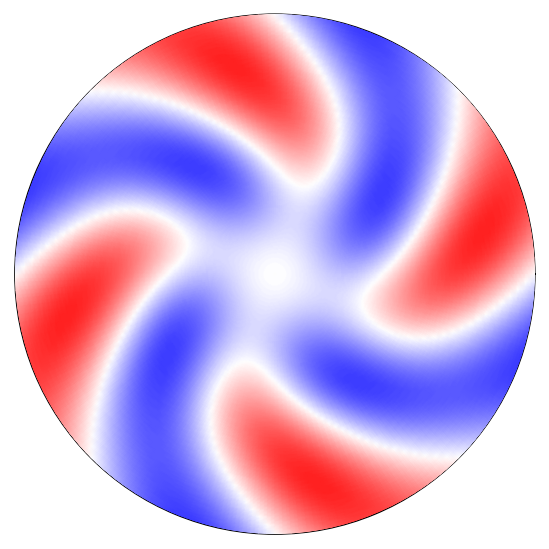} & \shshimage{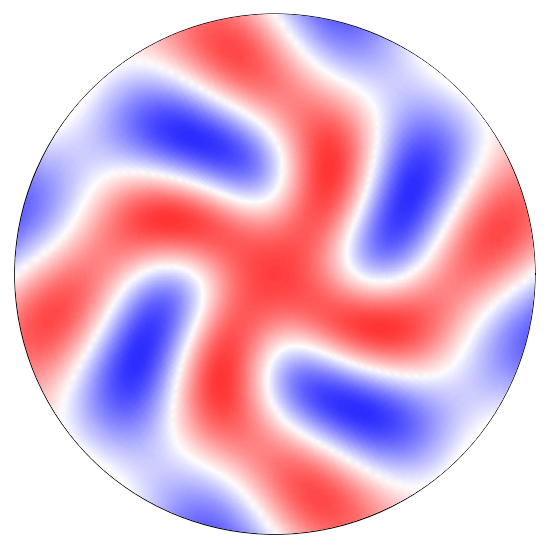} & \shshimage{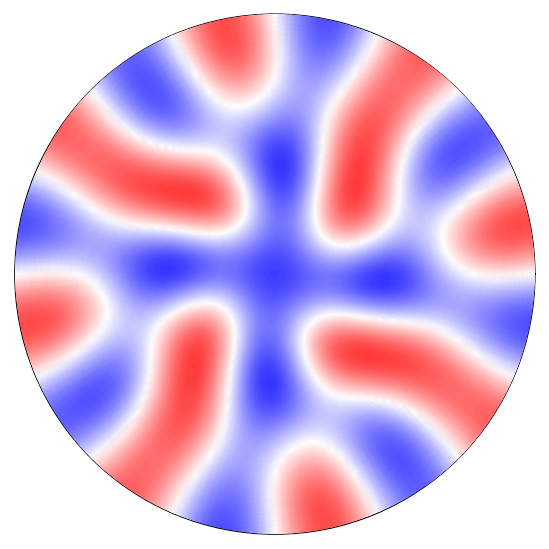} \\
\rotatebox[origin=c]{90}{$\lambda = 3$}  & \shshimage{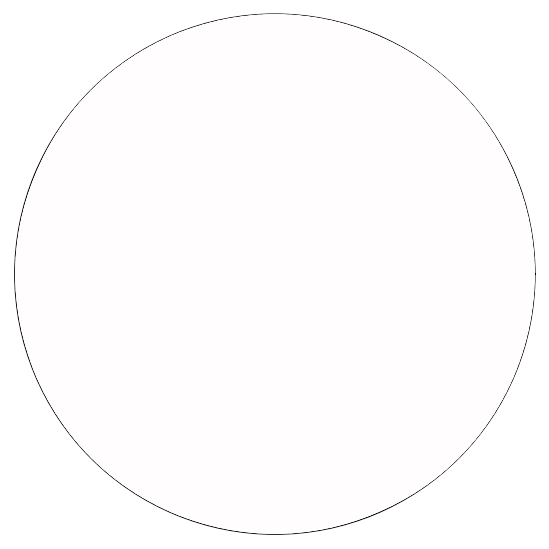} & \shshimage{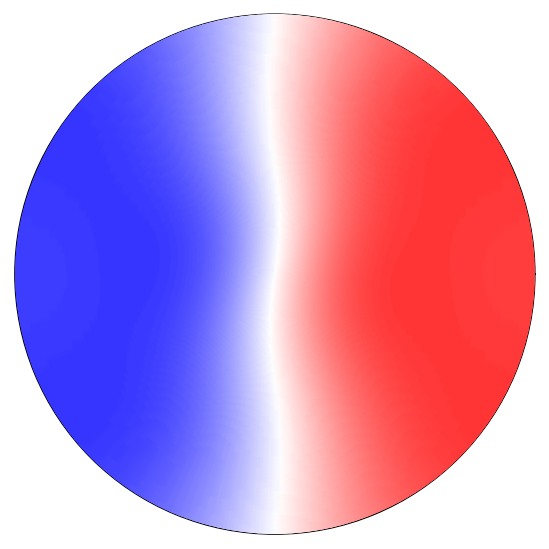} & \shshimage{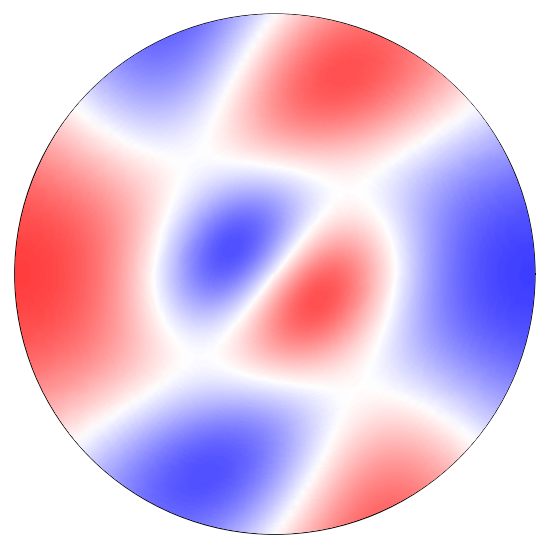} & \shshimage{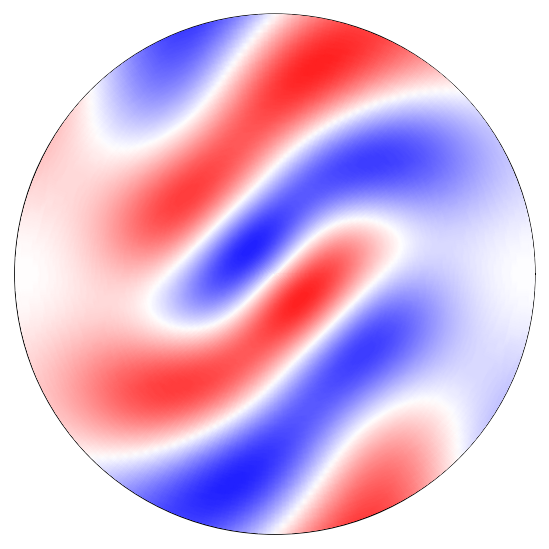} & \shshimage{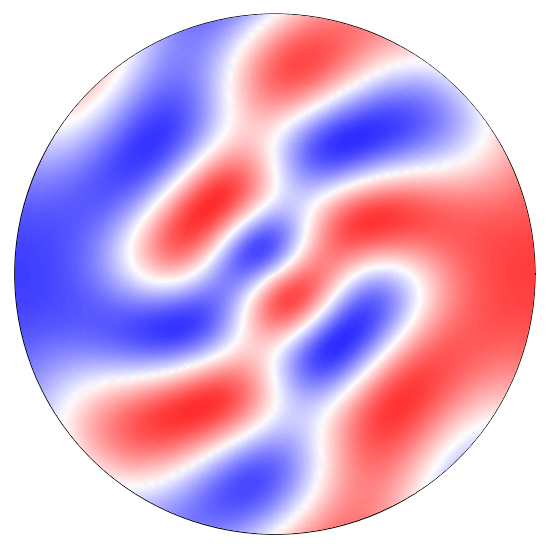} & \shshimage{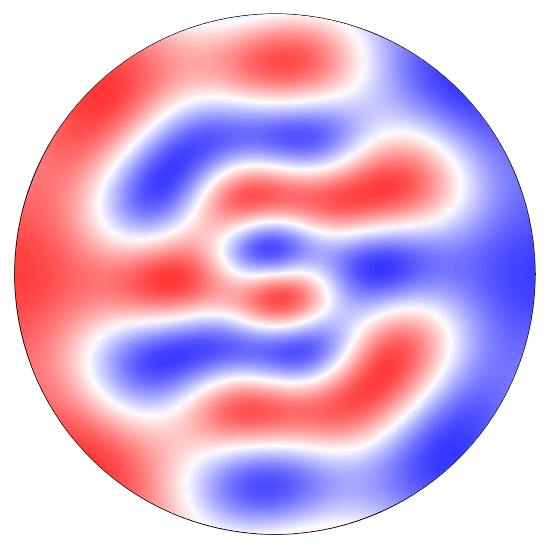} \\
\rotatebox[origin=c]{90}{$\lambda = 4$}  & \shshimage{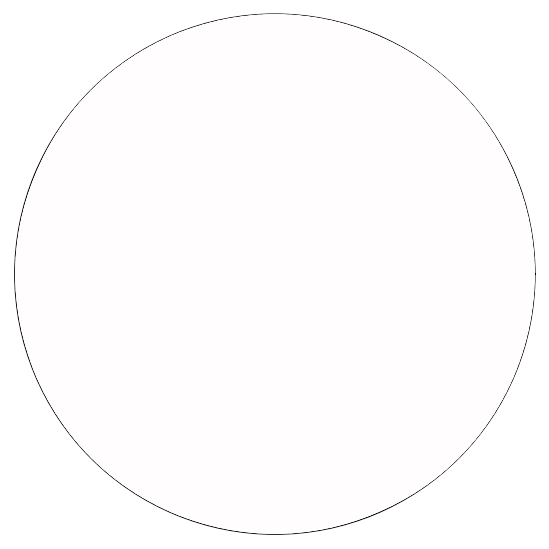} & \shshimage{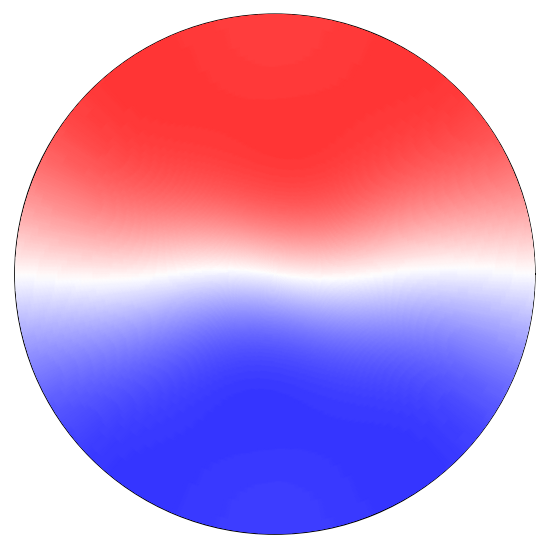} & \shshimage{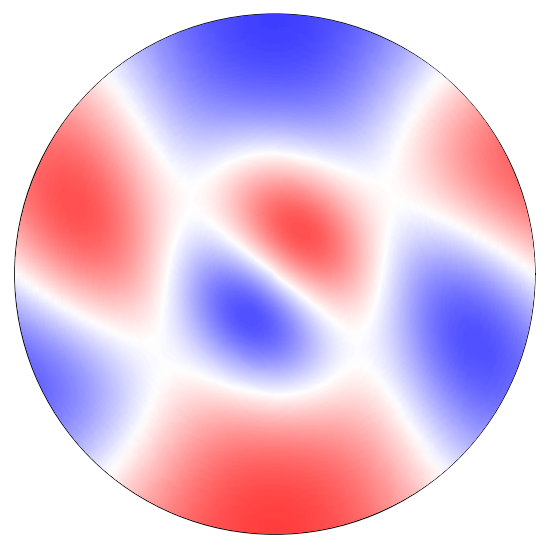} & \shshimage{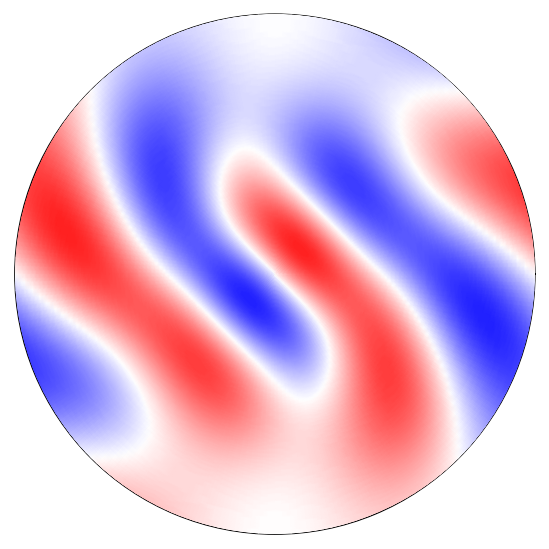} & \shshimage{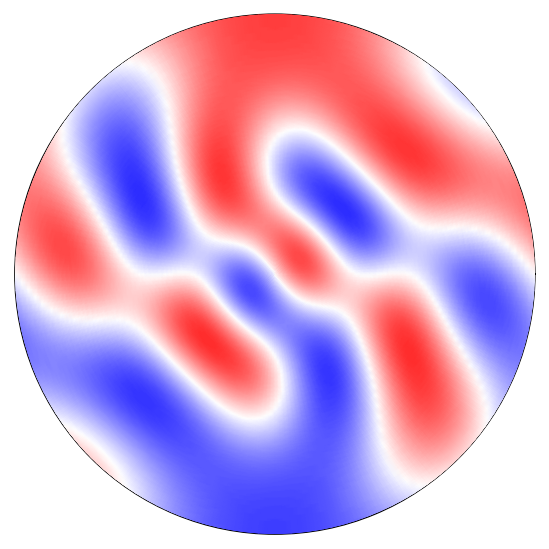} & \shshimage{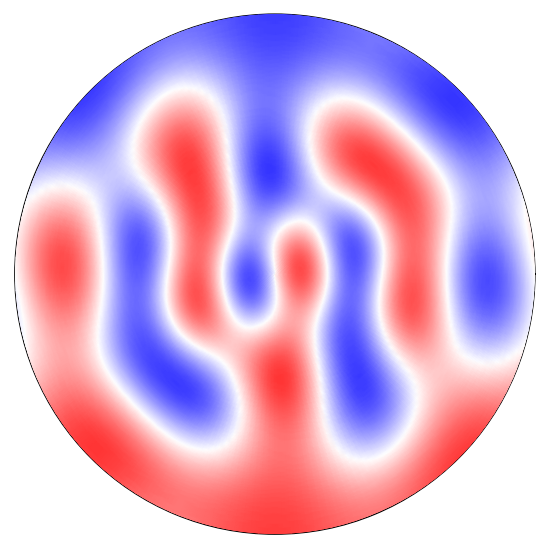} \\
\rotatebox[origin=c]{90}{$\lambda = 5$}  & \shshimage{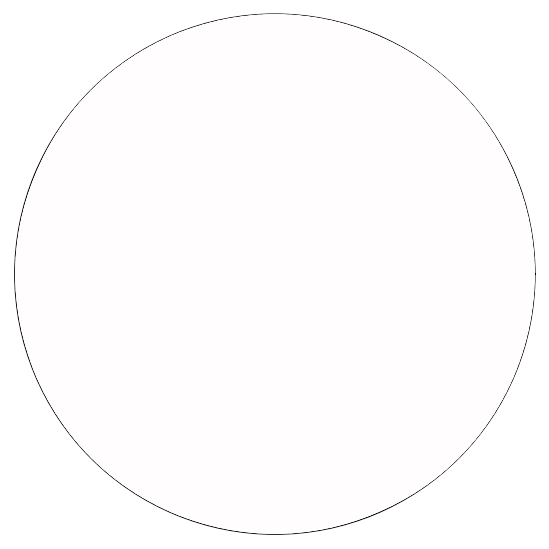} & \shshimage{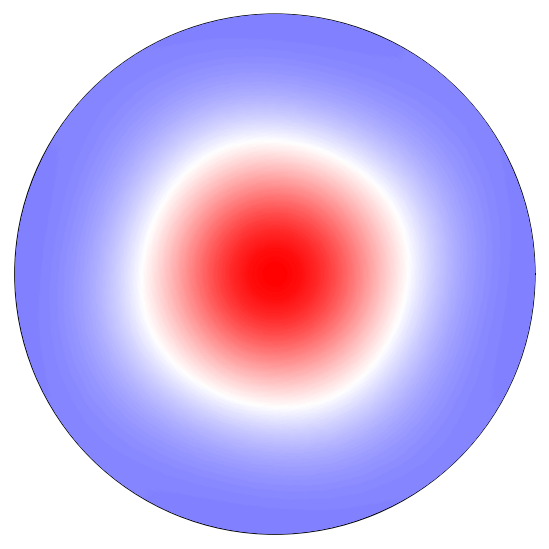} & \shshimage{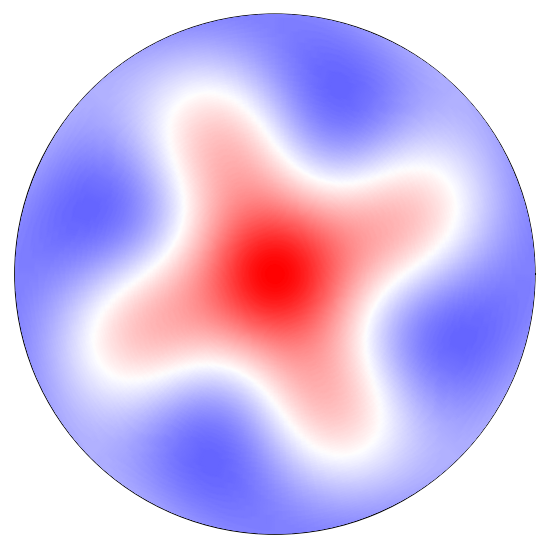} & \shshimage{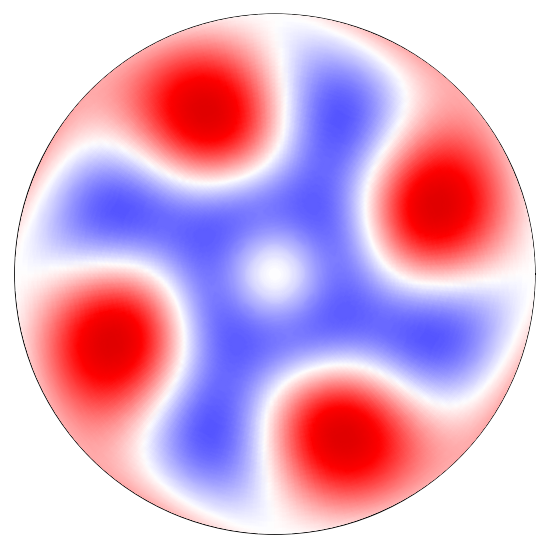} & \shshimage{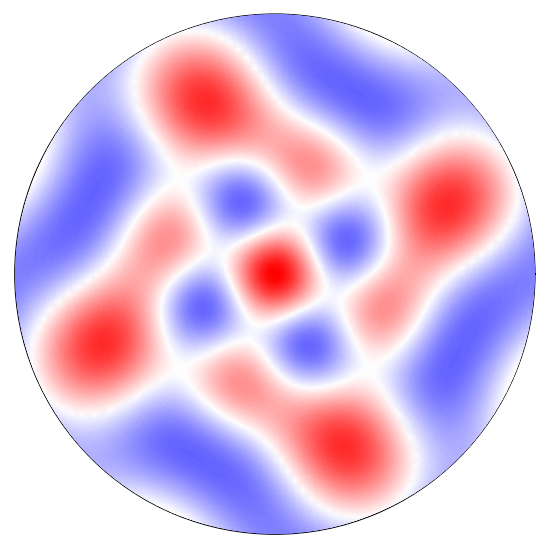} & \shshimage{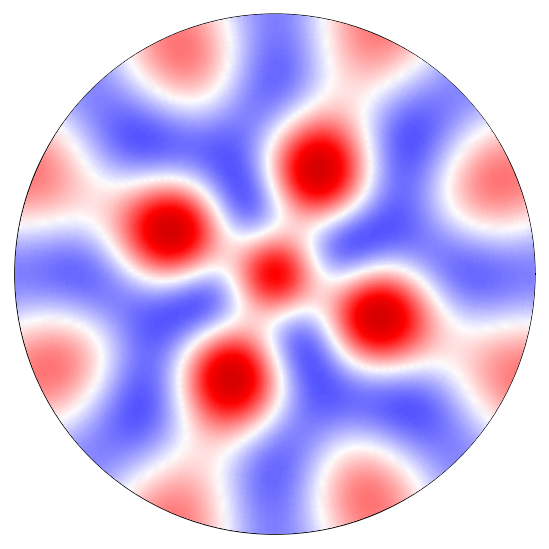} \\
\rotatebox[origin=c]{90}{$\lambda = 6$}  & \shshimage{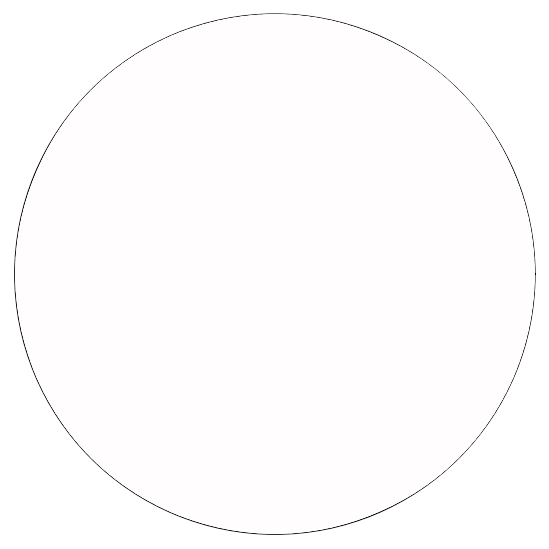} & \shshimage{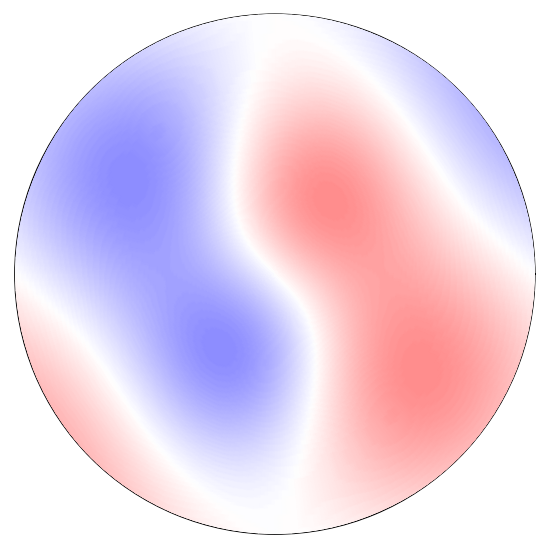} & \shshimage{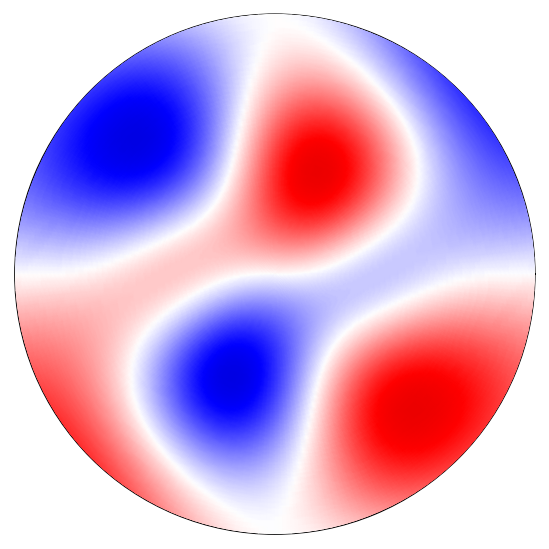} & \shshimage{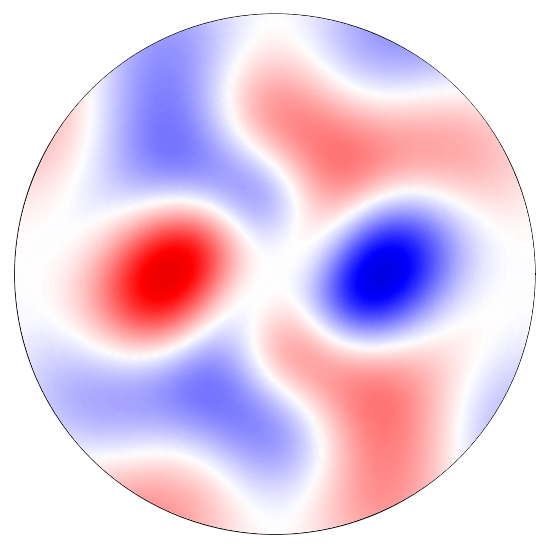} & \shshimage{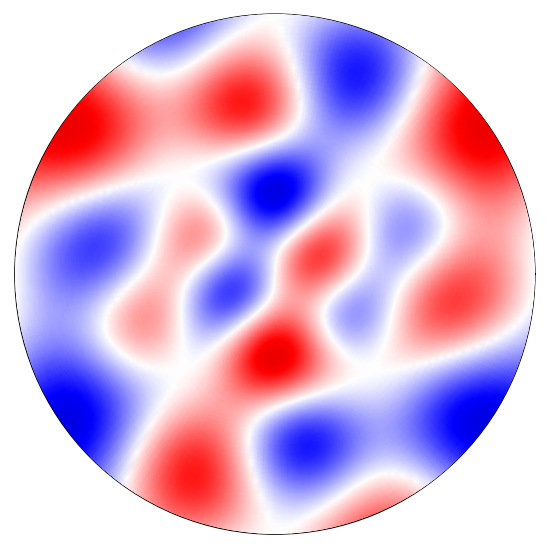} & \shshimage{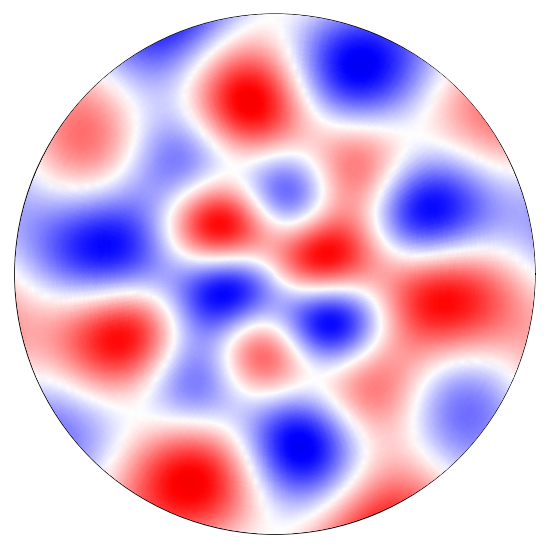} \\
\rotatebox[origin=c]{90}{$\lambda = 7$}  & \shshimage{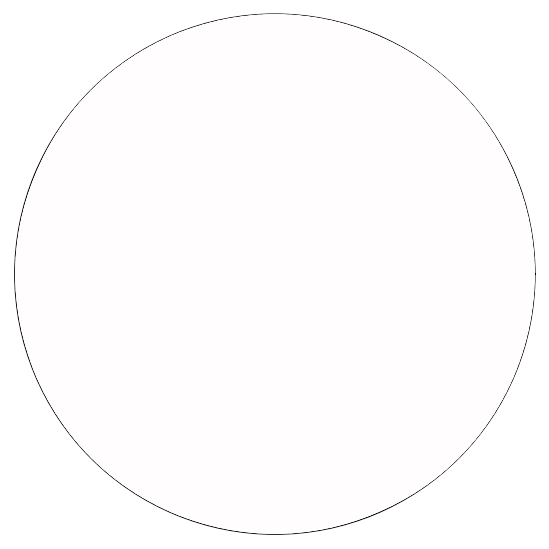} & \shshimage{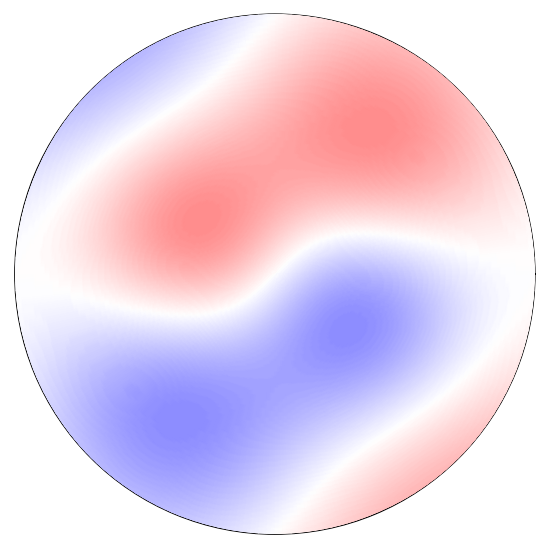} & \shshimage{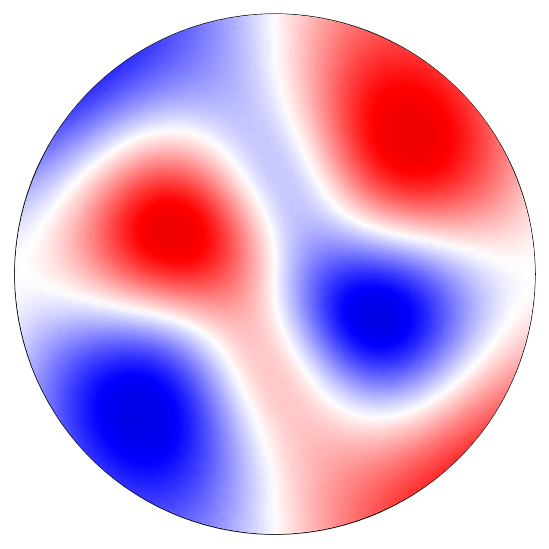} & \shshimage{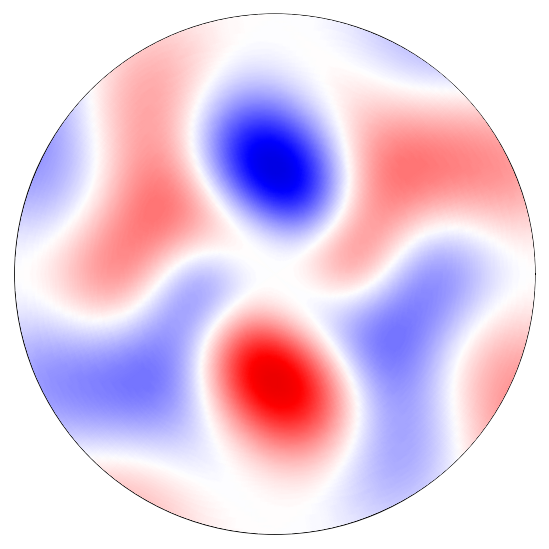} & \shshimage{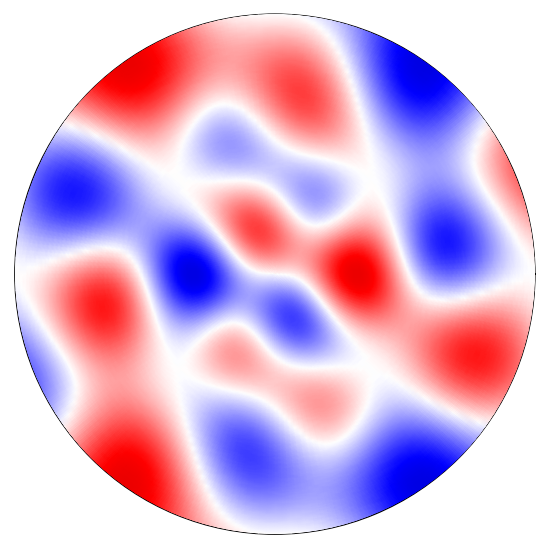} & \shshimage{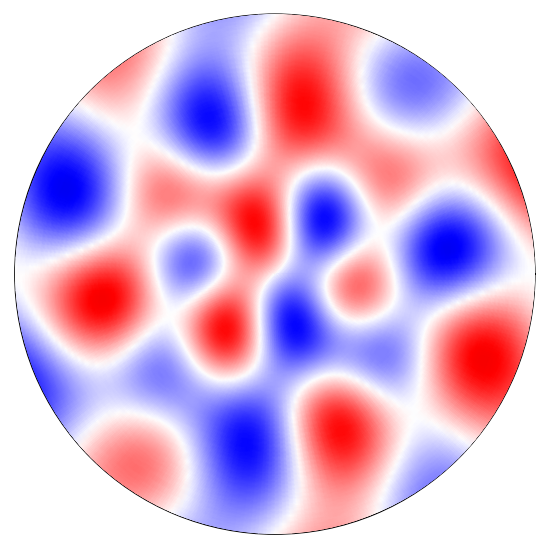} \\
\rotatebox[origin=c]{90}{$\lambda = 8$}  & \shshimage{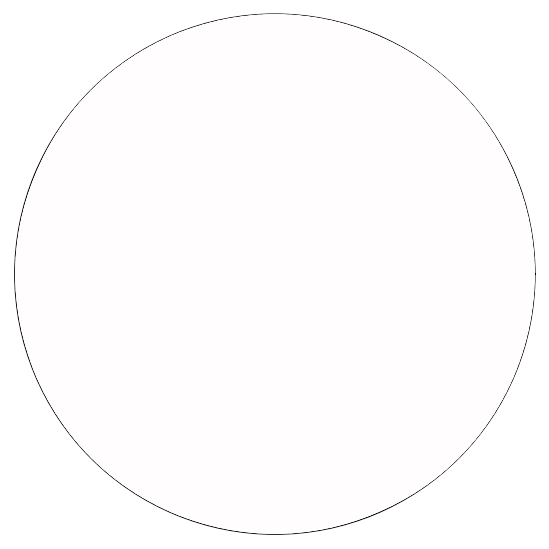} & \shshimage{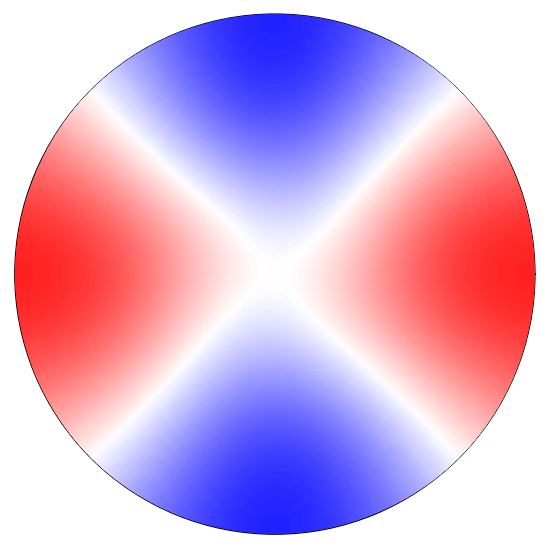} & \shshimage{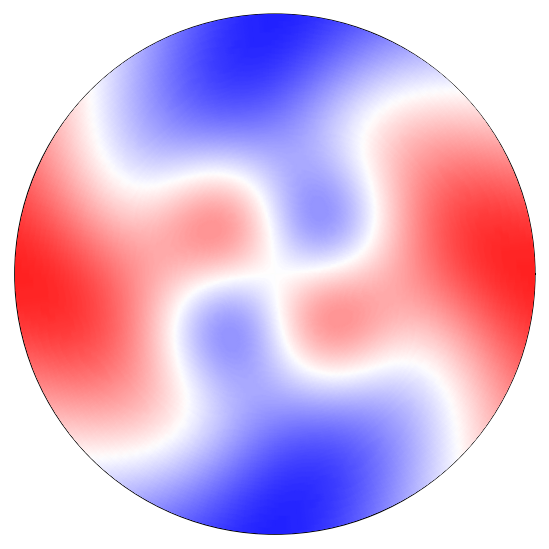} & \shshimage{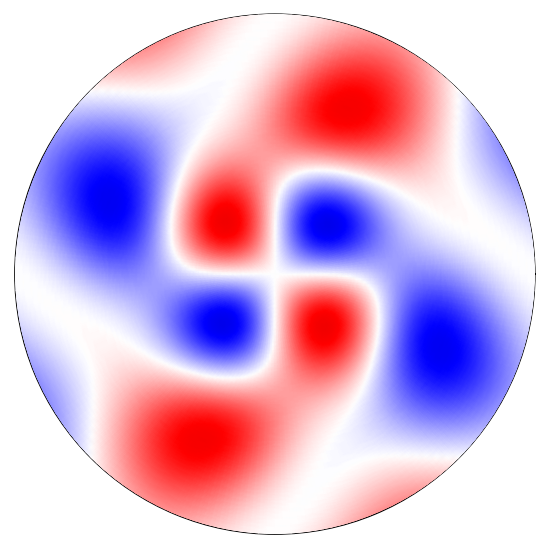} & \shshimage{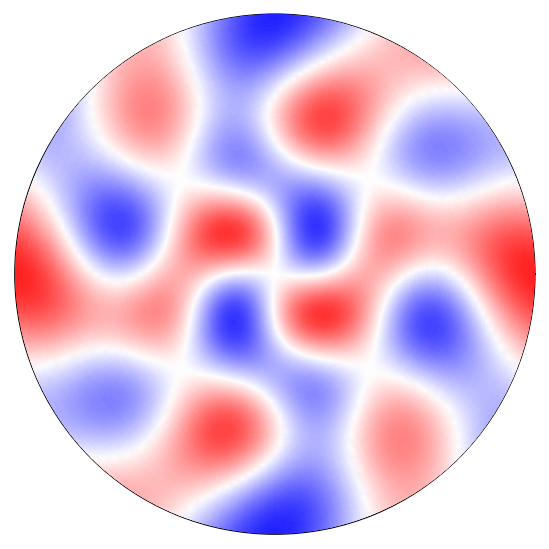} & \shshimage{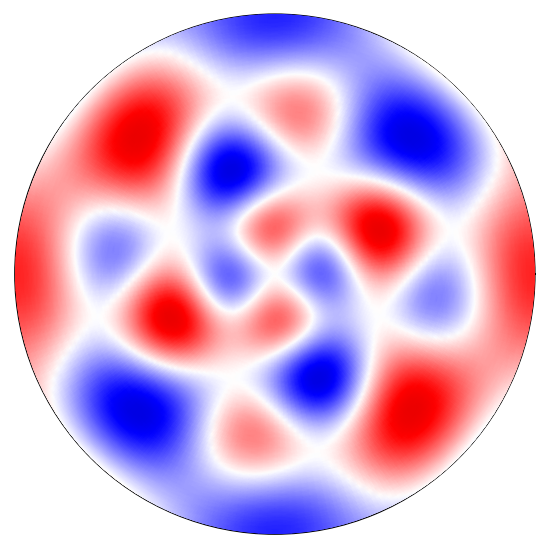} \\
\rotatebox[origin=c]{90}{$\lambda = 9$}  & \shshimage{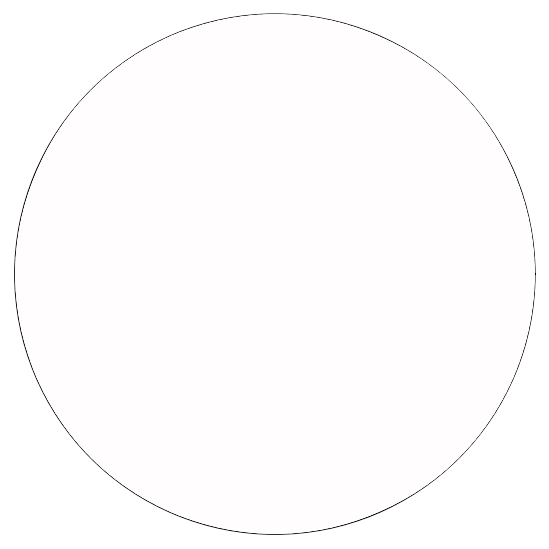} & \shshimage{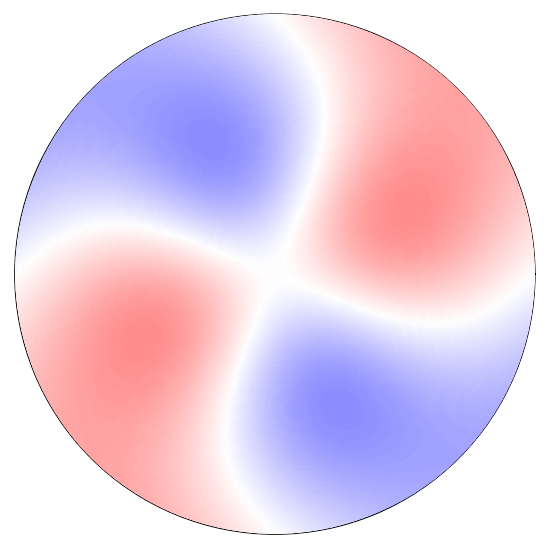} & \shshimage{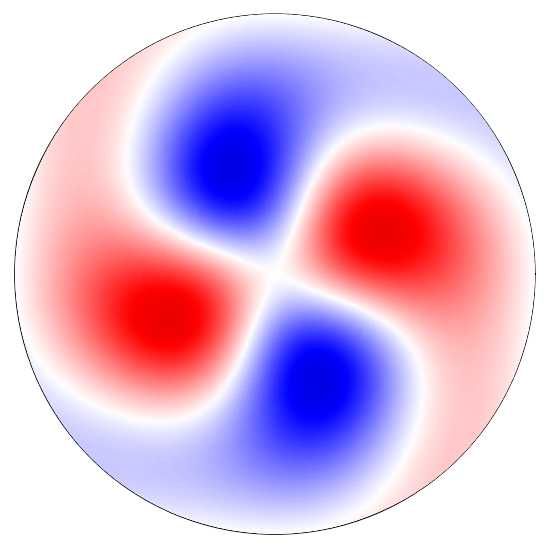} & \shshimage{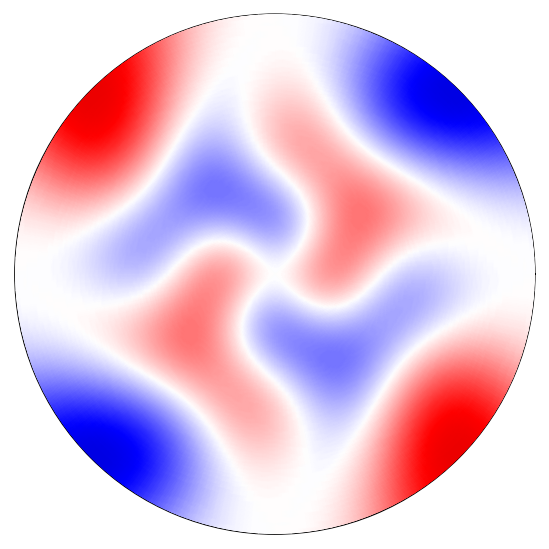} & \shshimage{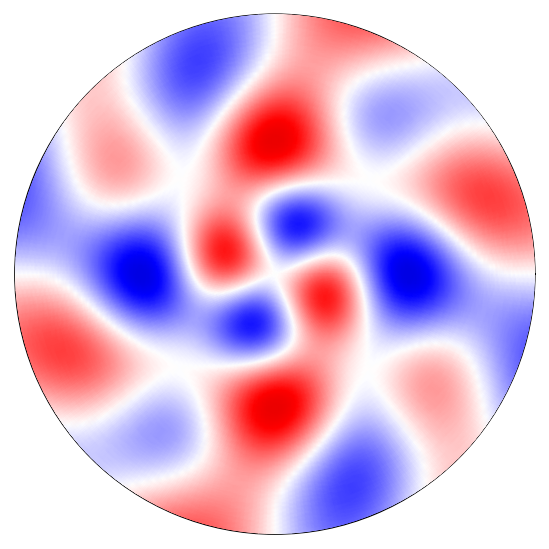} & \shshimage{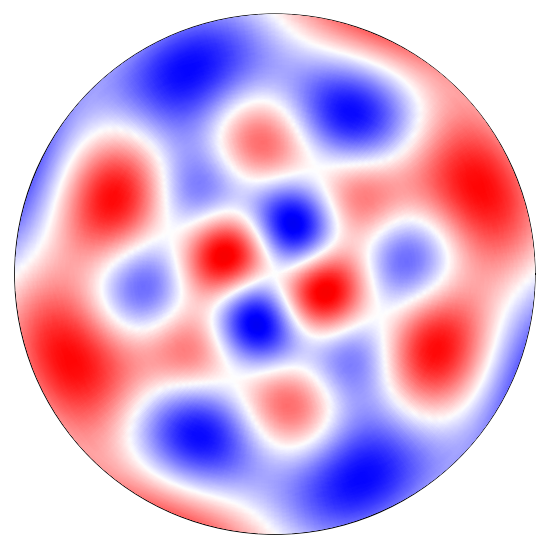} \\
\end{tabular}\\
\includegraphics[width=0.15\textwidth]{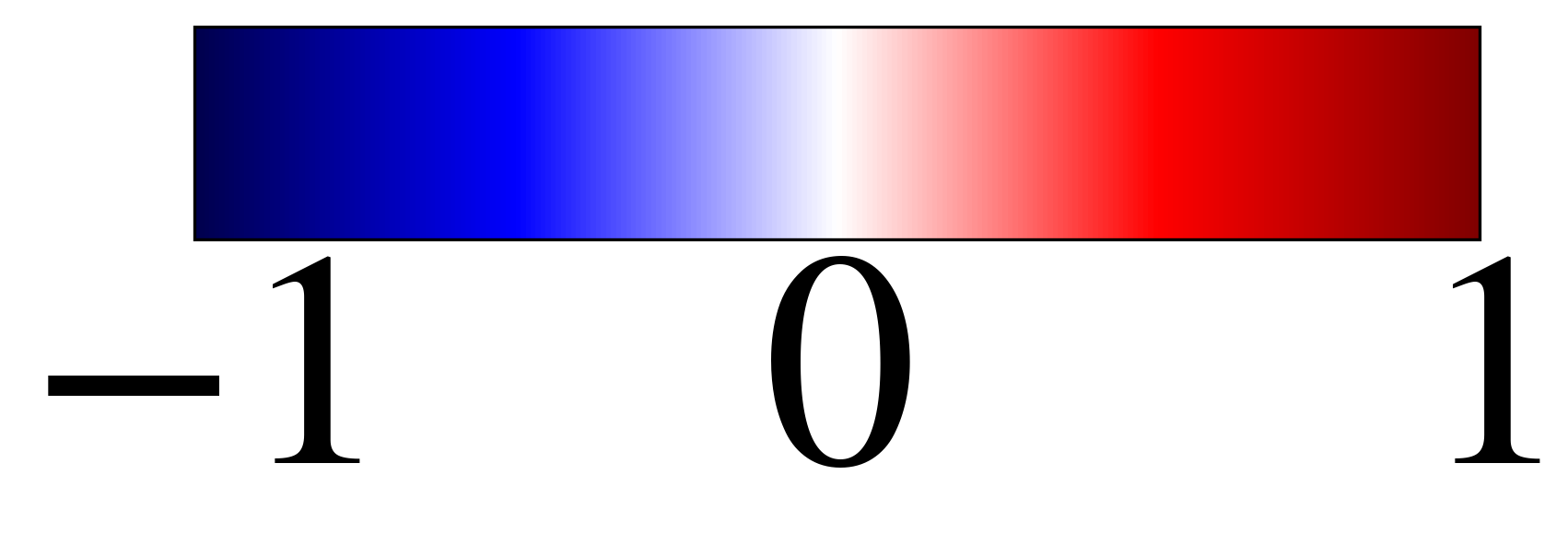}
\caption{Evaluation of the SHSH functions for the first nine $\lambda$ components at $n=8$ for cubic crystal symmetry \cite{mason_hyperspherical_2008, seibert_microstructure_2025}.}
\label{fig:SHSH_basefunction}
\end{figure}

The stereographic projections in Fig.~(\ref{fig:SHSH_basefunction}) illustrate how the SHSH basis functions vary for cubic symmetry. By showing the first nine SHSH components at harmonic degree \(n=8\), it can be seen that different basis functions capture distinct symmetry patterns while remaining smooth, and free of the discontinuities associated with conventional orientation parameterizations.

\begin{figure}[h]
\centering
\begin{tabular}{ccccccc}
       & $\omega = 180 ^\circ$ & $\omega = 210^\circ$ & $\omega = 240^\circ$ & $\omega = 270^\circ$ & $\omega = 300^\circ$ & $\omega = 330^\circ$ \\
\rotatebox[origin=c]{90}{$\lambda = 1$}  & \shshimage{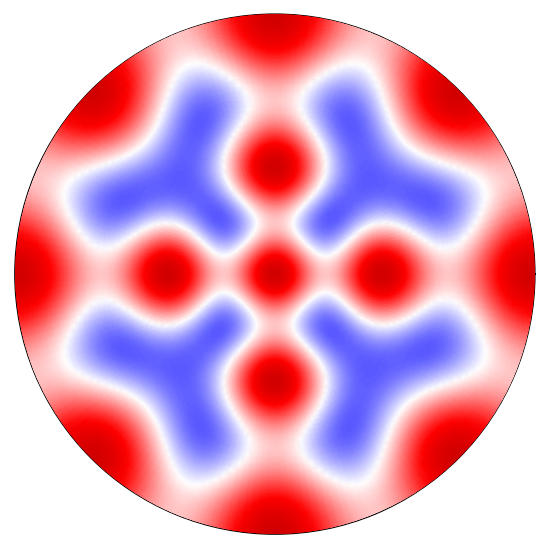} & \shshimage{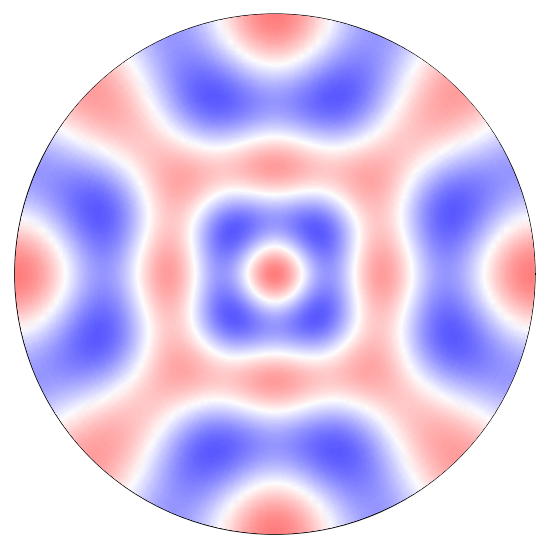} & \shshimage{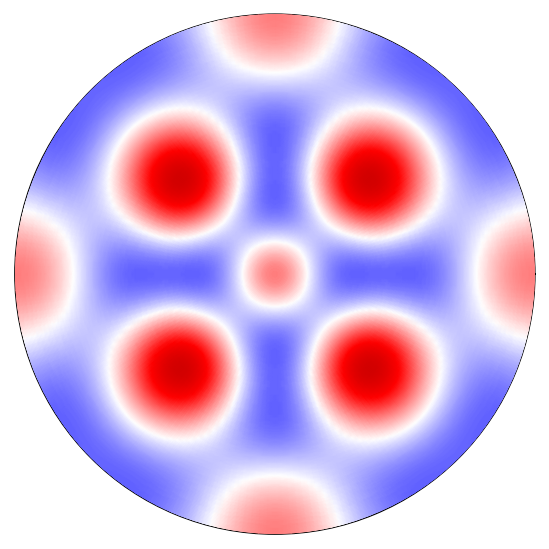} & \shshimage{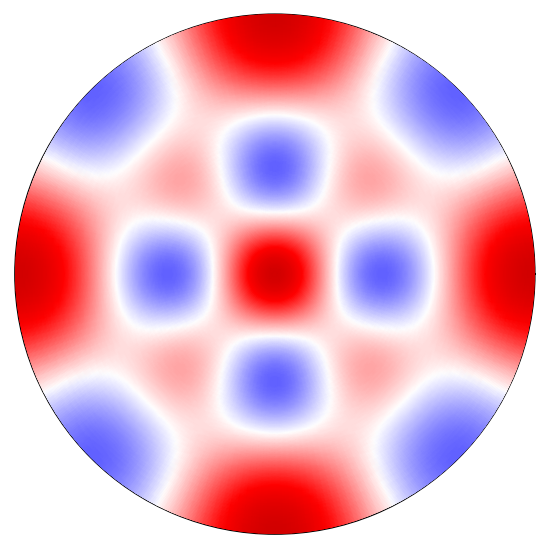} & \shshimage{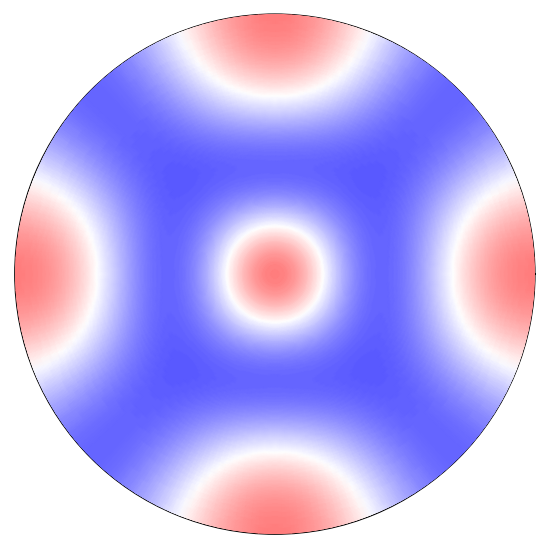} & \shshimage{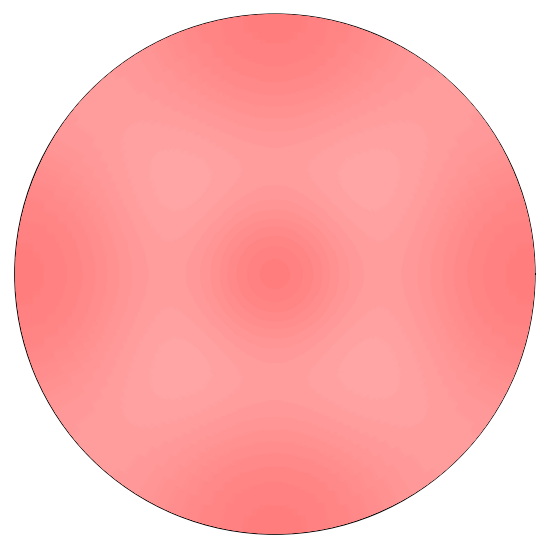} \\
\rotatebox[origin=c]{90}{$\lambda = 2$}  & \shshimage{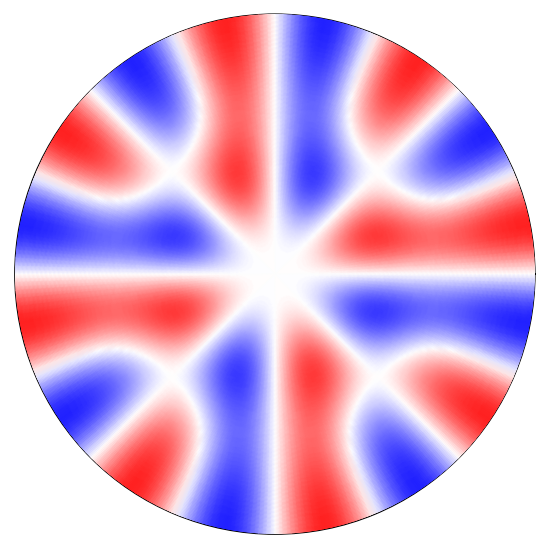} & \shshimage{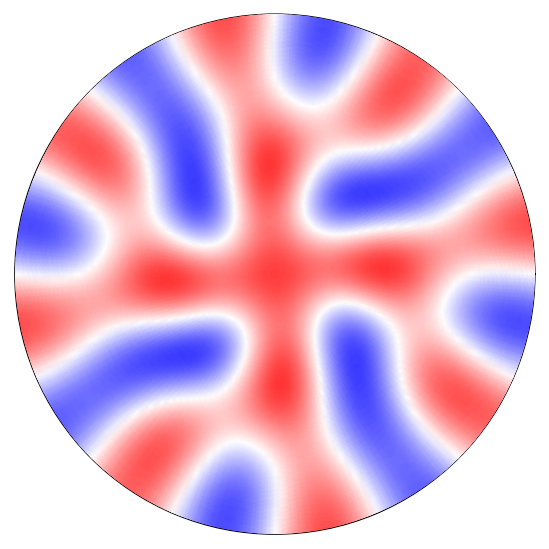} & \shshimage{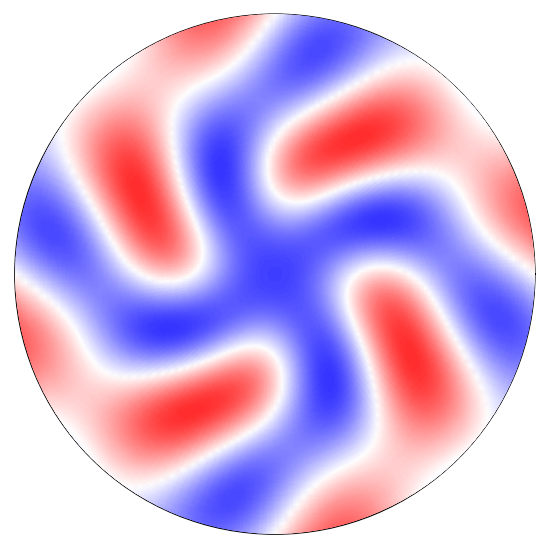} & \shshimage{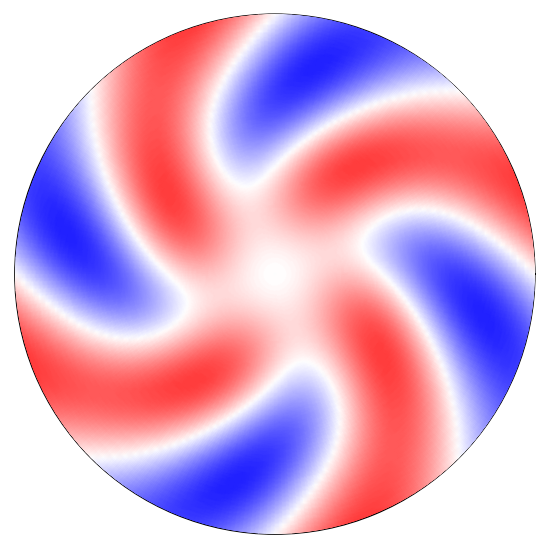} & \shshimage{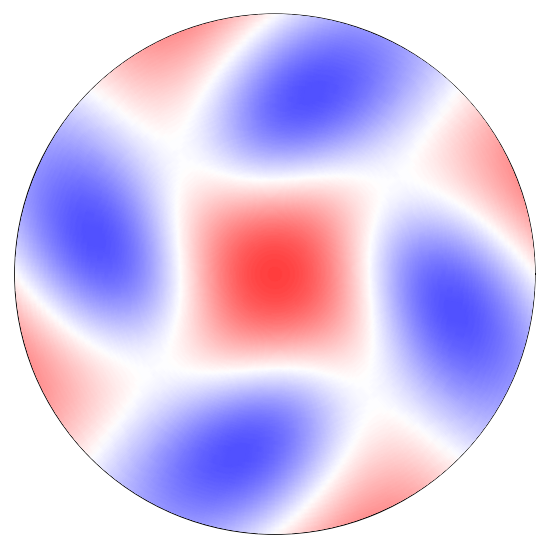} & \shshimage{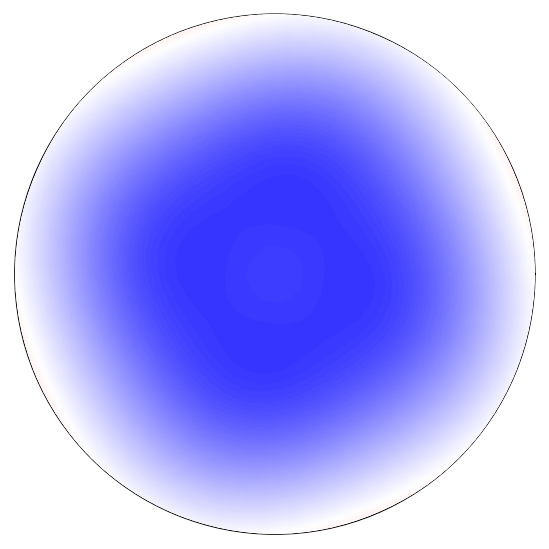} \\
\rotatebox[origin=c]{90}{$\lambda = 3$}  & \shshimage{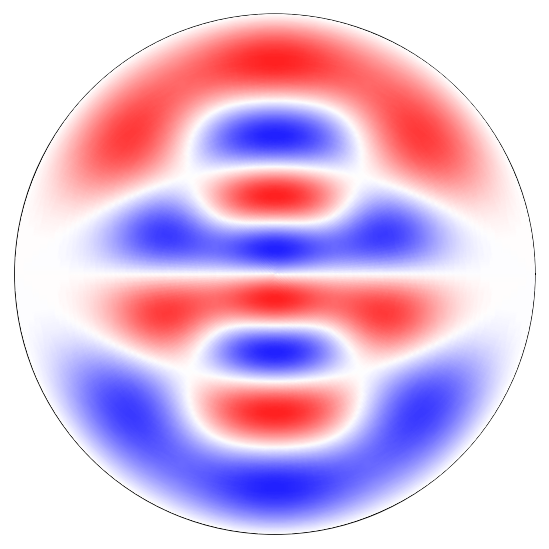} & \shshimage{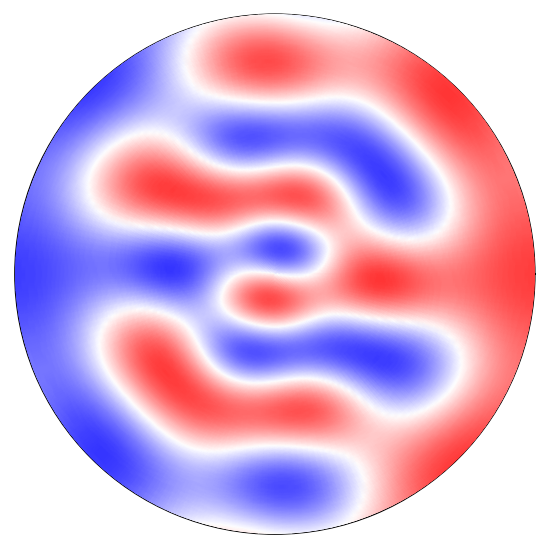} & \shshimage{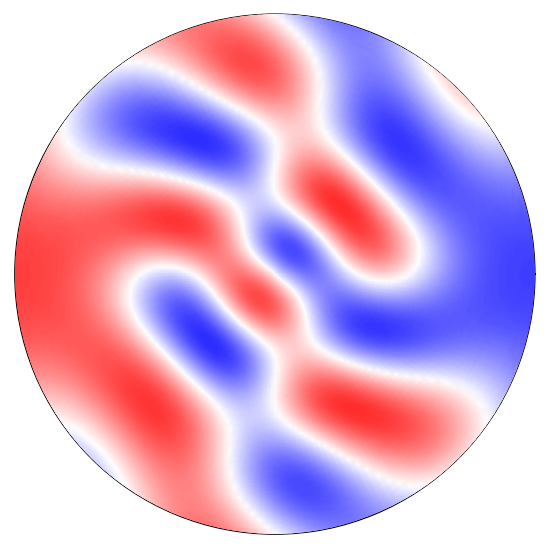} & \shshimage{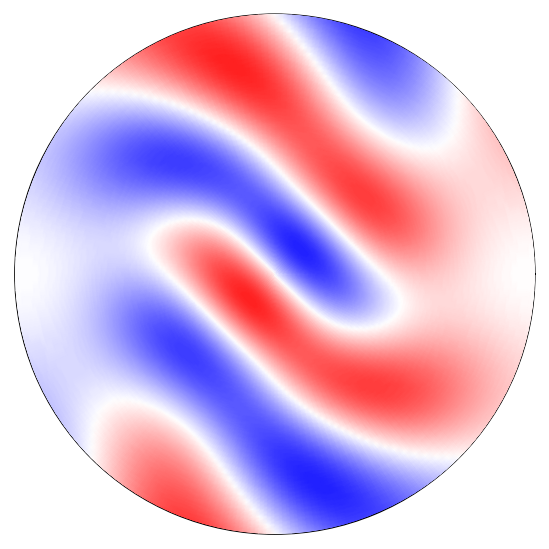} & \shshimage{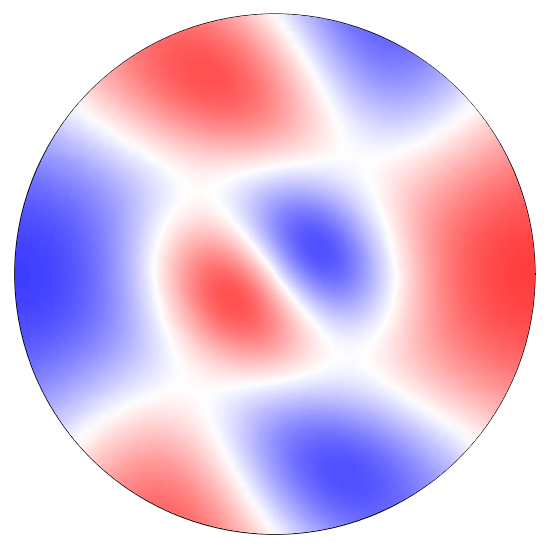} & \shshimage{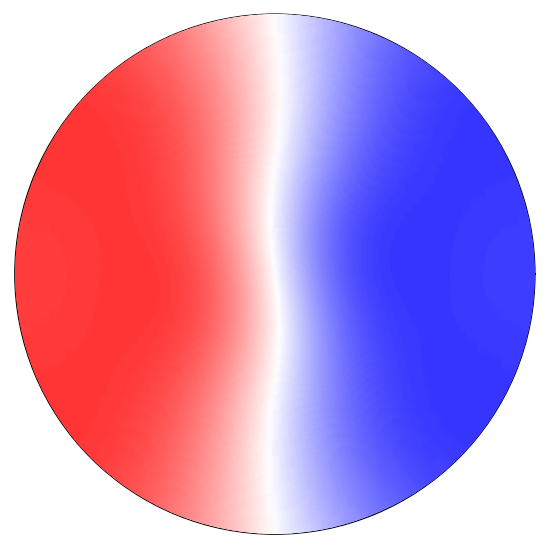} \\
\rotatebox[origin=c]{90}{$\lambda = 4$}  & \shshimage{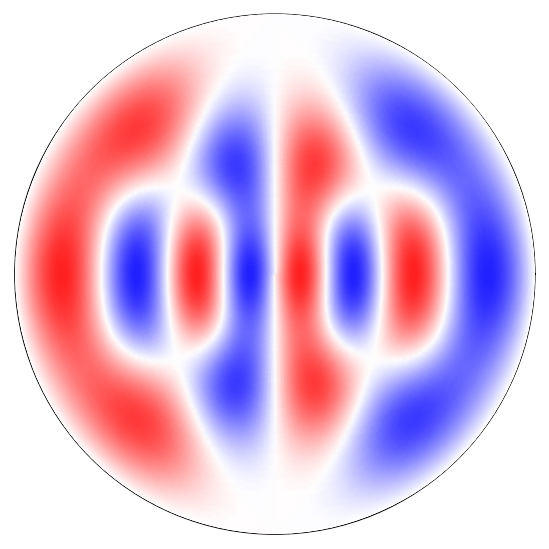} & \shshimage{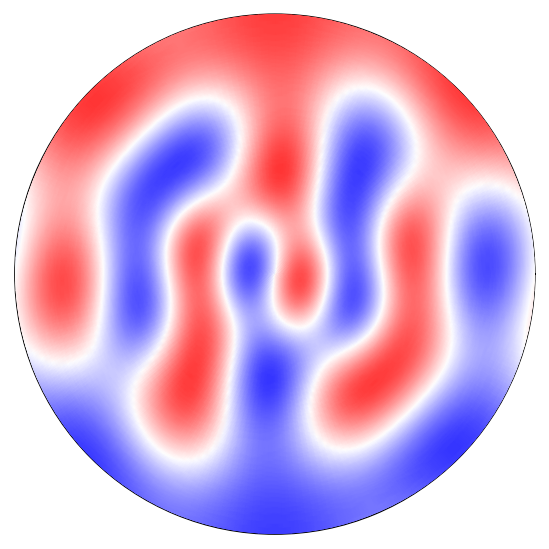} & \shshimage{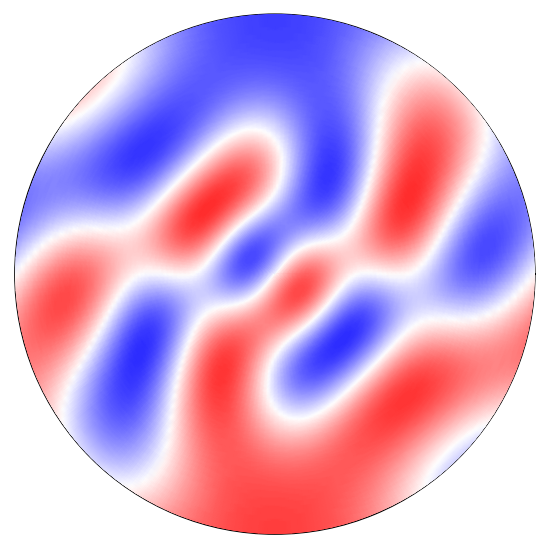} & \shshimage{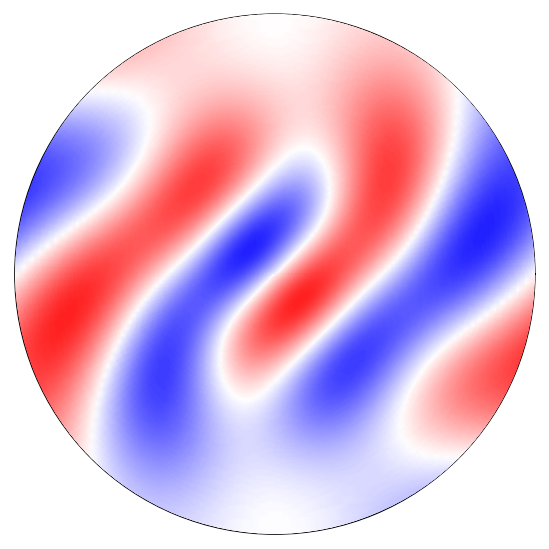} & \shshimage{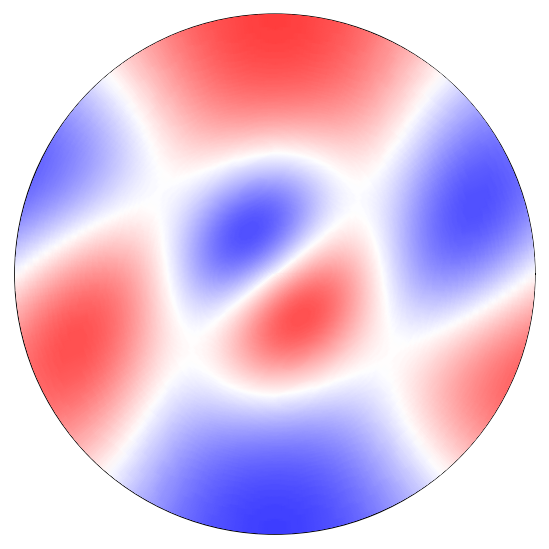} & \shshimage{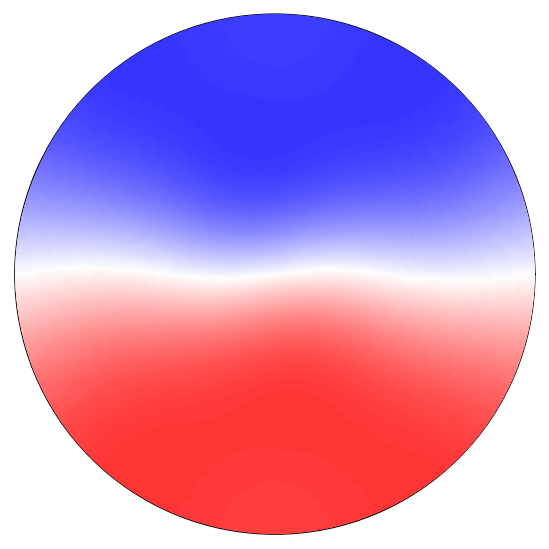} \\
\rotatebox[origin=c]{90}{$\lambda = 5$}  & \shshimage{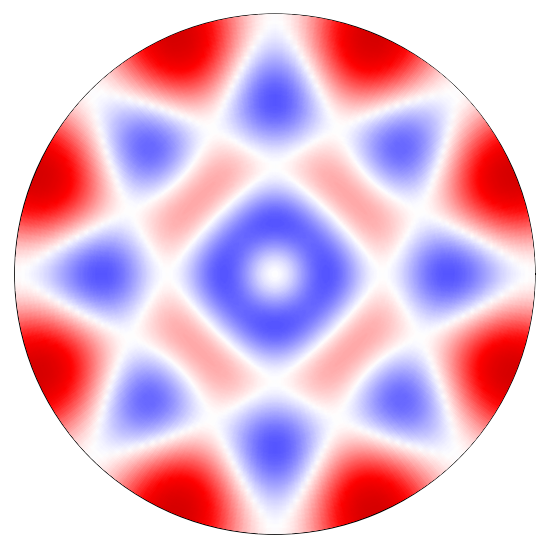} & \shshimage{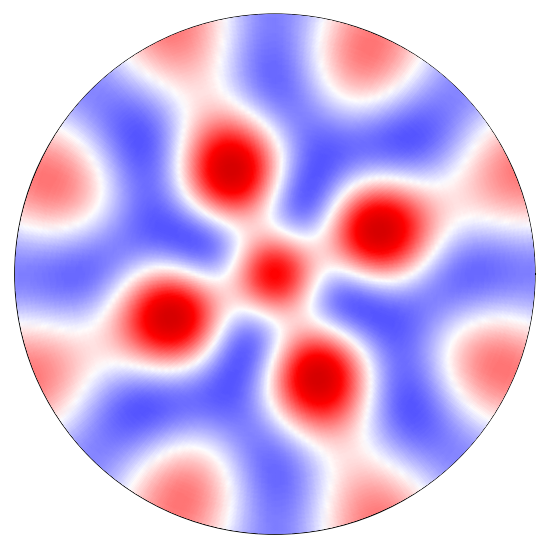} & \shshimage{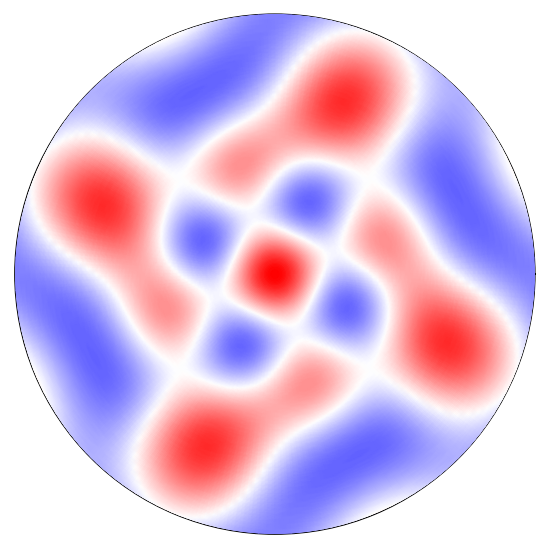} & \shshimage{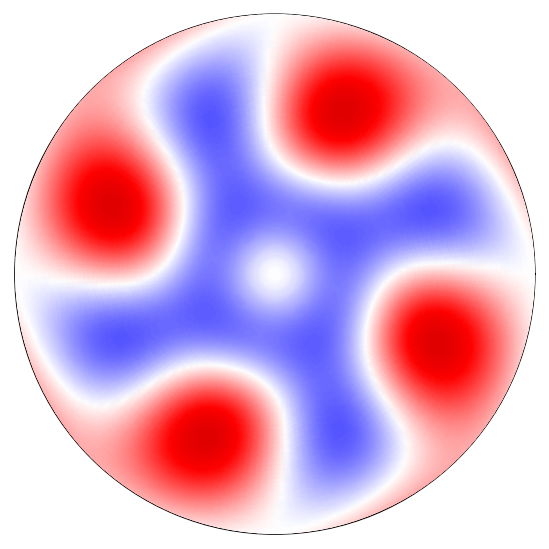} & \shshimage{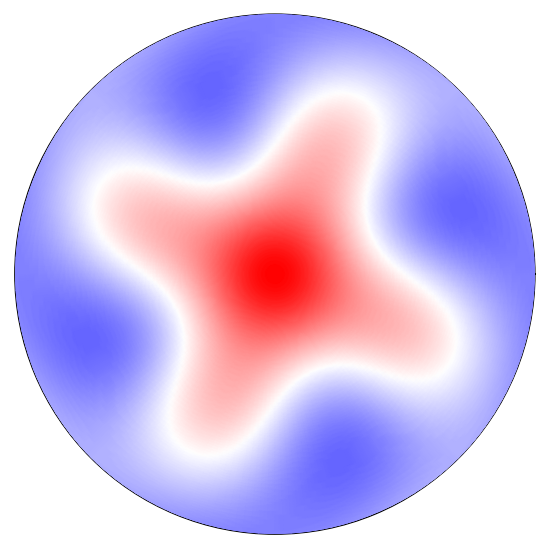} & \shshimage{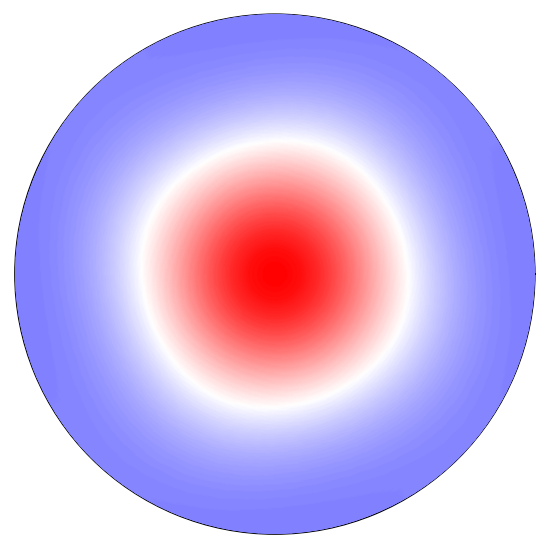} \\
\rotatebox[origin=c]{90}{$\lambda = 6$}  & \shshimage{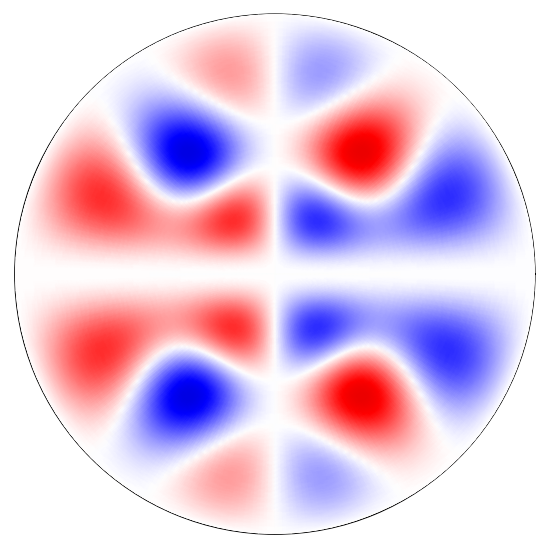} & \shshimage{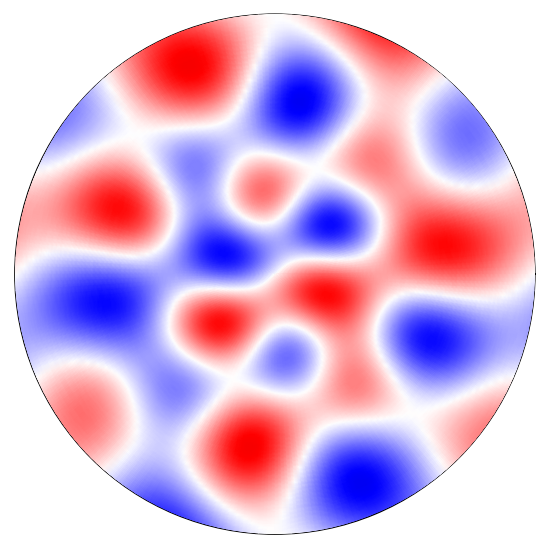} & \shshimage{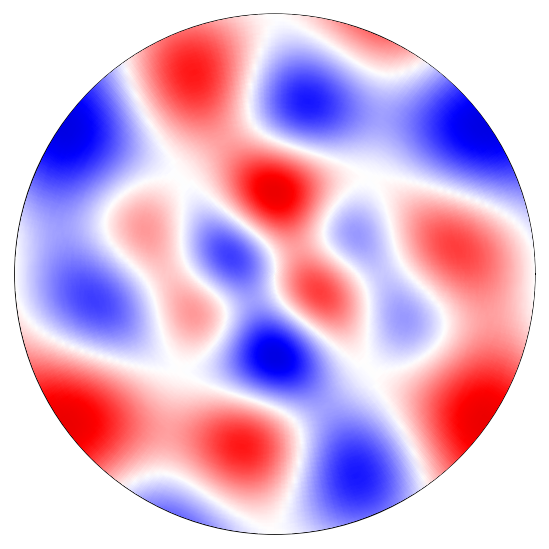} & \shshimage{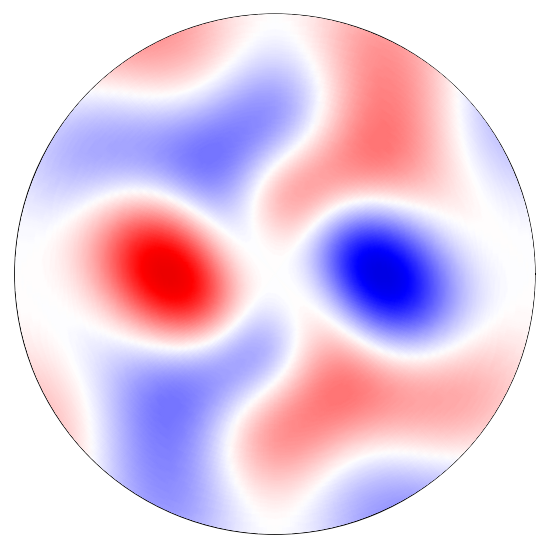} & \shshimage{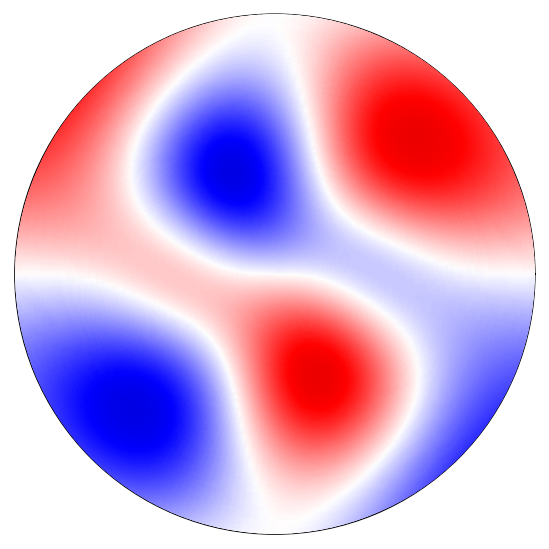} & \shshimage{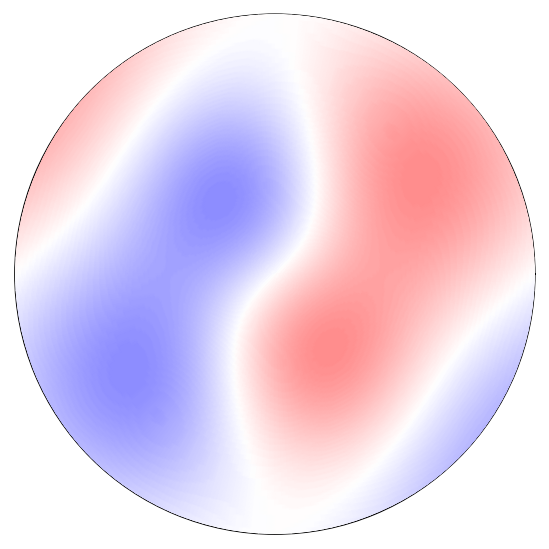} \\
\rotatebox[origin=c]{90}{$\lambda = 7$}  & \shshimage{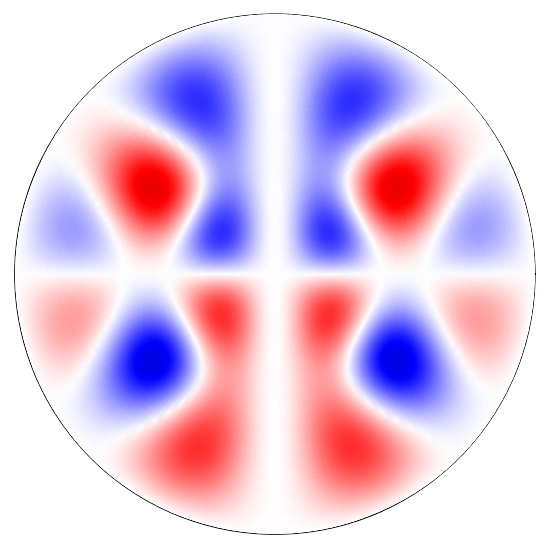} & \shshimage{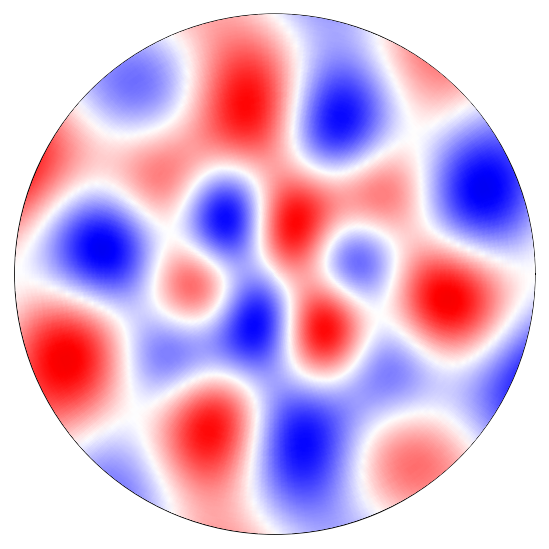} & \shshimage{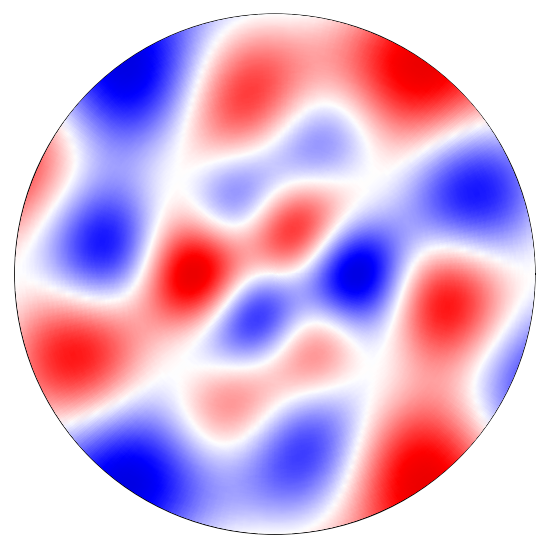} & \shshimage{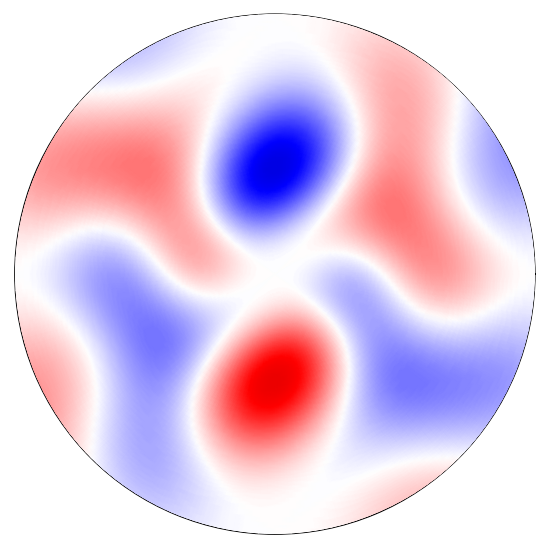} & \shshimage{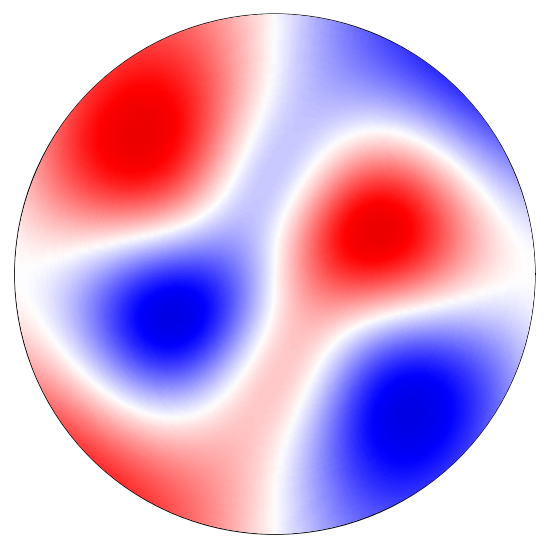} & \shshimage{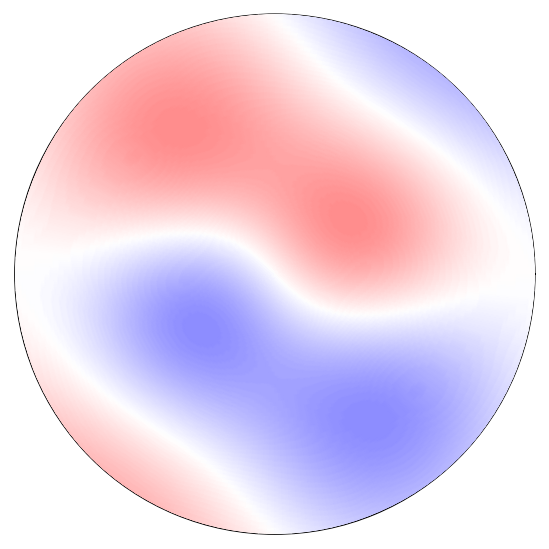} \\
\rotatebox[origin=c]{90}{$\lambda = 8$}  & \shshimage{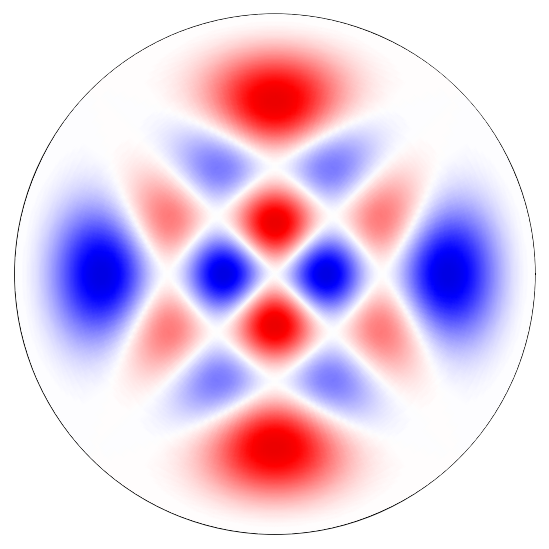} & \shshimage{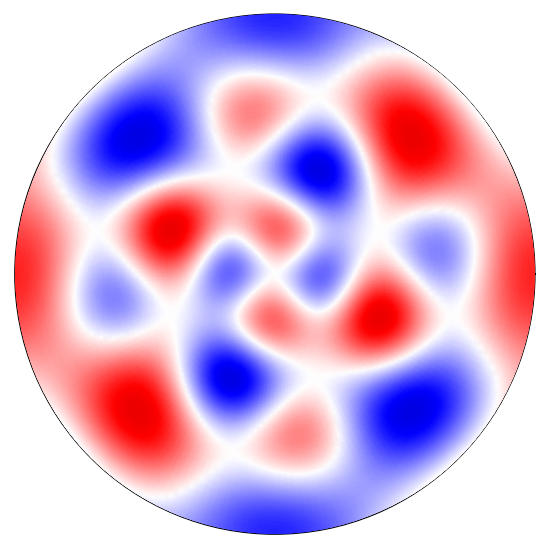} & \shshimage{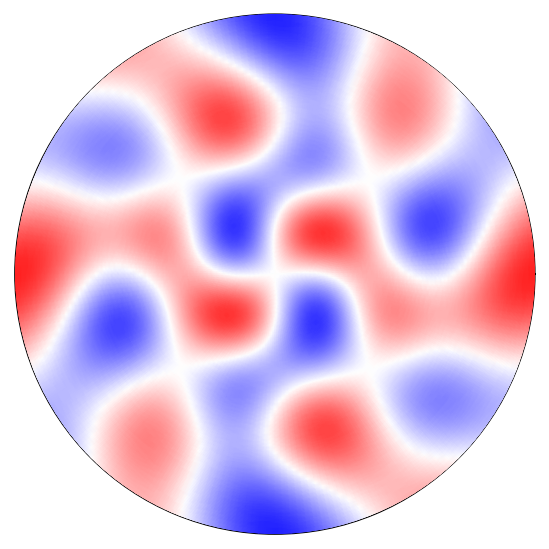} & \shshimage{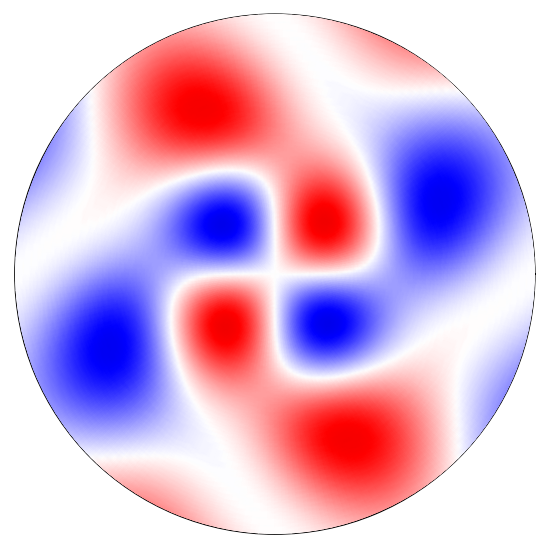} & \shshimage{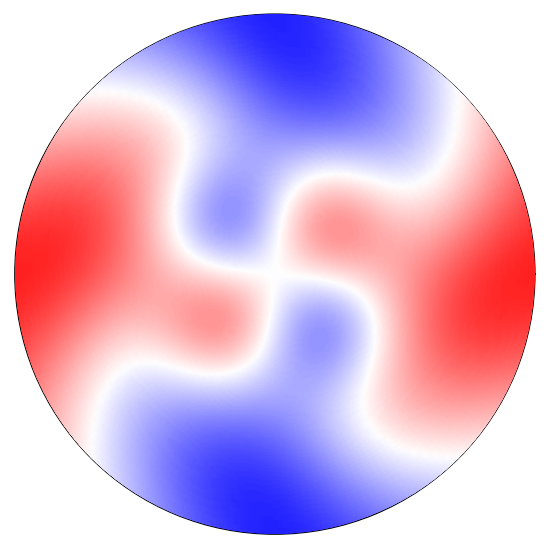} & \shshimage{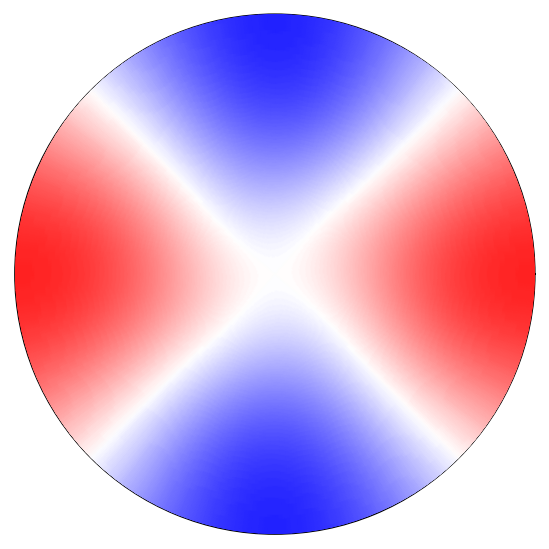} \\
\rotatebox[origin=c]{90}{$\lambda = 9$}  & \shshimage{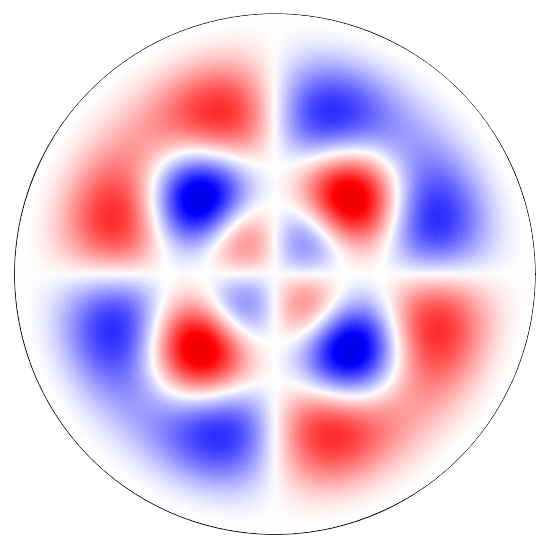} & \shshimage{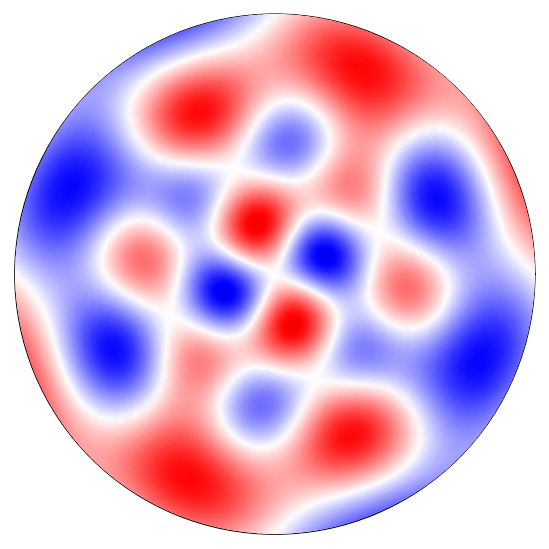} & \shshimage{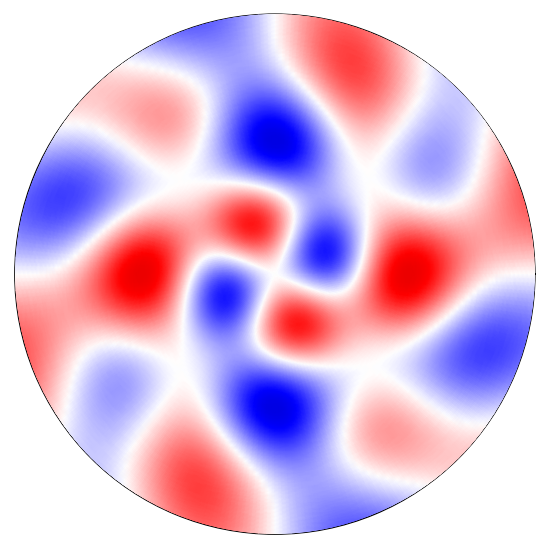} & \shshimage{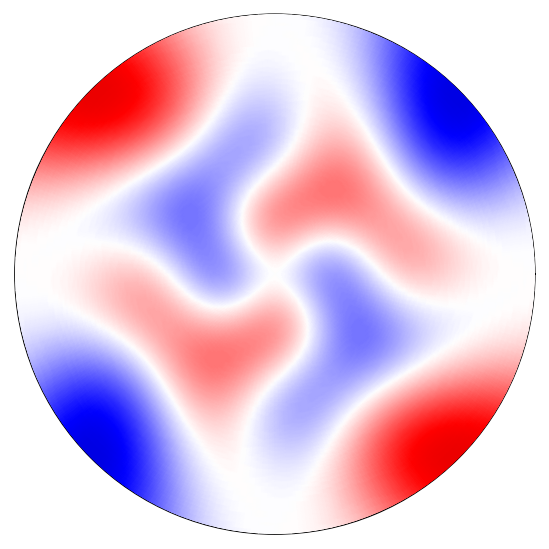} & \shshimage{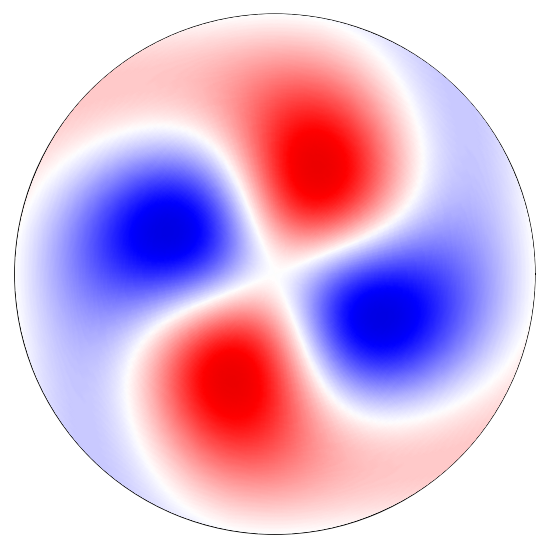} & \shshimage{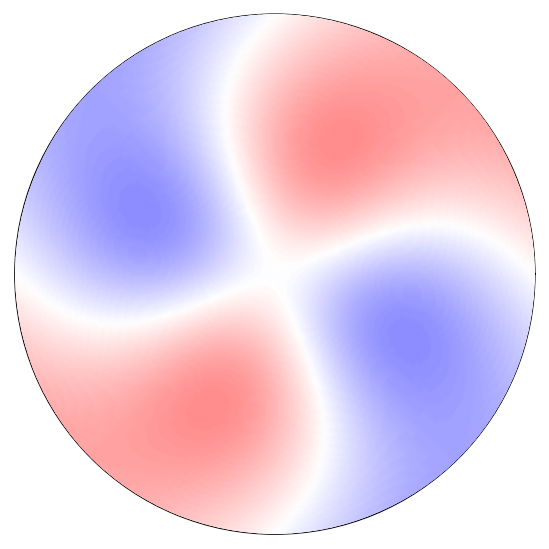} \\
\end{tabular}\\
\includegraphics[width=0.15\textwidth]{SHSHCbar.png}
\captionof*{figure}{Figure \ref*{fig:SHSH_basefunction}: (continued)}
\end{figure}

\clearpage
\section{Microstructure interpolation in orientation space}\label{app2}
\begin{figure*}[h]
  \centering
  \begin{subfigure}{0.25\textwidth}
    \includegraphics[width=1\textwidth]{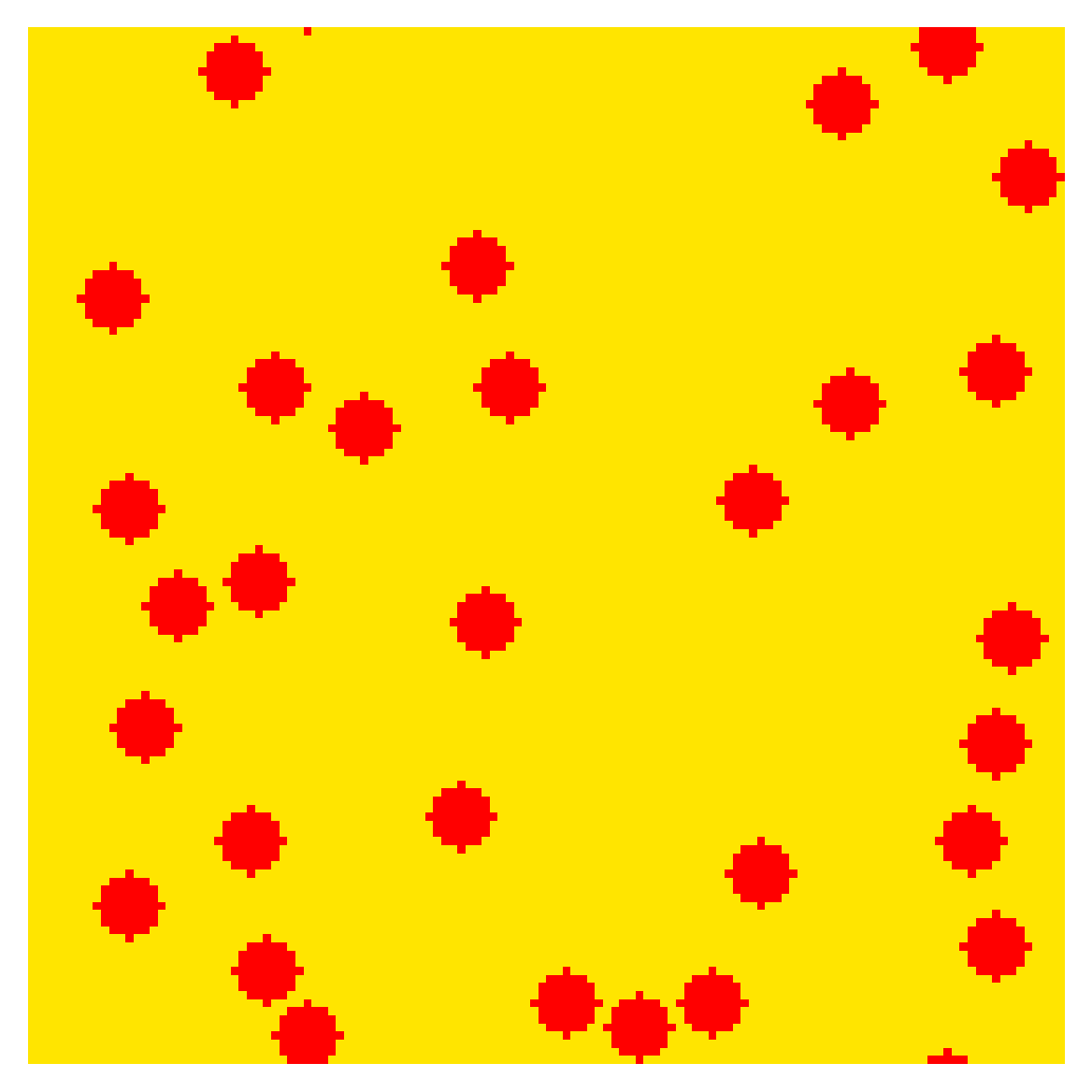}
    \caption{}
  \end{subfigure}%
  \begin{subfigure}{0.25\textwidth}
    \includegraphics[width=1\textwidth]{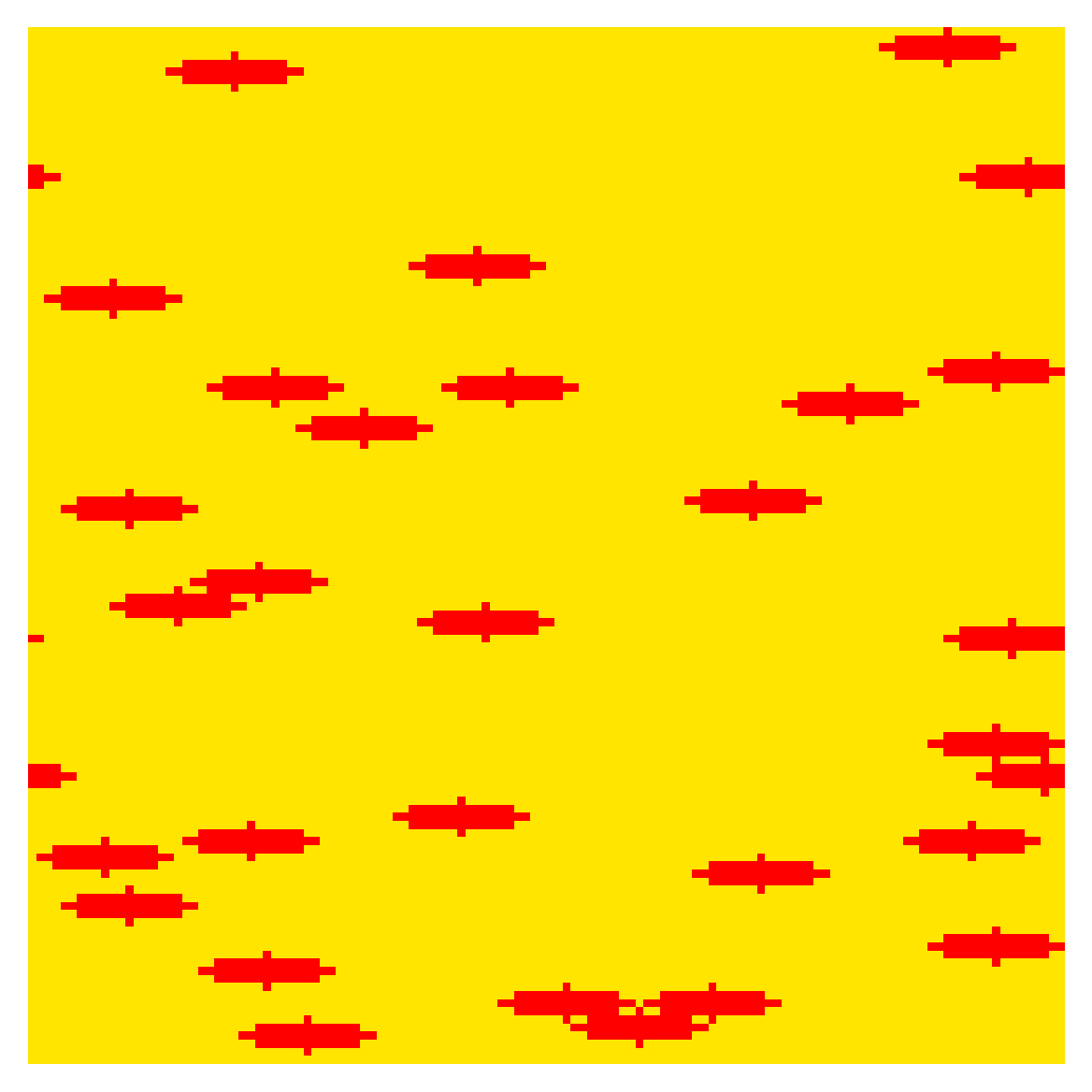}
    \caption{}
  \end{subfigure}%
  \\
  \begin{subfigure}{0.25\textwidth}
    \includegraphics[width=1\textwidth, trim={1.7 1cm 0.9cm 0.8cm},clip]{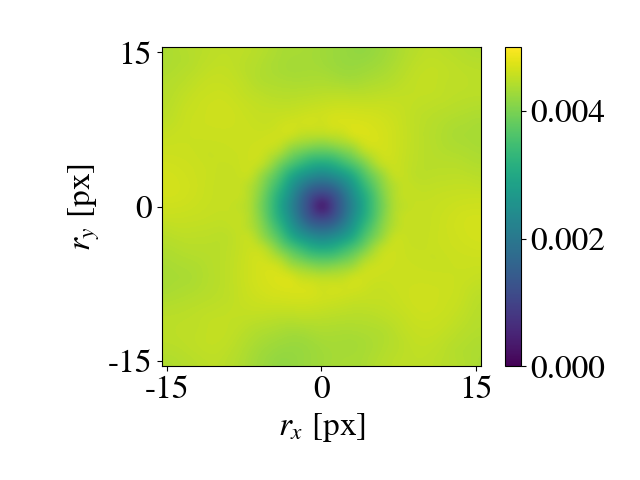}
    \caption{}
  \end{subfigure}%
  \begin{subfigure}{0.25\textwidth}
     \includegraphics[width=1\textwidth, trim={1.7 1cm 0.9cm 0.8cm},clip]{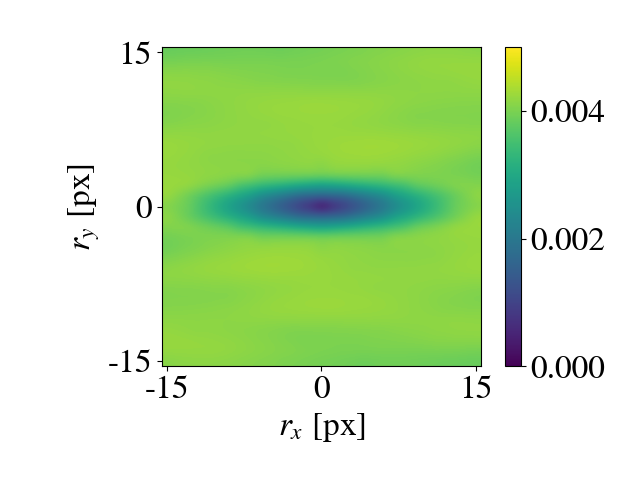}
    \caption{}
  \end{subfigure}%
  \caption{Two distinct hypothetical microstructural states for interpolation analysis with (a) an isotropic microstructure populated with equiaxed, circular inclusions, characterized by (c) the rotationally symmetric variogram, and (b) an anisotropic microstructure containing flattened, needle-shaped particles, corresponding to (d) the vertically compressed variogram.}
  \label{fig:Interpolation_study}
\end{figure*}

\begin{figure*}
  \centering
  \begin{subfigure}{0.19\textwidth}
    \includegraphics[width=1\textwidth]{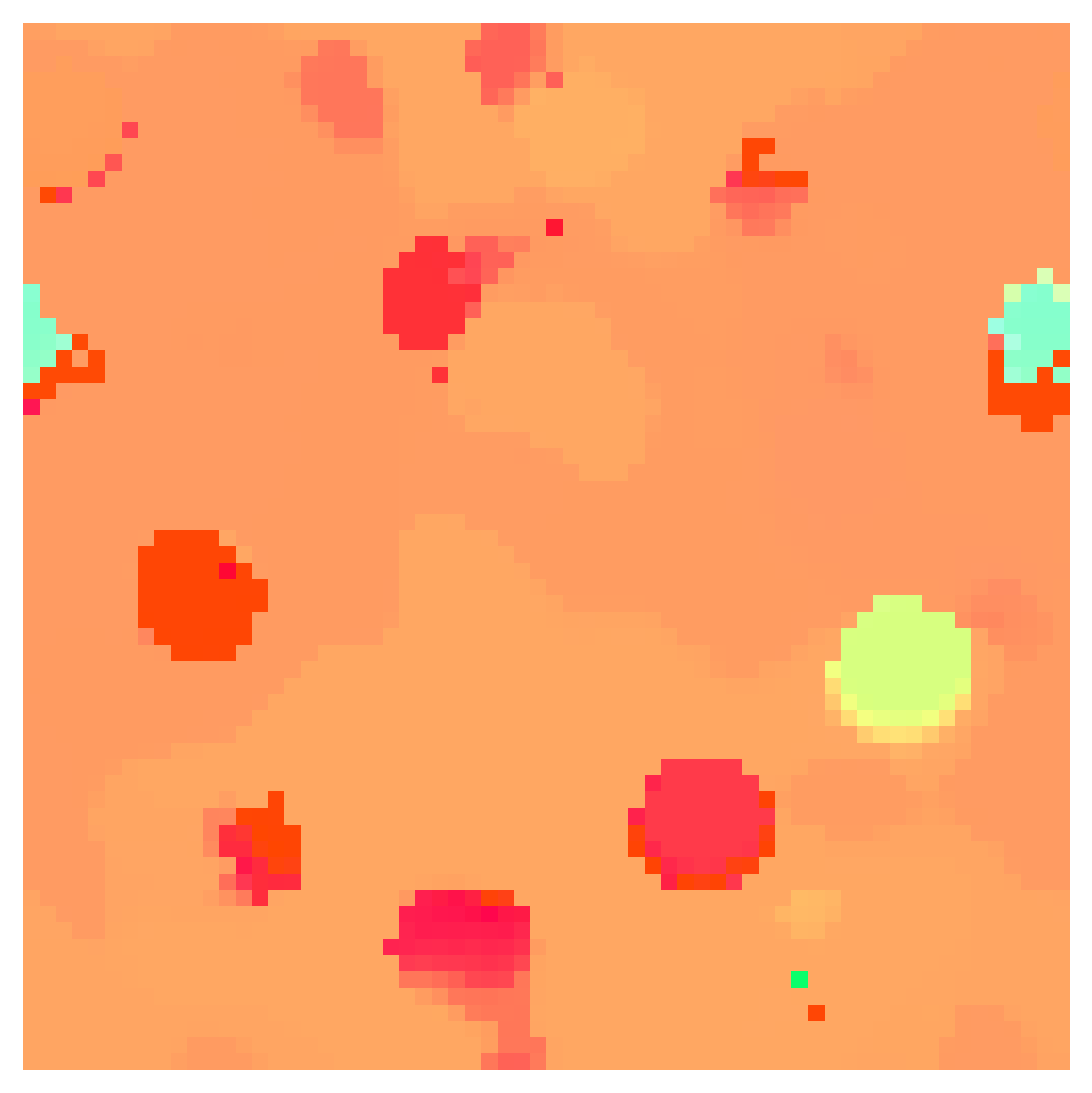}
    \caption{}
  \end{subfigure}%
  \begin{subfigure}{0.19\textwidth}
    \includegraphics[width=1\textwidth]{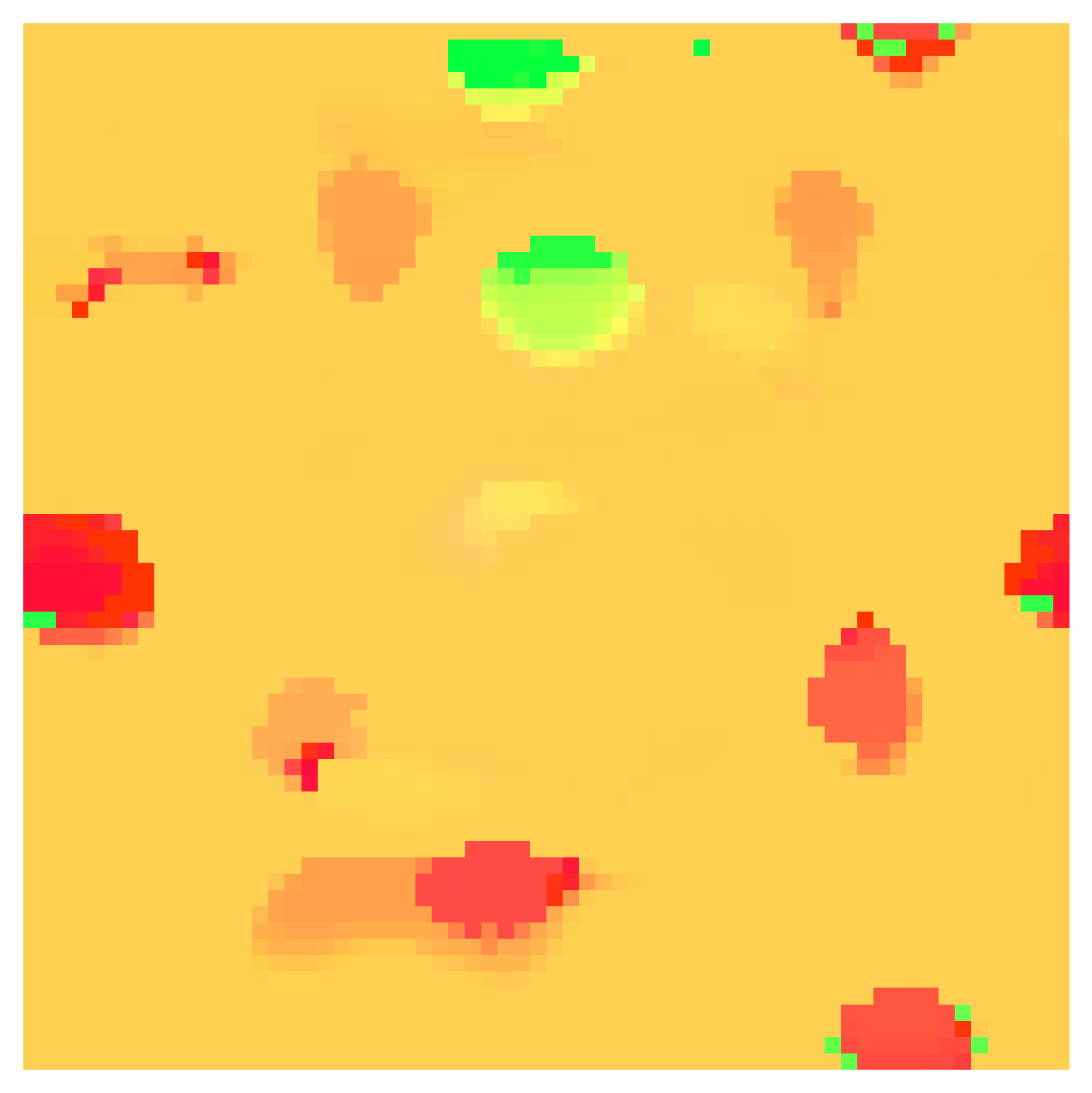}
    \caption{}
  \end{subfigure}%
  \begin{subfigure}{0.19\textwidth}
    \includegraphics[width=1\textwidth]{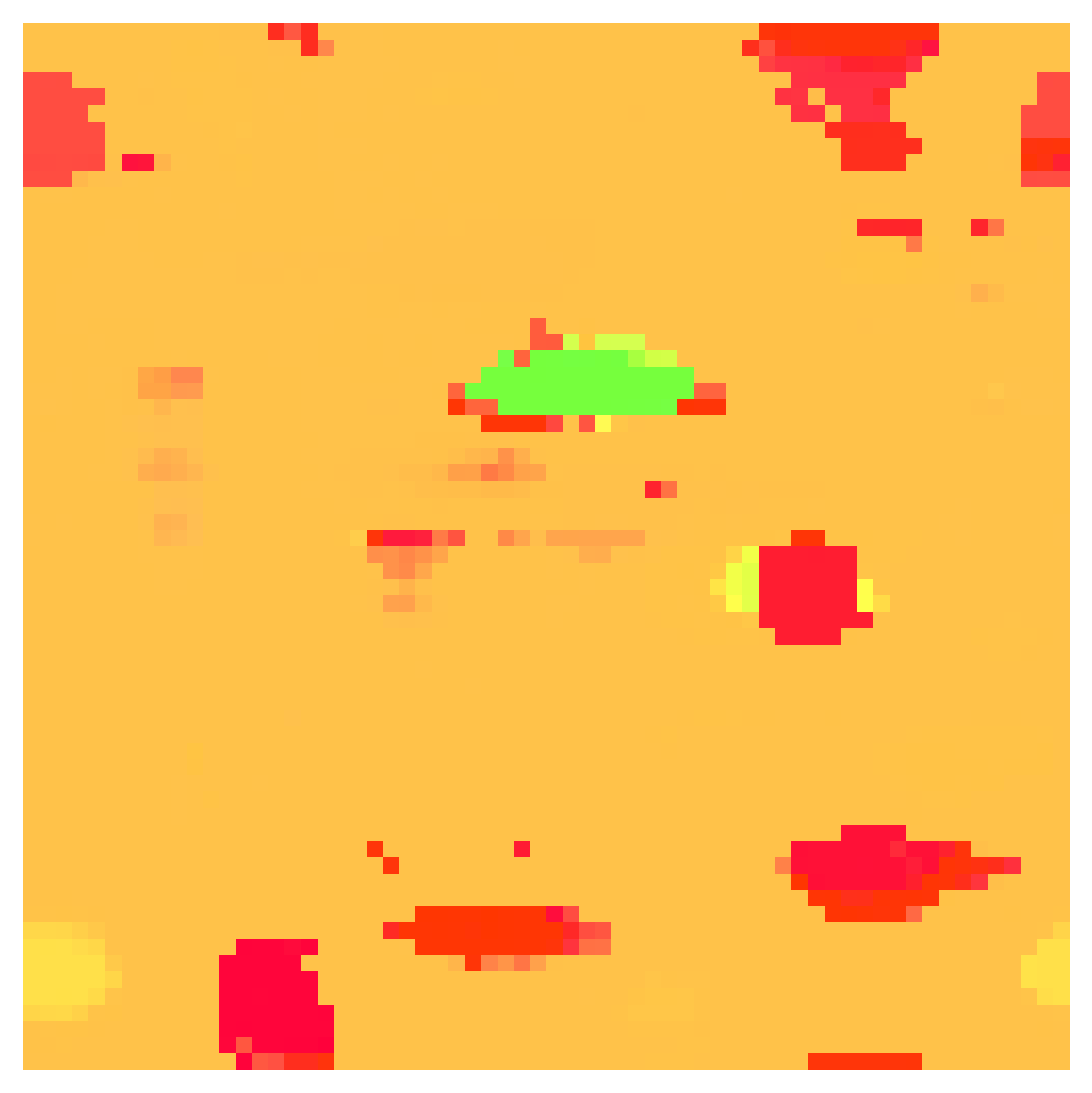}
    \caption{}
  \end{subfigure}%
  \begin{subfigure}{0.19\textwidth}
    \includegraphics[width=1\textwidth]{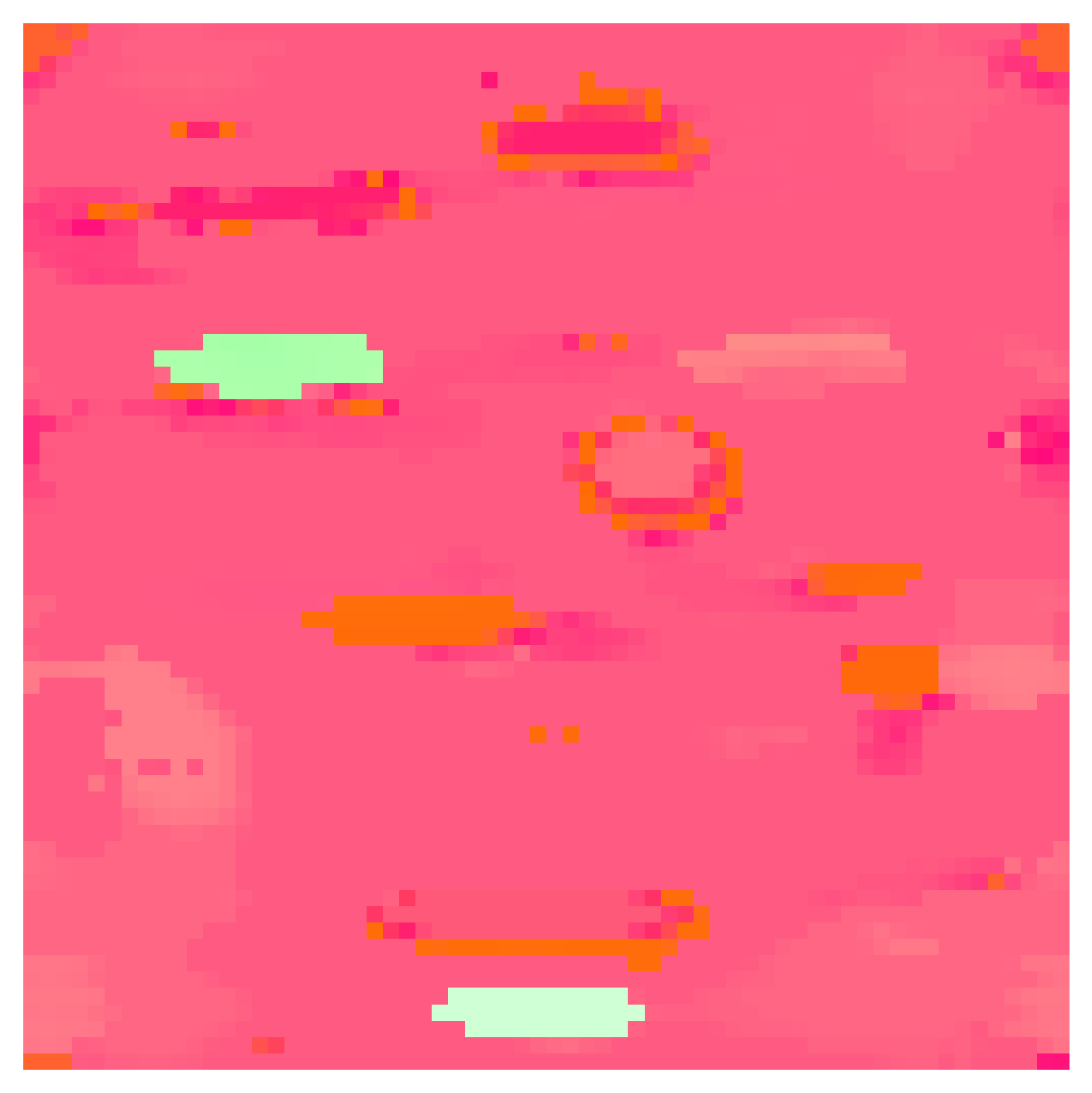}
    \caption{}
  \end{subfigure}%
  \begin{subfigure}{0.19\textwidth}
    \includegraphics[width=1\textwidth]{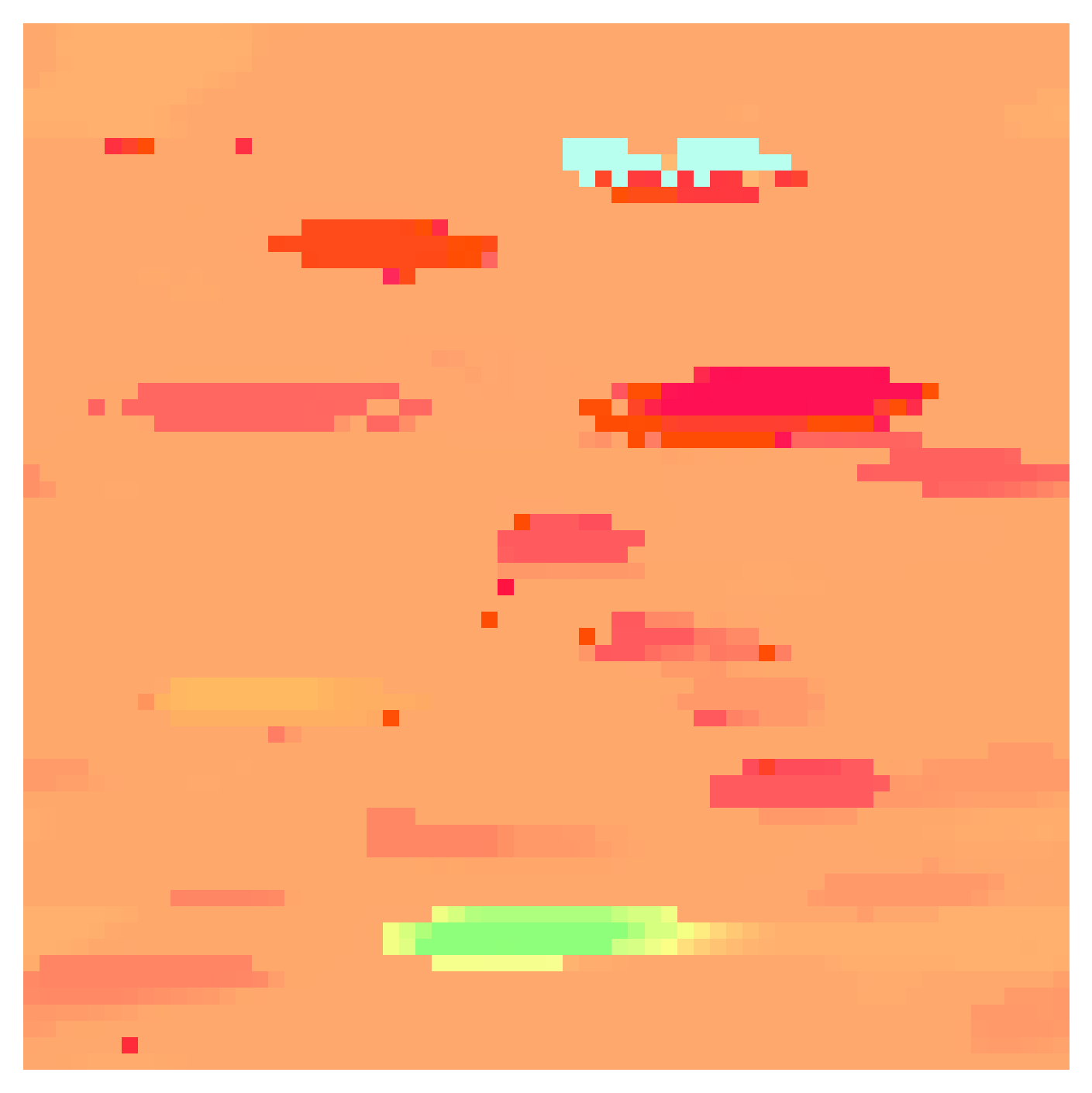}
    \caption{}
  \end{subfigure}%
  \\
  \begin{subfigure}{0.19\textwidth}
    \includegraphics[width=1\textwidth, trim={1.7 1cm 0.9cm 0.8cm},clip]{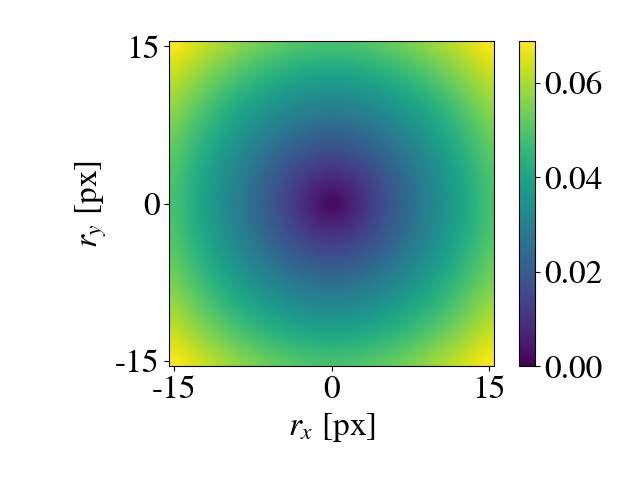}
    \caption{}
  \end{subfigure}%
  \begin{subfigure}{0.19\textwidth}
     \includegraphics[width=1\textwidth, trim={1.7 1cm 0.9cm 0.8cm},clip]{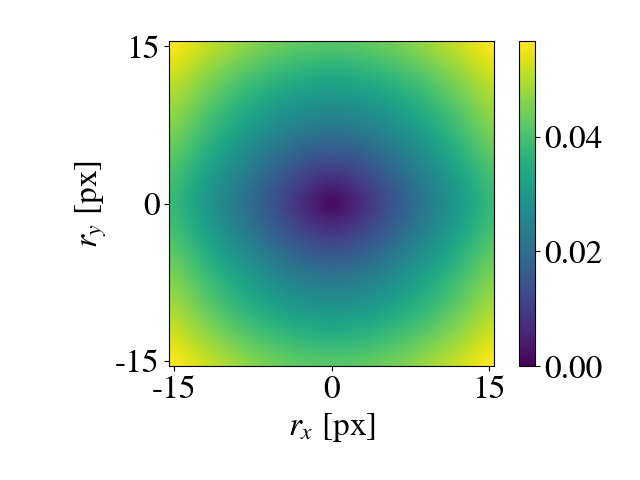}
    \caption{}
  \end{subfigure}%
  \begin{subfigure}{0.19\textwidth}
     \includegraphics[width=1\textwidth, trim={1.7 1cm 0.9cm 0.8cm},clip]{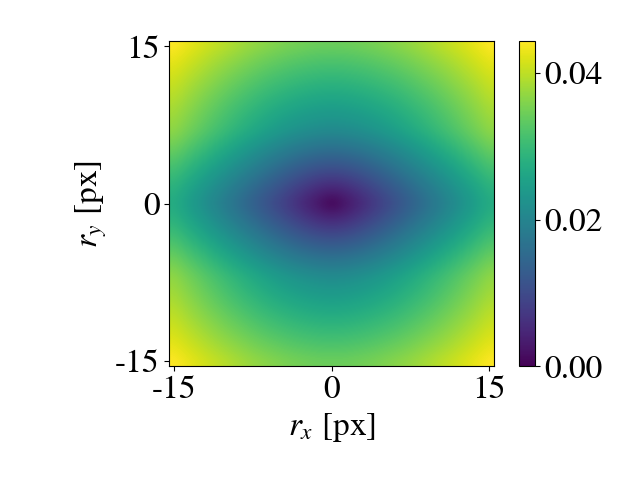}
    \caption{}
  \end{subfigure}%
  \begin{subfigure}{0.19\textwidth}
     \includegraphics[width=1\textwidth, trim={1.7 1cm 0.9cm 0.8cm},clip]{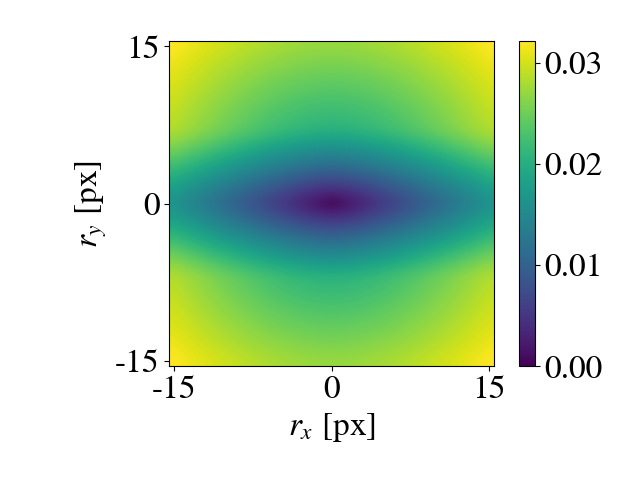}
    \caption{}
  \end{subfigure}%
  \begin{subfigure}{0.19\textwidth}
     \includegraphics[width=1\textwidth, trim={1.7 1cm 0.9cm 0.8cm},clip]{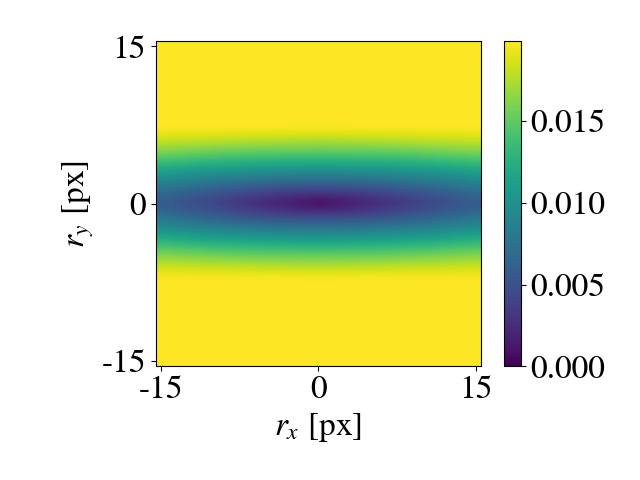}
    \caption{}
  \end{subfigure}%
  \caption{Non-linear evolution of microstructural morphology during statistical interpolation after 1000 iterations. The top row (a–e) illustrates the sequence of intermediate microstructures reconstructed from descriptors interpolated linearly between the isotropic and anisotropic limits defined in Figure 16. The bottom row (f–j) displays the corresponding target variograms for each step. The transition reveals significant topological complexity; rather than a smooth geometric elongation of grains, the intermediate states exhibit diverse artifacts such as shape fragmentation, irregular clustering, and the emergence of spurious orientation gradients (visible as non-physical color transitions in d). This demonstrates that linear interpolation in the high-dimensional descriptor space does not guarantee a physically continuous or unique morphological evolution in the spatial domain.}
  \label{fig:Interpolation_recon}
\end{figure*}

In the high-dimensional design space of the DMCR framework, the path between two statistical endpoints is neither linear nor unique. This is visually demonstrated in Fig.~\ref{fig:Interpolation_recon}, which explores the interpolation between the two limit states shown in Fig.~\ref{fig:Interpolation_study}, consisting of a microstructure characterized by isotropic, equiaxed inclusions (Fig.~\ref{fig:Interpolation_study}(a)) and one defined by highly anisotropic, flattened precipitates (Fig.~\ref{fig:Interpolation_study}(b)), with the particles being characterized with orientation $\boldsymbol{g}_0$ and the matrix with $\boldsymbol{g}_1$. Consistent with studies on descriptor interpolations using phase indicator functions \cite{seibert_microstructure_2025}, the intermediate realizations using the SHSH formulation (Fig.~\ref{fig:Interpolation_recon}(a-e)) reveal that satisfying the interpolated statistical descriptors does not result in a simple geometric morphing of the particles. Instead, the optimizer discovers a diverse manifold of valid microstructural solutions that satisfy the transitional statistics. For instance, the intermediate states do not merely elongate the spherical particles uniformly. They may manifest as mixtures of shapes, non-physical aspect ratios, or fragmented clusters that statistically average to the target intermediate variogram (Fig.~\ref{fig:Interpolation_recon}(j-f)).

\section{Gradient-based optimizer}\label{app3}

\begin{figure*}
  \centering
    \includegraphics[width=0.8\textwidth]{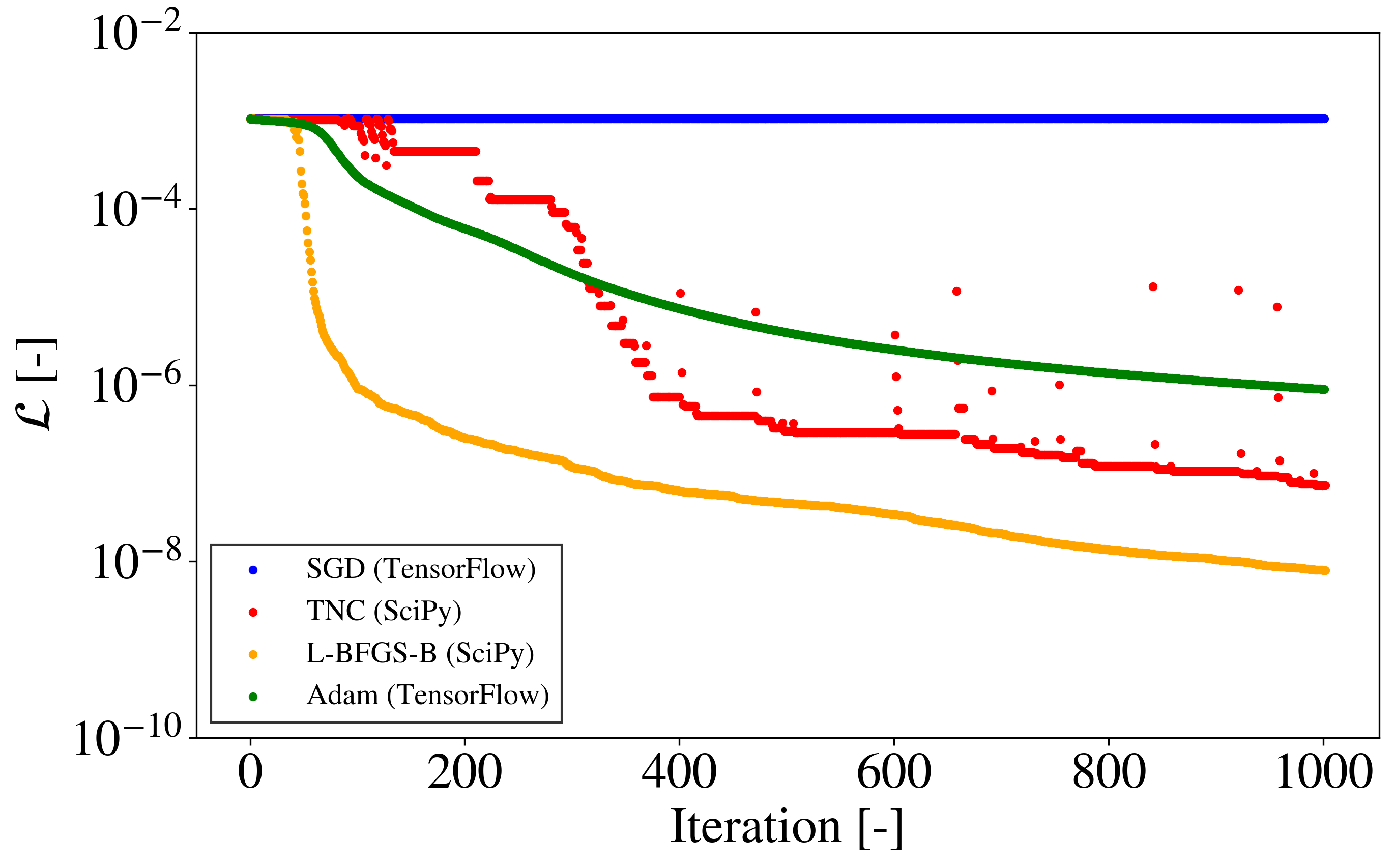}
    \caption{Comparison of convergence trajectories for various optimization algorithms. L-BFGS-B and Adam show the most efficient descent, while TNC exhibits a stepped convergence. Basic SGD struggles to move beyond the initial loss plateaus in this benchmark configuration.}
  \label{fig:Optimizer_comp}
\end{figure*}

The performance of various optimizers implemented in MCRpy is compared across five distinct solvers, using the configurations detailed in Table~\ref{tbl:Optimizer_settings}. As illustrated in Fig.~\ref{fig:Optimizer_comp}, which tracks the evolution of the descriptor loss $\mathcal{L}$ over 1000 iterations, the quasi-Newton L-BFGS-B algorithm \cite{fletcher_practical_2000} emerges as the superior optimizer for this class of inverse problems. It achieves rapid convergence, reaching a terminal loss below $10^{-8}$ and exhibiting a particularly steep descent within the first 100 iterations. The Adam optimizer \cite{kingma_adam_2017} also demonstrates robust performance, maintaining a smooth, monotonic descent to approximately $10^{-6}$, while the Truncated Newton (TNC) method \cite{dembo_truncated-newton_1983} eventually attains a loss below $10^{-7}$ despite a more irregular convergence profile. Conversely, basic first-order and stochastic methods exhibit significant limitations in this context. Standard Stochastic Gradient Descent (SGD) \cite{robbins_stochastic_1951} stagnates at a high loss plateau of $10^{-3}$. These findings establish L-BFGS-B as the currently preferred solver for high-fidelity reconstructions. Nevertheless, a trade-off with respect to computational resources remains. While second-order methods offer superior convergence rates, TensorFlow native solvers provide better potential scalability for large-scale, GPU-parallelized workloads.

\begin{table}[b]
\caption{Settings of the optimizers used in the comparison benchmark.}
\centering
\begin{tabular}{p{0.12\linewidth}p{0.43\linewidth}p{0.35\linewidth}}
\toprule
Optimizer & Primary parameter & Stopping criteria/Constraints \\
\midrule
L-BFGS-B  & $\textit{ftol}=0$                                                       & $\textit{maxiter}=1000$,\, $\textit{bounds}=(-1,1)$\\
Adam      & $\textit{learning rate}=0.001$,\, $\beta_1=0.9$,\,$\beta_2=0.999$ & $\textit{maxiter}=1000$\\
TNC       & $\textit{default}$                                                     & $\textit{maxiter}=1000$\\
SGD & $\textit{learning rate}=0.001$,\, $\textit{momentum}=0.0$ & $\textit{maxiter}=1000$ \\
\bottomrule
\end{tabular}
\label{tbl:Optimizer_settings}
\end{table}

\section{Stochastic variance}\label{app4}

\begin{figure}
    \centering
    \includegraphics[width=0.5\linewidth]{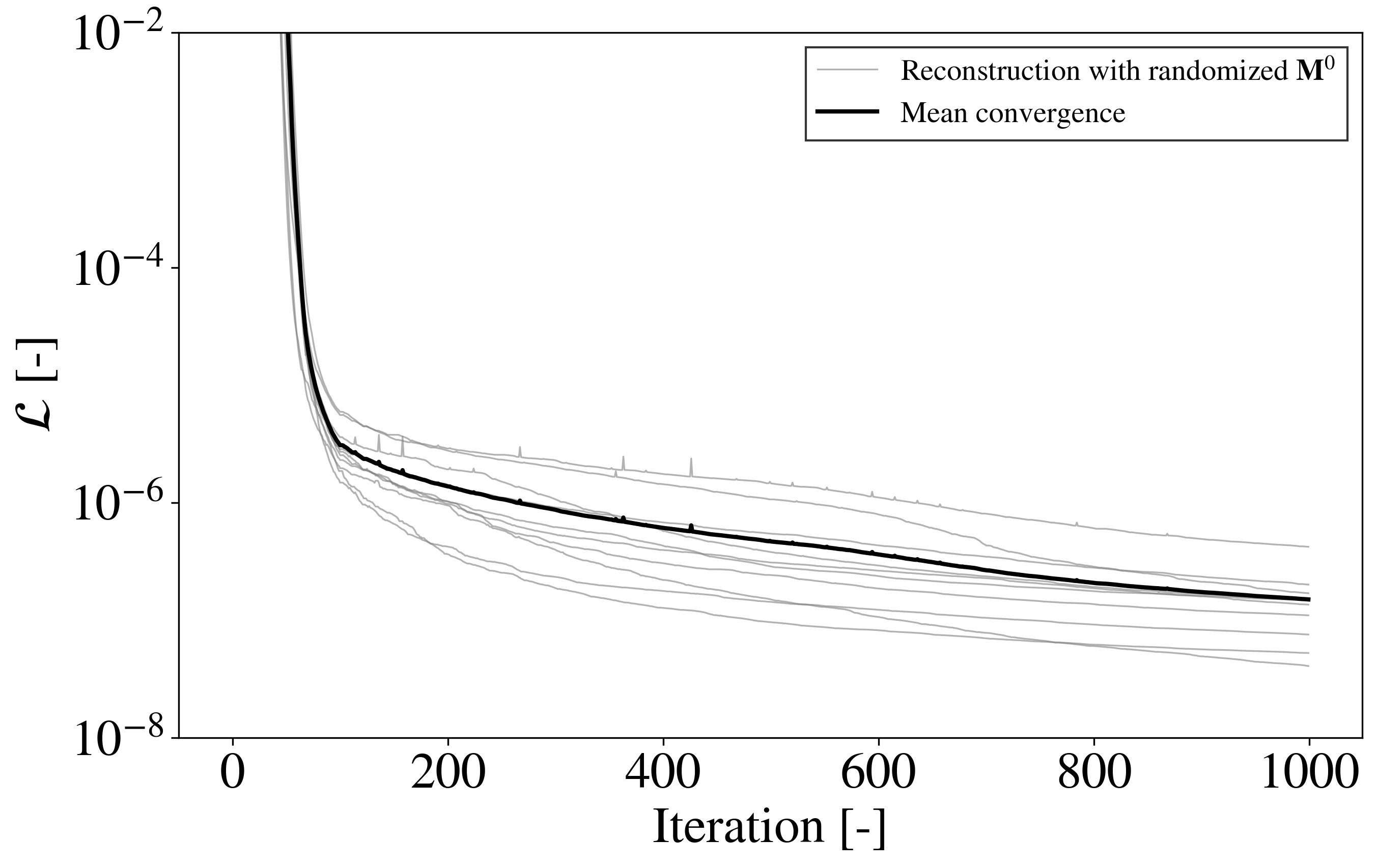}
    \caption{Convergence variability for ten independent stochastic realizations. Individual loss lines (grey) show distinct loss plateaus resulting from different initial random states ($\mathbf{M}^0$), highlighting the local minimum problem in high-dimensional orientation space. The mean convergence behavior is indicated by the solid black line.}
    \label{fig:Convergence_reproducibility}
\end{figure}

The susceptibility of the solver to become trapped in local minima within the high-dimensional descriptor landscape is analyzed by comparing different microstructure initializations. Fig.~\ref{fig:Convergence_reproducibility} quantifies this sensitivity by tracking the convergence variability across ten independent realizations, each initialized from a distinct randomized orientation field $\mathbf{M}^0$. While all trajectories exhibit a rapid initial descent, significant divergence emerges after approximately 100 iterations, with terminal loss values at iteration 1000 spanning nearly an order of magnitude (from $10^{-7}$ to $10^{-8}$). This stochastic spread confirms that the specific initialization vector $\mathbf{M}^0$ dictates the loss path into which the optimizer settles. Crucially, this variance correlates directly with microstructural fidelity as realizations that stagnate at moderate loss plateaus may typically recover global statistics like the ODF but fail to resolve fine-scale morphological features. As evidenced in Fig.~\ref{fig:micro_vs_loss}, interface migration mainly occurs in the lowest loss regimes, implying that solutions at local minima lack the gradient drive to overcome localized loss barriers necessary for refining detailed morphology. These findings demonstrate that a deterministic reliance on a single optimization run is insufficient for guaranteeing physical representativeness.

\end{document}